\documentclass[12pt]{article}
\usepackage{epsfig,latexsym,amssymb,amsmath}
\usepackage[arrow,matrix]{xy}
\usepackage{mythm}

\date{January 1, 2008}

\makeatletter

\newsavebox{\boxA}
\newsavebox{\boxB}
\newlength{\lenA}
\newlength{\lenB}

\newtheorem{Theorem}{Theorem}[section]
\newtheorem{Proposition}[Theorem]{Proposition}
\newtheorem{Lemma}[Theorem]{Lemma}
\newtheorem{Corollary}[Theorem]{Corollary}
\newtheorem{Conjecture}[Theorem]{Conjecture}

\newtheorem[definition]{Definition}[Theorem]{Definition}
\newtheorem[remark]{Example}[Theorem]{Example}

\newtheorem[remark]{Remark}[Theorem]{Remark}

\newtheorem*[definition]{definition}{Definition}
\newtheorem*[remark]{remark}{Remark}

\newcommand{\ZZ}{\mathbb{Z}}
\newcommand{\RR}{\mathbb{R}}
\newcommand{\CC}{\mathbb{C}}
\newcommand{\qq}{\mathbf{q}}
\newcommand{\pp}{\mathbf{p}}
\newcommand{\rr}{\mathbf{r}}
\newcommand{\nn}{\mathbf{n}}
\newcommand{\mm}{\mathbf{m}}
\newcommand{\calA}{\mathcal{A}}
\newcommand{\calB}{\mathcal{B}}
\newcommand{\calC}{\mathcal{C}}
\newcommand{\calD}{\mathcal{D}}
\newcommand{\calF}{\mathcal{F}}
\newcommand{\calG}{\mathcal{G}}
\newcommand{\calH}{\mathcal{H}}
\newcommand{\calL}{\mathcal{L}}
\newcommand{\calM}{\mathcal{M}}
\newcommand{\calS}{\mathcal{S}}
\newcommand{\calT}{\mathcal{T}}
\newcommand{\eps}{\varepsilon}
\newcommand{\ph}{\varphi}
\newcommand{\FS}{\varkappa}

\renewcommand{\le}{\leqslant}
\renewcommand{\ge}{\geqslant}
\newcommand{\bydef}{\stackrel{\mathrm{def}}{=}}
\newcommand{\dlangle}{\langle\mskip-3mu\langle}
\newcommand{\drangle}{\rangle\mskip-3mu\rangle}
\newcommand{\bdlangle}{\bigl\langle\mskip-4mu\bigl\langle}
\newcommand{\bdrangle}{\bigr\rangle\mskip-4mu\bigr\rangle}

\let\Re\relax\DeclareMathOperator{\Re}{Re}
\let\Im\relax\DeclareMathOperator{\Im}{Im}
\DeclareMathOperator{\Ker}{Ker}
\DeclareMathOperator{\Tr}{Tr}
\DeclareMathOperator{\tr}{tr}
\DeclareMathOperator{\QTr}{\widetilde{Tr}}
\DeclareMathOperator{\sgn}{sgn}

\DeclareMathOperator{\Pf}{Pf}
\DeclareMathOperator{\UU}{U}
\DeclareMathOperator{\SU}{SU}
\DeclareMathOperator{\OO}{O}
\DeclareMathOperator{\SO}{SO}
\DeclareMathOperator{\Spin}{Spin}
\DeclareMathOperator{\Hom}{Hom}
\DeclareMathOperator{\Rep}{\mathit{Rep}}

\newcommand{\HS}{\mathrm{HS}}
\newcommand{\eff}{\mathrm{eff}}
\newcommand{\std}{\mathrm{std}}

\newcommand{\spinup}{{\uparrow}}
\newcommand{\spindown}{{\downarrow}}
\newcommand{\id}{\mathrm{id}}
\newcommand*{\lefttens}[1]{[#1\,\otimes]}
\newcommand*{\refobj}[1]{{[#1]}}
\newcommand{\unit}{\mathbf{1}}
\newcommand{\tens}{\mathbin{\Box}}
\newcommand*{\ltens}[1]{[#1\,\Box]}
\newcommand*{\anti}[1]{\bar{#1}}
\newcommand{\Ver}{\mathit{Ver}}
\newcommand{\Hilb}{\mathit{Hilb}}
\newcommand{\Vect}{\mathit{Vec}}
\newcommand{\Matr}{\mathit{Mat}}
\newcommand{\Func}{\mathop{\mathit{Fun}}\nolimits}

\newcommand*{\latin}[1]{#1}

\newcommand{\afrac}[2]{\genfrac{}{}{0pt}{}{#1}{#2}}

\newcommand{\hexagon}{\text{\epsfysize=1.8ex\epsfbox{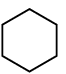}}}
\newcommand{\SolidLeftArrow}{\figbox{1.0}{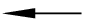}}
\newcommand{\DashedLeftArrow}{\figbox{1.0}{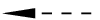}}

\newsavebox{\boxM}
\newcommand*{\saveM}[2]{\global\setbox\boxM=\hbox{$\m@th#1{#2}$}}

\newcommand*{\Copy}[1]{\mathpalette\saveM{#1}\copy\boxM}
\newcommand{\Paste}{\box\boxM}

\newcommand*{\rightAR}[2]{*i{\scriptstyle\Copy{#2}} \ar#1^-{\Paste}[l];[r]}
\newcommand*{\leftAR}[2]{*i{\scriptstyle\Copy{#2}} \ar#1_-{\Paste}[r];[l]}

\newlength{\fighskip} \fighskip=2pt
\newlength{\figvskip} \figvskip=4pt

\newcommand*{\figbox}[2]{{
  \def\figscale{#1}
  \def\arraystretch{0.8}
  \arraycolsep=0pt
  \begin{array}{c}
    \vbox{\vskip\figscale\figvskip
      \hbox{\hskip\figscale\fighskip
        \def\epsfsize##1##2{\figscale##1}%
        \epsfbox{#2}}}
  \end{array}}}

\newcommand{\specpar}[1]{{%
  \lenA=\parindent
  \parbox{0.5\textwidth}{\parindent=\lenA #1}}}

\newcommand{\figboxright}[3]{%
  \noindent\specpar{#1}%
  \hfil$\figbox{#2}{#3}$%
  \medskip}

\newcommand{\widefbox}[1]{%
  \fboxsep=0pt
  \lenA=\textwidth \advance\lenA-2\fboxrule \advance\lenA-2\fboxsep
  \lenB=\lenA \advance\lenB-20pt
  \framebox[\lenA]{\parbox{\lenB}{#1}}}

\def\sectModify#1#2#3#4{%
  \let\sectSavedFormat\@seccntformat
  \def\@seccntformat##1{#2{\csname the##1\endcsname}}%
  \let\sectOption\relax
  \def\sectDo##1{%
    \ifx\sectOption\relax \def\sectOption{#4{##1}}\fi
    \sectCommand{#3{##1}}
    \let\@seccntformat\sectSavedFormat}
  \@ifstar
    {\def\sectCommand{#1*}\sectDo}%
    {\def\sectCommand{#1[\sectOption]}\sectParse}}
\def\sectParse{\@ifnextchar[\sectParseOpt\sectDo}
\def\sectParseOpt[#1]{\def\sectOption{#1}\sectDo}

\newcommand*{\sectFormat}[1]{#1\quad}
\newcommand*{\AddDot}[1]{#1.}
\newcommand*{\AddColon}[1]{#1:}
\newcommand*{\Identity}[1]{#1}

\renewcommand{\subsubsection}{\sectModify\subsubsect
  \sectFormat\AddDot\Identity}
\newcommand{\subsubsect}{\@startsection{subsubsection}{3}{\z@}%
  {-2ex plus -0.7ex minus -0.15ex}%
  {-1em}%
  {\normalfont\normalsize\bfseries}%
  }

\renewcommand{\paragraph}{\sectModify\paragr
  \sectFormat\AddColon\Identity}
\newcommand{\paragr}{\@startsection{paragraph}{4}{\z@}%
  {2ex plus 0.7ex minus 0.15ex}%
  {-0.8em}%
  {\normalfont\normalsize\bfseries\itshape}%
  }

\let\stdsection\section
\let\stdappendix\appendix
\def\appFormat#1{Appendix #1:\quad}

\renewcommand{\appendix}{%
  \stdappendix
  \gdef\section{\sectModify\stdsection\appFormat\Identity\Identity}}

\makeatother

\title{Anyons in an exactly solved model and beyond}
\author{Alexei Kitaev
\smallskip\\ 
{\normalsize\it
California Institute of Technology, Pasadena, CA 91125, U.S.A.}\\
{\normalsize e-mail: kitaev@iqi.caltech.edu}}

\begin{document}
\maketitle

\begin{abstract}
A spin $1/2$ system on a honeycomb lattice is studied.  The interactions
between nearest neighbors are of XX, YY or ZZ type, depending on the direction
of the link; different types of interactions may differ in strength. The model
is solved exactly by a reduction to free fermions in a static $\mathbb{Z}_{2}$
gauge field.  A phase diagram in the parameter space is obtained. One of the
phases has an energy gap and carries excitations that are Abelian anyons. The
other phase is gapless, but acquires a gap in the presence of magnetic
field. In the latter case excitations are non-Abelian anyons whose braiding
rules coincide with those of conformal blocks for the Ising model. We also
consider a general theory of free fermions with a gapped spectrum, which is
characterized by a spectral Chern number $\nu$. The Abelian and non-Abelian
phases of the original model correspond to $\nu=0$ and $\nu=\pm 1$,
respectively. The anyonic properties of excitation depend on $\nu\bmod 16$,
whereas $\nu$ itself governs edge thermal transport. The paper also provides
mathematical background on anyons as well as an elementary theory of Chern
number for quasidiagonal matrices.
\end{abstract}

\tableofcontents

\subsection*{Comments to the contents: What is this paper about?}
Certainly, the main result of the paper is an exact solution of a particular
two-dimensional quantum model. However, I was sitting on that result for too
long, trying to perfect it, derive some \emph{properties} of the model, and
put them into a more general framework. Thus many ramifications have come
along. Some of them stem from the desire to avoid the use of conformal field
theory, which is more relevant to edge excitations rather than the bulk
physics.  This program has been partially successful, but some rudiments of
conformal field theory (namely, the topological spin $\theta_{a}=e^{2\pi i
(h_{a}-\overline{h}_{a})}$ and the chiral central charge
$c_{-}=c-\overline{c}$) are still used.

The paper is self-contained and provides an introduction into the subject. For
most readers, a good strategy is to follow the exposition through the
beginning of Sec.~\ref{sec_nonabelian}, \emph{Nonabelian anyons} and take a
glance at the rest of that section, where things become more technical. But
the mathematically inclined reader may be interested in those details, as well
as some of the appendices. I have tried to make the paper modular so that some
parts of it can be understood without detailed reading of the other. This has
caused some redundancy though.

Appendix~\ref{sec_algth}, \emph{Algebraic theory of anyons} is an elementary
introduction into unitary modular categories, which generalizes the discussion
in Sec.~\ref{sec_nonabelian}.

Appendix~\ref{sec_locmatr}, \emph{Quasidiagonal matrices} is also mostly
expository but some of the arguments may be new.  It begins with a simplified
treatment of ``operator flow'' and ``noncommutative Chern number'' (the latter
has been used to prove the quantization of Hall conductivity in disordered
systems~\cite{BES-B94}), but the main goal is to explain ``unpaired Majorana
modes'', a certain parity phenomenon related to the Chern number.

Appendix~\ref{sec_anomaly} on the \emph{chiral central charge} and
Appendix~\ref{sec_symbreak} on \emph{weak symmetry breaking} contain some raw
ideas that might eventually develop into interesting theories.

\section*{Introduction}
\addcontentsline{toc}{section}{Introduction}

\subsubsection*{Overview of the subject}
\emph{Anyons} are particles with unusual statistics (neither Bose nor Fermi),
which can only occur in two dimensions. Quantum statistics may be understood
as a special kind of interaction: when two particles interchange along some
specified trajectories, the overall quantum state is multiplied by
$e^{i\ph}$. In three dimensions, there is only one topologically distinct way
to swap two particles. Two swaps are equivalent to the identity
transformation, hence $e^{i\ph}=\pm 1$.  On the contrary, in two dimensions
the double swap corresponds to one particle making a full turn around the
other; this process is topologically nontrivial. Therefore the exchange phase
$\ph$ can, in principle, have any value --- hence the name
``anyon''. (However, a stability consideration requires that $\ph$ be a
rational multiple of $2\pi$.) Of course, the real question is whether such
particles exist in nature or can be built somehow, but we will follow the
historic path, approaching the problem from the mathematical end.

The study of anyons was initiated by Wilczek~\cite{Wilczek-1,Wilczek-2} in
early 1980's. He proposed a simple but rather abstract model, which was based
on ($2+1$)-dimensional electrodynamics. This theory has integer electric
charges and vortices carrying magnetic flux (which is a real number defined up
to an integer). Considered separately, both kinds of particles are bosons. But
when a charge $q$ goes around a vortex $v$, it picks up the phase $2\pi qv$,
known as the Aharonov-Bohm phase. Thus, charges and vortices have nontrivial
\emph{mutual} statistics and therefore must be called anyons when considered
together. Moreover, composite objects $(q,v)$ consisting of a charge and a
vortex are anyons by themselves because they have nontrivial exchange phase
$\ph_{(q,v)}=2\pi qv$.

A general way to describe quantum statistics is to consider particle
worldlines in the ($2+1$)-dimensional space-time. Such worldlines form a
braid, therefore the statistics is characterized by a representation of the
braid group. In the preceding discussion we assumed that braiding is
characterized just by phase factors, i.e., that the representation is
one-dimensional. The corresponding anyons are called \emph{Abelian}. But one
can also consider multidimensional representations of the braid group; in this
case the anyons are called \emph{non-Abelian}. Actually, it may not be so
important how the braid group acts, but the very existence of a
multidimensional space associated with several particles is a key
feature. Vectors in this space are quantum states that have almost the same
energy (see discussion of topological quantum computation below).

Historically, the theory of non-Abelian anyons emerged from conformal theory
(CFT). However, only topological and algebraic structure in CFT is relevant to
anyons. Different pieces of this structure were discovered in a colossal work
of many people, culminating in the paper by Moore and
Seiberg~\cite{MooreSeiberg}. Witten's work on quantum Chern-Simons
theory~\cite{Witten} was also very influential. A more abstract approach
(based on local field theory) was developed by Fredenhagen, Rehren, and
Schroer~\cite{FRS89} and by Frohlich and Gabbiani~\cite{FrohlichGabbiani90}.

The most amazing thing about anyons is that they actually exist as excitations
in some condensed matter systems. Such systems also have highly nontrivial
ground states that are described as possessing \emph{topological order}. The
best studied example (both theoretically and experimentally) is the Laughlin
state~\cite{Laughlin83} in the fractional quantum Hall system at the filling
factor $\nu=1/3$. It carries Abelian anyons with exchange phase $\ph=\pi/3$
and electric charge $\pm 1/3$. It is the fractional value of the charge that
was predicted in original Laughlin's paper and confirmed by several methods,
in particular by a shot noise
measurement~\cite{Saminadayar_atal,dePicciotto_atal}. The statistical phase is
a subtler property which is deduced theoretically~\cite{Halperin84,ASW84}; a
nontrivial experimental test has been performed recently using quasiparticle
tunneling~\cite{CaminoZhouGoldman}.

A different kind of state is observed at the filling factor $\nu=5/2$, though
it is more fragile and less studied experimentally. There is much evidence
suggesting that this system is described by a beautiful theory proposed by
Moore and Read~\cite{MooreRead-1,MooreRead-2}. The Moore-Read state admits
non-Abelian anyons with charge $\pm 1/4$. If $2n$ such particles are present,
the associated Hilbert space has dimension $2^{n-1}$. (The non-Abelian anyons
studied in this paper have similar properties, though there is no
electric charge.)

The notion of anyons assumes that the underlying state has an energy gap (at
least for topologically nontrivial quasiparticles). Otherwise excitations are
not localizable and braiding may not be defined. Note that if all excitations
are gapped, then all equal-time correlators decay exponentially with
distance~\cite{Hastings04}.

An example of anyons in a spin-$1/2$ system originates from the theory of
resonating valence bond (RVB). The idea of RVB was put forward by
Anderson~\cite{Anderson73} and used later as a model of the undoped insulating
phase in high-$T_{c}$ cuprate superconductors~\cite{Anderson87}. Without
electrically charged holes, the problem seems to be described adequately by a
Heisenberg-like Hamiltonian, but its solution has proved very
difficult. Several variants of an RVB state have been proposed, both gapless
and gapped. Here we discuss a particular gapped RVB phase, namely the one
which is realized on the triangular lattice~\cite{MoessnerSondhi00}, but which
apparently exists on the square lattice as well. This phase admits
quasiparticles of four types: trivial excitations (such as spin waves),
\emph{spinons} (spin-$1/2$ fermions, which are conserved \latin{modulo}~$2$),
$\ZZ_{2}$-vortices (spinless bosons, also called \emph{visons}), and
spinon-vison composites~\cite{ReadChakraborty89}. The mutual statistics of
spinons and visons is characterized by the Aharonov-Bohm factor $-1$,
therefore the composite particles are bosons. Note that the relevance of this
theory to cuprate superconductors is under debate. Senthil and Fisher proposed
an interesting way to detect visons in these materials~\cite{SenthilFisher01},
but the experiment gave a negative result~\cite{no_visons}. However, some kind
of RVB state is likely to realize in a different material,
$\mathrm{Cs_{2\mathstrut}CuCl_{4\mathstrut}}$. This conclusion is drawn from
neutron-scattering experiments that have shown the presence of spin-$1/2$
excitations~\cite{CTTT01}.
 
\begin{figure}
\centerline{\begin{tabular}{c@{\qquad\qquad}c}
$\figbox{0.5}{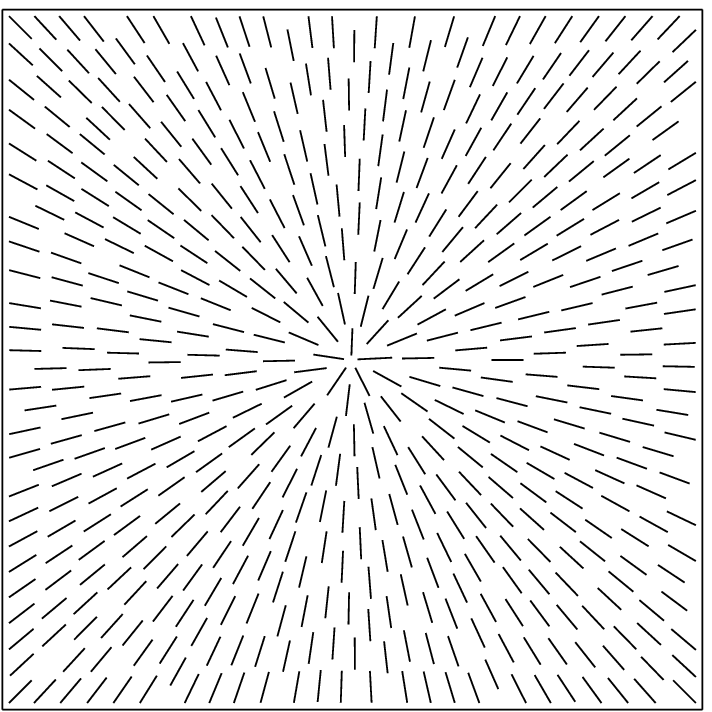}$ & $\figbox{0.5}{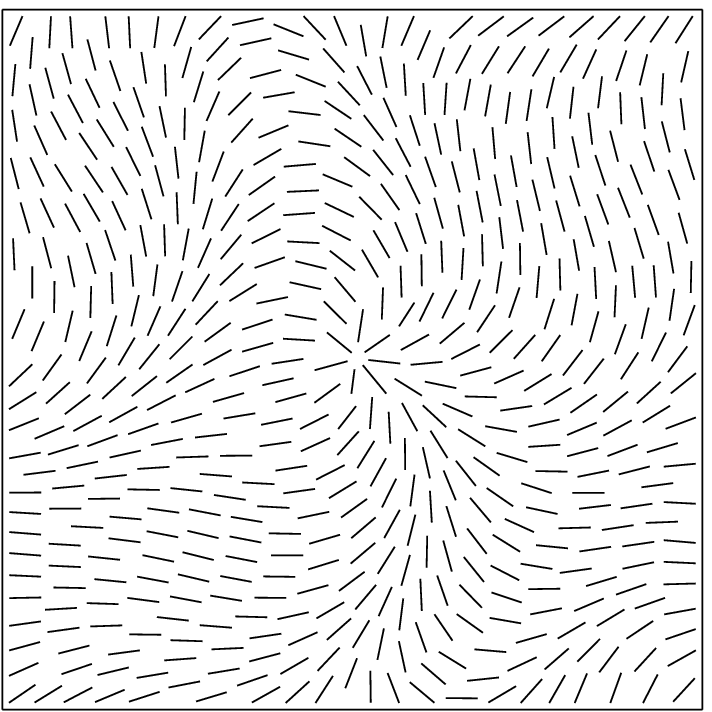}$ \\
a) & b)
\end{tabular}}
\caption{A classical vortex (a) distorted by fluctuations (b).}
\label{fig_vcart}
\end{figure}

Anyonic particles are best viewed as a kind of topological defects that reveal
non-trivial properties of the ground state. Thus anyons carry some topological
quantum numbers which make them stable: a single particle cannot be
annihilated locally but only through the fusion with an antiparticle.  An
intuitive way to picture an anyon is to imagine a vortex in a medium with a
local order parameter (see Fig.~\ref{fig_vcart}a). Now suppose that quantum
fluctuations are so strong that the order parameter is completely washed out
and only the topology remains (see Fig.~\ref{fig_vcart}b). Of course, that is
only a rough illustration. It resembles the Kosterlitz-Thouless phase with
power-law correlation decay, while in anyonic systems correlations decay
exponentially due to the energy gap.

A real example can be constructed with spins on the edges of a square
lattice. Basis states of the spins are described by the variables $s_{j}=\pm
1$, which may be regarded as a $\ZZ_{2}$ gauge field (i.e., ``vector
potential''), whereas the ``magnetic field intensity'' on plaquette $p$ is
given by
\begin{equation}
w_{p}\,=\!\prod_{j\in\text{boundary}(p)}\!\!s_{j}.
\end{equation}
We say that there is a \emph{vortex} on plaquette $p$ if $w_{p}=-1$. Now we
may define the vortex-free state:
\begin{equation}\label{Z_2vac}
|\Psi\rangle\,=\,
c \!\!\sum_{\def\arraystretch{0.7}\arraycolsep=2pt \begin{array}{ll}
\scriptstyle \mathbf{s}: & \scriptstyle w_{p}(\mathbf{s})=1\\
& \scriptstyle\text{for all $p$}
\end{array}} \!\!\! |\mathbf{s}\rangle,
\qquad\quad \text{where}\,\ \mathbf{s}=(s_{1},\dots,s_{N})
\end{equation}
($c$ is a normalization factor). The state with a single vortex on a given
plaquette is defined similarly. It is clear that the vortex can be detected by
measuring the observable $\prod_{j\in l}\sigma^{z}_{j}$ for any enclosing path
$l$, though no local order parameter exists.

The state~(\ref{Z_2vac}) can be represented as the ground state of the
following Hamiltonian with four-body interaction~\cite{Kitaev97}:
\begin{equation} \label{torcode}
H \,=\, -J_{e} \sum_{\text{vertices}}A_{s}
-J_{m}\sum_{\text{plaquettes}}B_{p},
\qquad\text{where}\quad\
A_{s}=\prod_{\text{star}(s)}\sigma^{x}_{j}, \quad\
B_{p}=\!\prod_{\text{boundary}(p)}\!\!\sigma^{z}_{j}.
\end{equation}
Its elementary excitations are $\ZZ_{2}$-charges with energy $2J_{e}$ and
vortices with energy $2J_{m}$. Certain essential features of this model are
stable to small local perturbations (such as external magnetic field or
Heisenberg interaction between neighboring spins). Note that the robust
characteristic of excitations is not the energy or the property of being
elementary, but rather \emph{superselection sector}. It is defined as a class
of states that can be transformed one to another by local operators. This
particular model has the vacuum sector $1$, the charge sector $e$, the vortex
sector $m$, and a charge-vortex composite $\eps$. Particles of type $e$ and
$m$ are bosons with nontrivial mutual statistics, whereas $\eps$ is a
fermion. Thus, the model represents a universality class of topological order
--- actually, the same class as RVB.

Anyonic superselection sectors may or may not be linked to conventional
quantum numbers, like spin or electric charge. Most studies have been focused
on the case where anyons carry fractional electric charge or half-integer
spin. Such anyons are potentially easier to find experimentally because they
contribute to collective effects (in particular, electric current) or have
characteristic selection rules for spin-dependent scattering. Chargeless and
spinless quasiparticles are generally harder to identify. But anyons, by
virtue of their topological stability, must have some observable
signatures. For example, anyons can be trapped by impurities and stay there
for sufficiently long time, modifying the spectrum of local modes (magnons,
excitons, etc.). However, effective methods to observe anyons are yet to be
found.

Thus, the hunt for anyons and topological order is a difficult endeavor. Why
do we care? First, because these are conceptually important phenomena,
breaking some paradigms. In particular, consider these principles (which work
well and provide important guidance in many cases):
\begin{enumerate}
\item Conservation laws come from symmetries (by Noether's theorem or its
  quantum analogue);
\item Symmetries are initially present in the Hamiltonian (or Lagrangian),
  but may be spontaneously broken.
\end{enumerate}
Let us limit our discussion to the case of gauge symmetries and local
conservation laws, which are described by \emph{fusion rules} between
superselection sectors. A profound understanding of the first principle and
its underlying assumptions is due to Doplicher and
Roberts~\cite{DoplicherRoberts90,DoplicherRoberts89}.  They proved that any
consistent system of fusion rules for bosons is equivalent to the
multiplication rules for irreducible representations of some compact group.
Fermions also fit into this framework. However, anyonic fusion rules are not
generally described by a group! As far as the second principle is concerned,
topological order does not require any preexisting symmetry but leads to new
conservation laws. Thus, the formation of topological order is exactly the
opposite of symmetry breaking!

\subsubsection*{Topological quantum computation}
A more practical reason to look for anyons is their potential use in quantum
computing. In Ref.~\cite{Kitaev97} I suggested that topologically ordered
states can serve as a physical analogue of error-correcting quantum
codes. Thus, anyonic systems provide a realization of quantum memory that is
\emph{protected from decoherence}. Some quantum gates can be implemented by
braiding; this implementation is \emph{exact} and does not require explicit
error correction. Freedman, Larsen, and Wang~\cite{FLW00} proved that for
certain types of non-Abelian anyons braiding enables one to perform universal
quantum computation. This scheme is usually referred to as \emph{topological
quantum computation} (TQC).

Let us outline some basic principles of TQC. First, topologically ordered
systems have degenerate ground states under certain circumstances. In
particular, the existence of Abelian anyons implies the ground state
degeneracy on the torus~\cite{Einarsson}. Indeed, consider a process in which
a particle-antiparticle pair is created, one of the particles winds around the
torus, and the pair is annihilated. Such a process corresponds to an operator
acting on the ground state. If $A$ and $B$ are such operators corresponding to
two basic loops on the torus, then $ABA^{-1}B^{-1}$ describes a process in
which none of the particles effectively crosses the torus, but one of them
winds around the other. If the Aharonov-Bohm phase is nontrivial, then $A$ and
$B$ do not commute. Therefore they act on a multidimensional space.

Actually, the degeneracy is not absolute but very precise. It is lifted due to
virtual particle tunneling across the torus, but this process is exponentially
suppressed. Therefore the distance between ground energy levels is
proportional to $\exp(-L/\xi)$, where $L$ the linear size of the torus and
$\xi$ is some characteristic length, which is related to the gap in the
excitation spectrum.

In non-Abelian systems, degeneracy occurs even in the planar geometry when
several anyons are localized in some places far apart from each other (it is
this space of quantum states the braid group acts on). The underlying
elementary property may be described as follows: \emph{two given non-Abelian
particles can fuse in several ways} (like multi-dimensional representations of
a non-Abelian group). For example, the non-Abelian phase studied in this paper
has the following fusion rules: 
\[
\eps\times\eps=1,\qquad\quad \eps\times\sigma=\sigma,\qquad\quad
\sigma\times\sigma=1+\eps,
\]
where $1$ is the vacuum sector, and $\eps$ and $\sigma$ are some other
superselection sectors. The last rule is especially interesting: it means that
two $\sigma$-particles may either annihilate or fuse into an
$\eps$-particle. But when the $\sigma$-particles stay apart, $1$ and $\eps$
correspond to two quantum states, $|\psi_{1}^{\sigma\sigma}\rangle$ and
$|\psi_{\eps}^{\sigma\sigma}\rangle$. These states are \emph{persistent}. For
example, if we create $|\psi_{\eps}^{\sigma\sigma}\rangle$ by splitting an
$\eps$ into two $\sigma$'s, wait some time, and fuse the $\sigma$-particles
back, we will still get an $\eps$-particle.

Here is a subtler property: the fusion states
$|\psi_{1}^{\sigma\sigma}\rangle$ and $|\psi_{\eps}^{\sigma\sigma}\rangle$ are
practically \emph{indistinguishable} and have almost the same energy. In fact,
a natural process that ``distinguishes'' them by multiplying by different
factors is tunneling of a virtual $\eps$-particle between the fixed
$\sigma$-particles (which is possible since $\sigma\times\eps=\sigma$).
However, $\eps$-particles are gapped, therefore this process is exponentially
suppressed. Of course, this explanation depends on many details, but it is a
general principle that \emph{different fusion states can only be distinguished
by transporting a quasiparticle}. Such processes are unlikely even in the
presence of thermal bath and external noise, as long as the temperature and
the noise frequency are much smaller than the gap.

\begin{figure}
\centerline{\begin{tabular}%
{c@{\qquad\qquad}c@{\qquad\qquad\quad}c@{\qquad\qquad}c}
$\figbox{1}{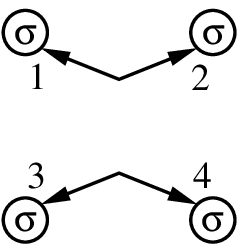}$  &
$\figbox{1}{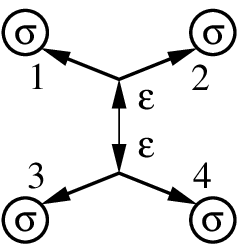}$  &
$\figbox{1}{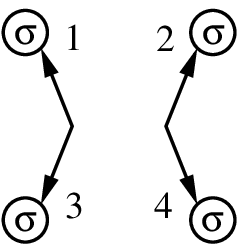}$  &
$\figbox{1}{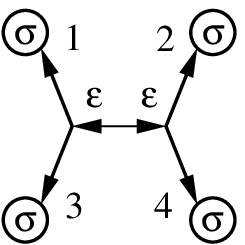}$ \\[5pt]
$|\xi_{1}\rangle$ & $|\xi_{\eps}\rangle$ &
$|\eta_{1}\rangle$ & $|\eta_{\eps}\rangle$
\end{tabular}}
\caption{Four ways to initialize an anyonic qubit.}
\label{fig_qubit}
\end{figure}

In the above example, the two-particle fusion states
$|\psi_{1}^{\sigma\sigma}\rangle$ and $|\psi_{\eps}^{\sigma\sigma}\rangle$
cannot form coherent superpositions because they belong to different
superselection sectors ($1$ and $\eps$, resp.). In order to implement a qubit,
one needs four $\sigma$-particles. A logical $|0\rangle$ is represented by the
quantum state $|\xi_{1}\rangle$ that is obtained by creating the pairs $(1,2)$
and $(3,4)$ from the vacuum (see Fig.~\ref{fig_qubit}). A logical $|1\rangle$
is encoded by the complementary state $|\xi_{\eps}\rangle$: we first create a
pair of $\eps$-particles, and then split each of them into a
$\sigma\sigma$-pair. Note that both states belong to the vacuum sector and
therefore \emph{can} form superpositions. Also shown in Fig.~\ref{fig_qubit}
are two alternative ways to initialize the qubit, $|\eta_{1}\rangle$ and
$|\eta_{\eps}\rangle$. The detailed analysis presented in
Secs.~\ref{sec_assoc} and~\ref{sec_algcons} implies that
\[
|\eta_{1}\rangle =
\frac{1}{\sqrt{2}}\Bigl(|\xi_{1}\rangle+|\xi_{\eps}\rangle\Bigr),\qquad
|\eta_{\eps}\rangle =
\frac{1}{\sqrt{2}}\Bigl(|\xi_{1}\rangle-|\xi_{\eps}\rangle\Bigr).
\]
Therefore we can perform the following gedanken experiment. We create the
state $|\xi_{1}\rangle$ and then measure the qubit in the
$\bigl\{|\eta_{1}\rangle,\,|\eta_{\eps}\rangle\}$-basis by fusing the pairs
$(1,3)$ and $(2,4)$. With probability $1/2$ both pairs annihilate, and with
probability $1/2$ we get two $\eps$-particles. One can also think of a simple
robustness test for quantum states: if there is no decoherence, then
\emph{both} $|\xi_{1}\rangle$ and $|\eta_{1}\rangle$ are persistent.

As already mentioned, braiding is described by operators that are \emph{exact}
(up to virtual quasiparticle tunneling). Indeed, the operators of
counterclockwise exchange between two particles ($R$-matrices) are related to
the fusion rules by so-called \emph{hexagon equations} and \emph{pentagon
equation}. We will see on concrete exapmles that these equations have only a
finite number of solutions and therefore do not admit small deformations. In
general, it is a nontrivial theorem known as \emph{Ocneanu
rigidity}~\cite{Ocneanu,ENO}, see Sec.~\ref{sec_Ocneanu}.

Thus, we have all essential elements of a quantum computer implemented in a
robust fashion: an initial state is made by creating pairs and/or by splitting
particles, unitary gates are realized by braiding, and measurements are
performed by fusion. This ``purely topological'' scheme is universal for
sufficiently complicated phases such as the $k=3$ parafermion
state~\cite{ReadRezayi}, lattice models based on some finite groups (e.g.\
$S_{5}$~\cite{Kitaev97}, $A_{5}$~\cite{OgburnPreskill,Preskill97} and
$S_{3}$~\cite{Mochon03}), and double Chern-Simons
models~\cite{Freedman01,FNSWW03,FNS03}. Unfortunately, the model studied in
this paper is not universal in this sense.  One can, however, combine a
topologically protected quantum memory with a nontopological realization of
gates (using explicit error correction). Note that some weak form of
topological protection is possible even in one-dimensional Josephson junction
arrays~\cite{DoucotVidal}, which is due to the build-in
$\UU(1)$-symmetry. Several other schemes of Josephson junction-based
topological quantum memory have been proposed
recently~\cite{IF02,DIF02,DIF03}.

Unlike many other quantum computation proposals, TQC should not have serious
scalability issues. What is usually considered an initial step, i.e.,
implementing a single gate, may actually be close to the solution of the whole
problem. It is an extremely challenging task, though. It demands the ability
to control individual quasiparticles, which is beyond the reach of present
technology. One should however keep in mind that the ultimate goal is to build
a practical quantum computer, which will contain at least a few hundred
\emph{logical} qubits and involve error-correcting coding: either in software
(with considerable overhead) or by topological protection or maybe by some
other means. At any rate, that is a task for the technology of the future. But
for the meantime, finding and studying topological phases seems to be a very
reasonable goal, also attractive from the fundamental science perspective.

\subsubsection*{Comparison with earlier work and a summary of the results}
In this paper we study a particular exactly solvable spin model on a
two-dimensional lattice. It only involves two-body interactions and therefore
is simpler than Hamiltonian~(\ref{torcode}) considered in~\cite{Kitaev97}, but
the solution is less trivial. It is not clear how to realize this model in
solid state, but an optical lattice implementation has been
proposed~\cite{DuanDemlerLukin}.

The model has two phases (denoted by $A$ and $B$) which occur at different
values of parameters. The exact solution is obtained by a reduction to free
real fermions. Thus quasiparticles in the system may be characterized as
fermions and $\ZZ_{2}$-vortices. Vortices and fermions interact by the
Aharonov-Bohm factor equal to $-1$. In phase $A$ the fermions have an energy
gap, and the vortices are bosons that fall into two distinct superselection
sectors. (Interestingly enough, the two types of vortices have identical
physical properties and are related to each other by a lattice translation.)
The overall particle classification, fusion rules, and statistics are the same
as in the model~(\ref{Z_2vac}) or RVB. In phase $B$ the fermions are gapless
and there is only one type of vortices with undefined statistics.  Adding a
magnetic field to the Hamiltonian opens a gap in the fermionic spectrum, and
the vortices become non-Abelian anyons. The difference between the vortex
statistics in phase $A$ and phase $B$ with the magnetic field may be
attributed to different topology of fermionic pairing.

Topological properties of Fermi-systems were first studied in the theory of
integer quantum Hall effect~\cite{TKNN82,ASS83}. Let us outline the main
result. To begin with, the Hall conductivity of noninteracting electrons in a
periodic potential (e.g., in the Hofstadter model with $m/n$ flux quanta per
plaquette) is expressed in terms of a single-electron Hamiltonian in the
Fourier basis. Such a Hamiltonian is an $n\times n$ matrix that depends on the
momentum $\mathbf{q}$. For each value of $q$ one can define a subspace
$\calL(\mathbf{q})\subseteq\CC^{n}$ that is associated with negative-energy
states, i.e., ones that are occupied by electrons. Thus, a vector bundle over
the momentum space is defined. The quantized Hall conductivity is proportional
to the Chern number of this bundle. Bellissard \latin{at al}~\cite{BES-B94}
have generalized this theory to disordered systems by using a powerful
mathematical theory called noncommutative geometry~\cite{Connes}.

Even more interesting topological phenomena occur when the number of particles
is not conserved (due to the presence of terms like
$a_{j}^{\dag}a_{k}^{\dag}$, as in the mean-field description of
superconductors). In this case the single-electron Hamiltonian is replaced by
a more general object, the Bogolyubov-Nambu matrix. It also has an associated
Chern number $\nu$, which is twice the number defined above when the previous
definition is applicable. But in general $\nu$ is an arbitrary integer. The
first physical example of this kind, the $\mathrm{{}^{3}He}$-$\mathrm{A}$
film, was studied by Volovik~\cite{Volovik88}. Volovik and
Yakovenko~\cite{VolovikYakovenko89} showed that the Chern number in this
system determines the statistics of solitons. More recently, Read and
Green~\cite{ReadGreen00} considered BCS pairing of spinless particles with
angular momentum $l=-1$. They identified a ``strong pairing phase'' with zero
Chern number and a ``weak pairing phase'' with $\nu=1$. The latter is closely
related to the Moore-Read state and has non-Abelian vortices and chiral edge
modes.

In the present paper, these results are generalized to an arbitrary
Fermi-system described by a quadratic Hamiltonian on a two-dimensional
lattice. We show that $\ZZ_{2}$-vortices are Abelian particles when the Chern
number $\nu$ is even and non-Abelian anyons when $\nu$ is odd. The non-Abelian
statistics is due to \emph{unpaired Majorana modes} associated with vortices.
Our method relies on a \emph{quasidiagonal matrix formalism} (see
Appendix~\ref{sec_locmatr}), which is similar to, but more elementary than,
noncommutative geometry. It can also be applied to disordered systems.

Furthermore, we find that there are actually $16$ ($8$ Abelian and $8$
non-Abelian) types of vortex-fermion statistics, which correspond to different
values of $\nu\bmod 16$. Only three of them (for $\nu=0,\pm 1$) are realized in
the original spin model.  We give a complete algebraic description of all $16$
cases, see tables on pages~\pageref{tab_nonabelian}, \pageref{tab_0mod4}, and
\pageref{tab_2mod4}.

\section{The model}

\begin{figure}[ht]
\centerline{\begin{tabular}{c@{\qquad}c}
\epsfbox{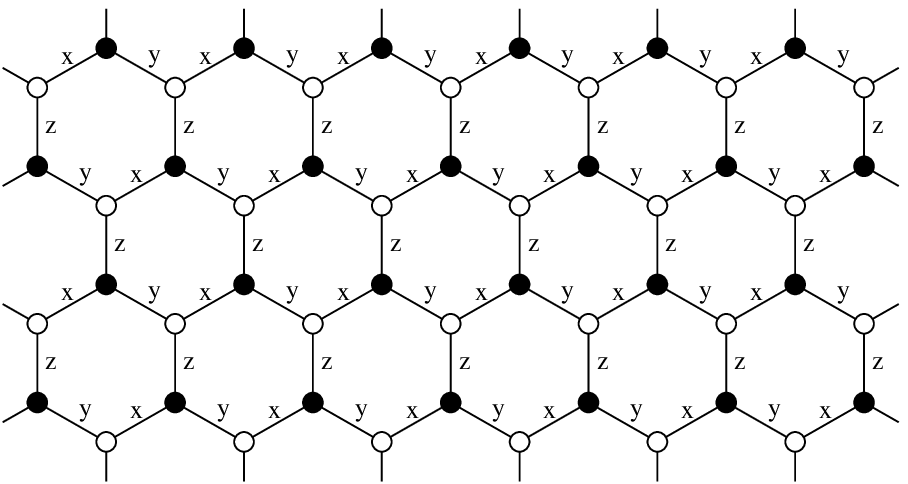} & \raisebox{1cm}{\epsfbox{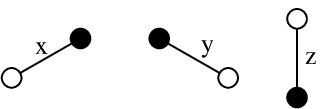}} \\
a) & b)
\end{tabular}}
\caption{Three types of links in the honeycomb lattice.}
\label{fig_honeycomb}
\end{figure}

We study a spin-$1/2$ system in which spins are located at the vertices
of a honeycomb lattice, see Fig.~\ref{fig_honeycomb}a.
This lattice consists of two equivalent simple sublattices, referred to as
``even'' and ``odd'' (they are shown by empty and full circles in the
figure). A unit cell of the lattice contains one vertex of each kind.  Links
are divided into three types, depending on their direction (see
Fig.~\ref{fig_honeycomb}b); we call them ``$x$-links'', ``$y$-links'', and
``$z$-links''. The Hamiltonian is as follows:
\begin{equation} \label{Hamiltonian}
H\,=\,
-J_{x} \sum_{\text{$x$-links}} \sigma_{j}^{x}\sigma_{k}^{x}
-J_{y} \sum_{\text{$y$-links}} \sigma_{j}^{y}\sigma_{k}^{y}
-J_{z} \sum_{\text{$z$-links}} \sigma_{j}^{z}\sigma_{k}^{z},
\end{equation}
where $J_{x}$, $J_{y}$, $J_{z}$ are model parameters.

Let us introduce a special notation for the individual terms in the
Hamiltonian:
\begin{equation} \label{terms}
K_{jk}= \left\{
\begin{array}{l@{\quad}l}
\sigma_{j}^{x}\sigma_{k}^{x}, & \text{if $(j,k)$ is an $x$-link;} \\
\sigma_{j}^{x}\sigma_{k}^{y}, & \text{if $(j,k)$ is an $y$-link;} \\
\sigma_{j}^{x}\sigma_{k}^{z}, & \text{if $(j,k)$ is an $z$-link.} 
\end{array}
\right.
\end{equation}
Remarkably, \emph{all} operators $K_{jk}$ commute with the following operators
$W_{p}$, which are associated to lattice plaquettes (i.e., hexagons):
\begin{equation} \label{invariant}
\begin{array}{c} \epsfbox{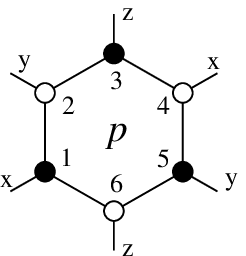} \end{array} \qquad\quad
W_{p}=\sigma_{1}^{x}\sigma_{2}^{y}\sigma_{3}^{z}
\sigma_{4}^{x}\sigma_{5}^{y}\sigma_{6}^{z}
=K_{12}K_{23}K_{34}K_{45}K_{56}K_{61}.
\end{equation}
Here $p$ is a label of the plaquette. Note that different operators $W_{p}$
commute with each other.

Thus Hamiltonian~(\ref{Hamiltonian}) has the set of ``integrals of motion''
$W_{p}$, which greatly simplifies the problem. In order to find eigenstates of
the Hamiltonian, we first divide the total Hilbert space $\calL$ into
\emph{sectors} --- eigenspaces of $W_{p}$, which are also invariant subspaces
of $H$. This can be written as follows:
\begin{equation} \label{v-sectors}
\calL =\bigoplus_{w_{1},\dots,w_{m}} \calL_{w_{1},\dots,w_{m}},
\end{equation}
where $m$ is the number of plaquettes. Each operator $W_{p}$ has eigenvalues
$+1$ and $-1$, therefore each sector corresponds to a choice of $w_{p}=\pm 1$
for each plaquette $p$. Then we need to solve for the eigenvalues of the
Hamiltonian restricted to a particular sector $\calL_{w_{1},\dots
w_{m}}$.

The honeycomb lattice has $1/2$ plaquette per vertex, therefore $m\approx
n/2$, where $n$ is the number of vertices. It follows that the dimension of
each sector is $\sim 2^{n}/2^{m}\sim 2^{n/2}$ (we will in fact see that
all these dimensions are equal). Thus splitting into sectors does not solve
the problem yet. Fortunately, it turns out that the degrees of freedom within
each sector can be described as real (Majorana) fermions, and the restricted
Hamiltonian is simply a quadratic form in Majorana operators. This makes an
exact solution possible.

\section{Representing spins by Majorana operators} \label{sec_spinferm}

\subsection{A general spin-fermion transformation}

Let us remind the reader some general formalism pertaining to Fermi systems. A
system with $n$ fermionic modes is usually described by the annihilation and
creation operators $a_{k}$, $a_{k}^\dag$\, ($k=1,\dots,n$).  Instead, one can
use their linear combinations,
\[
c_{2k-1}=a_{k}+a_k^\dag, \qquad\quad c_{2k}=\frac{a_k-a_k^\dag}{i},
\]
which are called Majorana operators. The operators $c_{j}$\,
($j=1,\dots,2n$) are Hermitian and obey the following relations:
\begin{equation}
c_{j}^{2} = 1,\qquad\quad
c_j c_l = -c_lc_j\ \text{ if } j\not=l.
\end{equation}
Note that all operators $c_{j}$ can be treated on equal basis.

We now describe a representation of a spin by two fermionic modes, i.e., by
four Majorana operators. Let us denote these operators by $b^{x}$, $b^{y}$,
$b^{z}$, and $c$ (instead of $c_{1}$, $c_{2}$, $c_{3}$, $c_{4}$).  The
Majorana operators act on the $4$-dimensional Fock space
$\widetilde{\calM}$, whereas the Hilbert space of a spin is identified
with a two-dimensional subspace $\calM\subset\widetilde{\calM}$
defined by this condition:
\begin{equation} \label{spin_rep}
|\xi\rangle\in\calM\quad \text{if and only if}\quad
D|\xi\rangle=|\xi\rangle,\qquad \text{where}\,\ D=b^{x}b^{y}b^{z}c.
\end{equation}
We call $\calM$ and $\widetilde{\calM}$ the \emph{physical
subspace} and the \emph{extended space}, respectively; the operator $D$ may be
thought of as a \emph{gauge transformation} for the group $\ZZ_2$.

The Pauli operator $\sigma^{x}$, $\sigma^{y}$, $\sigma^{z}$ can be represented
by some operators $\widetilde{\sigma}^{x}$, $\widetilde{\sigma}^{y}$,
$\widetilde{\sigma}^{z}$ acting on the extended space. Such a representation
must satisfy two conditions: (1) $\widetilde{\sigma}^{x}$,
$\widetilde{\sigma}^{y}$, $\widetilde{\sigma}^{z}$ preserve the subspace
$\calM$;\, (2) when restricted to $\calM$, the operators
$\widetilde{\sigma}^{x}$, $\widetilde{\sigma}^{y}$, $\widetilde{\sigma}^{z}$
obey the same algebraic relations as $\sigma^{x}$, $\sigma^{y}$,
$\sigma^{z}$. We will use the following particular representation:
\begin{equation} \label{XYZ}
\begin{array}[c]{c} \epsfbox{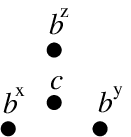} \end{array} \qquad\qquad
\widetilde{\sigma}^{x}=ib^{x}c,\quad\ \widetilde{\sigma}^{y}=ib^{y}c,\quad\ 
\widetilde{\sigma}^{z}=ib^{z}c.
\end{equation}
(We have associated the Majorana operators with four dots for a reason that
will be clear later.) This representation is correct since
$\widetilde{\sigma}^{\alpha}$ ($\alpha=x,y,z$) commutes with $D$ (so that
$\calM$ is preserved),
$(\widetilde{\sigma}^{\alpha})^\dag=\widetilde{\sigma}^{\alpha}$,\,
$(\widetilde{\sigma}^{\alpha})^{2}=1$, and
\[
\widetilde{\sigma}^{x}\widetilde{\sigma}^{y}\widetilde{\sigma}^{z}
=ib^{x}b^{y}b^{z}c=iD.
\]
The last equation is consistent with the formula
$\sigma^{x}\sigma^{y}\sigma^{z}=i$ because $D$ acts as the identity operator
on the subspace $\calM$.

A multi-spin system is described by four Majorana operators per spin.  The
corresponding operators $\widetilde{\sigma}^{\alpha_{j}}$, $D^{\alpha}_{j}$
and the physical subspace $\calL\subset\widetilde{\calL}$ are
defined as follows:
\begin{equation} \label{multispin_rep}
\begin{array}{c}
\widetilde{\sigma}^{\alpha_{j}}=ib^{\alpha}_{j}c_{j},\qquad\
D_{j}=b^{x}_{j}b^{y}_{j}b^{z}_{j}c_{j};
\bigskip\\
|\xi\rangle\in\calL\quad \text{if and only if}\quad
D_{j}|\xi\rangle=|\xi\rangle\ \text{for all $j$}.
\end{array}
\end{equation}
Any spin Hamiltonian $H\{\sigma^{\alpha}_{j}\}$ can be replaced by the
fermionic Hamiltonian $\widetilde{H}\{b^{\alpha}_{j},c_{j}\}
=H\{\widetilde{\sigma}^{\alpha_{j}}\}$ the action of which is restricted to
the physical subspace. (The resulting Hamiltonian $\widetilde{H}$ is rather
special; in particular, it commutes with the operators $D_{j}$.)

\begin{Remark}
The substitution $\sigma^{\alpha}_{j}\mapsto \widetilde{\sigma}^{\alpha}_{j}
=ib^{\alpha}_{j}c_{j}$ is gauge-equivalent to a more familiar one
(see \cite{spinMaj} and references therein):
\begin{equation} \label{multispin_rep1}
\sigma^{\alpha}_{j}\mapsto D_{j}\widetilde{\sigma}^{\alpha}_{j}, \qquad\quad
\text{i.e.,}\quad
\sigma^{x}_{j} \mapsto -ib^{y}_{j}b^{z}_{j}, \quad\
\sigma^{y}_{j} \mapsto -ib^{z}_{j}b^{x}_{j}, \quad\
\sigma^{z}_{j} \mapsto -ib^{x}_{j}b^{y}_{j}.
\end{equation}
Thus one can represent a spin by only 3 Majorana operators without imposing
gauge constraints. However, this is not sufficient for our purposes.
\end{Remark}

\subsection{Application to the concrete model}

\begin{figure}
\centerline{\epsfbox{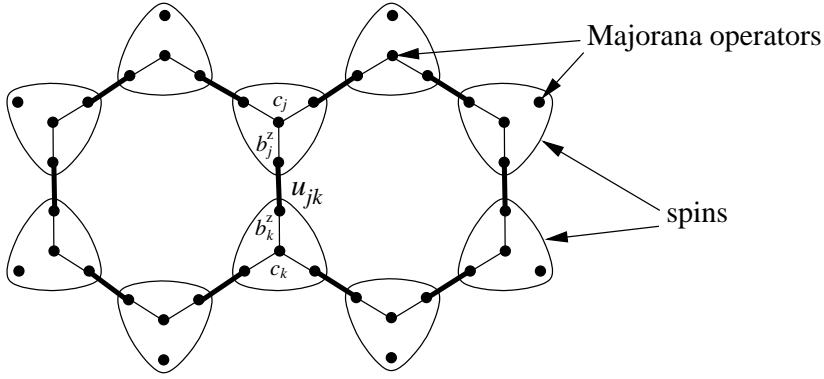}}
\caption{Graphic representation of Hamiltonian~(\ref{Htilde}).}
\label{fig_linkvar}
\end{figure}

Let us apply the general procedure to the the spin
Hamiltonian~(\ref{Hamiltonian}). Each term
$K_{jk}=\sigma^{\alpha}_{j}\sigma^{\alpha}_{k}$ becomes
$\widetilde{K}_{jk}=(ib^{\alpha}_{j}c_{j})
(ib^{\alpha}_{k}c_{k})=-i\,(ib^{\alpha}_{j}b^{\alpha}_{k})\,c_{j}c_{k}$. The
operator in parentheses, $\hat{u}_{jk}=ib^{\alpha}_{j}b^{\alpha}_{k}$, is
Hermitian; we associated it with the link $(j,k)$. (The index $\alpha$ takes
values $x$, $y$ or $z$ depending on the direction of the link, i.e.,
$\alpha=\alpha_{jk}$.)  Thus we get:
\begin{equation} \label{Htilde}
\begin{array}{c}
\displaystyle
\widetilde{H} = \frac{i}{4}\sum_{j,k} \hat{A}_{jk}c_{j}c_{k},
\qquad
\hat{A}_{jk} = \left\{ \begin{array}{cl}
2J_{\alpha_{jk}}\hat{u}_{jk} & \text{if $j$ and $k$ are connected},\\ 
0 & \text{otherwise}, \end{array} \right.
\medskip\\
\hat{u}_{jk}=ib^{\alpha_{jk}}_{j}b^{\alpha_{jk}}_{k}.
\end{array}
\end{equation}
Note that each pair of connected sites is counted twice, and
$\hat{u}_{kj}=-\hat{u}_{jk}$. The structure of this Hamiltonian is shown in
Fig.~\ref{fig_linkvar}.

Remarkably, the operators $\hat{u}_{jk}$ commute with the Hamiltonian and with
each other. Therefore the Hilbert space $\widetilde{\calL}$ splits into
common eigenspaces of $\hat{u}_{jk}$, which are indexed by the corresponding
eigenvalues $u_{jk}=\pm 1$. Similarly to~(\ref{v-sectors}) we may write
$\widetilde{\calL} = \bigoplus_{u} \widetilde{\calL}_{u}$, where
$u$ stands for the collection of all $u_{jk}$. The restriction of
Hamiltonian~(\ref{Htilde}) to the subspace $\widetilde{\calL}_{u}$ is
obtained by ``removing hats'', i.e., replacing operators by numbers. This
procedure results in the Hamiltonian $\widetilde{H}_{u} =
\frac{i}{4}\sum_{j,k} A_{jk}c_{j}c_{k}$, which corresponds to free
fermions. The ground state of $\widetilde{H}_{u}$ can be found exactly; let us
denote it by $|\widetilde{\Psi}_{u}\rangle$.

Note, however, that the subspace $\widetilde{\calL}_{u}$ is not
gauge-invariant: applying the gauge operator $D_{j}$ changes the values of
$u_{jk}$ on the links connecting the vertex $j$ with three adjacent vertices
$k$.  Therefore the state $|\widetilde{\Psi}_{u}\rangle$ does not belong to
the physical subspace. To obtain a physical space we must symmetrize over all
gauge transformations. Specifically, we construct the following state:
\begin{equation}
|\Psi_{w}\rangle\,=\,\prod_{j}\left(\frac{1+D_{j}}{2}\right)
|\widetilde{\Psi}_{u}\rangle\, \in\,\calL.
\end{equation}
Here $w$ denotes the equivalence class of $u$ under the gauge transformations.
For the planar lattice (but not on the torus) $w$ is characterized by numbers
$w_{p}=\pm 1$ defined as products of $u_{jk}$ around hexagons. To avoid
ambiguity (due to the relation $u_{kj}=-u_{jk}$), we choose a particular
direction for each link:
\begin{equation} \label{invariant1}
w_{p}=\prod_{(j,k)\in\text{boundary}(p)} u_{jk} \qquad
\bigl(j \in\text{even sublattice},\ k \in\text{odd sublattice}\bigr).
\end{equation}
The corresponding operator $\widetilde{W}_{p}=\prod\hat{u}_{jk}$ commutes with
the gauge transformations as well as the Hamiltonian. The restriction of this
operator to the physical subspace coincides with the integral of motion
$W_{p}$ defined earlier (see~(\ref{invariant})).

\begin{remark}[Notation change:]
From now on, we will not make distinction between operators acting in the
extended space and their restrictions to the physical subspace e.g.,
$\widetilde{W}_{p}$ versus $W_{p}$. The tilde mark will be used for other
purposes.
\end{remark}

\subsection{Path and loop operators}

One may think of the variables $u_{kj}$ as a $\ZZ_{2}$ gauge field. The number
$w_{p}$ is interpreted as the magnetic flux through the plaquette $p$. If
$w_{p}=-1$, we say that the plaquette carries a vortex.

The product of $\hat{u}_{jk}$ along an arbitrary path corresponds to the
transfer of a fermion between the initial and the final point. However, this
product is not gauge-invariant. One can define an invariant \emph{fermionic
path operator} in terms of spins or in terms of fermions:
\begin{equation} \label{pathop}
W(j_{0},\dots,j_{n}) \,=\,
K_{j_{n}j_{n-1}} \dots K_{j_{1}j_{0}} \,=\,
\left(\prod_{s=1}^{n} -i\hat{u}_{j_{s}j_{s-1}}\right) c_{n}c_{0},
\end{equation}
where $K_{jk}$ is given by~(\ref{terms}). If the path is closed, i.e.,
$j_{n}=j_{0}$, the factors $c_{n}$ and $c_{0}$ cancel each other. In this case
the path operator is called the \emph{Wilson loop}; it generalizes the notion
of magnetic flux.

On the honeycomb lattice all loops have even length, and
formula~(\ref{pathop}) agrees with the sign convention based on the partition
into the even and odd sublattice. However, the spin model can be generalized
to any trivalent graph, in which case the loop length is arbitrary. For an odd
loop $l$ the operator $W(l)$ has eigenvalues $w_{l}=\pm i$. This is
not just an artifact of the definition: odd loops are special in that they
cause spontaneous breaking of the time-reversal symmetry.

The time-reversal operator is a conjugate-linear unitary operator $T$ such
that
\begin{equation} \label{timerev}
T\sigma^{\alpha}_{j}T^{-1}=-\sigma^{\alpha}_{j},\qquad\quad
Tb_{j}T^{-1}=b_{j},\quad\ Tc_{j}T^{-1}=c_{j}.
\end{equation}
(The first equation is a physical requirement; the other two represent the
action of $T$ in the extended space.) Therefore $T$ commutes with the
Hamiltonian~(\ref{Hamiltonian}) and the Wilson loop. Multiplying the equation
$W_{l}|\Psi\rangle=w_{l}|\Psi\rangle$ by $T$, we get $W_{l}T|\Psi\rangle
=w_{l}^{*}T|\Psi\rangle$. Thus the time-reversal operator changes $w_{l}$ to
$w_{l}^{*}$. For a bipartite graph $w_{l}^{*}=w_{l}$ for all loops, therefore
fixing the variables $w_{l}$ does not break the time-reversal symmetry. On the
contrary, for a similar model on a non-bipartite graph (e.g., a lattice
containing triangles) the operator $T$ does not preserve the field
configuration, which is defined by the values of $w_{l}$ on all loops. But $T$
is a symmetry operator, therefore all Hamiltonian eigenstates are (at least)
two-fold degenerate.

\section{Quadratic Hamiltonians} \label{sec_quadratic}

In the previous section we transformed the spin model~(\ref{Hamiltonian}) to a
quadratic fermionic Hamiltonian of the general form
\begin{equation} \label{quadH}
H(A) = \frac{i}{4}\sum_{j,k} A_{jk}c_{j}c_{k},
\end{equation}
where $A$ is a real skew-symmetric matrix of size $n=2m$. Let us briefly state
some general properties of such Hamiltonians and fix the terminology.

First, we comment on the normalization factor $1/4$ in Eq.~(\ref{quadH}). It
is chosen so that
\begin{equation}
[-iH(A),-iH(B)]=-iH\bigl([A,B]\bigr).
\end{equation}
Thus the Lie algebra of quadratic operators $-iH(A)$ (acting on the
$2^{m}$-dimensional Fock space) is identified with $\mathfrak{so}(2m)$.
Operators of the form $e^{-iH(A)}$ constitute the Lie group $\Spin(2m)$. The
center of this group consists of phase factors $\pm 1$ (e.g., $e^{\pi
c_{1}c_{2}}=-1$).  The quotient group $\Spin(2m)/\{+1,-1\} =\SO(2m)$ describes
the action of $e^{-iH(A)}$ on Majorana operators by conjugation:
\begin{equation} \label{linact}
e^{-iH(A)}\, c_{k}\, e^{iH(A)} = \sum_{j}Q_{jk}c_{j},\qquad
\text{where}\,\  Q=e^{A}
\end{equation}
($Q$ is a real orthogonal matrix with determinant $1$).

Note that the sum in Eq.~(\ref{linact}) corresponds to the multiplication of
the row vector $(c_{1},\dots,c_{2m})$ by the matrix $Q$. On the other hand,
when we consider a linear combination of Majorana operators,
\begin{equation}
F(x)=\sum_{j}x_{j}c_{j},
\end{equation}
we prefer to view $x$ as a column vector. If the coefficients $x_{j}$ are
real, we call $F(x)$ (or $x$ itself) a \emph{Majorana mode}.

To find the ground state of the Hamiltonian~(\ref{quadH}), one needs to
reduce it to a canonical form
\begin{equation}\label{canonical_form}
H_{\text{canonical}} \,=\,\frac{i}{2}\sum_{k=1}^{m}\eps_k b_k'b_k'' \,=\,
\sum_{k=1}^{m}\eps_{k}(a_k^\dag a_k
- {\textstyle\frac{1}{2}}), \qquad\ \eps_k\ge 0,
\end{equation}
where $b_k'$, $b_k''$ are \emph{normal modes}, and
$a_k^\dag=\frac{1}{2}(b_k'-ib_k'')$,\, $a_k=\frac{1}{2}(b_k'+ib_k'')$ are the
corresponding creation and annihilation operators. The ground state of
$H_{\text{canonical}}$ is characterized by the condition $a_k|\Psi\rangle=0$
for all $k$. The reduction to the canonical form is achieved by the
transformation
\begin{equation}\label{reduction_to_canonical}
(b_1', b_1'',\dots,b_m',b_m'')
\,=\, (c_1,c_2,\dots,c_{2m-1},c_{2m})\,Q,
\qquad\ Q\in\OO(2m),
\end{equation}
such that
\begin{equation}
A \,=\; Q\begin{pmatrix}
  0 & \eps_1 &&&\\ -\eps_1 & 0 &&&\\ &&\ddots&&\\
  &&& 0 & \eps_m\\ &&& -\eps_m & 0 \end{pmatrix}Q^{T}.
\end{equation}
The numbers $\pm\eps_{k}$ constitute the spectrum of the Hermitian matrix
$iA$, whereas odd (even) columns of $Q$ are equal to the real (resp.,
imaginary) part of the eigenvectors. The ground state of the Majorana system
has energy
\begin{equation}\label{quadHenergy}
E=-\frac{1}{2}\sum_{k=1}^{m}\eps_{k}
=-\frac{1}{4}\Tr|iA|,
\end{equation}
where the function $|\cdot|$ acts on the eigenvalues, the eigenvectors being
fixed. (In fact, any function of a real variable can be applied to Hermitian
matrices.)

Note that different quadratic Hamiltonians may give rise to the same ground
state. The latter actually depends on
\begin{equation} \label{matrixB}
B\,=\,-i\sgn(iA)\,=\;
Q \begin{pmatrix}
  0 & 1 &&&\\ -1 & 0 &&&\\ &&\ddots&&\\
  &&& 0 & 1 \\ &&& -1 & 0 \end{pmatrix} Q^{T},
\end{equation}
(We assume that $A$ is not degenerate.) The matrix $B$ is real skew-symmetric
and satisfies $B^{2}=-1$. It determines the ground state through the condition
\begin{equation} \label{specproj}
\sum_{j}P_{jk}c_{j}\,|\Psi\rangle=0 \quad \text{for all}\ k,\qquad
\text{where}\ P_{jk}=\frac{1}{2}(\delta_{jk}-iB_{jk}).
\end{equation}
Loosely speaking, $B$ corresponds to a pairing\footnote{Note the nonstandard
use of terminology: in our sense, electron half-modes in an insulator are
paired as well as in a superconductor. The only difference is that the
insulating pairing commutes with the number of electrons while the
superconducting one doesn't. This distinction is irrelevant to our model
because the Hamiltonian doesn't preserve any integral charge, and the number
of fermions is only conserved \latin{modulo} $2$.} between Majorana modes. The
operators $b'=F(x')$ and $b''=F(x'')$ are \emph{paired} if $x''=\pm Bx'$.

The matrix $P$ in Eq.~(\ref{specproj}) is called the \emph{spectral
projector}. It projects the $2m$-dimensional complex space $\CC^{2m}$ onto the
$m$-dimensional subspace $L$ spanned by the eigenvectors of $iA$ corresponding
to negative eigenvalues. For any vector $z\in L$ the corresponding operator
$F(z)$ annihilates the ground state, so we may call $L$ the \emph{space of
annihilation operators}. Note that if $z,z'\in L$ then
$\sum_{j}z_{j}z'_{j}=0$. The choice of an $m$-dimensional subspace
$L\subseteq\CC^{2m}$ satisfying this condition is equivalent to the choice of
matrix $B$.

The ground state of a quadratic Hamiltonian can also be characterized by
correlation functions. The second-order correlator is
$\langle\Psi|c_{j}c_{k}|\Psi\rangle=2P_{kj}$; higher-order correlators can be
found using Wick's formula.

\section{The spectrum of fermions and the phase diagram} \label{sec_spectrum}

We now study the system of Majorana fermions on the honeycomb lattice. It is
described by the quadratic Hamiltonian $H_{u}=H(A)$, where
$A_{jk}=2J_{\alpha_{jk}}u_{jk}$,\, $u_{jk}=\pm 1$. Although the Hamiltonian is
parametrized by $u_{jk}$, the corresponding gauge-invariant state (or the
state of the spin system) actually depends on the variables $w_{p}$,
see~(\ref{invariant1}).

First, we remark that the global ground state energy does not depend on the
signs of the exchange constants $J_{x}$, $J_{y}$, $J_{z}$ since changing the
signs can be compensated by changing the corresponding variables $u_{jk}$. We
further notice that the ground state energy for $H_{u}$ does not depend on
these signs even if $u$ is fixed. Suppose, for instance, that we replace
$J_{z}$ by $-J_{z}$. Such a change is equivalent to altering $u_{jk}$ for all
$z$-links. But the gauge-invariant quantities $w_{p}$ remain constant, so we
may apply a gauge transformation that returns $u_{jk}$ to their original
values.  The net effect is that the Majorana operators at some sites are
transformed as $c_{j}\mapsto-c_{j}$. Specifically, the transformation acts on
the set of sites $\Omega_{z}$ that lie in the shaded area in the picture
below. In terms of spins, this action is induced by the unitary operator
\begin{equation} \label{jzflip}
\begin{array}{c} \epsfbox{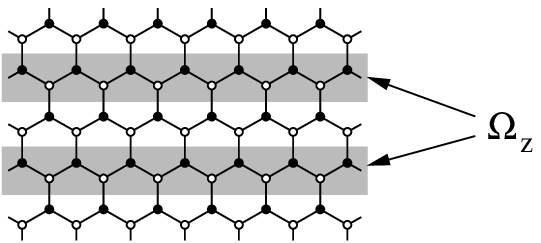} \end{array} \qquad\qquad
R_{z}=\prod_{j\in\Omega_{z}}\sigma^{z}_{j} \hspace{20pt}
\end{equation}
So, for the purpose of finding the ground state energy and the excitation
spectrum, the signs of exchange constants do not matter (but other physical
quantities may depend on them).

The most interesting choice of $u_{jk}$ is the one that minimizes the ground
state energy. It turns out that the energy minimum is achieved by the
vortex-free field configuration, i.e., $w_{p}=1$ for all plaquettes $p$. This
statement follows from a beautiful theorem proved by Lieb~\cite{Lieb94}. (Not
knowing about Lieb's result, I did some numerical study suggesting the same
answer, see Appendix~\ref{sec_numerics}.) Thus we may assume that $u_{jk}=1$
for all links $(j,k)$, where $j$ belongs to the even sublattice, and $k$
belongs to the odd sublattice. This field configuration (denoted by
$u_{jk}^{\std}$) possesses a translational symmetry, therefore the fermionic
spectrum can be found analytically using the Fourier transform.

The general procedure is as follows. Let us represent the site index $j$ as
$(s,\lambda)$, where $s$ refers to a unit cell, and $\lambda$ to a position
type inside the cell (we choose the unit cell as shown in the
figure accompanying Eq.~(\ref{spectrum})). The Hamiltonian becomes
$H=(i/4)\sum_{s,\lambda,t,\mu}A_{s\lambda,t\mu} c_{s\lambda}c_{t\mu}$, where
$A_{s\lambda,t\mu}$ actually depends on $\lambda$, $\mu$, and $t-s$. Then we
pass to the momentum representation:
\begin{gather}
H=\frac{1}{2}\sum_{\qq,\lambda,\mu}i\widetilde{A}_{\lambda\mu}(\qq)
a_{-\qq,\lambda}a_{\qq,\mu}, \qquad\quad
\widetilde{A}_{\lambda\mu}(\qq)
=\sum_{t}e^{i(\qq,\rr_{t})} A_{0\lambda,t\mu},
\\
a_{\qq,\lambda}
=\frac{1}{\sqrt{2N}}\sum_{s}e^{-i(\qq,\rr_{s})} c_{s\lambda},
\end{gather}
where $N$ is the total number of the unit cells. (Here and on, operators in
the momentum representation are marked with tilde.) Note that
$a_{\qq,\lambda}^{\dag} =a_{\mathbf{-q},\lambda}^{\phantom{\dag}}$ and
$a_{\pp,\mu}^{\phantom{\dag}}a_{\qq,\lambda}^{\dag}
+a_{\qq,\lambda}^{\dag}a_{\pp,\mu}^{\phantom{\dag}}
=\delta_{\pp\qq}\delta_{\mu\lambda}$. The spectrum $\eps(\qq)$ is given by the
eigenvalues of the matrix $i\widetilde{A}(\qq)$. One may call it a ``double
spectrum'' because of its redundancy: $\eps(-\qq)=-\eps(-\qq)$. The ``single
spectrum'' can be obtained by taking only positive eigenvalues (if none of the
eigenvalues is zero).

We now apply this procedure to the concrete Hamiltonian
\begin{equation} \label{Hvf}
H_{\text{vortex-free}}= \frac{i}{4}\sum_{j,k} A_{jk}c_{j}c_{k},\qquad\quad
A_{jk}=2J_{\alpha_{jk}}u_{jk}^{\std}.
\end{equation}
We choose a basis $(\nn_{1},\nn_{2})$ of the translation group and obtain the
following result:
\begin{equation} \label{spectrum}
\begin{array}{c} \epsfbox{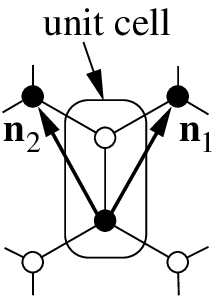} \end{array}\qquad
\begin{array}{c}
i\widetilde{A}(\qq)=\left(\begin{array}{@{}c@{\,}c@{}}
0 & i f(\qq)\\ -i f(\qq)^{*} & 0
\end{array}\right), \quad\ \eps(\qq)=\pm|f(\qq)|,
\bigskip\\
f(\qq)=2\bigl(J_{x}e^{i(\qq,\nn_{1})} +J_{y}e^{i(\qq,\nn_{2})} +J_{z}\bigr),
\bigskip\\
\end{array}
\end{equation}
where $\nn_{1}=(\frac{1}{2},\frac{\sqrt{3}}{2})$,\,
$\nn_{2}=(-\frac{1}{2},\frac{\sqrt{3}}{2})$ in the standard $xy$-coordinates.

\begin{figure}
\centerline{\epsfbox{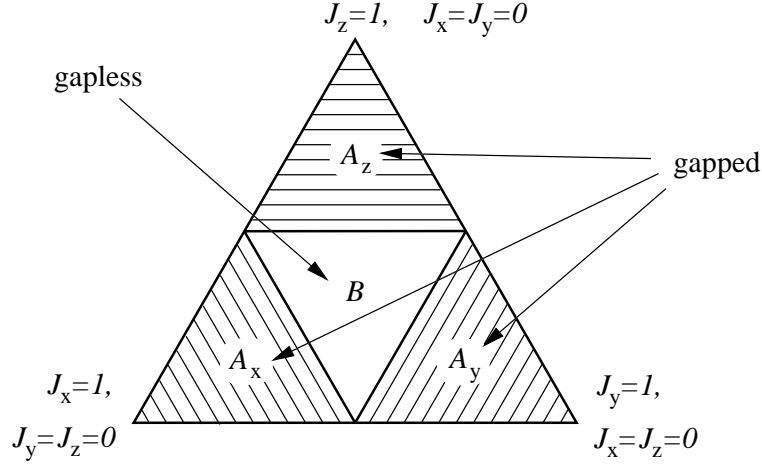}}
\caption{Phase diagram of the model. The triangle is the section of the
positive octant ($J_{x},J_{y},J_{z}\ge 0$) by the plane
$J_{x}+J_{y}+J_{z}=1$. The diagrams for the other octants are similar.}
\label{fig_phdiag}
\end{figure}

An important property of the spectrum is whether it is gapless, i.e., whether
$\eps(\qq)$ is zero for some $\qq$. The equation
$J_{x}e^{i(\qq,\nn_{1})} +J_{y}e^{i(\qq,\nn_{2})} +J_{z}=0$ has
solutions if and only if $|J_{x}|$, $|J_{y}|$, $|J_{z}|$ satisfy the
triangle inequalities:
\begin{equation} \label{trineq}
|J_{x}|\le |J_{y}|+|J_{z}|,\quad\
|J_{y}|\le |J_{x}|+|J_{z}|,\quad\
|J_{z}|\le |J_{x}|+|J_{y}|.
\end{equation}
If the inequalities are strict (``$<$'' instead of ``$\le$''), there are
exactly $2$ solutions: $\qq=\pm\qq_{*}$. The region defined by
inequalities~(\ref{trineq}) is marked by $B$ in Fig.~\ref{fig_phdiag}; this
phase is gapless. The gapped phases, $A_{x}$, $A_{y}$, and $A_{z}$, are
algebraically distinct, though related to each other by rotational
symmetry. They differ in the way lattice translations act on anyonic states
(see Section~\ref{sec_Abelian}). Therefore a continuous transition from one
gapped phase to another is impossible, even if we introduce new terms in the
Hamiltonian. On the other hand, the 8 copies of each phase (corresponding to
different sign combinations of $J_{x}$, $J_{y}$, $J_{z}$) have the same
translational properties. It is unknown whether the 8 copies of the gapless
phase are algebraically different.

We now consider the zeros of the spectrum that exist in the gapless phase. The
momentum $\qq$ is defined \latin{modulo} the reciprocal lattice, i.e., it
belongs to a torus. We represent the momentum space by the parallelogram
spanned by $(\qq_{1},\qq_{2})$ --- the basis dual to $(\nn_{1},\nn_{2})$. In
the symmetric case ($J_{x}=J_{y}=J_{z}$) the zeros of the spectrum are given
by
\begin{equation} \label{qstar}
\begin{array}{c} \epsfbox{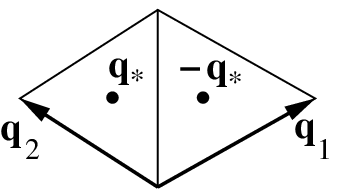} \end{array} \qquad\
\begin{array}{r@{}l}
\qq_{*} & {}\equiv \frac{1}{3}\qq_{1}+\frac{2}{3}\qq_{2}
\pmod{\qq_{1},\qq_{2}}
\bigskip\\
-\qq_{*} & {}\equiv \frac{2}{3}\qq_{1}+\frac{1}{3}\qq_{2}
\pmod{\qq_{1},\qq_{2}}
\end{array}
\end{equation}
If $|J_{x}|$ and $|J_{y}|$ decrease while $|J_{z}|$ remains constant,
$\qq_{*}$ and $-\qq_{*}$ move toward each other (within the parallelogram)
until they fuse and disappear. This happens when $|J_{x}|+|J_{y}|=|J_{z}|$.
The points $\qq_{*}$ and $-\qq_{*}$ can also effectively fuse at opposite
sides of the parallelogram. (Note that the equation $\qq_{*}=-\qq_{*}$ has
three nonzero solutions on the torus).

At the points $\pm \qq_{*}$ the spectrum has conic singularities (assuming
that $\qq_{*}\not=-\qq_{*}$):
\begin{equation} \label{cone}
\begin{array}{c} \epsfbox{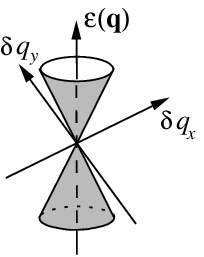} \end{array} \qquad\
\begin{array}{c}
\eps(\qq) \approx
\pm\sqrt{g_{\alpha\beta}\,\delta q_{\alpha}\,\delta q_{\beta}},
\bigskip\\
\text{where}\,\ \delta\qq=\qq-\qq_{*}\,\ \text{or}\,\ \delta\qq=\qq+\qq_{*}.
\end{array}
\end{equation}

\section{Properties of the gapped phases}
\label{sec_gapped}

In a gapped phase, spin correlations decay exponentially with distance,
therefore spatially separated quasiparticles cannot interact directly. That
is, a small displacement or another local action on one particle does not
influence the other. However, the particles can interact topologically if they
move around each other. This phenomenon is described by \emph{braiding rules}.
(We refer to braids that are formed by particle worldlines in the
3-dimensional space-time.) In our case the particles are vortices and
fermions. When a fermion moves around a vortex, the overall quantum state is
multiplied by $-1$. As mentioned in the introduction, such particles (with
braiding characterized simply by phase factors) are called \emph{Abelian
anyons}.

The description of anyons begins with identifying \emph{superselection
sectors}, i.e., excitation types defined up to local operations. (An
``excitation'' is assumed to be localized in space, but it may have uncertain
energy or be composed of several unbound particles.)  The trivial
superselection sector is that of the vacuum; it also contains all excitations
that can be obtained from the vacuum by the action of local operators.

At first sight, each gapped phase in our model has three superselection
sectors: a fermion, a vortex and the vacuum. However, we will see that there
are actually \emph{two} types of vortices that live on different subsets of
plaquettes. They have the same energy and other physical characteristics, yet
they belong to different superselection sectors: to transform one type of
vortex into the other one has to create or annihilate a fermion.

To understand the particle types and other algebraic properties of the gapped
phases, we will map our model to an already known one~\cite{Kitaev97}. Let us
focus on the phase $A_{z}$, which occurs when $|J_{x}|+|J_{y}|<|J_{z}|$. Since
we are only interested in discrete characteristics of the phase, we may set
$|J_{x}|,|J_{y}|\ll|J_{z}|$ and apply the perturbation theory.

\subsection{Perturbation theory study}

The Hamiltonian is $H=H_{0}+V$, where $H_{0}$ is the main part and $V$ is the
perturbation:
\[
H_{0} = -J_{z} \sum_{\text{$z$-links}} \sigma_{j}^{z}\sigma_{k}^{z},
\qquad\quad
V = -J_{x} \sum_{\text{$x$-links}} \sigma_{j}^{x}\sigma_{k}^{x}
-J_{y} \sum_{\text{$y$-links}} \sigma_{j}^{y}\sigma_{k}^{y}
\]
Let us assume that $J_{z}>0$ (the opposite case is studied analogously).

\begin{figure}
\centerline{\begin{tabular}{c@{\qquad\quad}c@{\qquad\quad}c}
$\figbox{1}{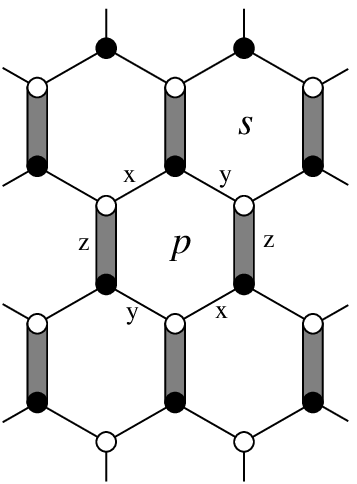}$ & $\figbox{1}{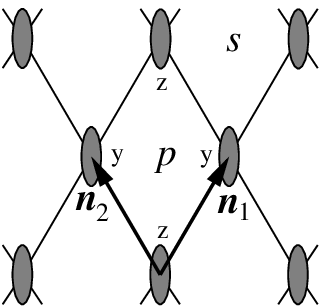}$ &
$\figbox{1}{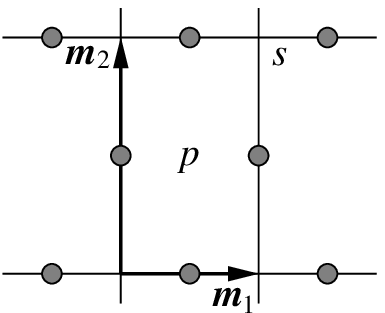}$\\
a) & b) & c)
\end{tabular}}
\caption{Reduction of the model. Strong links in the original model (a) become
effective spins (b), which are associated with the links of a new lattice
(c).}
\label{fig_perturb}
\end{figure}

We first set $J_{x}=J_{y}=0$ and find the ground state. It is highly
degenerate: each two spins connected by a $z$-link are aligned
($\spinup\spinup$ or $\spindown\spindown$), but their common direction is
not fixed. We regard each such pair as an effective spin.  The transition from
physical spins to effective spins is shown in
Fig.~\ref{fig_perturb}a,b. The ground state energy is $E_{0}=-NJ_{z}$,
where $N$ is the number of unit cells, i.e., half the number of spins.

Our goal is to find an effective Hamiltonian that would act in the space of
effective spins $\calL_{\eff}$. One way to solve the problem is
to choose a basis in $\calL_{\eff}$ and compute the matrix
elements
\[
\langle a|H_{\eff}^{(1)}|b\rangle
=\langle a|V|b\rangle,\quad\
\langle a|H_{\eff}^{(2)}|b\rangle
=\mathop{{\sum}'}_{j}\frac{\langle a|V|j\rangle\langle j|V|b\rangle}
{E_{0}-E_{j}},\quad \text{etc.}
\]
However, we will use the more general Green function formalism. 

Let $\Upsilon:\calL_{\eff}\to\calL$ be the embedding that maps the effective
Hilbert space onto the ground subspace of $H_{0}$. The map $\Upsilon$ simply
doubles each spin: $\Upsilon|m\rangle=|mm\rangle$, where $m=\spinup$ or
$m=\spindown$. The eigenvalues of the ``effective Hamiltonian'' (if one
exists) are supposed to be the energy levels of $H$ that originate from ground
states of $H_{0}$. These levels can be unambiguously defined as poles of the
\emph{Green function} $G(E)=\Upsilon^{\dag}(E-H)^{-1}\Upsilon$, which is an
operator acting on $\calL_{\eff}$ and depending on the parameter $E$. The
Green function is conventionally expressed as
$\bigl(E-E_{0}-\Sigma(E)\bigr)^{-1}$, where $\Sigma(E)$ is called
\emph{self-energy}, so the energy levels in question are the values of $E$ for
which the operator $E-E_{0}-\Sigma(E)$ is degenerate. Neglecting the the
dependence of $\Sigma(E)$ on $E$ (for $E\approx E_{0}$), we define the
effective Hamiltonian as $H_{\eff}=E_{0}+\Sigma(E_{0})$.

The self-energy is computed by the standard method. Let
$G_{0}'(E)={\bigl((E-H_{0})^{-1}\bigr)'}$ be the unperturbed Green function
for exited states of $H_{0}$. The ${}'$ sign indicates that the operator
$\bigl((E-H_{0})^{-1}\bigr)'$ acts on excited states in the natural way but
vanishes on ground states. Then
\begin{equation} \label{self-energy}
\Sigma(E)= \Upsilon^{\dag}
\Bigl(V+VG_{0}'(E)V+VG_{0}'(E)VG_{0}'(E)V+\cdots\Bigr) \Upsilon,
\end{equation}
We set $E=E_{0}$ and compute $H_{\eff}=E_{0}+\Sigma$ in the zeroth order
($H_{\eff}^{(0)}=E_{0}$), first order, second order, and on, until we find a
nonconstant term.\footnote{Higher order terms may be less significant than the
dependence of $\Sigma(E)$ on $E$ in the first nonconstant term.} The
calculation follows.

\begin{enumerate}
\item\, $H_{\eff}^{(1)}=\Upsilon^{\dag}V\Upsilon=0$.
\item\, $\displaystyle H_{\eff}^{(2)} =\Upsilon^{\dag}VG_{0}'V\Upsilon
=-\!\!\sum_{\text{$x$-links}}\!\frac{J_{x}^{2}}{4J_{z}}
-\!\!\sum_{\text{$y$-links}}\!\frac{J_{y}^{2}}{4J_{z}}
=-N\frac{J_{x}^{2}+J_{y}^{2}}{4J_{z}}$. Indeed, consider the action of the
second $V$ in the expression $\Upsilon^{\dag}VG_{0}'V\Upsilon$. Each term
$\sigma_{j}^{x}\sigma_{k}^{x}$ or $\sigma_{j}^{y}\sigma_{k}^{y}$ flips two
spins, increasing the energy by $4J_{z}$. The other $V$ must flip them back.
\item\, $H_{\eff}^{(3)}=\Upsilon^{\dag}VG_{0}'VG_{0}'V\Upsilon=0$.
\item\, $\displaystyle H_{\eff}^{(4)}
=\Upsilon^{\dag}VG_{0}'VG_{0}'VG_{0}'V\Upsilon =\mathrm{const}
-\frac{J_{x}^{2}J_{y}^{2}}{16J_{z}^{3}}\sum_{p}Q_{p}$,\, where
$Q_{p}=(W_{p})_{\eff}$ is the effective spin representation of the
operator~(\ref{invariant}). The factor $\frac{1}{16}$ is obtained by summing
$24$ terms, each of which corresponds to flipping $4$ spin pairs in a
particular order:
\[
\frac{1}{16}=8\cdot\frac{1}{64}+8\cdot\frac{-1}{64}+8\cdot\frac{1}{128}.
\]
\end{enumerate}

The above arguments can easily be adapted to the case $J_{z}<0$. Now we have
$\Upsilon:\ |\spinup\rangle\mapsto|\spinup\spindown\rangle,\:\
|\spindown\rangle\mapsto|\spindown\spinup\rangle$. The result turns out
to be the same, with $J_{z}$ replaced by $|J_{z}|$.

Thus the effective Hamiltonian has the form
\begin{equation} \label{HAz}
H_{\eff}=-\frac{J_{x}^{2}J_{y}^{2}}{16|J_{z}|^{3}}\sum_{p}Q_{p},\qquad
Q_{p} = \sigma_{\text{left}(p)}^{y} \sigma_{\text{right}(p)}^{y}
\sigma_{\text{up}(p)}^{z} \sigma_{\text{down}(p)}^{z}
\end{equation}
(the geometric arrangement of the spins corresponds to
Fig.~\ref{fig_perturb}b).

\subsection{Abelian anyons} \label{sec_Abelian}

Hamiltonian~(\ref{HAz}) already lends itself to direct analysis. However, let
us first reduce it to the more familiar form~(\ref{torcode}). We construct a
new lattice $\Lambda'$ so that the effective spins lie on its links (see
Fig.~\ref{fig_perturb}c). This is a sublattice of index $2$ in the original
lattice $\Lambda$ (here ``lattice'' means ``translational group''). The basis
vectors of $\Lambda'$ are $\mathbf{m}_{1}=\mathbf{n}_{1}-\mathbf{n}_{2}$ and
$\mathbf{m}_{2}=\mathbf{n}_{1}+\mathbf{n}_{2}$. The plaquettes of the
effective spin lattice become plaquettes and vertices of the new lattice, so
the Hamiltonian can be written as follows:
\[
H_{\eff} \,=\, -J_{\eff} 
\left(  \sum_{\text{vertices}}Q_{s} + \sum_{\text{plaquettes}}\!Q_{p} \right),
\]
where $J_{\eff}=J_{x}^{2}J_{y}^{2}/(16|J_{z}|^{3})$.

Now we apply the unitary transformation
\begin{equation} \label{transbreak}
U\,=\!\prod_{\text{horizontal links}}\!\!X_{j}\,
\prod_{\text{vertical links}}\!\!Y_{k}
\end{equation}
for suitably chosen spin rotations $X$ and $Y$ so that the Hamiltonian
becomes
\begin{equation}
H_{\eff}' \,=\, UH_{\eff}U^{\dag} \,=\, -J_{\eff}
\left( \sum_{\text{vertices}}A_{s} + \sum_{\text{plaquettes}}B_{p} \right),
\end{equation}
where $A_{s}$ and $B_{p}$ are defined in Eq.~(\ref{torcode}).
(Caution: transformation~(\ref{transbreak}) breaks the translational symmetry
of the original model.)

The last Hamiltonian has been studied in detail~\cite{Kitaev97}. Its key
properties are that all the terms $A_{s}$, $B_{p}$ commute, and that the
ground state minimizes each term separately. Thus the ground state satisfies
these conditions:
\begin{equation}
A_{s}|\Psi\rangle=+|\Psi\rangle,\qquad B_{p}|\Psi\rangle=+|\Psi\rangle.
\end{equation}
Excited states can be obtained by replacing the $+$ sign to a $-$ sign for a
few vertices and plaquettes. Those vertices and plaquettes are the locations
of anyons. We call them ``electric charges'' and ``magnetic vortices'', or
$e$-particles and $m$-particles, respectively. When an $e$-particle moves
around a $m$-particle, the overall state of the system is multiplied by
$-1$. This property is stable with respect to small local perturbations of the
Hamiltonian. (A local operator is a sum of terms each of which acts on a small
number of neighboring spins.) It is also a robust property that the number of
particles of each type is conserved \latin{modulo}~$2$.

The model has $4$ superselection sectors: $1$ (the vacuum), $e$, $m$, and
$\eps=e\times m$. The latter expression denotes a composite object consisting
of an ``electric charge'' and a ``magnetic vortex''. These are the fusion
rules:
\begin{equation} \label{Abfusion}
\begin{array}{c}
e\times e
=m\times m
=\eps\times\eps=1,
\medskip\\
e\times m=\eps, \quad
e\times\eps=m, \quad
m\times\eps=e.
\end{array}
\end{equation}
(In general, fusion rules must be supplemented by associativity relations, or
$6j$-symbols, but they are trivial in our case.)

Let us discuss the braiding rules. One special case has been mentioned: moving
an $e$-particle around an $m$-particle yields $-1$. This fact can be
represented pictorially:
\begin{equation}
\figbox{1}{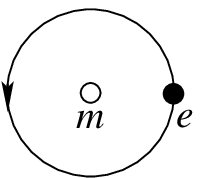}\ =\ -1\ ,\qquad \text{or} \qquad
\figbox{1}{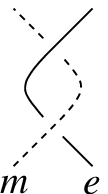}\ =\ -\ \figbox{1}{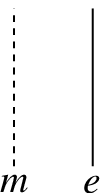}\ .
\end{equation}
The fist diagram shows the ``top view'' of the process. The diagrams in the
second equation correspond to the ``front view'': the ``up'' direction is
time.

It is easy to show that $e$-particles are bosons with respect to themselves
(though they clearly do not behave like bosons with respect to $m$-particles);
$m$-particles are also bosons. However $\eps$-particles are fermions. To see
this, consider two processes. In the first process two $\eps\eps$-pairs are
created, two of the four $\eps$-objects are exchanged by a $180^{\circ}$
counterclockwise rotation, then the pairs are annihilated. (Each $\eps$-object
is represented by $e$ and $m$, so there are 8 elementary particles involved.)
In the second process the two pairs are annihilated immediately. It does not
matter how exactly we create and annihilate the pairs, but we should do it the
same way in both cases. For example, we may use this definition:
\begin{equation}
\text{creation} =\, \figbox{0.8}{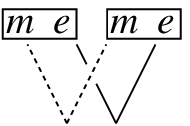}\ , \qquad\
\text{annihilation} =\, \figbox{0.8}{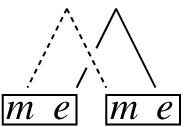}\ . 
\end{equation}
Now we compare the two processes. Each one effects the multiplication of the
ground state by a number, but the two numbers differ by $-1$:
\begin{equation}
\figbox{1}{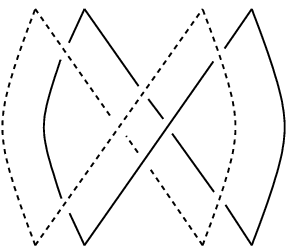}\,\ =\,\ -\ \figbox{1}{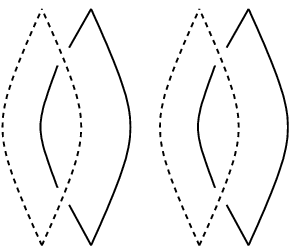}\ .
\end{equation}
Indeed, in the left diagram the dashed line is linked with the solid line.
This corresponds to an $e$-particle going around an $m$ particle, which yields
the minus sign.

Braiding an $\eps$-particle with an $e$- or $m$-particle also gives $-1$.
This completes the description of braiding rules.

\begin{figure}
\centerline{\epsfbox{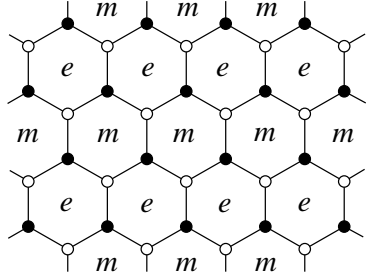}}
\caption{Weak breaking of the translational symmetry: $e$-vortices and
$m$-vortices live on alternating rows of hexagons.}
\label{fig_trbreak}
\end{figure}

It remains to interpret these properties in terms of vortices and fermions in
the original model. Tracing sites and plaquettes of the reduced
model~(\ref{torcode}) back to the original model, we conclude that
$e$-particles and $m$-particles are vortices that live on alternating rows of
hexagons, see Fig.~\ref{fig_trbreak}.  Note that the two types of vortices
have the same energy and other physical properties, yet they cannot be
transformed one to another without creating or absorbing a fermion. A general
view on this kind of phenomenon is given in Appendix~\ref{sec_symbreak}.

The fermions in the original model belong to the superselection sector $\eps$,
although they are not composed of $e$ and $m$. In the perturbation-theoretic
limit, the energy of a fermion is about $2|J_{z}|$ whereas an $em$-pair has
energy $4J_{\eff}\ll|J_{z}|$. The fermions are stable due to the conservation
of $W_{p}$ (and also due to the conservation of energy). However, they will
decay into $e$ and $m$ if we let the spins interact with a zero-temperature
bath, i.e., another system that can absorb the energy released in the decay.

\section{Phase $B$ acquires a gap in the presence of magnetic field}

\subsection{The conic singularity and the time-reversal symmetry}

Phase $B$ (cf.\ Fig.~\ref{fig_phdiag}) carries gapped vortices and gapless
fermions. Note that vortices in this phase do not have well-defined
statistics, i.e., the effect of transporting one vortex around the other
depends on details of the process. Indeed, a pair of vortices separated by
distance $L$ is strongly coupled to fermionic modes near the singularity of
the spectrum, $|\qq-\qq_{*}|\sim L^{-1}$. This coupling results in effective
interaction between the vortices that is proportional to $\eps(\qq)\sim
L^{-1}$ and oscillates with characteristic wavevector $2\qq_{*}$. When one
vortex moves around the other, the quantum state picks up a nonuniversal
phase $\ph\sim L^{-1}t$, where $t$ is the duration of the process. Since the
vortex velocity $v=L/t$ must be small to ensure adiabaticity (or, at least, to
prevent the emission of fermionic pairs), the nonuniversal phase $\ph$ is
always large.

The conic singularity in the spectrum is, in fact, a robust feature that is
related to time-reversal symmetry. As discussed in the end of
Section~\ref{sec_spinferm}, this symmetry is not broken by fixing the gauge
sector (i.e., the variables $w_{p}$) since the honeycomb lattice is
bipartite. Let us show that a small perturbation commuting with the
time-reversal operator $T$ can not open a spectral gap.

We may perform the perturbation theory expansion relative to the gauge sector
of the ground state (i.e., any term that changes the field configuration is
taken into account in the second and higher orders). This procedure yields an
effective Hamiltonian which acts in the fixed gauge sector and therefore can
be represented in terms of $c_{j}$ and $\hat{u}_{jk}$. The operator $T$
defined by~(\ref{timerev}) changes the sign of
$\hat{u}_{jk}=ib^{\alpha_{jk}}_{j}b^{\alpha_{jk}}_{k}$, but this change can be
compensated by a gauge transformation. Thus we get a physically equivalent
operator $T'$ such that
\[
T'\hat{u}_{jk}(T')^{-1}=\hat{u}_{jk},\qquad\quad
T'c_{j}(T')^{-1}=
\left\{\begin{array}{c@{\ \text{if}\ }l}
c_{j} & j\in\text{even sublattice},
\\
-c_{j} & j\in\text{odd sublattice}.
\end{array}\right.
\]
A $T'$-invariant perturbation to the fermionic Hamiltonian~(\ref{Hvf})
(corresponding to a fixed gauge) can not contain terms like $ic_{j}c_{k}$,
where $j$ and $k$ belong to the same sublattice. Thus the perturbed matrix
$\widetilde{A}(\qq)$ in Eq.~(\ref{spectrum}) still has zeros on the diagonal,
though the exact form of the function $f(\qq)$ may be different. However, a
zero of a complex-valued function in two real variables is a topological
feature, therefore it survives the perturbation.

\subsection{Derivation of an effective Hamiltonian}

What if the perturbation does not respect the time-reversal symmetry? We will
now show that the simplest perturbation of this kind,
\begin{equation} \label{magperturb}
V = -\sum_{j}
\bigl( h_{x}\sigma^{x}_{j}+h_{y}\sigma^{y}_{j}+h_{z}\sigma^{z}_{j} \bigl),
\end{equation}
does open a spectral gap. (Physically, the vector
$\mathbf{h}=(h_{x},h_{y},h_{z})$ is an external magnetic field acting on all
spins.) For simplicity we will assume that $J_{x}=J_{y}=J_{z}=J$.

Let us use the perturbation theory to construct an effective Hamiltonian
$H_{\eff}$ acting on the vortex-free sector. One can easily see that
$H_{\eff}^{(1)}=0$. Although the second-order term $H_{\eff}^{(2)}$ does not
vanish, it preserves the time-reversal symmetry. Therefore we must consider
the third-order term, which can be written as follows:
\[
H_{\eff}^{(3)}= \Pi_{0}VG_{0}'(E_{0})VG_{0}'(E_{0})V\Pi_{0},
\]
where $\Pi_{0}$ is the projector onto the vortex-free sector, and $G_{0}'$ is
the unperturbed Green function with the vortex-free sector excluded. In
principle, the Green function can be computed for each gauge sector using the
formula $G_{0}(E)=-i\int_{0}^{\infty}e^{i(E-H_{0}+i\delta)t}\,dt$ (where
$\delta$ is an infinitely small number). For fixed values of the field
variables $u_{jk}$ the unperturbed Hamiltonian may be represented in the
form~(\ref{quadH}) and exponentiated implicitly by exponentiating the
corresponding matrix $A$; the final result may be written as a normal-ordered
expansion up to the second order. However, it is a rather difficult
calculation, so we will use a qualitative argument instead.

Let us assume that all intermediate states involved in the calculation have
energy $\Delta E\sim|J|$ above the ground state. (Actually, $\Delta E\approx
0.27|J|$ for the lowest energy state with two adjacent vortices, see
Appendix~\ref{sec_numerics}.) Then $G_{0}'(E_{0})$ can be replaced by
$-(1-\Pi_{0})/|J|$. The effective Hamiltonian becomes
\begin{equation}
H_{\eff}^{(3)}\,\sim\, -\frac{h_{x}h_{y}h_{z}}{J^{2}}
\sum_{j,k,l} \sigma^{x}_{j}\sigma^{y}_{k}\sigma^{z}_{l},
\end{equation}
where the summation takes place over spin triples arranged as follows:
\begin{equation}
\text{a)}\ \begin{array}{c} \epsfbox{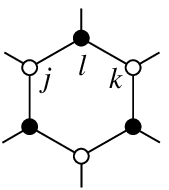} \end{array}
\ \text{(and symmetry-equivalent),}
\qquad \text{or} \qquad
\text{b)}\ \begin{array}{c} \epsfbox{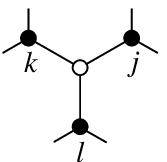} \end{array} \ .
\end{equation}
Configuration (a) corresponds to the term
$\sigma^{x}_{j}\sigma^{y}_{k}\sigma^{z}_{l}
=-iD_{l}\hat{u}_{jl}\hat{u}_{kl}c_{j}c_{k}$ (where $D_{l}$ may be omitted as
we work in the physical subspace), or simply $-ic_{j}c_{k}$ in the standard
gauge. Configuration (b) corresponds to a four-fermion term and therefore does
not directly influence the spectrum. Thus we arrive at this effective
Hamiltonian:
\begin{equation} \label{Hmageff}
\begin{array}{c} \epsfbox{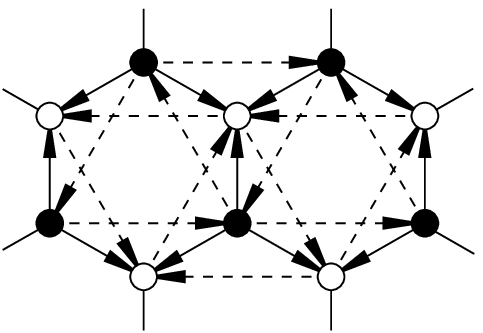} \end{array} \qquad\qquad
\begin{array}{l}
\displaystyle
H_{\eff} = \frac{i}{4}\sum_{j,k} A_{jk}c_{j}c_{k},
\medskip\\
\displaystyle
A=2J\,(\SolidLeftArrow)\,+\,2\kappa\,(\DashedLeftArrow), \qquad
\medskip\\
\displaystyle
\kappa\sim\frac{h_{x}h_{y}h_{z}}{J^{2}}.
\end{array}
\end{equation}
Here $(\SolidLeftArrow)$ is just another notation for $u^{\std}$, i.e., the
matrix whose entry $(\SolidLeftArrow)_{jk}$ is equal to $1$ if there is a
solid arrow from $k$ to $j$ in the figure, $-1$ if an arrow goes from $j$ to
$k$, and $0$ otherwise. $(\DashedLeftArrow)$ is defined similarly.

\subsection{The spectrum and the Chern number}

The fermionic spectrum $\eps(\qq)$ of the Hamiltonian~(\ref{Hmageff}) is given
by the eigenvalues of a modified matrix $i\widetilde{A}(\qq)$ (cf.\
Eq.~(\ref{spectrum})):
\begin{equation} \label{spectrum1}
i\widetilde{A}(\qq)=\left(\begin{array}{@{}c@{\;\,}c@{}}
\Delta(\qq) & i f(\qq)\\
-i f(\qq)^{*} & -\Delta(\qq)
\end{array}\right),
\qquad\quad \eps(\qq)=\pm\sqrt{|f(\qq)|^{2}+\Delta(\qq)^{2}},
\end{equation}
where $f(\qq)=2J\bigl(e^{i(\qq,\nn_{1})}+e^{i(\qq,\nn_{2})}+1\bigr)$ and
$\Delta(\qq)=4\kappa\,\bigl(\sin(\qq,\nn_{1}) +\sin(\qq,-\nn_{2})
+\sin(\qq,\nn_{2}{-}\nn_{1})\bigr)$. Actually, the exact form of the
function $\Delta(\qq)$ does not matter; the important parameter is
\begin{equation} \label{maggap}
\Delta = \Delta(\qq_{*})= -\Delta(-\qq_{*})=6\sqrt{3}\,\kappa
\sim \frac{h_{x}h_{y}h_{z}}{J^{2}},
\end{equation}
which determines the energy gap. The conic singularities are resolved as
follows:
\begin{equation} \label{cone1}
\begin{array}{c} \epsfbox{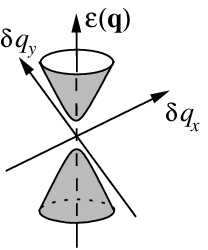} \end{array} \qquad\
\qquad\quad
\begin{array}{c}
\eps(\qq) \approx\ \pm\sqrt{3J^{2}\,|\delta\qq|^{2}+\Delta^{2}},
\bigskip\\
\text{where}\,\ \delta\qq=\qq-\qq_{*}\,\ \text{or}\,\ \delta\qq=\qq+\qq_{*}.
\end{array}
\end{equation}

\begin{Remark}
The magnetic field also gives nontrivial dispersion to vortices.  Indeed, the
operators $W_{p}$ are no longer conserved, therefore a vortex can hop to an
adjacent hexagon. Thus the vortex energy depends on the momentum. This effect
is linear in $h$, but it is not so important as the change in the fermionic
spectrum.
\end{Remark}

Let us also find the fermionic spectral projector, which determines the ground
state. The global spectral projector $P$ is defined by Eq.~(\ref{specproj});
we now consider its Fourier component:
\begin{equation} \label{spectrum1a}
\widetilde{P}(\qq)=\frac{1}{2}\bigl(1-\sgn(i\widetilde{A}(\qq)\bigr)
=\frac{1}{2}
\bigl(1+m_{x}(\qq)\sigma^{x} +m_{y}(\qq)\sigma^{y}+m_{z}(\qq)\sigma^{z}\bigr),
\end{equation}
\begin{equation} \label{spectrum1b}
\mm(\qq) \approx
\left\{\begin{array}{l@{\quad}l}
\frac{1}{\sqrt{(\delta q_{x})^{2}+(\delta q_{y})^{2}+\Delta^{2}/(3J^{2})}}
\left(-\delta q_{y},\, -\delta q_{x},\, -\frac{\Delta}{\sqrt{3}J}\right)
& \text{if}\ \qq \approx \qq_{*},
\bigskip\\
\frac{1}{\sqrt{(\delta q_{x})^{2}+(\delta q_{y})^{2}+\Delta^{2}/(3J^{2})}}
\left(-\delta q_{y},\, \delta q_{x},\, \frac{\Delta}{\sqrt{3}J}\right)
& \text{if}\ \qq \approx -\qq_{*}.
\end{array}\right.
\end{equation}
The function $\mm$ maps the torus to the unit sphere. If $\Delta>0$, then this
map has degree $1$. Indeed, the neighborhood of $\qq_{*}$ is mapped onto the
lower hemisphere, the neighborhood of $-\qq_{*}$ is mapped onto the upper
hemisphere; in both cases the orientation is preserved. (The rest of the torus
is mapped onto the equator.) For negative $\Delta$ the map has degree $-1$.

An important topological quantity characterizing a two-dimensional system of
noninteracting (or weakly interacting) fermions with an energy gap is the
\emph{spectral Chern number}. It plays a central role in the theory of the
integer quantum Hall effect~\cite{TKNN82,ASS83,BES-B94}. In our model there is
no analogue of Hall conductivity (because the number of fermions is not
conserved), but the Chern number determines the edge mode chirality and
anyonic properties of vortices (cf.~\cite{ReadGreen00}).

The spectral Chern number is defined as follows. For each value of the
momentum $\qq$ we consider the space $\widetilde{L}(\qq)$ of annihilation
operators, i.e., fermionic modes with negative energy; this is the subspace
the matrix $\widetilde{P}(\qq)$ projects onto. Thus we obtain a complex vector
bundle over the momentum space. (In our case $\widetilde{L}(\qq)$ is a
one-dimensional subspace of $\CC^{2}$, so the bundle is one-dimensional.) The
first Chern number of this bundle is denoted by $\nu$ and can be expressed as
follows~(cf.~\cite{ASS83}):
\begin{equation} \label{nu}
\nu\,=\, \frac{1}{2\pi i}\int\Tr
\bigl(\widetilde{P}\,d\widetilde{P}\wedge d\widetilde{P}\bigr)
\,=\, \frac{1}{2\pi i}\int\Tr
\biggl(\widetilde{P}\biggl(
\frac{\partial\widetilde{P}}{\partial q_{x}}
\frac{\partial\widetilde{P}}{\partial q_{y}} -
\frac{\partial\widetilde{P}}{\partial q_{y}}
\frac{\partial\widetilde{P}}{\partial q_{x}}
\biggl)\biggl)\,dq_{x}\,dq_{y} .
\end{equation}

The Chern number is always an integer. If the spectral projector
$\widetilde{P}(\qq)$ is given by Eqs.~(\ref{spectrum1a}), (\ref{spectrum1b}),
then
\begin{equation}
\nu\,=\, \frac{1}{4\pi}\int
\left(\frac{\partial\mm}{\partial q_{x}}\times
\frac{\partial\mm}{\partial q_{y}},\:\mm\right)dq_{x}\,dq_{y}\,=\,
\sgn\Delta\,=\,\pm 1.
\end{equation}
We will use the notation $B_{\nu}$ (where $\nu=\pm 1$) to designate phase $B$
in the magnetic field. In the Abelian phases $A_{x}$, $A_{y}$, $A_{z}$ the
Chern number is zero.

\section{Edge modes and thermal transport}

\begin{figure}
\centerline{\begin{tabular}{c@{\qquad\qquad\qquad}c}
$\figbox{0.8}{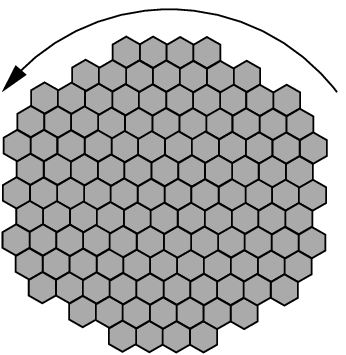}$ & $\figbox{0.8}{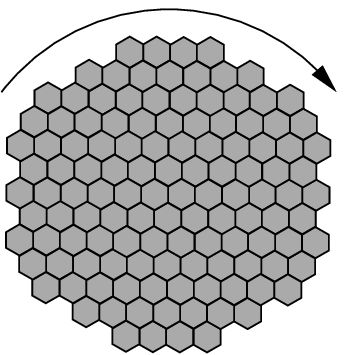}$
\\
a) & b)
\end{tabular}}
\caption{Chiral edge modes: left-moving (a) and right-moving (b).} 
\label{fig_lrwave}
\end{figure}

Remarkably, any system with nonzero Chern number possesses gapless edge modes.
Such modes were first discovered in the integer quantum Hall
effect~\cite{Halperin83}; they are \emph{chiral}, i.e., propagate only in one
direction (see Fig.~\ref{fig_lrwave}). In fact, left-moving and right-moving
modes may coexist, but the following relation holds~\cite{Hatsugai93}:
\begin{equation} \label{numedgemod}
\nu_{\text{edge}}\;\bydef\
\bigl(\text{\# of left-movers} \,-\, \text{\# of right-movers}\bigr)
\ =\; \nu.
\end{equation}
In the absence of special symmetry, counterpropagating modes usually cancel
each other, so the surviving modes have the same chirality.  A calculation of
the edge spectrum in phases $B_{\nu}$ (for some specific boundary conditions)
and a simple proof of Eq.~(\ref{numedgemod}) are given in
Appendix~\ref{sec_edge}. More rigorous and general results, which even apply
to disordered systems, can be found in Refs.~\cite{S-BKR00,KRS-B02}.

It is important to note that the analogy to the quantum Hall effect is not
exact. In our model (like in two-dimensional superfluid and superconducting
systems~\cite{Volovik_edge,ReadGreen00}) edge modes are described by real
fermions, as contrasted to complex fermions in the quantum Hall
effect. Therefore each quantum Hall mode is equivalent to two modes in our
system.

Chiral edge modes can carry energy, leading to potentially measurable thermal
transport. (The temperature $T$ is assumed to be much smaller than the energy
gap in the bulk, so that the effect of bulk excitations is negligible.) For
quantum Hall systems, this phenomenon was discussed in
Refs.~\cite{KaneFisher97,CHZ01}. The energy current along the edge in the left
(counterclockwise) direction is given by the following formula:
\begin{equation}\label{heatcurrent}
I=\frac{\pi}{12}c_{-}T^{2},
\end{equation}
where $c_{-}$ is some real number. (The factor $\pi/12$ is introduced to make
a connection to conformal field theory, see below.) It is remarkable that
$c_{-}$ does not depend on particular conditions at the edge, but rather on
the bulk state. Indeed, the energy current is conserved, therefore it remains
constant even if some conditions change along the edge. The effect is
invariant with respect to time rescaling. Since the energy current has
dimension $(\mathrm{time})^{-2}$, it must be proportional to $T^{2}$. But the
value of the dimensionless proportionality coefficient cannot be found using
such simple arguments.

There are two standard ways to calculate the coefficient $c_{-}$. They both
rely on certain assumptions but can be applied to our model, yielding this
result:
\begin{equation}\label{anomnu}
c_{-}=\frac{\nu}{2}.
\end{equation}
The first argument~\cite{SivanImry86} (adapted to real fermions) assumes
translational invariance and the absence of interaction. Each edge mode is
described by a free fermion with an energy spectrum $\eps(q)$ such that
$\eps(-q)=-\eps(q)$ and $\eps(q)\to\pm\infty$ as $q\to\pm\infty$. The signs in
the last two expressions agree if the mode propagates in the direction of
positive $q$ (for simplicity we may assume that $\eps(q)>0$ when $q$ is
positive and $\eps(q)<0$ when $q$ is negative). Thus the Hamiltonian has the
form
\[
H\,=\,\frac{1}{2}\sum_{q}\eps(q)a_{-q}a_{q}
\,=\,\sum_{q:\,\eps(q)>0}\eps(q)a_{-q}a_{q}.
\]
If $\eps(q)>0$, then $a_{q}$ is an annihilation operator and
$a_{-q}=a_{q}^{\dag}$ is the corresponding creation operator. The mode
propagates with group velocity $v(q)=d\eps/dq$, and the occupation number
$n(q)$ is given by the Fermi distribution. The energy flow due to each mode
propagating in the positive direction can be calculated as follows:
\[
I_{1}\,=\,\int_{\eps(q)>0}n(q)\,\eps(q)\,v(q)\,\frac{dq}{2\pi}\,=\,
\int_{\eps(q)>0}\frac{\eps(q)}{1+e^{\eps(q)/T}}\,
\frac{d\eps}{dq}\,\frac{dq}{2\pi}
\,=\, \frac{1}{2\pi}\int_{0}^{\infty}\frac{\eps\,d\eps}{1+e^{\eps/T}}
\,=\,\frac{\pi}{24}T^{2}.
\]
Each mode propagating in the opposite direction contributes $-I_{1}$,
therefore $I=\frac{\pi}{24}\nu T^{2}$.

The second derivation~\cite{CHZ01} is based on the assumption that the edge
modes can be described by a conformal field theory (CFT). In this case,
\begin{equation}
c_{-}=c-\overline{c},
\end{equation}
where $c$ and $\overline{c}$ are the Virasoro central charges. Thus, $c_{-}$
is called the \emph{chiral central charge}. Left-moving fermions have
$(c,\overline{c})=\bigl(\frac{1}{2},0\bigr)$ whereas right-moving fermions
have $(c,\overline{c})=\bigl(0,\frac{1}{2}\bigr)$, which implies
Eq.~(\ref{anomnu}). More generally, $c$ and $\overline{c}$ are some rational
numbers, and so is $c_{-}$. The chiral central charge parametrizes a
two-dimensional gravitational anomaly~\cite{gravanomaly} of the corresponding
CFT; it can also be identified with the coefficient in front of the
gravitational Chern-Simons action in a three-dimensional theory~\cite{Witten}.
Volovik~\cite{Volovik90} suggested that for $\mathrm{{}^{3}He}$-$\mathrm{A}$
films, the role of gravitational field is played by the order parameter
interacting with fermions. However, it is not obvious how to define a
``gravitational field'' for lattice models.

It remains a bit mysterious how the chiral central charge is related to the
ground state and spin correlators in the bulk. This question is partially
answered in Appendix~\ref{sec_anomaly}, but the obtained expression for
$c_{-}$ is not easy to use, nor can we demonstrate that $c_{-}$ is
rational. Note that there is a beautiful relation between the chiral central
charge and algebraic properties of anyons~\cite{FrohlichGabbiani90,Rehren90},
which does imply the rationality of $c_{-}$. We discuss that relation in
Appendix~\ref{sec_algth} (see Eq.~(\ref{anomaly}) on page~\pageref{anomaly}),
though it is unclear how to deduce it only considering the bulk. The only
known argument is to assume that edge modes are described by a CFT, then one
can use modular invariance~\cite{Verlinde}. In fact, the modular invariance
alone would suffice. In Appendix~\ref{sec_anomaly} we try to derive it from
general principles, but again, encounter a problem.

\section{Non-Abelian anyons} \label{sec_nonabelian}

We continue the study of phase $B$ in the magnetic field. Now that all bulk
excitations are gapped, their braiding rules must be well-defined. Of course,
this is only true if the particles are separated by distances that are much
larger than the correlation length associated with the
spectrum~(\ref{cone1}). The correlation length may be defined as follows:
$\xi=|\Im\qq|^{-1}$, where $\qq$ is a complex solution to the equation
$\eps(\qq)=0$. Thus
\begin{equation}
\xi=\biggl|\frac{\sqrt{3}J}{\Delta}\biggr|
\sim \biggl|\frac{J^{3}}{h_{x}h_{y}h_{z}}\biggr|.
\end{equation}

The braiding rules for vortices depend on the spectral Chern number
$\nu$. Although $\nu$ is actually equal to $+1$ or $-1$ (depending on the
direction of the magnetic field), one may formally consider a model with an
arbitrary gapped fermionic spectrum, in which case $\nu$ may take any integer
value. We will see that vortices behave as non-Abelian anyons for any odd
value of $\nu$, but their exact statistics depends on $\nu\bmod 16$.

The properties of the anyons are summarized in Table~\ref{tab_nonabelian} on
page~\pageref{tab_nonabelian}. The notation and underlying concepts are
explained below; see also Appendix~\ref{sec_algth}. Let us first show a quick
way of deriving those properties from conformal filed theory (CFT) in the most
important case, $\nu=\pm 1$. (For a general reference on CFT, see~\cite{CFT}.)
Then we will give an alternative derivation, which uses only rudimentary CFT
and refers to the operational meaning of braiding and fusion.

\begin{table}[p]
\widefbox{\vskip 10pt
\noindent\textbf{Superselection sectors:}\quad
$1$ (vacuum),\quad $\eps$ (fermion),\quad $\sigma$ (vortex).
\[
\begin{array}{l@{\qquad}l@{\qquad}l@{\qquad}l}
\text{Quantum dimension:} &
d_{1}=1, & d_{\eps}=1 &  d_{\sigma}=\sqrt{2};
\smallskip\\
\text{Topological spin:} &
\theta_{1}=1, & \theta_{\eps}=-1,
& \theta_{\sigma}=\theta=\exp\bigl(\frac{\pi}{8}i\nu\bigr);
\smallskip\\
\text{Frobenius-Schur indicator:} &
\FS_{1}=1, & \FS_{\eps}=1,
& \FS_{\sigma}=\FS=(-1)^{(\nu^{2}-1)/8}.
\bigskip\\
\multicolumn{4}{l}{\text{Global dimension:}\quad 
\calD^{2}\bydef\sum\nolimits_{u}d_{u}^{2}=4.}
\end{array}
\]
\smallskip

\noindent\textbf{Fusion rules:}
$\qquad\quad \eps\times\eps=1,\qquad\quad
\eps\times\sigma=\sigma,\qquad\quad
\sigma\times\sigma=1+\eps.$
\medskip

\noindent\textbf{Associativity relations:}
\[
\figbox{1.0}{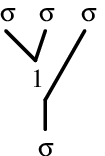} \ =\,\ 
\frac{\FS}{\sqrt{2}}\figbox{1.0}{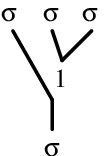} \;+\ 
\frac{\FS}{\sqrt{2}}\figbox{1.0}{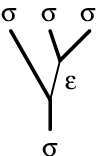}\,,\qquad\qquad
\figbox{1.0}{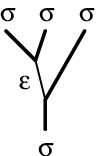} \ =\,\
\frac{\FS}{\sqrt{2}}\figbox{1.0}{ssss3.eps} \;-\ 
\frac{\FS}{\sqrt{2}}\figbox{1.0}{ssss4.eps}\,,
\]
\[
\figbox{1.0}{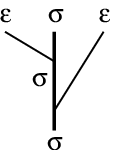}\,=\; -\ \figbox{1.0}{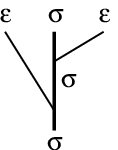}\,,\qquad\qquad
\figbox{1.0}{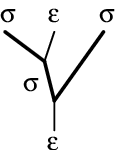}\,=\; -\figbox{1.0}{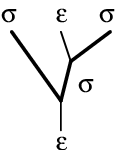}\,,\qquad\qquad
\figbox{1.0}{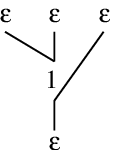}\;=\; \figbox{1.0}{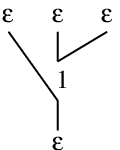}\,,
\]
\[
\figbox{1.0}{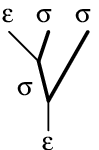}\; =\; \figbox{1.0}{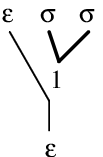}\,,\qquad\quad
\figbox{1.0}{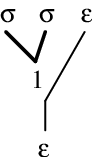}\; =\; \figbox{1.0}{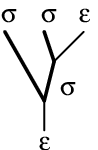}\,,\qquad\quad
\figbox{1.0}{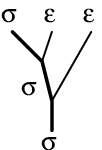}\; =\; \figbox{1.0}{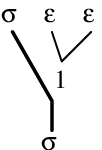}\,,\qquad\quad
\figbox{1.0}{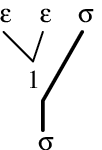}\; =\; \figbox{1.0}{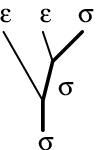}\,,
\]
\[
\figbox{1.0}{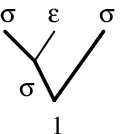}\; =\; \figbox{1.0}{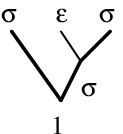}\,,\qquad\qquad
\figbox{1.0}{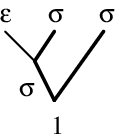}\; =\; \figbox{1.0}{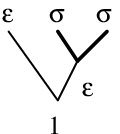}\,,\qquad\qquad
\figbox{1.0}{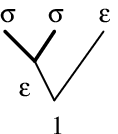}\; =\; \figbox{1.0}{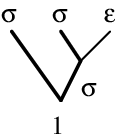}\,.
\]

\noindent\textbf{Braiding rules:}
\[
\hskip-35pt
\begin{array}{ll}
\text{Definition of $R^{xy}_{z}$:}\\[3pt]
\figbox{1.0}{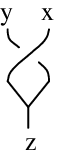}\ =\ R^{xy}_{z}\figbox{1.0}{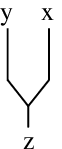}\,;
\end{array}
\qquad\qquad
\begin{array}{l@{\qquad\quad}l}
R^{\eps\eps}_{1}=-1, &
R^{\sigma\sigma}_{1}=\FS\exp\bigl(-\frac{\pi}{8}i\nu\bigr),
\bigskip\\
R^{\eps\sigma}_{\sigma}=R^{\sigma\eps}_{\sigma}=-i^{\nu}, &
R^{\sigma\sigma}_{\eps}=\FS\exp\bigl(\frac{3\pi}{8}i\nu\bigr).
\end{array}
\]

\noindent\textbf{Topological $S$-matrix:}
\[
\bigl(S_{z}\bigr)_{xy}\ \bydef\ \frac{1}{\calD}\ \figbox{1.0}{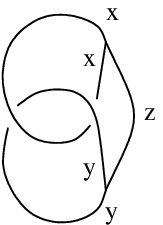}\;;
\qquad\quad
S_{1}\,=\,\left(\begin{array}{ccc}
\frac{1}{2} & \frac{1}{2} & \frac{1}{\sqrt{2}} \smallskip\\
\frac{1}{2} & \frac{1}{2} & -\frac{1}{\sqrt{2}} \smallskip\\
\frac{1}{\sqrt{2}} & -\frac{1}{\sqrt{2}} & 0
\end{array}\right), \qquad
{\textstyle\bigl(S_{\eps}\bigr)_{\sigma\sigma}
=\exp\bigl(-\frac{\pi}{4}i\nu\bigr)}.
\]
\vskip -5pt}
\caption{Algebraic properties of anyons in non-Abelian phases ($\nu$ is odd).}
\label{tab_nonabelian}
\end{table}

\subsection{Bulk-edge correspondence}

\begin{figure}
\centerline{\begin{tabular}{c@{\qquad\qquad\qquad}c}
$\figbox{1.0}{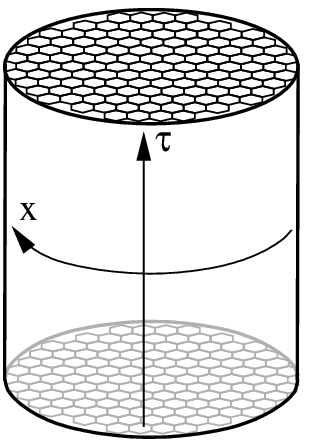}$ & $\figbox{1.0}{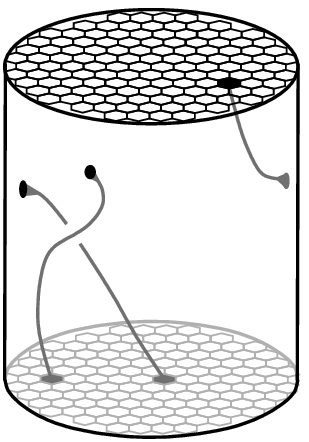}$
\\
a) & b)
\end{tabular}}
\caption{The space-time coordinates (a) and particle worldlines (b).}
\label{fig_cylinder}
\end{figure}

The properties listed in Table~\ref{tab_nonabelian} form the same type of
algebraic structure that was described by Moore and Seiberg~\cite{MooreSeiberg}
in the CFT context.  However, the actual connection to CFT is indirect: anyons
are related to edge modes, which in turn can be described by a field theory in
$1+1$ dimensions. More concretely, the space-time may be represented as a
cylinder (see Fig.~\ref{fig_cylinder}). It is convenient to use the imaginary
time formalism ($t=-i\tau$, where $\tau\in\RR$), so that we have a
two-dimensional Euclidean field theory on the side surface of the
cylinder. The surface may be parametrized by a complex variable $z=\tau+ix$,
where $x$ is the spatial coordinate.

The two-dimensional field theory describes the physics of the edge, which is
generally richer than that of the bulk. The theory possesses both local and
nonlocal fields. The insertion of a nonlocal field $\phi(\tau+ix)$ corresponds
to an anyonic particle emerging on the edge or sinking into the bulk at point
$(\tau,x)$. The correlation function of several nonlocal fields has nontrivial
monodromy which coincides with the anyonic braiding. Specifically, the
counterclockwise exchange of anyons in the bulk is equivalent to moving the
fields counterclockwise on surface (if we look at the cylinder from outside).
Moreover, the value of the correlator is not a number, but rather an operator
transforming the initial anyonic state (on the bottom of the cylinder) into
the final one (on the top of the cylinder). For non-Abelian anyons, the
space of such operators is multidimensional.

The anyon-CFT correspondence has been successfully used in the study of
quantum Hall systems~\cite{MooreRead-1,MooreRead-2,ReadRezayi}. The
correspondence is well-understood if all boundary fields are either
holomorphic or antiholomorphic, which is the case for our model.  We have seen
that the edge carries a left-moving (holomorphic) fermion for $\nu=1$, or a
right-moving (antiholomorphic) fermion for $\nu=-1$. A vortex emerging on the
surface corresponds to a twist field $\sigma$. The correlation functions for
such fields are given by holomorphic (resp.\ antiholomorphic) conformal blocks
for the Ising model.

A partial bulk-edge correspondence can be established at a more elementary
level. Let $\nu=1$. A particularly important parameter of the edge theory is
the conformal weight of the twist field,
$(h_{\sigma},\overline{h}_{\sigma})=\bigl(\frac{1}{16},0\bigr)$. The related
bulk parameter is the topological spin of a vortex, $\theta_{\sigma}$; it
corresponds to the counterclockwise rotation by angle $2\pi$. Since the
rotations in the bulk and on the surface must agree, the following equation
holds for an arbitrary field $a$:
\begin{equation} \label{confspin}
\theta_{a}=e^{2\pi i(h_{a}-\overline{h}_{a})}.
\end{equation}
Thus
$\theta_{\sigma}=e^{i\pi/8}$. Similarly, if $\nu=-1$, then
$\theta_{\sigma}=e^{-i\pi/8}$. In the general case, there are $|\nu|=2n+1$
species of the free fermion; the twist field also comes with some
multiplicity.\footnote{Specifically, there are $2^{n}$ copies of the twist
field, which are transformed according to the fundamental representation of
$\Spin(2n+1)$.}  One may argue that the conformal weight of the twist field is
proportional to the number of fermionic species. Hence
\begin{equation} \label{theta}
\theta_{\sigma}=e^{i\pi\nu/8}.
\end{equation}

\subsection{Unpaired Majorana modes}

We now begin a rather lengthy derivation of the properties listed in
Table~\ref{tab_nonabelian}, dealing only with anyons in the bulk. First, we
give a crude description of vortices by Majorana operators and find the fusion
rules. The braiding rules and so-called associativity relations (also known as
crossing symmetry, or $6j$-symbols) are determined up to several free
parameters; we discuss what combinations of these parameters have invariant
meaning. Solving the so-called pentagon and hexagon equations, we reduce the
continuum of possibilities down to eight consistent theories. The right theory
is selected using Eq.~(\ref{theta}).

The starting point is this: \emph{if $\nu$ is odd, then each vortex carries
an unpaired Majorana mode}. A sufficiently rigorous proof of this statement
can be found in Appendix~\ref{sec_locmatr}. Here we give a rough explanation
based on the quantum Hall analogy.

It is known that the quantized Hall conductivity for noninteracting electrons
(in units of $e^{2}/h$) is equal to the Chern number of the projector onto the
occupied electron states~\cite{ASS83}. The essential difference from our case
is that electrons are ordinary fermions, not Majorana fermions. However, we
can concoct an analogue of an electron system from two copies of the Majorana
system, which may be pictured as two layers. The Hamiltonian is as follows:
\begin{equation}
H_{\text{electron}}=
\frac{i}{4}\sum_{j,k} A_{jk} (c_{j}'c_{k}'+c_{j}''c_{k}'')
=\sum_{j,k} iA_{jk} a_{j}^{\dag}a_{k}, \qquad
\text{where}\ a_{j}=\frac{1}{2}(c_{j}'+ic_{j}'').
\end{equation}
(We forget about the original spin model for the purpose of this
construction.) Note that the Hamiltonian possesses a global $\OO(2)$ symmetry
that consists of orthogonal linear transformations $c_{j}'\mapsto \alpha
c_{j}'+\beta c_{j}''$,\, $c_{j}''\mapsto \gamma c_{j}'+\delta c_{j}''$. The
rotational subgroup $\UU(1)\cong\SO(2)\subset\OO(2)$ corresponds to the
conservation of electric charge, whereas reflections (i.e., transformations
with determinant $-1$) change the sign of the charge.

Assuming that the Fermi energy is zero, the projector onto the occupied
electron states coincides with the spectral projector for the Majorana
system. Thus the Hall conductivity of the electron system is $\nu$.

A vortex piercing both Majorana layers corresponds to half-integer magnetic
flux. Such a vortex carries excessive charge $q=\nu/2+n$, where $n$ is an
arbitrary integer. In particular, if $\nu$ is odd, a state with $q=1/2$
exist. By an $O(2)$ reflection, it is related to a $q=-1/2$ state with the
same energy. The doublet of states with unit charge difference may be
attributed to a zero-energy electron mode that can be empty or occupied. This
mode can be represented by two Majorana modes. Since the layers are
independent, we conclude that each layer contains one zero-energy Majorana
mode.

It has been previously shown that zero-energy Majorana modes exist in some
exotic one-dimensional systems~\cite{qwires} as well as vortices in
two-dimensional $p$-wave superconductors~\cite{ReadGreen00,Ivanov}. A pair of
Majorana modes at two vortices constitutes a full fermionic mode with a
two-dimensional Fock space. The quantum state of such a pair is virtually
inaccessible to measurements or perturbations as long as the vortices stay far
apart from each other. A system of $2n$ vortices possesses a protected space
of dimensionality $2^{n}$ (or $2^{n-1}$, if we require that the whole system
have even fermionic parity).

\subsection{Fusion and braiding rules} \label{sec_fubr}

In our model a vortex carrying an unpaired Majorana mode is just one of the
superselection sectors, denoted by $\sigma$. The other sectors are $1$ (the
vacuum) and $\eps$ (a fermion). If two vortices fuse, they either annihilate
completely or leave a fermion behind: $\sigma\times\sigma=1+\eps$. The actual
fusion outcome depends on the initial quantum state. Hence the protected space
of the vortex pair has two basis vectors: $|\psi_{1}^{\sigma\sigma}\rangle$
and $|\psi_{\eps}^{\sigma\sigma}\rangle$. (The upper indices indicate the
particle types before the fusion whereas the subscript indicates the resulting
particle.) The complete set of fusion rules is as
follows:
\begin{equation}
\eps\times\eps=1,\qquad\quad \eps\times\sigma=\sigma,\qquad\quad
\sigma\times\sigma=1+\eps,
\end{equation}
plus trivial rules of the form $1\times x=x$. Read backwards, these relations
are understood as splitting rules: for example, an $\eps$-particle can split
into two $\sigma$-particles. However, these rules do not capture more subtle
aspects of fusion and splitting, which will be discussed later.

Braiding rules for Majorana half-vortices in a spin-triplet superconductor
have been derived by D.\,Ivanov~\cite{Ivanov}. Here we follow the main idea of
Ivanov's work. We should, however, keep in mind one important difference
between his setting and our model. A spin-triplet superconductor has a locally
measurable vector order parameter, which contributes to vortex-vortex
interaction and can interact with impurities. One vortex making a full turn
around another may pick up a nonuniversal phase, hence the non-Abelian
statistics is defined up to arbitrary phase factors. That is not the case for
our model (or for spinless superconductors~\cite{ReadGreen00}), so additional
arguments are required to find the Abelian part of the vortex statistics.

Once again, we use the fact that each vortex $p$ carries an unpaired Majorana
mode $C_{p}$, which is a linear combination of the operators $c_{j}$ on
neighboring sites. The operators $c_{j}$ do not commute with gauge
transformations and therefore should be used with care. The gauge can be fixed
in a neighborhood of each vortex, so constructing the linear combination is
not a problem. However, the overall sign of $C_{p}$ does not have invariant
meaning. This ambiguity is avoided if we consider fermionic path
operators~(\ref{pathop}), which are gauge-invariant. A suitable linear
combination of elementary paths constitutes a path that begins or ends at a
vortex.

Let us choose some reference path $l_{p}$ connecting each vortex $p=1,2\dots$
to a reference point $0$. We will assume that the vortices lie on the
horizontal axis whereas the reference point $0$ is located in the lower
half-plane. Fixing the gauge along the paths, we may write
\begin{equation}
W(l_{p})=C_{p}c_{0}.
\end{equation}

Let us exchange vortices~$1$ and~$2$ by moving them
\emph{counterclockwise}. The exchange process is described by a unitary
operator $R$ acting on the physical Hilbert space. It also acts on operators
by conjugation: $X\mapsto RXR^{\dag}$.  Clearly,
\begin{equation}
\setbox\boxB\vbox{\hbox{\epsfbox{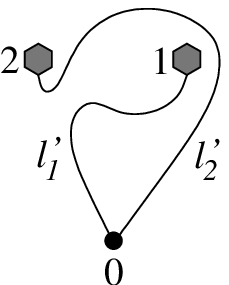}}}\lenB=\ht\boxB
\begin{array}{c} \vbox to \lenB{\vfill\hbox{\epsfbox{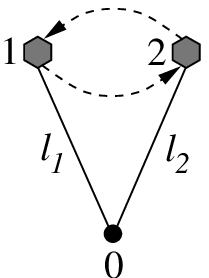}}}
\end{array}
\quad\longmapsto\quad
\begin{array}{c} \vbox to \lenB{\vfill\hbox{\epsfbox{refpath2.eps}}}
\end{array}
\,\ ; \qquad\quad
\begin{array}{r@{}l}
RW(l_{1})R^{\dag} &{}\,=\; W(l_{1}')\,=\,W(l_{2}),
\bigskip\\
RW(l_{2})R^{\dag} &{}\,=\; W(l_{2}')\,=\,-W(l_{1}).
\end{array}
\end{equation}
(We have used the fact that $W(l_{2}')=-W(l_{1})$. Indeed, the paths
$l_{2}'$ and $l_{1}$ differ by a loop enclosing a vortex; transporting
a fermion around the vortex gives rise to the minus sign.) Hence
\begin{equation} \label{RMaj}
\begin{array}{r@{}l}
RC_{1}R^{\dag} &{}\,=\,C_{2},
\smallskip\\
RC_{2}R^{\dag} &{}\,=\,-C_{1},
\end{array}
\qquad\qquad
R=\,\theta\,\exp\Bigl(-\frac{\pi}{4}C_{1}C_{2}\Bigr),
\end{equation}
where $\theta$ is a phase factor. (We will see that $\theta$ is actually the
topological spin of a vortex; for now it is just an unknown parameter).

At first sight, it is not clear whether the number $\theta$ has an invariant
meaning. Indeed, the operator that moves a vortex along a given path may be
defined up to an arbitrary phase. However, the ambiguity can be avoided by a
careful definition of the particle exchange process. The following argument is
completely general; it is not based on adiabaticity or translational
invariance.

\begin{figure}
\[
\figbox{1.0}{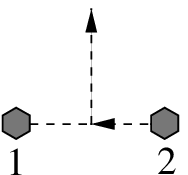}\qquad \longrightarrow\qquad
\figbox{1.0}{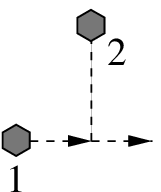}\qquad \longrightarrow\qquad
\figbox{1.0}{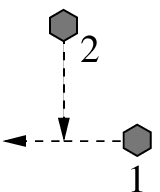}\qquad \longrightarrow\qquad
\figbox{1.0}{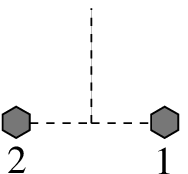}
\]
\caption{A realization of the particle exchange in which all local
contributions cancel, revealing the topological effect in a pure form.}
\label{fig_exch}
\end{figure}

Let us consider a \emph{vortex path operator} $W_{\sigma}(l)$ that is composed
of elementary steps, i.e., displacements of the vortex from hexagon to
hexagon.\footnote{The displacement may be realized as the action of
$\sigma^{\alpha}$ on one of the two spins at the boundary between the hexagons
(which changes the field configuration) followed by an operator of the form
$\exp\bigl(\sum F_{km}c_{k}c_{m}\bigr)$ (which adjusts the fermionic
subsystem).} Each elementary displacement is defined up to a phase, but the
path operator satisfies these equations:
\begin{equation} \label{Wsigma}
W_{\sigma}(\overline{l})=W_{\sigma}(l)^{\dag},\qquad\quad
W_{\sigma}(l_{1}l_{2})=W_{\sigma}(l_{2})W_{\sigma}(l_{1}).
\end{equation}
Here $\overline{l}$ denotes the reverse path, whereas $l_{1}l_{2}$ is the
composite path (the vortex goes along $l_{1}$ and then along $l_{2}$).
Remarkably, two vortices can be exchanged in such a way that the arbitrary
phases cancel each other, see Fig.~\ref{fig_exch}.
The vortices move along the indicated lines; each line segment is passed by a
vortex in both directions.

As mentioned above, the vortex pair has two states corresponding to the
possible fusion outcomes, $|\psi_{1}^{\sigma\sigma}\rangle$ and
$|\psi_{\eps}^{\sigma\sigma}\rangle$. They should be identified with the
eigenvectors of $C_{1}C_{2}$, but we do not know which is which. We can only
write:
\begin{equation} \label{defalpha}
C_{1}C_{2}|\psi_{1}^{\sigma\sigma}\rangle
=i\alpha|\psi_{1}^{\sigma\sigma}\rangle,\qquad\quad
C_{1}C_{2}|\psi_{\eps}^{\sigma\sigma}\rangle
=-i\alpha|\psi_{\eps}^{\sigma\sigma}\rangle,
\end{equation}
where $\alpha=\pm 1$ is unknown. Thus the braiding operator $R$ acts as
follows:
\begin{equation} \label{Rinv}
\begin{array}{r@{}l}
R|\psi_{1}^{\sigma\sigma}\rangle
&{}\,=\,R_{1}^{\sigma\sigma}|\psi_{1}^{\sigma\sigma}\rangle,
\medskip\\
R|\psi_{\eps}^{\sigma\sigma}\rangle
&{}\,=\,R_{\eps}^{\sigma\sigma}|\psi_{\eps}^{\sigma\sigma}\rangle,
\end{array}
\qquad\quad \text{where}\qquad
\begin{array}{r@{}l}
R_{1}^{\sigma\sigma} &{}\,=\, \theta\,e^{-i\alpha\pi/4},
\medskip\\
R_{\eps}^{\sigma\sigma} &{}\,=\, \theta\,e^{i\alpha\pi/4}.
\end{array}
\end{equation}

\begin{Remark}
One may wonder why formula~(\ref{RMaj}) contains the minus sign in front of
$C_{1}$, but not in front of $C_{2}$. This is, in fact, a consequence of the
convention that the reference point $0$ lies in the lower half-plane. If we
move it to the upper half-plane, the signs will change, $\alpha$ will turn
into $-\alpha$, and the signs in front of $\alpha$ in Eq.~(\ref{Rinv}) will
change too. The numbers $R_{1}^{\sigma\sigma}$, $R_{\eps}^{\sigma\sigma}$ and
$\theta$ are invariant though.
\end{Remark}

\subsection{Associativity relations}\label{sec_assoc}

The fusion rules only indicate what fusion and splitting events are possible.
Nontrivial relations arise if we consider a sequence of such events.

\subsubsection{Relations based on the anticommutativity of fermionic operators}
Let us consider the splitting of a $\sigma$-particle into $\eps$, $\sigma$,
and $\eps$. This can be done in two different ways, depending on the order the
two $\eps$-particles are created. Let us suppose that the $\sigma$-particle is
located at point~$2$ in the middle; the $\eps$-particles will appear at
points~$1$ and~$3$ on the left and on the right, respectively. The particles
(or their future locations) are connected by the paths $l_{12}$ and $l_{23}$.

The $\sigma$-particle is described by the associated Majorana mode
$C_{2}$. The $\eps$-particles can be created from the vacuum by operators
$a_{1}^{\dag}$ and $a_{3}^{\dag}$. However, these operators are not physical
by themselves. To actually split the first $\eps$-particle off the
$\sigma$-particle, one needs to apply a fermionic path operator $W(l_{12})$,
which equals $a_{1}^{\dag}C_{2}$ in a suitable gauge. The second
$\eps$-particle is created by the operator
$W(l_{23})=a_{3}^{\dag}C_{2}$. Since $W(l_{12})W(l_{23})=-W(l_{23})W(l_{12})$,
we obtain the following \emph{associativity relation} between the two
splitting processes:
\begin{equation} \label{as_sese}
\figbox{1.0}{sese1.eps}\ =\ -\ \figbox{1.0}{sese2.eps}\, .
\end{equation}

Similarly, we can consider the splitting of one $\eps$-particle into three.
For this, we use two conventions:\footnote{Although these conventions are
somewhat arbitrary, the result is invariant (see
subsection~\ref{sec_fusgen}).} (i) moving an $\eps$-particle from place $j$ to
place $k$ is described by the operator $a_{k}^{\dag}a_{j}$; (ii) creating a
fermionic pair from the vacuum is described by $a_{j}^{\dag}a_{k}^{\dag}$,
where $j$ is located left of $k$. Thus we get:
\begin{equation}  \label{as_eeee}
\figbox{1.0}{eeee1.eps}\ =\ \figbox{1.0}{eeee2.eps}
\end{equation}
because $(a_{1}^{\dag}a_{2}^{\dag})(a_{3}^{\dag}a_{2})
=(a_{2}^{\dag}a_{3}^{\dag})(a_{1}^{\dag}a_{2})$.  (In these graphs the ``up''
direction is time. The intermediate state of the middle particle is shown in
all cases, but lines corresponding to label $1$, i.e., the vacuum
superselection sector, are suppressed.)

\subsubsection{Some generalities}\label{sec_fusgen} The analysis of other
splitting processes (e.g., $\sigma\to\sigma\sigma\sigma$) is more
complicated. Ideally, we would use the vortex path operator, but its exact
form is unknown. Therefore we have to rely on more abstract arguments. Before
we proceed, let us clarify some points pertaining to anyons in general.

\begin{enumerate}
\item Speaking about anyons, we are interested in particle types (i.e.,
superselection sectors) and topology of the particle worldlines (braids,
trees, etc.). An actual particle is also characterized by position and local
degrees of freedom; for example, an $\eps$-particle may be ``elementary'' or
consist of two adjacent vortices. We generally ignore such details though.

\item In the study of splitting and fusion, we consider particles located on
the horizontal axis. In this case, we only care about the order of the
particles on the line, but not about their positions. All configurations with
the same particle types and order can be identified with each other. Unlike in
the two-dimensional case, cycling through several configurations has trivial
effect on the quantum state.\footnote{This claim can be justified as
follows. To move a particle $z$ from one place to another, we apply a path
operator $W_{z}(r,r')$, which acts on the spins in some neighborhood of the
interval $[r,r']$. These operators are defined in such a way that it does not
matter whether the particle is moved at once or in several steps (cf.\
Eq.~(\ref{Wsigma})). When several particles are present, we do not let them
pass through each other, which imposes a certain restriction on the sequence
of operators applied. All such sequences are actually equivalent since the
path operators for nonoverlapping intervals commute.}

\item ``Quantum state of a particle'' is a rather subtle notion. It may be
understood as a projector that enforces certain spin correlations in some
neighborhood of the supposed particle location. (The neighborhood radius must
be much larger than the correlation length). Such an object has no ``overall
phase''. Superpositions of states from different superselection sectors cannot
be constructed either.

\item In spite of that, the state $|\psi^{\sigma\sigma}_{1}\rangle$ has a
well-defined phase if we consider it relative to the vacuum. A similar
argument holds for an arbitrary state $|\psi^{xy}_{z}\rangle$, which describes
particles $x$ and $y$ with total ``anyonic charge'' $z$. For a general anyonic
system, the $xy$-pair may have several distinct states that belongs to the
same superselection sector $z$; in other words, there may be several ways to
split $z$ into $x$ and $y$. Such states form a finite-dimensional Hilbert
space $V^{xy}_{z}$, which is called \emph{fusion space}.\footnote{Technically,
it should be called ``splitting space'', whereas the fusion space is its dual,
$V_{xy}^{z}=\bigl(V^{xy}_{z}\bigr)^{*}$.}  In our model, all such spaces have
dimension one or zero. For example, $|\psi^{\sigma\sigma}_{1}\rangle$ is a
unit vector in the one-dimensional space $V^{\sigma\sigma}_{1}$.
\end{enumerate}

For a more rigorous definition of the space $V^{xy}_{z}$, consider the
splitting of $z$ into $x$ and $y$ by an operator $L$ that acts on the spins in
some fixed finite region $\Omega$. (We call such operators \emph{local}.) Let
us also fix a quantum state $|\Psi_{z}\rangle$ that has a $z$-particle at a
given place and no other particles in $\Omega$ or within the correlation
length from $\Omega$. Finally, we consider the set of local operators $L$ for
which the state $L|\Psi_{z}\rangle$ has an $x$-particle and an $y$-particle at
the required places. By definition, these are all possible states of the
$xy$-pair that can be obtained from $z$. On the other hand, such states are in
one-to-one correspondence with \emph{equivalence classes} of operators $L$: we
say that $L$ and $L'$ are \emph{equivalent} if
$L|\Psi_{z}\rangle=L'|\Psi_{z}\rangle$. For local operators, the equivalence
relation does not depend on the choice of $|\Psi_{z}\rangle$. Thus we arrive
at the following definition:
\begin{quote}
\noindent $V^{xy}_{z}$ is the set of equivalence classes of local
operators that split $z$ into $x$ and $y$.
\end{quote}

Each vertex in a splitting graph (as in Eqs.~(\ref{as_sese}),
(\ref{as_eeee}) above) designates an equivalence class of local
operators. For example,
$\figbox{1.0}{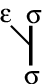}=|\psi^{\eps\sigma}_{\sigma}\rangle\in
V^{\eps\sigma}_{\sigma}$. One may choose an arbitrary basis in the space
$V^{xy}_{z}$. In our case, this amounts to fixing the phase of the vectors
$|\psi^{\sigma\sigma}_{1}\rangle$, $|\psi^{\sigma\sigma}_{\eps}\rangle$,
$|\psi^{\eps\sigma}_{\sigma}\rangle$, $|\psi^{\sigma\eps}_{\sigma}\rangle$,
and $|\psi^{\eps\eps}_{1}\rangle$. Different bases are related by a
transformation of the form
\begin{equation} \label{baschng}
|\psi^{xy}_{z}\rangle'\,=\,u^{xy}_{z}|\psi^{xy}_{z}\rangle,\qquad\quad
|u^{xy}_{z}|=1,
\end{equation}
which generally affects relations like~(\ref{as_sese})
and~(\ref{as_eeee}). However, these particular relations are invariant since
they have the same set of basis vectors on both sides:
$|\psi^{\eps\sigma}_{\sigma}\rangle$, $|\psi^{\sigma\eps}_{\sigma}\rangle$ in
Eq.~(\ref{as_sese}) and $|\psi^{\eps\eps}_{1}\rangle$ in Eq.~(\ref{as_eeee}).

\subsubsection{More relations} Let us write some other possible
associativity relations in a general form:
\begin{equation} \label{as_1}
\figbox{1.0}{eess1.eps}\;=\; \beta_{1}\figbox{1.0}{eess2.eps}\,,\qquad\qquad
\figbox{1.0}{esse1.eps}\;=\; \beta_{2}\figbox{1.0}{esse2.eps}\,,
\end{equation}
\begin{equation} \label{as_2}
\figbox{1.0}{ssee1.eps}\;=\; \gamma_{1}\figbox{1.0}{ssee2.eps}\,,\qquad\qquad
\figbox{1.0}{sees1.eps}\;=\; \gamma_{2}\figbox{1.0}{sees2.eps}\,,
\end{equation}
\begin{equation} \label{as_3}
\figbox{1.0}{eses1.eps}\;=\; \delta\figbox{1.0}{eses2.eps}\,,\qquad\qquad
\figbox{1.0}{ises1.eps}\;=\; \tau\figbox{1.0}{ises2.eps}\,,
\end{equation}
\begin{equation} \label{as_4}
\figbox{1.0}{iess1.eps}\;=\; \mu_{1}\figbox{1.0}{iess2.eps}\,,\qquad\qquad
\figbox{1.0}{isse1.eps}\;=\; \mu_{2}\figbox{1.0}{isse2.eps}\,.
\end{equation}
The numbers $\beta_{1}$, $\beta_{2}$, $\gamma_{1}$, $\gamma_{2}$, $\delta$,
$\tau$, $\mu_{1}$, $\mu_{2}$ are equal to $1$ in magnitude. These and similar
coefficients are called \emph{structural parameters}.

Using the transformation~(\ref{baschng}), we can eliminate some of the
parameters, e.g., $\beta_{1}$, $\beta_{2}$, and $\gamma_{1}$. Indeed, they are
transformed as follows:
\begin{equation} \label{baschng1}
\beta_{1}'\,=\,
\frac{u^{\eps\sigma}_{\sigma}u^{\sigma\sigma}_{\eps}}{u^{\sigma\sigma}_{1}}\,
\beta_{1},\qquad\quad
\beta_{2}'\,=\,
\frac{u^{\sigma\sigma}_{1}}{u^{\sigma\eps}_{\sigma}u^{\sigma\sigma}_{\eps}}\,
\beta_{2},\qquad\quad
\gamma_{1}'\,=\,
\frac{(u^{\sigma\eps}_{\sigma})^{2}}{u^{\eps\eps}_{1}}\, \gamma_{1}.
\end{equation}
We require that $\beta_{1}'=\beta_{2}'=\gamma_{1}'=1$ and solve for
$u^{xy}_{z}$.  Note that the solution is unique up to trivial variations not
affecting the structural parameters, namely
\begin{equation} \label{trivbaschng}
u^{xy}_{z}\mapsto \zeta_{x}\zeta_{y}\zeta_{z}^{-1}u^{xy}_{z}.
\end{equation}

Finally, let us consider the splitting of one $\sigma$-particle into three
$\sigma$-particles. The space $V^{\sigma\sigma\sigma}_{\sigma}$ is
two-dimensional. A basis in this space can be chosen in two ways:
\begin{equation} \label{basis_ssss}
\left\{\,|\xi_{1}\rangle\; =\figbox{1.0}{ssss1.eps}, \quad\
|\xi_{\eps}\rangle\; =\figbox{1.0}{ssss2.eps}\;\right\}
\qquad \text{or} \qquad
\left\{\,|\eta_{1}\rangle\; =\figbox{1.0}{ssss3.eps}, \quad\
|\eta_{\eps}\rangle\; =\figbox{1.0}{ssss4.eps}\;\right\}.
\end{equation}
Of course, one basis can be expressed in terms of the other:
\begin{equation} \label{as_ssss}
|\xi_{1}\rangle \,=\,
a_{11}|\eta_{1}\rangle + a_{\eps 1}|\eta_{\eps}\rangle,\qquad
|\xi_{\eps}\rangle \,=\,
a_{1\eps}|\eta_{1}\rangle + a_{\eps\eps}|\eta_{\eps}\rangle,\qquad\
\text{where}\quad a_{xy}=\langle\eta_{x}|\xi_{y}\rangle.
\end{equation}
The coefficients $a_{xy}$ and the other structural parameters will be found
later. At this point, we can only tell that the matrix $(a_{xy})$ is unitary.

To illustrate the physical meaning of the structural parameters, let us return
to the gedanken experiment considered in the introduction. Suppose we create
two pairs of $\sigma$-particles from the vacuum.  Then we take one particle
from each pair and fuse them. With probability $|a_{11}|^{2}$ they will
annihilate, and with probability $|a_{\eps 1}|^{2}$ they will fuse into an
$\eps$-particle.  (We will see that $|a_{11}|^{2}=|a_{\eps 1}|^{2}=1/2$.)

\subsection{Algebraic consistency} \label{sec_algcons}

All the previous arguments were based on the result that vortices carry
unpaired Majorana modes. Not surprisingly, this property alone is not
sufficient to fully characterize the fusion and braiding of anyons. We now
invoke some additional principles. The first one is consistency: successive
fusion and braiding events must commute with each other in certain cases.  A
more careful statement of this requirement amounts to the formulation of an
\emph{algebraic theory of anyons}.

Anyons may be described in the framework of topological quantum field theory
(TQFT), which originates from Witten's paper on quantum Chern-Simons
fields~\cite{Witten} and the work of Moore and Seiberg on conformal field
theory~\cite{MooreSeiberg}. Important mathematical studies in this area were
done by Reshetikhin and Turaev~\cite{ReshTur} and Walker~\cite{Walker}. For
our purposes, it suffices to use a construction called \emph{unitary modular
category} (UMC), which constitutes the algebraic core of TQFT~\cite{Turaev}.
This construction will be outlined in Appendix~\ref{sec_algth}. Actually, the
full theory is not necessary to understand the calculations below. On the
contrary, one may use these calculations to motivate some of the UMC axioms.

One of the axioms is known as the \emph{pentagon equation}, see
Fig.~\ref{fig_pentquadr}a on page~\pageref{fig_pentquadr}. It deals with the
five ways to split a particle $u$ into $x$, $y$, $z$, $w$, or five
representation of the space $V^{xyzw}_{u}$. The arrows in the figure may be
regarded as equality signs, therefore the diagram must commute. For example,
consider the splitting process $\sigma\to\eps\sigma\sigma\sigma$ via
intermediate states $q=1$ and $p=\sigma$. The upper path across the diagram
looks like this:
\[
\figbox{1.0}{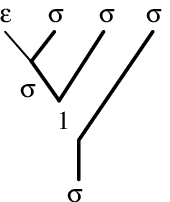} \,\ =\,\
a_{11}\figbox{1.0}{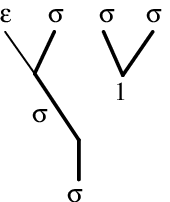} \,+\; a_{\eps 1}\figbox{1.0}{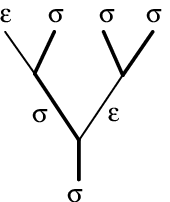}
\,\ =\,\
a_{11}\!\figbox{1.0}{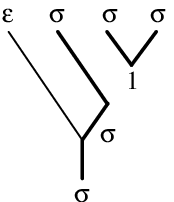} \,-\; a_{\eps 1}\!\figbox{1.0}{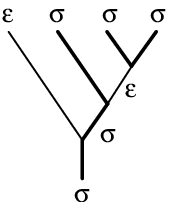},
\]
whereas the lower path is as follows:
\[
\figbox{1.0}{sesss1.eps} \,\ =\,\
\mu_{1}\!\figbox{1.0}{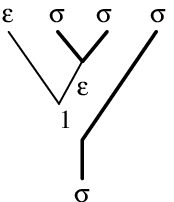} \,\ =\,\
\mu_{1}\gamma_{2}\!\figbox{1.0}{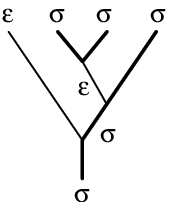} \,\ =\,\
\mu_{1}\gamma_{2}a_{1\eps}\!\figbox{1.0}{sesss41.eps} \,+\;
\mu_{1}\gamma_{2}a_{\eps\eps}\!\figbox{1.0}{sesss4e.eps}.
\]
Thus $\mu_{1}\gamma_{2}a_{1\eps}=a_{11}$,\,
$\mu_{1}\gamma_{2}a_{\eps\eps}=-a_{\eps 1}$.  One can also start with the
splitting graph that has intermediate state $\eps$ instead of $1$; this gives
another pair of relations between the structural parameters.

Overall, there are $23$ spaces $V^{xyzw}_{u}$ with nontrivial upper
indices. Thus we obtain $23$ pentagon equations, $5$ of which are satisfied
automatically.  The remaining $18$ equations imply the following relations
between the structural parameters:
\begin{equation} \label{pent-first}
V^{\eps\sigma\sigma\sigma}_{\sigma}:
\left\{\begin{array}{r@{}l}
\mu_{1}\gamma_{2}a_{1\eps}&{}=a_{11},\\
\mu_{1}\gamma_{2}a_{\eps\eps}&{}=-a_{\eps 1},\\
\beta_{1}a_{11}&{}=a_{1\eps},\\
\beta_{1}a_{\eps 1}&{}=-a_{\eps\eps};
\end{array}\right.
\qquad\qquad
V^{\sigma\sigma\sigma\eps}_{\sigma}:
\left\{\begin{array}{r@{}l}
\mu_{2}\gamma_{1}a_{\eps 1}&{}=a_{11},\\
\beta_{2}a_{11}&{}=a_{\eps 1},\\
\mu_{2}\gamma_{1}a_{\eps\eps}&{}=-a_{1\eps},\\
\beta_{2}a_{1\eps}&{}=-a_{\eps\eps};
\end{array}\right.
\end{equation}
\smallskip
\begin{equation}
V^{\sigma\eps\sigma\sigma}_{\sigma}:
\left\{\begin{array}{r@{}l}
\tau\beta_{1}a_{\eps 1}&{}=a_{11},\\
\tau\mu_{1}a_{11}&{}=\gamma_{1}a_{\eps 1},\\
\delta\beta_{1}a_{\eps\eps}&{}=a_{1\eps},\\
\delta\mu_{1}a_{1\eps}&{}=\gamma_{1}a_{\eps\eps};
\end{array}\right.
\qquad\qquad
V^{\sigma\sigma\eps\sigma}_{\sigma}:
\left\{\begin{array}{r@{}l}
\tau\mu_{2}a_{11}&{}=\gamma_{2}a_{1\eps},\\
\delta\mu_{2}a_{\eps 1}&{}=\gamma_{2}a_{\eps\eps},\\
\tau\beta_{2}a_{1\eps}&{}=a_{11},\\
\delta\beta_{2}a_{\eps\eps}&{}=a_{\eps 1};
\end{array}\right.
\end{equation}
\smallskip
\begin{equation}
V^{\sigma\sigma\sigma\sigma}_{1}:
\left\{\begin{array}{r@{}l}
a_{11}^{2}+\tau a_{1\eps}a_{\eps 1}&{}=1,\\
a_{\eps 1}a_{11}+\tau a_{\eps\eps}a_{\eps 1}&{}=0,\\
a_{11}a_{1\eps}+\tau a_{1\eps}a_{\eps\eps}&{}=0,\\
a_{\eps 1}a_{1\eps}+\tau a_{\eps\eps}^{2}&{}=1;
\end{array}\right.
\qquad\qquad
V^{\sigma\sigma\sigma\sigma}_{\eps}:
\left\{\begin{array}{r@{}l}
a_{11}^{2}+\delta a_{1\eps}a_{\eps 1}&{}=0,\\
a_{\eps 1}a_{11}+\delta a_{\eps\eps}a_{\eps 1}&{}=\beta_{2},\\
a_{11}a_{1\eps}+\delta a_{1\eps}a_{\eps\eps}&{}=\beta_{1},\\
a_{\eps 1}a_{1\eps}+\delta a_{\eps\eps}^{2}&{}=0;
\end{array}\right.
\end{equation}
\smallskip
\begin{equation} \label{pent-last}
\begin{array}{rl@{\qquad\qquad}rl@{\qquad\quad}}
V^{\sigma\eps\eps\sigma}_{1}: & \tau^{2}=\gamma_{1}\gamma_{2}, &
V^{\eps\eps\sigma\sigma}_{1},V^{\eps\eps\sigma\sigma}_{\eps}: &
\mu_{1}\beta_{1}\gamma_{2}=1,
\smallskip\\
V^{\sigma\eps\eps\sigma}_{\eps}: & \delta^{2}=\gamma_{1}\gamma_{2}, &
V^{\sigma\sigma\eps\eps}_{1},V^{\sigma\sigma\eps\eps}_{\eps}: &
\mu_{2}\beta_{2}\gamma_{1}=1,
\smallskip\\
V^{\eps\sigma\eps\sigma}_{1},V^{\eps\sigma\eps\sigma}_{\eps},
V^{\sigma\eps\sigma\eps}_{1},V^{\sigma\eps\sigma\eps}_{\eps}: &
\delta=-\tau, &
V^{\eps\sigma\sigma\eps}_{1},V^{\eps\sigma\sigma\eps}_{\eps}: &
\mu_{1}\mu_{2}=\beta_{1}\beta_{2}.
\end{array}
\end{equation}

Equations~(\ref{pent-first}--\ref{pent-last}) have only two solutions
satisfying the convention that $\beta_{1}=\beta_{2}=\gamma_{1}=1$:
\begin{equation} \label{structparam}
\begin{array}{c}
\beta_{1}=\beta_{2}=\gamma_{1}=\gamma_{2}=\mu_{1}=\mu_{2}=\tau=1,\qquad
\delta=-1,
\bigskip\\
\displaystyle
\begin{pmatrix}a_{11}&a_{1\eps}\\a_{\eps 1}&a_{\eps\eps}\end{pmatrix} \,=\,
\frac{\FS}{\sqrt{2}}
\begin{pmatrix}1&1\\1&-1\end{pmatrix},\qquad
\text{where}\ \FS=\pm 1.
\end{array}
\end{equation}

\begin{Remark}
The number of independent equations can be reduced by using a symmetry between
upper and lower indices: $V^{xy}_{z}\cong V^{\anti{z}x}_{\anti{y}}$, where
$\anti{z}$ denotes the antiparticle for $z$ (in our case,
$\anti{\eps}=\eps$,\, $\anti{\sigma}=\sigma$). However this symmetry generally
involves nontrivial phase factors, so we find the brute-force calculation a
safer approach.
\end{Remark}

Let us now examine consistency between fusion and braiding. Braiding is fully
characterized by the action of the counterclockwise rotation on fusion spaces:
$\widehat{R}^{xy}_{z}:\,V^{xy}_{z}\to V^{yx}_{z}$. Since in our case the
spaces $V^{xy}_{z}$ and $V^{yx}_{z}$ are one-dimensional, the linear map
$\widehat{R}^{xy}_{z}$ is given by a single matrix element:\footnote{In
Appendix~\ref{sec_algth}, we take the liberty to omit the hat from the
notation $\widehat{R}^{xy}_{z}$. This should not cause confusion because we do
not consider matrix elements there.}
 
\begin{equation} \label{numberR}
\widehat{R}^{xy}_{z}|\psi^{xy}_{z}\rangle=R^{xy}_{z}|\psi^{yx}_{z}\rangle,
\qquad\quad \text{where}\quad
R^{xy}_{z}=\langle\psi^{yx}_{z}|\widehat{R}^{xy}_{z}|\psi^{xy}_{z}\rangle.
\end{equation}
In graphic notation,\,
$\widehat{R}^{xy}_{z}|\psi^{xy}_{z}\rangle=\figbox{1.0}{zyxr.eps}\,$,\,
therefore Eq.~(\ref{numberR}) reads:
\begin{equation}
\qquad\figbox{1.0}{zyxr.eps}\ =\ R^{xy}_{z}\figbox{1.0}{zyxi.eps}.
\end{equation}

Nontrivial relations arise if we consider the action of braiding operators on
the fusion space of three particles. These relations can be expressed by two
commutative diagrams called the \emph{hexagon equations}, see
Fig.~\ref{fig_hexagon} on page~\pageref{fig_hexagon}. The arrows in those
diagrams may be understood as equalities of vectors in the space
$V^{yzx}_{u}$. Let us consider the following example of the first equation
(the first line below corresponds to the upper path across the hexagon, the
second to the lower path):
\[
\figbox{1.0}{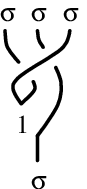} \,\ =\,\
R^{\sigma\sigma}_{1}\figbox{1.0}{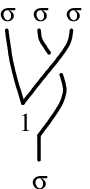} \,\ =\,\
\frac{\FS R^{\sigma\sigma}_{1}}{\sqrt{2}}\figbox{1.0}{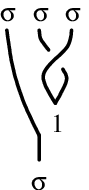} \,+\;
\frac{\FS R^{\sigma\sigma}_{1}}{\sqrt{2}}\figbox{1.0}{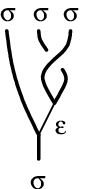} \,\ =\,\
\frac{\FS(R^{\sigma\sigma}_{1})^{2}}{\sqrt{2}}
\figbox{1.0}{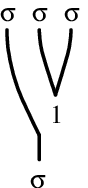} \,+\;
\frac{\FS R^{\sigma\sigma}_{1}R^{\sigma\sigma}_{\eps}}{\sqrt{2}}
\figbox{1.0}{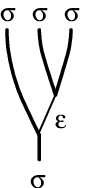};
\]
\[
\figbox{1.0}{s1sssrr.eps} \,\ =\,\
\frac{\FS}{\sqrt{2}}\figbox{1.0}{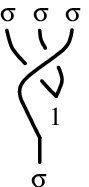}\,+\;
\frac{\FS}{\sqrt{2}}\figbox{1.0}{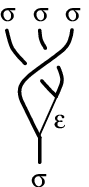}=\,\
\frac{\FS}{\sqrt{2}}\figbox{1.0}{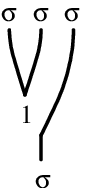}\,+\;
\frac{\FS R^{\sigma\eps}_{\sigma}}{\sqrt{2}}\figbox{1.0}{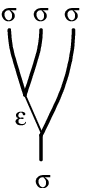}
\,\ =\,\
\frac{1+R^{\sigma\eps}_{\sigma}}{2}\figbox{1.0}{ss1ss.eps} \,+\;
\frac{1-R^{\sigma\eps}_{\sigma}}{2}\figbox{1.0}{ssess.eps}.
\]
Thus we get:\, $1+R^{\sigma\eps}_{\sigma}
=\FS\sqrt{2}(R^{\sigma\sigma}_{1})^{2}$,\,\,
$1-R^{\sigma\eps}_{\sigma}
=\FS\sqrt{2}R^{\sigma\sigma}_{1}R^{\sigma\sigma}_{\eps}$.

The full set of equations arising from the first hexagon is as follows:
\begin{equation}
V^{\sigma\sigma\sigma}_{\sigma}:
\left\{\begin{array}{r@{}l}
1+R^{\sigma\eps}_{\sigma}&{}=\FS\sqrt{2}(R^{\sigma\sigma}_{1})^{2},\\
1-R^{\sigma\eps}_{\sigma}
&{}=\FS\sqrt{2}R^{\sigma\sigma}_{1}R^{\sigma\sigma}_{\eps},\\
1+R^{\sigma\eps}_{\sigma}&{}=-\FS\sqrt{2}(R^{\sigma\sigma}_{\eps})^{2},
\end{array}\right.
\qquad\qquad
V^{\eps\eps\eps}_{\eps}:\ (R^{\eps\eps}_{\eps})^{2}=1;
\end{equation}
\smallskip
\begin{equation}
\begin{array}{rl@{\qquad\qquad}rl}
V^{\eps\sigma\sigma}_{1},V^{\sigma\eps\sigma}_{1}: &
R^{\sigma\eps}_{\sigma}R^{\sigma\sigma}_{\eps}=R^{\sigma\sigma}_{1},
&
V^{\sigma\sigma\eps}_{1}: &
(R^{\eps\sigma}_{\sigma})^{2}=R^{\eps\eps}_{1},
\medskip\\
V^{\eps\sigma\sigma}_{\eps},V^{\sigma\eps\sigma}_{\eps}: &
R^{\sigma\eps}_{\sigma}R^{\sigma\sigma}_{1}=-R^{\sigma\sigma}_{\eps},
&
V^{\sigma\sigma\eps}_{\eps}: &
-(R^{\eps\sigma}_{\sigma})^{2}=1,
\medskip\\
V^{\sigma\eps\eps}_{\sigma},V^{\eps\sigma\eps}_{\sigma}: &
R^{\eps\eps}_{1}=-1,
&
V^{\eps\eps\sigma}_{\sigma}: &
-(R^{\sigma\eps}_{\sigma})^{2}=1.
\end{array}
\end{equation}
The second hexagon is obtained from the first by replacing
$R^{xy}_{z}$ with $(R^{-1})^{xy}_{z}=(R^{yx}_{z})^{-1}$. For example, the
equation $R^{\sigma\eps}_{\sigma}R^{\sigma\sigma}_{\eps}=R^{\sigma\sigma}_{1}$
becomes $(R^{\eps\sigma}_{\sigma})^{-1}(R^{\sigma\sigma}_{\eps})^{-1}
=(R^{\sigma\sigma}_{1})^{-1}$.

Eliminating redundancies, we get this system of equations:
\begin{equation} \label{Req}
R^{\eps\eps}_{1}=-1,\qquad
R^{\sigma\eps}_{\sigma}=R^{\eps\sigma}_{\sigma}=\pm i,\qquad
(R^{\sigma\sigma}_{1})^{2}
=\frac{\FS(1+R^{\sigma\eps}_{\sigma})}{\sqrt{2}},\qquad
R^{\sigma\sigma}_{\eps}=-R^{\sigma\eps}_{\sigma}R^{\sigma\sigma}_{1},
\end{equation}
where $\FS=\pm 1$. The system has eight solutions, which all fit
Eq.~(\ref{Rinv}). Specifically, the solutions have the form
\begin{equation} \label{Rsol1}
R^{\eps\eps}_{1}=-1,\qquad
R^{\sigma\eps}_{\sigma}=R^{\eps\sigma}_{\sigma}= -i\alpha,\qquad
R^{\sigma\sigma}_{1}=\theta\,e^{-i\alpha\pi/4},\qquad
R^{\sigma\sigma}_{\eps}=\theta\,e^{i\alpha\pi/4},
\end{equation}
where the following combinations of $\theta$, $\alpha$, and $\FS$ are
possible:
\begin{equation} \label{Rsol2}
\def\arraystretch{0}
\begin{array}{|c||c|c|c|c|c|c|c|c|}
\hline
&&&&&&&&\\[2pt]
\strut \theta & e^{-7i\pi/8} & e^{-5i\pi/8} & e^{-3i\pi/8} & e^{-i\pi/8}
& e^{i\pi/8} & e^{3i\pi/8} & e^{5i\pi/8} & e^{7i\pi/8} \\
\hline
\strut \alpha & +1 & -1 & +1 & -1 & +1 & -1 & +1 & -1\\
\hline
\strut \FS & +1 & -1 & -1 & +1 & +1 & -1 & -1 & +1\\
\hline
\end{array}\ .
\end{equation}

Thus the properties of anyons are defined by the value of $\theta$, which
satisfies $\theta^{8}=-1$.

\subsection{Final details}

To choose the the right solution and to complete the description of
non-Abelian anyons, we use Eq.~(\ref{theta}) in conjunction with some general
results from Appendix~\ref{sec_algth}.

The matrix element $a_{11}=\langle\eta_{1}|\xi_{1}\rangle$ (cf.\
Eqs.~(\ref{basis_ssss}), (\ref{as_ssss}), (\ref{structparam})) is equal to
$\FS/\sqrt{2}$. But according to Eq.~(\ref{qdimFS}) on
page~\pageref{qdimFS}, $a_{11}=\FS_{\sigma}/d_{\sigma}$, where
$d_{\sigma}$ is the quantum dimension of the $\sigma$-particle, and
$\FS_{\sigma}$ is the Frobenius-Schur indicator. Therefore
\begin{equation}
d_{\sigma}=\sqrt{2},\qquad\quad \FS_{\sigma}=\FS.
\end{equation}

The topological spin of the $\sigma$-particle can be computed using
Eq.~(\ref{topspin}) on page~\pageref{topspin}:
\begin{equation}
\theta_{\sigma}
=d_{\sigma}^{-1}\bigl(R^{\sigma\sigma}_{1}+R^{\sigma\sigma}_{\eps}\bigr)
=\theta.
\end{equation}
This result also follows from Eq.~(\ref{tsFS}). Matching it with the
expression of $\theta_{\sigma}$ in terms of the Chern number, we represent all
eight cases in~(\ref{Rsol2}) by these formulas:
\begin{equation}
\theta=e^{i\pi\nu/8},\qquad
\alpha=(-1)^{(\nu-1)/2},\qquad \FS=(-1)^{(\nu^{2}-1)/8}.
\end{equation}

Thus we have obtained almost all properties in Table~\ref{tab_nonabelian}. The
remaining structure, namely the topological $S$-matrix, can be found using
general rules (see Appendix~\ref{sec_algth}).

\section{The sixteen-fold way} \label{sec_16}

Let us again consider the theory with $\ZZ_{2}$-vortices and free fermions
whose spectrum is gapped and characterized by the Chern number $\nu$. The
properties of anyons in this model depend on $\nu\bmod 16$. In the previous
section we studied the case of odd $\nu$; now we assume that $\nu$ is even.

\begin{table}[p]
\widefbox{\vskip10pt
\noindent\textbf{Superselection sectors:}\quad
$1$ (vacuum),\quad $\eps$ (fermion),\quad $e,m$ (vortices).
\[
\begin{array}{l@{\qquad}l@{\qquad}l@{\qquad}l}
\text{Quantum dimension:} &
d_{1}=1, & d_{\eps}=1 &  d_{e}=d_{m}=1;
\smallskip\\
\text{Topological spin:} &
\theta_{1}=1, & \theta_{\eps}=-1,
& \theta_{e}=\theta_{m}=\theta=\exp\bigl(\frac{\pi}{8}i\nu\bigr);
\smallskip\\
\text{Frobenius-Schur indicator:} &
\FS_{1}=1, & \FS_{\eps}=1,
& \FS_{e}=\FS_{m}=\FS=\exp\bigl(\frac{\pi}{4}i\nu\bigr).
\end{array}
\]
\noindent\textbf{Fusion rules:}
\[
\eps\times\eps=1,\qquad \eps\times e=m,\qquad
\eps\times m=e,\qquad\quad
e\times e=m\times m=1,\qquad e\times m=\eps.
\]

\underline{\textbf{Case 1}}:\; $\nu\equiv 0,8\pmod{16}$;\qquad
$\theta=\pm 1$,\; $\FS=1$.
\begin{itemize}
\item[]\textbf{Associativity relations}:\;
all associativity relations are trivial.

\item[]\textbf{Braiding rules:}
\vspace{-10pt}
\[
\begin{array}{c}
R^{\eps\eps}_{1}=-1,
\\[8pt]
R^{ee}_{1}=R^{mm}_{1}=\theta,
\end{array}
\hskip40pt
\begin{array}{l@{\qquad}l}
R^{e\eps}_{m}=1, & R^{\eps e}_{m}=-1,
\\[5pt]
R^{\eps m}_{e}=1, & R^{m\eps}_{e}=-1,
\\[5pt]
R^{em}_{\eps}=\theta, & R^{me}_{\eps}=-\theta,
\end{array}
\]
\end{itemize}
\medskip

\underline{\textbf{Case 2}}:\; $\nu\equiv\pm 4\pmod{16}$;\qquad
$\theta=\pm i$,\; $\FS=-1$.
\begin{itemize}
\item[]\textbf{Nontrivial associativity relations:}
\[
\figbox{1.0}{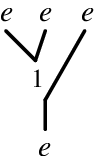} \;=\; -\ \figbox{1.0}{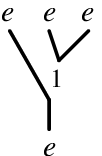}\,, \qquad\qquad
\figbox{1.0}{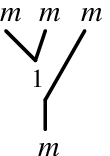} \;=\; -\ \figbox{1.0}{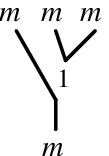}\,,
\]
\[
\figbox{1.0}{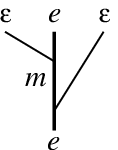}\,=\; -\ \figbox{1.0}{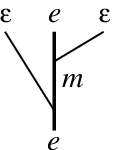}\,,\qquad\qquad
\figbox{1.0}{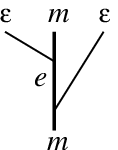}\,=\; -\figbox{1.0}{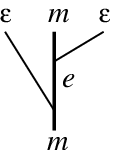}\,,
\]
\[
\figbox{1.0}{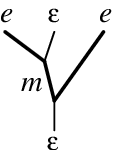}\,=\; -\ \figbox{1.0}{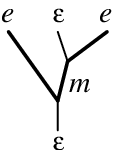}\,,\qquad\qquad
\figbox{1.0}{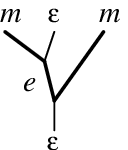}\,=\; -\figbox{1.0}{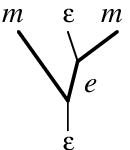}\,.
\]

\item[]\textbf{Braiding rules:}
\[
\begin{array}{c@{\hskip50pt}c}
R^{\eps\eps}_{1}=-1, &
R^{e\eps}_{m}=R^{\eps e}_{m}=R^{\eps m}_{e}=R^{m\eps}_{e}=\theta,
\\[6pt]
R^{ee}_{1}=R^{mm}_{1}=\theta, &
R^{em}_{\eps}=R^{me}_{\eps}=1.
\end{array}
\]
\end{itemize}
}
\caption{Properties of anyons for $\nu\equiv 0\pmod{4}$.}
\label{tab_0mod4}
\end{table}

\begin{table}[t]
\widefbox{\vskip10pt
\noindent\textbf{Superselection sectors:}\quad
$1$ (vacuum),\quad $\eps$ (fermion),\quad $a,\anti{a}$ (vortices).
\[
\begin{array}{l@{\qquad}l@{\qquad}l@{\qquad}l}
\text{Quantum dimension:} &
d_{1}=1, & d_{\eps}=1 &  d_{a}=d_{\anti{a}}=1;
\smallskip\\
\text{Topological spin:} &
\theta_{1}=1, & \theta_{\eps}=-1,
& \theta_{a}=\theta_{\anti{a}}=\theta=\exp\bigl(\frac{\pi}{8}i\nu\bigr)
=\frac{\pm1\pm i}{\sqrt{2}};
\smallskip\\
\text{Frobenius-Schur indicator:} &
\FS_{1}=1, & \FS_{\eps}=1. & 
\end{array}
\]
\noindent\textbf{Fusion rules:}
\[
a\times\eps=\anti{a}, \qquad \anti{a}\times\eps=a, \qquad \eps\times\eps=1,
\qquad\quad a\times a=\anti{a}\times\anti{a}=\eps, \qquad a\times\anti{a}=1.
\]

\noindent\textbf{Nontrivial associativity relations:}
\[
\figbox{1.0}{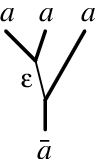} \;=\; -\ \figbox{1.0}{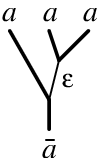}\,, \qquad\qquad
\figbox{1.0}{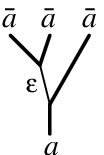} \;=\; -\ \figbox{1.0}{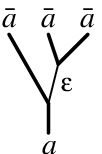}\,,
\]
\[
\figbox{1.0}{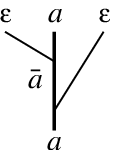}\,=\; -\ \figbox{1.0}{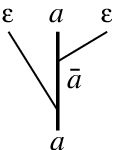}\,,\qquad\qquad
\figbox{1.0}{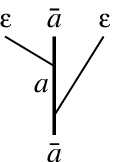}\,=\; -\figbox{1.0}{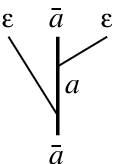}\,,
\]
\[
\figbox{1.0}{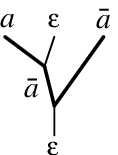}\,=\; -\ \figbox{1.0}{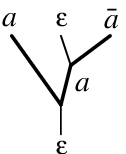}\,,\qquad\qquad
\figbox{1.0}{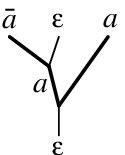}\,=\; -\figbox{1.0}{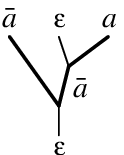}\,.
\]

\noindent\textbf{Braiding rules:}
\[
\hskip-35pt
\begin{array}{c@{\hskip50pt}c}
R^{\eps\eps}_{1}=-1, &
R^{a\eps}_{\anti{a}}=R^{\eps a}_{\anti{a}}
=R^{\anti{a}\eps}_{a}=R^{\eps\anti{a}}_{a}=\theta^{-2},
\\[6pt]
R^{aa}_{\eps}=R^{\anti{a}\anti{a}}_{\eps}=\theta, &
R^{a\anti{a}}_{1}=R^{\anti{a}a}_{1}=\theta^{-1}.
\end{array}
\]
}
\caption{Properties of anyons for $\nu\equiv 2\pmod{4}$.}
\label{tab_2mod4}
\end{table}

For even $\nu$, the vortices do not carry unpaired Majorana modes (see
Appendix~\ref{sec_locmatr}), therefore a vortex cannot absorb a fermion while
remaining in the same superselection sector. Thus, there are actually two
types of vortices, which are transformed one to another by adding a
fermion. Two vortices of the same type may either annihilate or fuse into a
fermion. In the first case the vortices are denoted by $e$ and $m$; they obey
the fusion rules~(\ref{Abfusion}). In the second case, we denote the vortices
by $a$ and $\anti{a}$; the fusion rules are as follows:
\begin{equation} \label{fusion2mod4}
a\times\eps=\anti{a}, \quad\ \anti{a}\times\eps=a, \quad\ \eps\times\eps=1,
\qquad\quad a\times a=\anti{a}\times\anti{a}=\eps, \qquad a\times\anti{a}=1.
\end{equation}
Both sets of rules are Abelian, i.e., they do not contain ``$+$'' on the
right-hand side.

The associativity relations can be found along the same lines as in the
non-Abelian case. First, the fermions must obey the trivial
relation~(\ref{as_eeee}). The argument used in the derivation of
Eq.~(\ref{as_sese}) remains valid, but we obtain two relations instead of one.
Note that they are not invariant individually; for example, the sign in both
relations may be changed to ``$+$''. Finally, we solve the pentagon equation.
The fusion rules with $e$ and $m$ admit two nonequivalent solutions, one of
which is trivial; see Table~\ref{tab_0mod4}. Both solutions are invariant
under the transformation
\begin{equation} \label{emfuseauto}
|\psi^{me}_{\eps}\rangle\mapsto -|\psi^{me}_{\eps}\rangle, \qquad
|\psi^{e\eps}_{m}\rangle\mapsto -|\psi^{e\eps}_{m}\rangle, \qquad
|\psi^{\eps m}_{e}\rangle\mapsto -|\psi^{\eps m}_{e}\rangle
\end{equation}
(where $|\psi^{xy}_{z}$ is the basis vector in the corresponding fusion
space), as well as trivial transformations
\begin{equation}
|\psi^{xy}_{z}\rangle\mapsto
\zeta_{x}\zeta_{y}\zeta_{z}^{-1}|\psi^{xy}_{z}\rangle.
\end{equation}
In the case of $a$ and $\anti{a}$, there is only one solution, which doesn't
admit any nontrivial symmetry; see Table~\ref{tab_2mod4}.

The braiding rules are found by solving the hexagon equations in conjunction
with the requirement that $R_{1}^{\eps\eps}=-1$ and that transporting a
fermion around a vortex is described by the multiplication by $-1$. Then we
compute the topological spin of the vortex and identify each solution with a
particular value of $\nu\bmod 16$. The results are summarized in
Tables~\ref{tab_0mod4} and~\ref{tab_2mod4}. (We have eliminated redundant
solutions that can be obtained by the transformation~(\ref{emfuseauto})).

Let us mention a couple of interesting properties of these Abelian theories.
For $\nu\equiv 8\pmod{16}$, all three nontrivial particles ($e$, $m$, and
$\eps$) are fermions. For $\nu\equiv\pm 4\pmod{16}$, the particles $e$ and $m$
are \emph{semions} with trivial mutual braiding. (In comparison, the
well-known Kalmeyer-Laughlin state~\cite{KalmeyerLaughlin} has one semionic
species.)
\enlargethispage{\baselineskip}

\section{Odds and ends}

What follows are some open questions, as well as thoughts of how the present
results can be extended.

\begin{enumerate}
\item Duan, Demler, and Lukin~\cite{DuanDemlerLukin} proposed an optical
lattice implementation of the Hamiltonian~(\ref{Hamiltonian}). It would be
interesting to find a solid state realization as well. For example, the
anisotropic exchange could be simulated by interaction of both lattice spins
with a spin-$1$ atom coupled to a crystal field.

\item The weak translational symmetry breaking in the Abelian phase has some
interesting consequences. A particularly unusual phenomenon takes place when
the lattice has a dislocation. A particle winding around the dislocation
changes its type: $e\leftrightarrow m$. Since $m=e\times\eps$, the fermionic
parity appears not to be conserved. To restore the conservation law, we must
assume that the dislocation carries an unpaired Majorana mode. Therefore,
Abelian phases can also be used for the implementation of quantum memory.

\item Chiral phases ($\nu\not=0$) require that the time-reversal symmetry be
broken. In the present model, this is achieved by applying a magnetic
field. However, a spontaneous breaking of time-reversal symmetry occurs in the
presence of odd cycles in the lattice. For example, one can replace each
vertex of the honeycomb lattice by a triangle. In this case, a gapped $\nu=\pm
1$ phase is realized without external magnetic field~\cite{FeigelmanKitaev}.

\item The representation of a spin by four Majorana fermions might be useful
for other models, even though it does not lead to an exact solution. In
particular, one can consider a variational mean-field state in which $c_{j}$
are decoupled from $b_{j}^{\alpha}$. It is unclear whether this type of states
occur in Heisenberg antiferromagnets. On the triangular lattice, such a state
has larger energy than the $120^{\circ}$ Neel-type state. It would be
interesting if the former could be stabilized by additional interactions and
quantum fluctuations.

\item Topological phases with free fermions coupled to an effective $\ZZ_2$
gauge field have been classified by $\nu\bmod 16$. However, this analysis does
not include multilayer systems. One can argue that if the interaction between
the layers is weak enough, topological particles cannot tunnel between the
layers. Thus, each layer is described by one of the $16$ theories studied
above. In mathematical terms, we have the direct product of several unitary
modular categories (UMCs). Strongly interacting layers are roughly described
by $n$ fermionic species interacting with $(\ZZ_{2})^{n}$-vortices, but a
complete classification of such phases is yet to be found.

\item In addition to the particle classification, the chiral central charge
$c_{-}=\nu/2$ is an important robust characteristic, though its topological
meaning is not so clear. It appears that in multilayer systems the total value
of $\nu$ is shared between the layers, i.e., increasing $\nu$ by $16$ in one
layer while decreasing it in another does make a different topological phase.
More generally, a topological phase is characterized by a UMC and a real
number $c_{-}$ satisfying the relation~(\ref{anomaly}). To prove or disprove
this statement, a mathematical notion of equivalence between topological
phases is necessary. It may be based on local (or quasilocal) isomorphisms
between operator algebras.

\item A related question is whether the space-time boundary of an arbitrary
topological phase can be described by a two-dimensional conformal field theory
(CFT) and when two such theories have the same topological content. Two CFTs
may be considered topologically equivalent if there is a consistent theory for
a one-dimensional boundary between them. Conjecturally, this is the case if
and only if both CFTs have the same chiral central charge
$c_{-}=c-\overline{c}$ and correspond to the same UMC.

\item Another topic to study is Bose-condensation of vortices, which occurs
when a vortex energy becomes negative due to some parameter change. The
condensation of $e$-particles in the $\nu=0$ phase is equivalent to the
confinement of $m$- and $\eps$-particles. Thus, the topological order is
destroyed. In the $\nu=16$ system, this process produces a phase without
anyons or fermions, but with the chiral central charge $c_{-}=8$. Under
special circumstances, the boundary of this phase is described by the $E_{8}$
CFT at level $1$, though a generic (nonconformal) perturbation drives it into
a state with $8$ chiral bosons propagating with different velocities.

\item\label{Boselaxfunc} The condensation of $\eps\eps$-pairs in two adjacent
$\nu=1$ layers leads to the binding of single-layer vortices into
$\sigma\sigma$ pairs, which are equivalent to $a$ or $\anti{a}$ in the $\nu=2$
phase. Thus the direct product of two $\nu=1$ theories is related to the
$\nu=2$ theory. It would be interesting to have a general mathematical
characterization of such relations. The notion of \emph{lax tensor functor}
(see Remark~\ref{laxfunc} on page~\pageref{laxfunc}) can be useful in this
regard.
\end{enumerate}

\pagebreak
\appendix

\section{Numerical results on the stability of the vortex-free phase}
\label{sec_numerics}

The goal of this study is to compare the energy of the vortex-free phase with
that of other phases. The case $J_{x}=J_{y}=J_{z}=1$ has been investigated
most carefully. The energy of the vortex-free phase equals $E_{0}\approx
-1.5746$ per unit cell (i.e., per two sites, or one hexagon). The actual
computation was done for tori with periodic or antiperiodic boundary
conditions in each direction. The vortex-free phase and other periodic phases
with small period are computationally very simple, so dealing with large tori
is not a problem. However, the energy calculation for an arbitrary vortex
configuration requires finding the singular values of an $N\times N$ matrix,
where $N$ is the number of unit cells. With our setup (MATLAB on a PC) we
could handle matrices of size $N\le 2500$, which corresponds to systems of
linear size $L=\sqrt{N}\le 50$. Although these numbers are pretty large,
finite-size effects are still appreciable due to the gapless nature of the
spectrum (see Section~\ref{sec_spectrum}).

\begin{figure}[h]
\centerline{\begin{tabular}{c@{\qquad}c}
{\def\epsfsize#1#2{0.5#1}\epsfbox{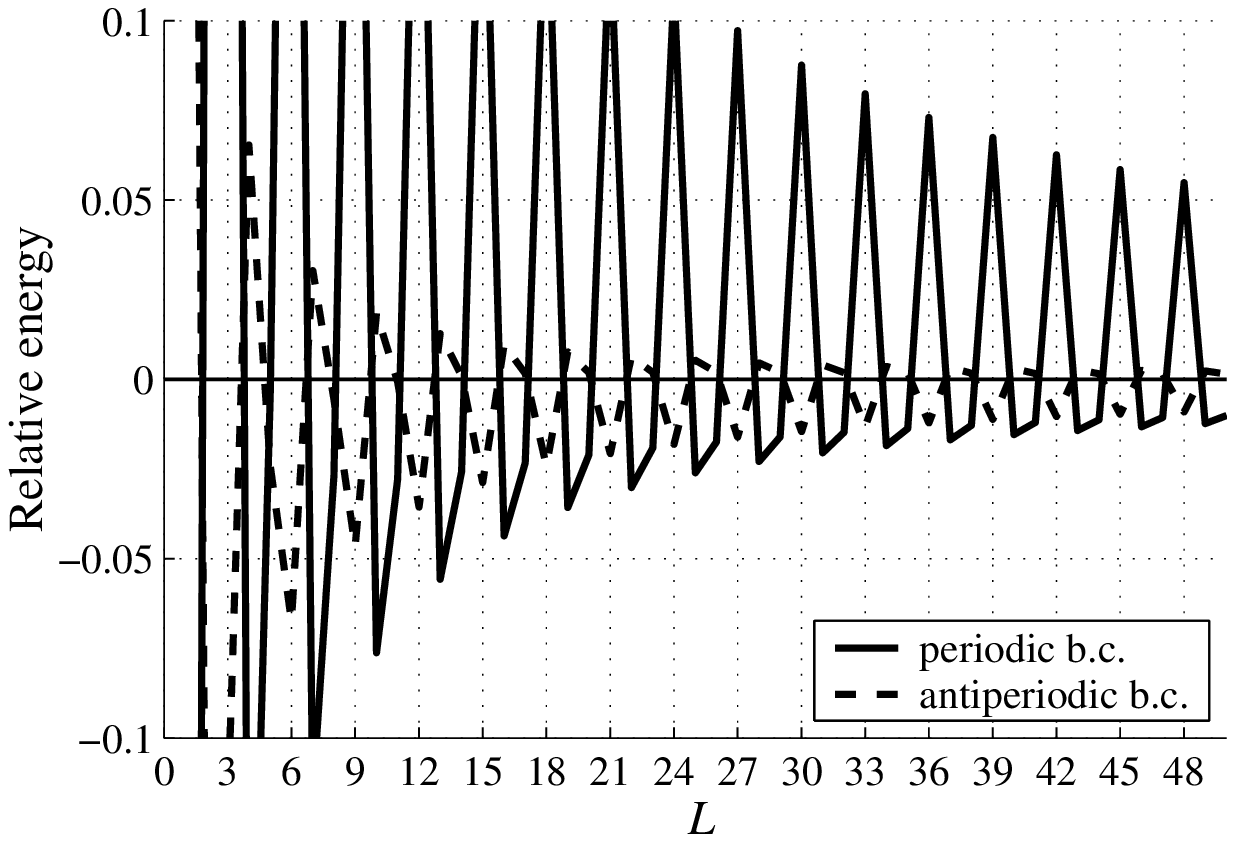}} &
{\def\epsfsize#1#2{0.5#1}\epsfbox{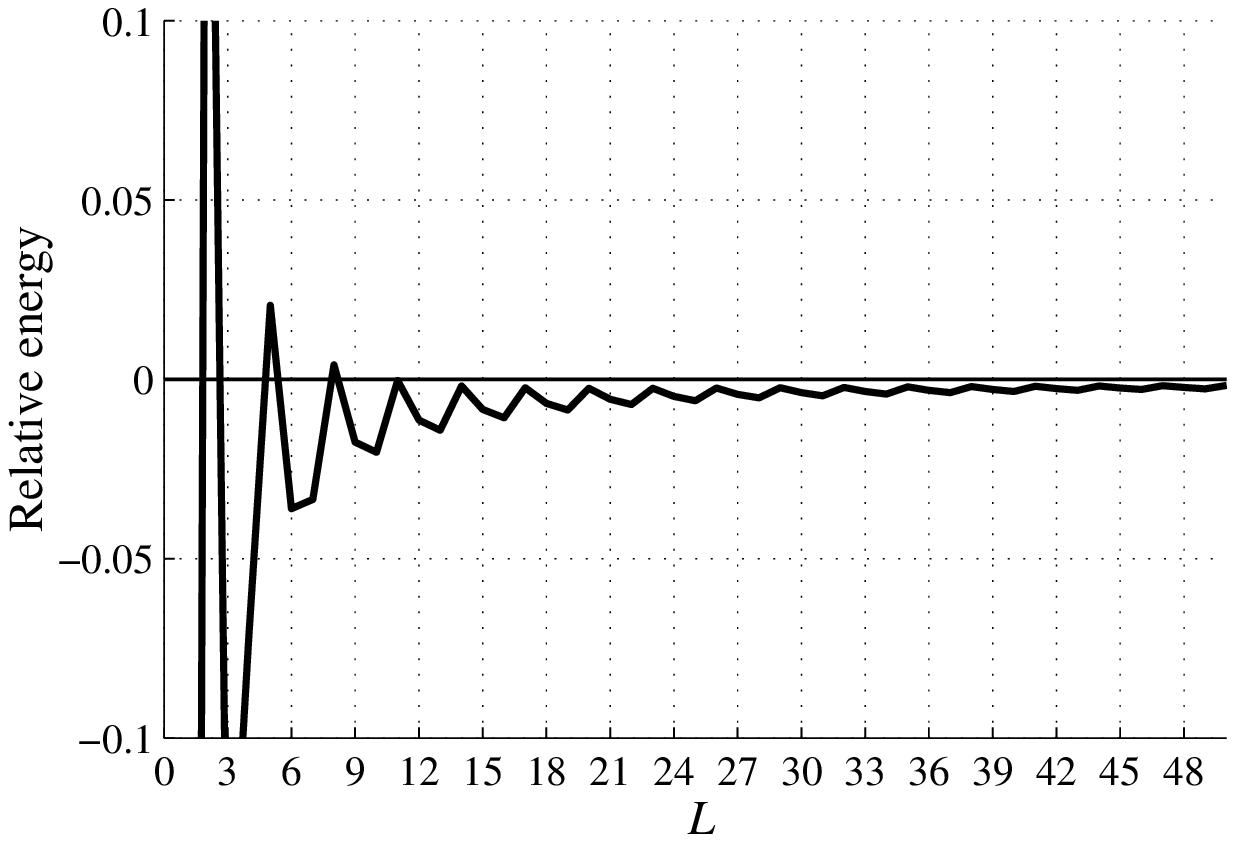}}
\\
a)\, tori with basis $(L\nn_{1},\,L\nn_{2})$ &
b)\, tori with basis $(L\nn_{1},\,L\nn_{2}+\nn_{1})$
\end{tabular}}
\caption{Finite size effects in the vortex-free phase.}
\label{fig_finsize}
\end{figure}

Let us first discuss the finite-size effects in the vortex-free phase. From
now on, we consider the relative energy, i.e., we subtract $NE_{0}$ from
actual results. The plot in Fig.~\ref{fig_finsize}a shows the relative
energy as a function of size for symmetric $L\times L$ tori. The oscillatory
behavior with period $3$ is due the fact that $\eps(\qq)$ vanishes at
$\qq=\qq_{*}$ (see~(\ref{qstar})). One may argue that each period $\rr$ of the
torus contributes $\sim|\rr|^{-1}\cos(2(\qq_{*},\rr))$ to the total energy
(there are infinitely many such terms since the periods form a
lattice). Interestingly enough, these contributions almost cancel each other
for periodic boundary conditions on the torus with basis
$(L\nn_{1},\,L\nn_{2}+\nn_{1})$, where $\nn_{1}$ and $\nn_{2}$ are the basis
vectors of the lattice (see figure in Eq.~(\ref{spectrum})). The
corresponding plot is shown in Fig.~\ref{fig_finsize}b.

\begin{figure}[p]
\centerline{\def\epsfsize#1#2{0.5#1}\epsfbox{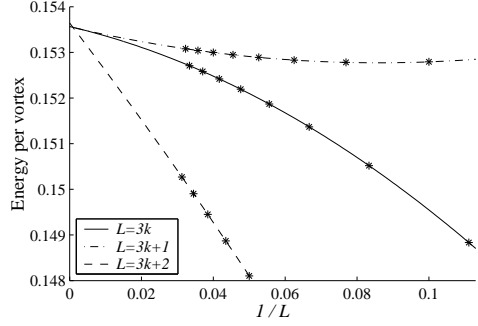}}
\caption{Extrapolation to infinite size: $4$ equally spaced vortices on the
torus with basis $(2L\nn_{1},\,2(L\nn_{2}+\nn_{1}))$.}
\label{fig_vortex}
\end{figure}

The energy of an isolated vortex is $E_{\text{vortex}}\approx 0.1536$ above
the ground state.\footnote{No rigorous precision analysis was attempted, but
the error in this and the other figures is expected not to exceed 1--2 units
of the least significant digit.} The calculation was done for tori with basis
$(L\nn_{1},\,L\nn_{2}+\nn_{1})$ for $L=9,\dots,32$. (We actually put $4$
vortices on the torus of twice this size because the number of vortices must
be even). Then the results were extrapolated to $L=\infty$ by fitting the
curve $E(L)=E_{\text{vortex}}+a_{1}L^{-1}+a_{2}L^{-2}$ to the data, separately
for $L=3k$, $L=3k+1$, and $L=3k+2$ (see Fig.~\ref{fig_vortex}).

\begin{table}[p]
\vspace{0.5cm}
\[
E_{\text{vortex}}\approx0.1536,\quad\ 
\Delta E\left( \figbox{0.6}{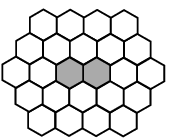} \right) \approx -0.04,\quad\
\Delta E\left( \figbox{0.6}{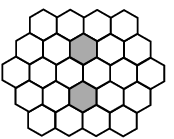} \right) \approx -0.07.
\]
\centerline{\begin{tabular}%
{|c|c|@{\ }c@{\ }|@{\ }c@{\ }|c|c|c|@{\ }c@{\ }|@{\ }c@{\ }|}
\cline{1-4}\cline{6-9}
& Phase &
\parbox{37pt}{\vskip3pt Vortex density \vskip3pt} &
\parbox{53pt}{\vskip3pt Energy per $\hexagon$ and per vortex \vskip3pt} &&
& Phase &
\parbox{37pt}{\vskip3pt Vortex density \vskip3pt} &
\parbox{53pt}{\vskip3pt Energy per $\hexagon$ and per vortex \vskip3pt} \\
\cline{1-4}\cline{6-9}
\cline{1-4}\cline{6-9}
1 & $\figbox{1}{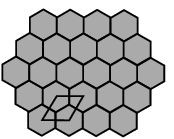}$ &
$\displaystyle\frac{1}{1}$ & $\displaystyle\afrac{0.067}{0.067}$ &&
8 & $\figbox{1}{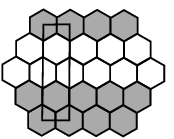}$ &
$\displaystyle\frac{2}{4}$ & $\displaystyle\afrac{0.042}{0.085}$ \\
\cline{1-4}\cline{6-9}
2 & $\figbox{1}{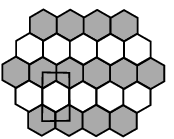}$ &
$\displaystyle\frac{1}{2}$ & $\displaystyle\afrac{0.052}{0.104}$ &&
9 & $\figbox{1}{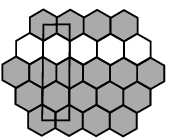}$ &
$\displaystyle\frac{3}{4}$ & $\displaystyle\afrac{0.059}{0.078}$ \\
\cline{1-4}\cline{6-9}
3 & $\figbox{1}{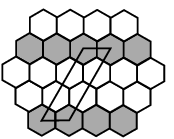}$ &
$\displaystyle\frac{1}{3}$ & $\displaystyle\afrac{0.041}{0.124}$ &&
10& $\figbox{1}{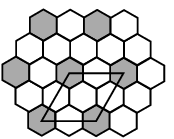}$ &
$\displaystyle\frac{1}{4}$ & $\displaystyle\afrac{0.042}{0.167}$ \\
\cline{1-4}\cline{6-9}
4 & $\figbox{1}{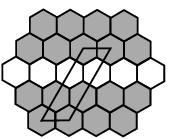}$ &
$\displaystyle\frac{2}{3}$ & $\displaystyle\afrac{0.054}{0.081}$ &&
11& $\figbox{1}{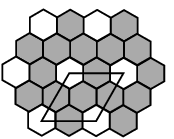}$ &
$\displaystyle\frac{3}{4}$ & $\displaystyle\afrac{0.074}{0.099}$ \\
\cline{1-4}\cline{6-9}
5 & $\figbox{1}{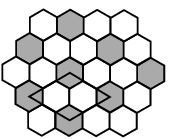}$ &
$\displaystyle\frac{1}{3}$ & $\displaystyle\afrac{0.026}{0.078}$ &&
12& $\figbox{1}{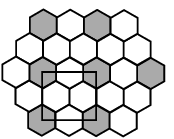}$ &
$\displaystyle\frac{1}{4}$ & $\displaystyle\afrac{0.025}{0.101}$ \\
\cline{1-4}\cline{6-9}
6 & $\figbox{1}{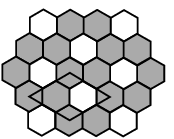}$ &
$\displaystyle\frac{2}{3}$ & $\displaystyle\afrac{0.060}{0.090}$ &&
13& $\figbox{1}{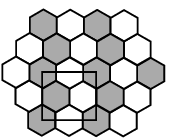}$ &
$\displaystyle\frac{2}{4}$ & $\displaystyle\afrac{0.046}{0.092}$ \\
\cline{1-4}\cline{6-9}
7 & $\figbox{1}{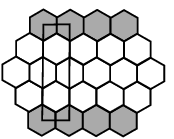}$ &
$\displaystyle\frac{1}{4}$ & $\displaystyle\afrac{0.034}{0.136}$ &&
14& $\figbox{1}{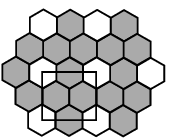}$ &
$\displaystyle\frac{3}{4}$ & $\displaystyle\afrac{0.072}{0.096}$ \\
\cline{1-4}\cline{6-9}
\end{tabular}}
\caption{Numerical results for $J_{x}=J_{y}=J_{z}=1$. Vortices are shown in
gray; for periodic phases the unit cell is indicated by a parallelogram.}
\label{tab_phases}
\end{table}

\pagebreak
The energy of a vortex pair is smaller than $2E_{vortex}$ if the vortices are
close to each other: by $\approx 0.04$ for nearest neighbors and by $\approx
0.07$ for next-nearest neighbors. (We didn't try to compensate the finite size
effects, so the precision is poor). These numbers suggest that inter-vortex
interaction is rather strong and could, in principle, result in some
configurations having negative energy. However, our further calculations give
evidence for the contrary.

We have checked all periodic phases with unit cell containing $1$, $2$, $3$ or
$4$ hexagons, see Table~\ref{tab_phases}. (As mentioned above, this
computation requires much less computer resources than the study of separate
vortices or vortex pairs). In all these cases the energy is positive and
increases as more vortices are added. The smallest energy per vortex is
achieved by phases $1$ and $5$ --- $0.067$ and $0.078$, respectively.

All $14$ phases have positive energies (relative to the vortex-free phase) for
all nonzero values of $J_{x}$, $J_{y}$, $J_{z}$.

\section{Edge modes in phases $B_{\nu}$} \label{sec_edge}

It is understood that the edge spectrum depends strongly on particular
conditions at the edge. The calculations below are only meant to illustrate
the universal feature of the spectrum --- the existence of a chiral gapless
mode.

Let us suppose that the honeycomb lattice fills the lower half-plane and is
cut as follows:
\[
\centerline{\epsfbox{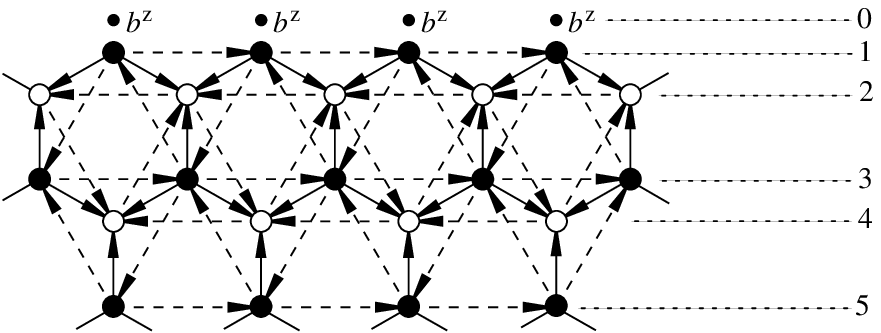}}
\]
Row $0$ consists of the Majorana operators $b^{z}_{j}$, which would be
decoupled from the rest of the system if not for the magnetic field.
Performing the Fourier transform in the horizontal direction, we compute
effective couplings between the rows as functions of $q_{x}$:
\begin{equation}
iA(q_{x})\,=\,
\begin{pmatrix}
0 & i\gamma & 0 &&&&\\
-i\gamma & \alpha & is & -\beta &&&\\
0 & -is & -\alpha & ir & \beta &&\\
& -\beta & -ir & \alpha & is & -\beta &\\
&& \beta & -is & -\alpha & ir & \ddots\\
&&& -\beta & -ir & \alpha & \ddots \\
&&&& \ddots & \ddots & \ddots
\end{pmatrix},
\qquad \text{where}\qquad
\begin{array}{r@{}l}
r &{}=2J, \smallskip\\
s &{}=-4J\cos\frac{q_{x}}{2}, \smallskip\\
\alpha &{}=4\kappa\sin q_{x}, \smallskip\\
\beta &{}=4\kappa\sin\frac{q_{x}}{2}, \smallskip\\
\gamma &{}=-2h_{z}.
\end{array}
\end{equation}

\begin{figure}
\centerline{\begin{tabular}{c@{\qquad\qquad}c}
$\figbox{1.0}{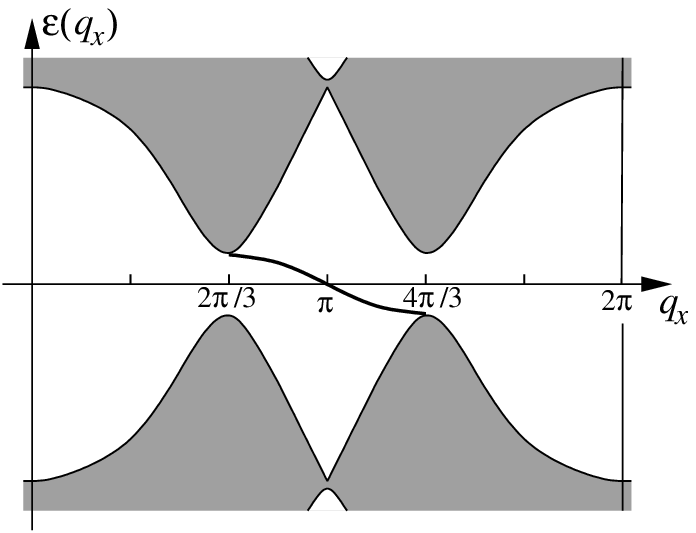}$ & $\figbox{1.0}{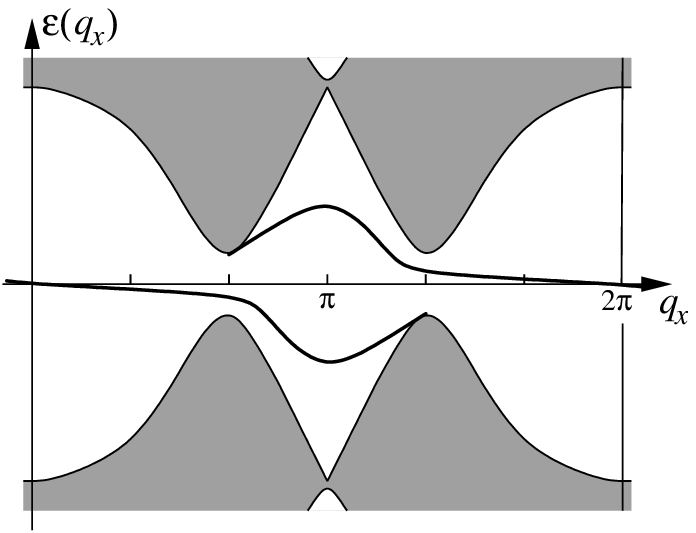}$
\\
a) & b)
\end{tabular}}
\caption{Schematic form of the edge spectrum in the simplest case (a) and with
the operators $b_{j}^{z}$ taken into account (b). The shaded area represents
the bulk spectrum.}
\label{fig_edgespec}
\end{figure}

Let us first find edge modes ignoring the operators $b^{z}_{j}$, i.e., the
first row and column in the above matrix. If $\kappa=0$, then for
$2\pi/3<q_{x}<4\pi/3$ the matrix has a null vector with elements $\psi(2j)=0$
and $\psi(2j+1)=\left(-2\cos\frac{q_{x}}{2}\right)^{j}$, which corresponds to
a zero-energy state localized near the edge. If $\kappa$ is not zero but still
small, we get this spectrum:
\begin{equation}
\eps(q_{x})\approx 12\kappa\sin q_{x},\qquad\quad q_{x}\in[2\pi/3,\,4\pi/3].
\end{equation}
It is shown in Fig.~\ref{fig_edgespec}a, assuming that $\kappa>0$, i.e.,
$\nu=+1$. The point where the energy curve crosses zero, $q_{x}=\pi$,
corresponds to a left-moving gapless mode.

Now we take the operators $b^{z}_{j}$ into account. If $\kappa=0$, then
a zero mode exists for $q_{x}\in[-2\pi/3,\,2\pi/3]$, whereas for
$q_{x}\in[2\pi/3,\,4\pi/3]$ there are two modes with energies
\begin{equation}
\eps(q_{x})\approx\pm\gamma\sqrt{1-4\cos^{2}\tfrac{q_{x}}{2}}.
\end{equation}
The spectrum for $\kappa>0$ is shown schematically in
Fig.~\ref{fig_edgespec}b. In this case a left-moving gapless mode occurs at
$q_{x}=0$.

In conclusion, let us give a simple (but not completely rigorous) proof of
Eq.~(\ref{numedgemod}). The idea is due to Laughlin~\cite{Laughlin81}: we put
the system on a cylinder and run magnetic flux through it. For simplicity, we
consider a cylinder of the smallest radius, which is equal to the lattice
period divided by $2\pi$. Thus the Hamiltonian is given by the matrix
$iA(q_{x})$; the variable $q_{x}$ plays the role of the magnetic flux.

Let $q_{x}$ vary from $0$ to $2\pi$. When the energy of an edge
state\footnote{We are using the first quantization formalism, therefore
``states'' are just superpositions of lattice points.}  $|\psi\rangle$ crosses
zero, the spectral projector $P(q_{x})$ changes by
$|\psi\rangle\langle\psi|$. For an arbitrary operator $Q$ the quantity
$\Tr(QP(q_{x}))$ changes by $\langle\psi|Q|\psi\rangle$. Let
\begin{equation}
Q_{jk}= g(j)\delta_{jk},\qquad\quad
g(j)=\left\{
\begin{array}{c@{\quad}l}
1 & \text{near the edge},
\\
0 & \text{far away from the edge}
\end{array}
\right.
\end{equation}
(the exact form of the function $g$ is not important). Then
$\langle\psi|Q|\psi\rangle\approx 1$. Since $\Tr(QP(q_{x}))$ is a periodic
function of $q_{x}$, the abrupt changes must be compensated by a continuous
variation, which we call ``adiabatic''. Thus
\begin{equation} \label{Iedge}
\nu_{\text{edge}}=\int I(q_{x})\,dq_{x}, \qquad\quad
\text{where}\quad I(q_{x})=-\Tr(Q\dot{P}),\quad
\dot{P}=\left(\frac{dP}{dq_{x}}\right)_{\text{adiabatic}}.
\end{equation}
  
The adiabatic evolution of the spectral projector may be represented by the
Heisenberg equation with a suitable Hamiltonian:
\begin{equation}
\dot{P}=i[H,P], \qquad\quad \text{where}\,\ H= i[P,\dot{P}].
\end{equation}
Indeed, $P^{2}=P$, hence $P\dot{P}+\dot{P}P=\dot{P}$ and $P\dot{P}P=0$,
implying that $-[[P,\dot{P}],P]=\dot{P}$. Therefore
\begin{equation} \label{Iedge1}
I(q_{x})=-i\Tr\bigl(QHP-QPH\bigr)
=-i\Tr\bigl(HPQ-HQP\bigr)=-i\Tr\bigl(H[P,Q]\bigr).
\end{equation}
(These transformations are valid because each trace is represented by a finite
sum.) The last expression in Eq.~(\ref{Iedge1}) may be calculated using the
spectral projector for the bulk. Indeed, if the function $g$ were constant,
the commutator $[P,Q]$ would vanish; thus the main contribution comes from the
region where $g$ changes from $1$ to $0$.

In the final part of the proof, we take into account the periodicity of the
bulk spectral projector, representing the site index $j$ as $(s,\lambda)$. As
is usual, lattice cells are indexed by $s$; it is an integer that increases in
the positive $y$-direction. We may assume that the function $g$ depends only
on $s$ and satisfies\, $\lim_{s\to\infty}g(s)=1$,\, $\lim_{s\to-\infty}g(s)=0$.
Then
\[
\bigl([P,Q]\bigr)_{t\mu,s\lambda}=\bigl(g(s)-g(t)\bigr)P_{t\mu,s\lambda},
\qquad\qquad
\sum_{t}\bigl(g(t+r)-g(t)\bigr)=r.
\]
Using these auxiliary identities, we pass to the momentum representation:
\[
\begin{array}{r@{}l}
\Tr\bigl(H[P,Q]\bigr)\;
& \displaystyle
=\; \sum_{s,\lambda,t,\mu}H_{s\lambda,t\mu}P_{t\mu,s\lambda}
\bigl(g(s)-g(t)\bigr)
\;=\; \sum_{r,\lambda,\mu}H_{0\lambda,-r\mu}P_{0\mu,r\lambda}
\sum_{t}\bigl(g(t+r)-g(t)\bigr)
\medskip\\
& \displaystyle
=\; \sum_{r,\lambda,\mu}H_{0\lambda,-r\mu}P_{0\mu,r\lambda}\,r
\;=\; -i\int\Tr\biggl(\widetilde{H}\,
\frac{\partial\widetilde{P}}{\partial q_{y}}\biggr)\frac{dq_{y}}{2\pi}
\;=\; \int\Tr\biggl(\biggl[\widetilde{P},\,
\frac{\partial\widetilde{P}}{\partial q_{x}}\biggr]\,
\frac{\partial\widetilde{P}}{\partial q_{y}}\biggr)\frac{dq_{y}}{2\pi}
\end{array}
\]
Substituting the result into~(\ref{Iedge1}) and~(\ref{Iedge}), we get
\[
\nu_{\text{edge}}\;=\; \int I(q_{x})\,dq_{x}
\;=\; \frac{-i}{2\pi}\int\Tr\biggl(\biggl[\widetilde{P},\,
\frac{\partial\widetilde{P}}{\partial q_{x}}\biggr]\,
\frac{\partial\widetilde{P}}{\partial q_{y}}\biggr)\,dq_{x}\,dq_{y}
\;=\; \nu.
\]

\section{Quasidiagonal matrices} \label{sec_locmatr}

The goal of this appendix is to provide a formal argument for the existence of
unpaired Majorana modes on vortices for odd values of the Chern number
$\nu$. However, the developed formalism may be interesting on its own right.
It suggests a rather efficient approach to problems like quantization of Hall
conductivity in disordered systems without the use of excessively heavy tools.
Some of results (in particular, the ones concerning the flow of a matrix and
the Chern number) are actually poor man's variants of known mathematical
theorems.\footnote{Note for experts: we are effectively trying to build a
K-theory on a manifold by considering functions that are not continuous, but
rather constant on cells that are dual to simplices. This seems to be an
awkward approach, but its possible advantage is the relation to a
second-quantized case, see Appendix~\ref{sec_anomaly}.} A powerful theory,
called noncommutative geometry~\cite{Connes} was used by Bellissard \latin{at
al}~\cite{BES-B94} to prove rigorously that the Hall conductivity is quantized
provided the electron are localized (see~\cite{ASS90,ASS94} for another
proof). We will use similar ideas, but focus mainly on providing the intuition
rather than mathematical rigor.

A physical example of a quasidiagonal matrix is an electron hopping matrix
$T=(t_{jk})$ on a $d$-dimensional lattice. It is a Hermitian matrix with the
property that $t_{jk}$ is bounded in magnitude and vanishes if the distance
$|j-k|$ between the sites $j$ and $k$ is greater that some constant
$L$. Furthermore, if $T$ has a spectral gap, i.e., if the eigenvalues are
bounded away from zero, then the corresponding spectral projector
$P=\frac{1}{2}\bigl(1-\sgn T\bigr)$ is also quasidiagonal. More specifically,
the matrix elements $P_{jk}$ decay exponentially with
distance.\footnote{Indeed, the function $f(x)=\sgn x$ can be approximated by a
sequence of polynomials $p_{n}(x)$ of degree $n\to\infty$ which converges
exponentially on the spectrum of $T$. Therefore $\|P-p_{n}(T)\|<ab^{n}$ for
some $b<1$ . On the other hand, the matrix elements
$\bigl(p_{n}(T)\bigr)_{jk}$ vanish if $|j-k|>nL$.}

In general, a \emph{quasidiagonal matrix} is a lattice-indexed matrix
$A=(A_{jk})$ with sufficiently rapidly decaying off-diagonal
elements. Technically, one requires that
\[
|A_{jk}|\le c|j-k|^{-\alpha},\qquad\quad \alpha>d,
\]
where $c$ and $\alpha$ are some constants, and $d$ is the dimension of the
space.  Note that ``lattice'' is simply a way to impose coarse $\RR^{d}$
geometry at large distances. We may think about the problem in these terms:
matrices are operators acting in some Hilbert space, and lattice points are
basis vectors. But the choice of the basis need not be fixed. One may safely
replace the basis vector corresponding to a given lattice point by a linear
combination of nearby points. One may also use some kind of coarse-graining,
replacing the basis by a decomposition into orthogonal subspaces corresponding
to groups of points, or regions in $\RR^{d}$.

Let us outline the main results. We first consider quasidiagonal unitary
matrices in one dimension and define an integral topological characteristic
called \emph{flow}. Then we study projection matrices in two dimensions. The
\emph{Chern number} $\nu(P)$ of a quasidiagonal projection matrix $P$ is
expressed directly in terms of the matrix elements $P_{jk}$, see
Eqs.~(\ref{Chern}) and~(\ref{2-current}). This definition does not rely on
translational invariance, but in the translationally invariant case we
reproduce Eq.~(\ref{nu}).

\begin{figure}[t]
\centerline{\begin{tabular}{c@{\qquad\qquad}c@{\qquad\qquad}c}
$\figbox{1.0}{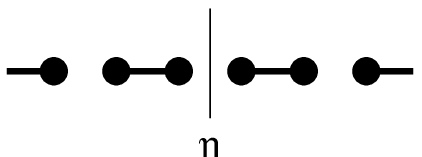}$
& $\figbox{1.0}{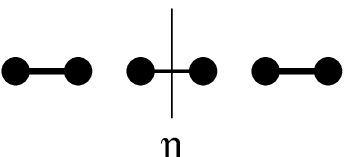}$
& $\figbox{1.0}{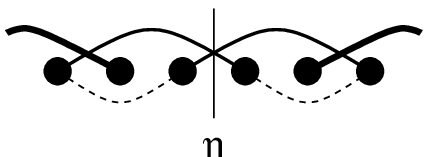}$
\smallskip\\
a) & b) & c)
\end{tabular}}
\caption{Cutting a Majorana chain: $\calM_{\eta}(B)=1$ in cases (a) and (c)
whereas $\calM_{\eta}(B)=-1$ in case (b). The dashed lines in (c) show a
possible way to reconnect the broken pairs.}
\label{fig_pairing}
\end{figure}

After those preliminaries, we switch to questions related to Majorana fermions
(all necessary background is given in Section~\ref{sec_quadratic}). From the
mathematical point of view, we study quasidiagonal real skew-symmetric
matrices $B$ satisfying the condition $B^{2}=-1$. In one dimension, such a
matrix is assigned a \emph{cutting obstruction} $\calM_{\eta}(B)=\pm 1$ with
respect to an arbitrary cut $\eta$ dividing the chain into two parts. Let us
explain the physical meaning of this number. The matrix $B$ defines a pairing
of Majorana modes (in the exact sense stated right after Eq.~(\ref{specproj}),
but we will use a cartoon description for illustration). If we cut the chain,
the resulting pieces may carry unpaired boundary modes~\cite{qwires}. This
happens when an odd number of pairs is broken, as in Fig.~\ref{fig_pairing}b;
otherwise one can modify the matrix near the cut so that to avoid broken
pairs, see Fig.~\ref{fig_pairing}c. More formally, let us consider
quasidiagonal real skew-symmetric matrices $B'$ that agree with $B$ at
infinity but have no nonzero elements across the cut. We prove that if
$\calM_{\eta}(B)=-1$, then no such $B'$ exists that satisfies the equation
$(B')^{2}=-1$. Note that while $\calM_{\eta}(B)$ depends on the cut, the
\emph{relative Majorana number} of two matrices on the same chain,
$\calM(A,B)=\calM_{\eta}(A)\calM_{\eta}(B)$ is invariant.

\begin{figure}[t]
\centerline{\begin{tabular}{c@{\qquad\qquad}c}
$\figbox{1.0}{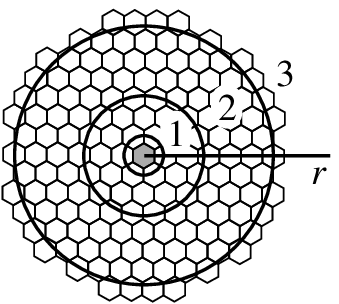}$ & $\figbox{1.0}{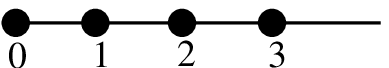}$
\\
a) & b)
\end{tabular}}
\caption{A map from two to one dimension: annular regions around the vortex
(a) are collapsed into points on a ray (b).}
\label{fig_annuli}
\end{figure}

The concept of Majorana number is also applicable in two dimensions. Let
$B^{2}=-1$, and let us construct another matrix $B'$ by inserting a
$\ZZ_{2}$-vortex at the origin. More exactly, $B'_{jk}=\pm B_{jk}$, where the
minus sign occurs when the link $(j,k)$ crosses some fixed ray $r$ (an
analogue of the Dirac string). Note that the condition $(B')^{2}=-1$ is only
true asymptotically (i.e., far away from the origin); it may or may not be
possible to satisfy this equation by altering the matrix elements near the
vortex. In fact, \emph{the presence of an unpaired Majorana mode at the vortex
is defined as locally unrepairable failure of the equation
$(B')^{2}=-1$}. Such a mode is detected by an \emph{absolute Majorana number}
$\calM(B')$. For its construction, let us regard $B$ and $B'$ as
one-dimensional by keeping track of the radial direction only: we divide the
plane into concentric rings and map all sites in each ring to a single
location on a ray (see Fig.~\ref{fig_annuli}). To simplify the calculation, we
cut out the interior of a sufficiently large circle $\eta$; let $|\eta|$ be
the number of sites removed. The presence or absence of an unpaired mode in
this case is given by $\calM_{\eta}(B')$, which is defined using some annular
neighborhood of $\eta$. Extrapolating to $|\eta|=0$, we get
$\calM(B')=(-1)^{|\eta|}\calM_{\eta}(B')=\calM(B,B')$. The last number only
depends on the matrix elements of $B$ in some neighborhood of the intersection
point between $\eta$ and $r$. Finally, we show that
$\calM(B,B')=(-1)^{\nu(P)}$, where $\nu(P)$ is the Chern number associated
with the projector $P=\frac{1}{2}(1-iB)$. Thus, vortices carry unpaired
Majorana modes if and only if $\nu(P)$ is odd.

\subsection{The flow of a unitary matrix}

\subsubsection{Definition} Let us consider an arbitrary (possibly infinite)
unitary matrix $U=(U_{jk})$. We refer to the values of $j$ and $k$ as
``sites'' and define a ``current'' flowing from $k$ to $j$:
\begin{equation}
f_{jk} = |U_{jk}|^{2}-|U_{kj}|^{2}.
\end{equation}
Since $U$ is unitary, $\sum_{j}U_{jl}^{*}U_{jk}
=\sum_{j}U_{lj}U_{kj}^{*}=\delta_{lk}$, therefore the current is conserved at
each site:
\begin{equation}
\sum_{j}f_{jk} = 0.
\end{equation}
Suppose that the sites are positioned on a line (more specifically, enumerated
by integers), and that the off-diagonal matrix elements of $U$ decay fast
enough. We will see later that the condition
\begin{equation} \label{decay1d}
|U_{jk}|\le c|j-k|^{-\alpha},\qquad
\text{where}\,\ \alpha>1,
\end{equation}
is sufficient for our purposes. For now, let us simply assume that $U_{jk}$
vanishes if $|j-k|$ is greater than some constant $L$.  Then it is obvious
that the total current through a ``cross section'' $\eta$,
\begin{equation} \label{flow}
\calF(U)=\sum_{j\ge\eta}\sum_{k<\eta}f_{jk},
\end{equation}
does not depend on the choice of $\eta$. The number $\calF(U)$ is called the
\emph{flow} of $U$. For example, the matrix with entries
$U_{jk}=\delta_{j,\,k+s}$ has flow $s$.

\subsubsection{Integrality of the flow} A nontrivial property is that the
flow is \emph{quantized}, i.e., has an integer value. To prove this statement,
let us introduce the projector onto the sites $M$ through $N$:
\begin{equation}\label{projMN}
\Pi^{[M,N]}_{jk} =
\left\{\begin{array}{ll}
1 & \text{if}\ M\le j=k\le N,\smallskip\\
0 & \text{otherwise}.
\end{array}\right.
\end{equation}
The projector onto an infinite interval is defined similarly. Let
$\Pi=\Pi^{[0,\infty)}$; then
\begin{equation}\label{flow1}
\calF(U)\,=\,
\Tr\bigl(U^{\dag}\Pi U(1-\Pi)\bigr)-\Tr\bigl(U^{\dag}(1-\Pi)U\Pi\bigr)
\,=\,\Tr\bigl(U^{\dag}\Pi U-\Pi\bigr).
\end{equation}
(\emph{Caution:} In the last expression, the order of operations is important:
we first compute the matrix $\Lambda=U^{\dag}\Pi U-\Pi$, and then its trace,
$\sum_{j}\Lambda_{jj}$. It is not possible to use the cyclic property of the
trace, $\Tr AB=\Tr BA$, since it is only valid if one of the matrices has a
finite number of nonzero elements or, more generally, if the sum involved in
the calculation of the trace converges absolutely.)

Note that the matrices $\Pi$ and $U^{\dag}\Pi U$ are orthogonal projectors. If
they were finite, their traces would be integers, and the difference would
also be an integer. In the infinite case, one may refer to the general notion
of the relative index of two projectors~\cite{ASS94,ASS94-1}. However, we will
proceed in a more pedestrian fashion and simply truncate the matrices.

Let $\Lambda=U^{\dag}\Pi U-\Pi$. It is clear that the matrix element
$\Lambda_{jk}$ vanishes if $|j|$ or $|k|$ is greater than $L$, so $\Lambda$ is
not changed by the truncation to the interval $[-L,L]$. Hence
\begin{equation}
\calF(U)=\Tr(U^{\dag}\Pi U)_{\mathrm{trunc}}-\Tr \Pi_{\mathrm{trunc}},
\qquad \text{where}\,\ A_{\mathrm{trunc}}\bydef\Pi^{[-L,L]}A\Pi^{[-L,L]}.
\end{equation}
If $A$ is an orthogonal projector and commutes with $\Pi^{[-L,L]}$, then
$A_{\mathrm{trunc}}$ is also an orthogonal projector. Obviously, $\Pi$ and
$\Lambda$ commute with $\Pi^{[-L,L]}$, and so does the matrix $U^{\dag}\Pi
U=\Pi+\Lambda$. Thus both $\Pi_{\mathrm{trunc}}$ and $(U^{\dag}\Pi
U)_{\mathrm{trunc}}$ are orthogonal projectors. It follows that
$\calF(U)$ is an integer.

In equation~(\ref{flow1}), the projector $\Pi$ may be replaced by another
operator with the same asymptotics at infinity. (In other words, we can
``blur'' the boundary between the left and the right half-line.) Indeed, the
expression $\Tr(U^{\dag}A U-A)$ vanishes if the cyclic property of the trace
is true, i.e., if the elements of $A$ decay at infinity fast enough. Adding
such a matrix $A$ to $\Pi$ will not change the result. Thus
\begin{equation}\label{flow2}
\calF(U)=\Tr\bigl(U^{\dag}QU-Q\bigr)=\Tr\bigl(U^{\dag}[Q,U]\bigr),
\qquad
\text{where}\quad
Q_{jk}=\left\{\begin{array}{ll}
\delta_{jk} & \text{for}\ j,k\to+\infty,\smallskip\\
0 & \text{for}\ j,k\to-\infty.
\end{array}\right.
\end{equation}
(Here we also assume that the matrix $Q$ is quasidiagonal, i.e, $|Q_{jk}|$ is
bounded by a rapidly decaying function of $|j-k|$, cf.~(\ref{decay1d}).)

\subsubsection{Translationally-invariant case} The flow can be easily
calculated if $U$ possesses a translational symmetry.  Let us group the sites
by unit cells and index them as $(s,\lambda)$, where $s$ is the number of the
cell, and $\lambda$ refers to a position type inside the cell. We assume that
$U_{s\lambda,t\mu}$ depends only on $t-s$, $\lambda$ and $\mu$. In this case,
we can express $\calF(U)$ in terms of the trace per unit cell (denoted
by $\tr$) and a position operator $X$:
\begin{equation}\label{flow3}
\calF(U)=\tr\bigl(U^{\dag}[X,U]\bigr),\qquad
\text{where}\,\ 
X_{s\lambda,t\mu}=s\,\delta_{st}\delta_{\lambda\mu}.
\end{equation}
(For a proof, replace the factor $s$ in the definition of $X$ by the function
$g(s)$ such that $g(s)=s$ for $|s|\le N$ and $g(s)=N\sgn s$ for $|s|>N$, where
$N$ is large; then use Eq.~(\ref{flow2})).

In the momentum representation, the operator $U$ becomes
$\widetilde{U}_{\mu\lambda}(q) =\sum_{t}e^{i(\qq,t)}U_{0\lambda,t\mu}$. We may
also use these simple rules:
\begin{equation}\label{Xqrules}
[X,A] \,\rightarrow\,
i\,\frac{d\widetilde{A}}{dq},\qquad
\tr A \,\rightarrow\,
\int_{-\pi}^{\pi}\frac{dq}{2\pi}\Tr\widetilde{A}.
\end{equation}
Thus
\begin{equation}\label{flow4}
\calF(U)\,=\, \frac{i}{2\pi}\int_{-\pi}^{\pi}
\Tr\biggl(\widetilde{U}^{\dag}\frac{d\widetilde{U}}{dq}\biggr)\,dq \,=\,
-\frac{1}{2\pi i}\int_{q=-\pi}^{q=\pi} d\bigl(\ln\det\widetilde{U}(q)\bigr).
\end{equation}

\subsection{General setting}

The result about the flow quantization can be extended to matrices satisfying
condition~(\ref{decay1d}), which guarantees that the sum in Eq.~(\ref{flow})
converges absolutely. 

Another generalization is coarse-graining, which means to allow multiple
states per site. The sites are now described by orthogonal subspaces of the
total Hilbert space. This generalization is useful if want to treat a
two-dimensional system as one-dimensional.

Finally, we may try to apply the notion of flow to finite systems. Let us
assume that the unitarity condition,
$\sum_{j}U_{jl}^{*}U_{jk}=\sum_{j}U_{lj}U_{kj}^{*}=\delta_{lk}$, is
approximate and holds only if $j$ and $k$ belong to some interval. In this
case, one needs to restrict the sum~(\ref{flow}) to the same interval, and the
result will not be an exact integer. But one can actually give an upper bound
for the deviation from the closest integer.

Setting the last generalization aside, let us put the integrality theorem
into a rigorous form. (The reader may safely skip this formalism and proceed
to the next subsection.)
\begin{Theorem}\label{th_flow}
Let $U$ be a unitary operator acting in the Hilbert space
$\calH=\bigoplus_{j=-\infty}^{\infty}\calH_{j}$ and represented by a matrix
whose entries $U_{jk}$ are linear maps from $\calH_{k}$ to
$\calH_{j}$. Suppose that the off-diagonal entries are Hilbert-Schmidt
operators satisfying the condition
\begin{equation}
\|U_{jk}\|_{\HS}\le c|j-k|^{-\alpha}
\end{equation}
where $c$ and $\alpha>1$ are some constants, and $\|\cdot\|_{\HS}$ is
the Hilbert-Schmidt norm. Then the sum in the expression for the flow,
\begin{equation}\label{flowHS}
\calF(U)=\sum_{j\ge 0}\sum_{k<0}
\bigl(\Tr(U_{jk}^{\dag}U_{jk})-\Tr(U_{kj}^{\dag}U_{kj})\bigr)
\end{equation}
converges absolutely and has an integer value.
\end{Theorem}

\begin{proof}[Proof sketch.] Let $\calH_{-}=\bigoplus_{j<0}\calH_{j}$ and
$\calH_{+}=\bigoplus_{j\ge 0}\calH_{j}$. An operator $A$ is said to be
\emph{quasi-trace-class} relative to the decomposition
$\calH=\calH_{-}\oplus\calH_{+}$ if the matrix elements $A_{++}$ and $A_{--}$
are trace class, whereas $A_{+-}$ and $A_{-+}$ are Hilbert-Schmidt. By
definition, $\Tr A=\Tr A_{++}+\Tr A_{--}$. An operator $B$ is said to be
\emph{quasi-bounded} if $B_{++}$, $B_{--}$ are bounded and $B_{+-}$, $B_{-+}$
are Hilbert-Schmidt. Both types of operators form Banach spaces with respect
to suitable norms; quasi-trace-class operators form an ideal in the algebra of
quasi-bounded operators.

One can show that the operator $\Lambda=U^{\dag}\Pi U-\Pi$ is
quasi-trace-class. Also, the operators $P(L)=\Pi^{[-L,L]}U^{\dag}\Pi
U\Pi^{[-L,L]}$ are almost projectors, in the sense that
$\lim_{L\to\infty}(P(L)^{2}-P(L))=0$ with respect to the quasi-trace
norm. Hence $P(L)$ can be approximated by a projector, which has an integer
trace.
\end{proof}

We omit the details and proceed to higher dimensions. Generally, our style
will not be very rigorous, but one can hopefully elaborate the results using
the following definition.
\begin{Definition}
Let a positive integer $d$ (the dimension) and a real number $\alpha>d$ be
fixed. A matrix $(A_{jk}:\, j,k\in\ZZ^{d})$ with operator entries is called
\emph{quasidiagonal} if all its off-diagonal entries are Hilbert-Schmidt, and
there are some constants $c$ and $c'$ such that
\begin{equation}
\label{loc1}
\|A_{jj}\|\le c,\qquad\quad
\|A_{jk}\|_{\HS}\le c'|j-k|^{-\alpha}\quad \text{for}\,\ j\not=k.
\end{equation}
\end{Definition}

Note that if $A$ is unitary, then $\|A_{jj}\|\le 1$
automatically. Quasidiagonal matrices form a Banach algebra with the norm
\begin{equation}
\label{loc2}
\|A\|_{\mathrm{qd}}\,=\,\sup_{j}\|A_{jj}\|
+\gamma\sup_{j\not=k}|j-k|^{\alpha}\|A_{jk}\|_{\HS}\,,
\end{equation}
where $\gamma$ is a sufficiently large constant.

\subsection{Chern number}\label{sec_locChern}

\subsubsection{General definition of $\nu(P)$}
Let $P$ be an orthogonal projector represented by a quasidiagonal matrix in
two dimensions. (For simplicity, we assume that the matrix elements $P_{jk}$
are scalars, though they may be operators as well.) For each triple $(j,k,l)$
we define a ``$2$-current'':\footnote{Another name of this object is
``simplicial $2$-chain''. More exactly, the $2$-chain is the formal sum
$\sum_{j<k<l}h_{jkl}\Delta_{jkl}$, where $\Delta_{jkl}$ is a combinatorial
simplex.}
\begin{equation}\label{2-current}
h_{jkl} \,=\, 12\pi i \bigl(P_{jk}P_{kl}P_{lj}-P_{jl}P_{lk}P_{kj}\bigr).
\end{equation}
It is clear that $h_{jkl}$ is antisymmetric in all three indices.
Since $P$ is Hermitian, $h_{jkl}$ is a real number. Moreover, since $P^{2}=P$,
\begin{equation}\label{curr2conserv}
(\partial h)_{kl} \bydef \sum_{j}h_{jkl}=0.
\end{equation}

Let us partition the plane into three sectors and define a quantity $\nu(P)$,
which will be shown to generalize the notion of the Chern number:
\begin{equation}\label{Chern}
\figbox{1.0}{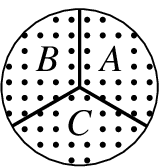}\qquad\qquad
\nu(P) \,=\,
h(A,B,C)\,\bydef\,\sum_{j\in A}\sum_{k\in B}\sum_{l\in C} h_{jkl}.
\end{equation}
This sum converges absolutely. On the other hand, its value does not change if
one reassigns any site from one sector to another (due to
Eq.~(\ref{curr2conserv})).  Therefore $\nu(P)$ is a topological invariant: it
is constant, provided that $A$, $B$, $C$ are arranged in the counterclockwise
order. We will see that $\nu(P)$ is actually an integer, and its value agrees
with Eq.~(\ref{nu}) in the translationally invariant case.

\subsubsection{Some properties and the translationally invariant case}
Let us rewrite Eq.~(\ref{Chern}) in an operator form using the relation
$A+B+C=1$, where the symbols $A$, $B$, $C$ designate the projectors on the
corresponding sectors:
\begin{align}
\nu(P)\,&=\, 12\pi i \bigl(\Tr(APBPCP)-\Tr(APCPBP)\bigr)
\nonumber\\[2pt]
&\hskip-5mm=\, 4\pi i\Tr\bigl(APBPCP+BPCPAP+CPAPBP-APCPBP-CPBPAP-BPAPCP\bigr)
\nonumber\\[2pt]
\label{Chern1}
&=\,4\pi i\Tr\bigl(PAPBP-PBPAP\bigr)\,=\, 4\pi i\Tr\bigl[PAP,PBP\bigr]
\,\bydef\,\nu(P,A,B).
\end{align}
Note that the value of $\nu(P,A,B)$ will not change if we add arbitrary finite
matrices to $A$ and $B$. Moreover, the commutator $[PAP,PBP]$ is nonzero only
in the region where the supports of $A$, $B$, and $C=1-A-B$ touch each other,
hence we can make arbitrary changes away from the triple contact point. Thus
the operators $A$, $B$, $C$ do not have to be projectors or even commute with
each other; it is only important that they are supported by regions with the
particular topological configuration (possibly with some overlap along the
boundaries and triple overlap at the center).

This freedom in the choice of $A$ and $B$ is very useful. As one application,
we show that $\nu(P)$ is additive. Let $P_{1}$ and $P_{2}$ be projectors onto
orthogonal subspaces (i.e., $P_{1}P_{2}=P_{2}P_{1}=0$) and let
$P_{3}=1-P_{1}-P_{2}$. We may replace $A$, $B$, and $C=1-A-B$ with
\[
A'=\sum_{k=1}^{3}P_{k}AP_{k},\qquad
B'=\sum_{k=1}^{3}P_{k}BP_{k},\qquad
C'=\sum_{k=1}^{3}P_{k}CP_{k},
\]
which differ from $A$, $B$, and $C$ only at the boundaries of the
corresponding regions. Then
\begin{gather*}
(P_{1}+P_{2})A'(P_{1}+P_{2})B'(P_{1}+P_{2})
=P_{1}AP_{1}BP_{1}+P_{2}AP_{2}BP_{2},
\\[2pt]
\nu\bigl(P_{1}+P_{2},A,B\bigr)\,=\,\nu\bigl(P_{1}+P_{2},A',B'\bigr)
\,=\,\nu(P_{1},A,B)+\nu(P_{2},A,B).
\end{gather*}
Thus
\begin{equation}
\nu(P_{1}+P_{2})=\nu(P_{1})+\nu(P_{2}),\qquad\quad \nu(1-P)=-\nu(P).
\end{equation}

Now, we calculate $\nu(P,A,B)$ for a different topological topological
configuration. Let $\Pi^{(x)}=\Pi^{(1)}+\Pi^{(4)}$ and
$\Pi^{(y)}=\Pi^{(2)}+\Pi^{(1)}$ be the projectors onto the right and the upper
half-plane, respectively (see the picture below). Then
\begin{equation}\label{Chern2}
\figbox{1.0}{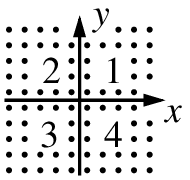}\quad
\begin{aligned}
&\nu(P,\Pi^{(x)},\Pi^{(y)})\,&
\\[2pt]
&\;=\, \nu(P,\Pi^{(1)},\Pi^{(2)})+\nu(P,\Pi^{(1)},\Pi^{(1)})
+\nu(P,\Pi^{(4)},\Pi^{(2)})+\nu(P,\Pi^{(4)},\Pi^{(1)})
\\[2pt]
&\;=\, \nu(P)+0+0+\nu(P)\,=\, 2\nu(P).
\end{aligned}
\end{equation}
It follows that
\begin{equation}\label{Chern3}
\nu(P)\,=\, 2\pi i\Tr\bigl[P\Pi^{(x)}P,\,P\Pi^{(y)}P\bigr]
\,=\, 2\pi i\Tr\Bigl(P\bigl[[\Pi^{(x)},P],[\Pi^{(y)},P]\bigr]+
P[\Pi^{(x)},\Pi^{(y)}]P\Bigr),
\end{equation}
where the last term vanishes because $\Pi^{(x)}$ and $\Pi^{(y)}$ commute. Of
course, $\Pi^{(x)}$, $\Pi^{(y)}$ may be replaced by any topologically
equivalent pair of operators, i.e., we may deform or blur the boundaries
of the corresponding half-planes.

In the translationally invariant case, the right-hand side of
Eq.~(\ref{Chern3}) can be calculated by widening the fuzzy boundaries so that
they turn into linear functions. Thus
\begin{equation}\label{Chern4}
\nu(P)\,=\, 2\pi i\tr\Bigl(P\bigl[[X,P],[Y,P]\bigr]\Bigr),
\end{equation}
where $X$ and $Y$ are the operators of $x$- and $y$-coordinate (resp.), and
$\,\tr\,$ is the trace per unit cell. Passing to the momentum representation
and using a two-dimensional analogue of the correspondence~(\ref{Xqrules}),
we recover Eq.~(\ref{nu}).

\subsubsection{The integrality of $\nu(P)$}
We have seen $\nu(P)$ has topological nature; let us show that it is an
integer. Let
\begin{equation}
Q^{(x)}=P\Pi^{(x)}P,\quad Q^{(y)}=P\Pi^{(y)}P,\qquad
U^{(x)}=\exp\bigl(2\pi iQ^{(y)}\bigr),\quad
U^{(y)}=\exp\bigl(-2\pi iQ^{(x)}\bigr).
\end{equation}
The operator $U^{(x)}$ coincides with the identity matrix away from the
$x$-axis, and $U^{(y)}$ is equal to the identity away from the $y$-axis.
We may proceed in two different ways.

\begin{proof}[First proof.]
Let us regard the two-dimensional lattice as one-dimensional by identifying
all sites in each vertical row. Then $U^{(x)}$ satisfies the hypothesis of
Theorem~\ref{th_flow}, hence it has an integer flow.  But the flow can be
expressed using Eq.~(\ref{flow2}) with $Q=Q^{(x)}$. Thus
\begin{equation}
\begin{aligned}
\calF(U^{(x)})
\,&=\,\Tr\Bigl(e^{-2\pi iQ^{(y)}}Q^{(x)}e^{2\pi iQ^{(y)}}-Q^{(x)}\Bigr)
\,=\,\int_{0}^{2\pi}\Tr\biggl(\frac{d}{d\ph}
\Bigl(e^{-i\ph Q^{(y)}}Q^{(x)}e^{i\ph Q^{(y)}}\Bigr)\biggr)d\ph
\\[2pt]
&=\,2\pi i\Tr\bigl[Q^{(x)},Q^{(y)}\bigr]\,=\, \nu(P).
\end{aligned}
\end{equation}
Note an analogy with the Laughlin argument: the integration over $\ph$
corresponds to the adiabatic insertion of a magnetic flux quantum.
\end{proof}

\begin{proof}[Second proof (sketch).]
Let us employ the notion of \emph{Fredholm determinant}. It is a
generalization of the determinant to infinite matrices that are close to the
identity, specifically of the form $1+K$, where $K$ has a well-defined
trace:\footnote{It is important that any matrix product that occurs in the
problem has a trace satisfying the cyclic property.}
\begin{equation}
\det(1+K)\, \bydef\, \exp\bigl(\Tr\ln(1+K)\bigr)
\,=\, 1+\Tr K+\left(-\frac{1}{2}\Tr K^{2}+\frac{1}{2}(\Tr K)^{2}\right)+\ldots.
\end{equation}
The exact meaning of this definition is as follows: we first obtain a formal
expression as a power series in $K$ and then evaluate each term. (If $K$ is a
matrix of size $n$, then all terms of degree higher than $n$ vanish.) We claim
that
\begin{equation}
\det\Bigl(U^{(x)}U^{(y)}
\bigl(U^{(x)}\bigr)^{-1}\bigl(U^{(y)}\bigr)^{-1}\Bigr)=1.
\end{equation}
Indeed, the identity $\det\bigl(ABA^{-1}B^{-1}\bigr)=1$ is true at the formal
power series level, assuming that the cyclic property of the trace holds for
any product of $A-1$, $A^{-1}-1$, $B-1$, $B^{-1}-1$ containing at least one
factor with $A$ and at least one factor with $B$. On the other hand, one can
show that
\begin{equation}
\det\Bigl(e^{i\ph_{y}Q^{(y)}}e^{i\ph_{x}Q^{(x)}}
e^{-i\ph_{y}Q^{(y)}}e^{-i\ph_{x}Q^{(x)}}\Bigr)
\,=\,\exp\Bigl(\ph_{x}\ph_{y}\Tr\bigl[Q^{(x)},Q^{(y)}\bigr]\Bigl).
\end{equation}
\end{proof}

\subsection{Majorana numbers}

In this section we study real skew-symmetric matrices satisfying the equation
$B^{2}=-1$. Recall that such a matrix defines the ground state $|\Psi\rangle$
of a Majorana system with a quadratic Hamiltonian (cf.\ Eqs.~(\ref{matrixB})
and~(\ref{specproj})).  First, we consider finite or general infinite matrices
(i.e., work in dimension zero), then proceed to quasidiagonal matrices in
dimension one, and finally to dimension two.

\subsubsection{Dimension zero: fermionic parity and the Pfaffian} 

A few more words about the Majorana system, then we will turn to matrices. Let
us group the sites (or the corresponding Majorana operators) into pairs
$(2k-1,2k)$ to form ``full'' fermionic modes with the occupation number
operators $a_{k}^{\dag}a_{k}=\frac{1}{2}(1+ic_{2k-1}c_{2k})$. Note that the
operator $-ic_{2k-1}c_{2k}$ has eigenvalue $1$ if the mode is empty and $-1$
if it is occupied. Thus the total \emph{fermionic parity} is characterized by
the operator $\prod_{k}(-ic_{2k-1}c_{2k})$. The fermionic parity of the ground
state is equal to the Pfaffian of the matrix $B$:
\begin{equation}
\prod_{k}(-ic_{2k-1}c_{2k})|\Psi\rangle=(\Pf B)|\Psi\rangle,\qquad\quad
\Pf B=\pm 1.
\end{equation}

Let us recall some standard definitions. The Pfaffian of a skew-symmetric
matrix is expressed as a sum over all partitions of the set $\{1,\dots,2N\}$
into pairs, or over all elements of the permutation group $S_{2N}$:
\begin{equation}\label{Pfaffian}
\Pf A \,=\, \frac{1}{2^{N}N!}\sum_{\tau\in S_{2N}}
\sgn(\tau)\,A_{\tau(1),\tau(2)}\cdots A_{\tau(2N-1),\tau(2N)}.
\end{equation}
For example,
\[
\Pf\begin{pmatrix}0&a_{12}\\-a_{12}&0\end{pmatrix}=a_{12},\qquad\quad
\Pf\begin{pmatrix} 0&a_{12}&a_{13}&a_{14}\\ -a_{12}&0&a_{23}&a_{24}\\
-a_{13}&-a_{23}&0&a_{34}\\ -a_{14}&-a_{24}&-a_{34}&0\end{pmatrix}
=a_{12}a_{34}+a_{14}a_{23}-a_{13}a_{24}.
\]
The Pfaffian satisfies the equations
\begin{equation}
(\Pf A)^{2} =\det A,\qquad\quad \Pf(WAW^T)\,=\,\Pf(A)\,\det(W).
\end{equation}

We will also need a generalization of the Pfaffian to infinite matrices, which
is called \emph{Fredholm Pfaffian}~\cite{Rains00}. Let $A$ and $B$ be real
skew-symmetric, $B$ invertible, and $A-B$ have a well-defined trace. Then
\begin{equation}
\begin{aligned}
\Pf(A,B)\,\bydef\,& \sqrt{\det(1+K)}=\exp\left(\frac{1}{2}\Tr\ln(1+K)\right)
\\[2pt]
=\,& 1+\frac{1}{2}\Tr K
+\left(-\frac{1}{4}\Tr K^{2}+\frac{1}{8}(\Tr K)^{2}\right) +\ldots,\qquad
\text{where}\,\ K=AB^{-1}-1.
\end{aligned}
\end{equation}
This definition is understood as that of the Fredholm determinant: we first
obtain a formal expression as a power series in $K$ and then evaluate each
term. (For finite matrices of size $2N$ we get $\Pf(A,B)=(\Pf A)(\Pf B)^{-1}$;
note that the terms of degree higher than $N$ vanish). The Fredholm Pfaffian
has the following properties:
\begin{gather}
\Pf(B,B)=1,\qquad\quad \Pf(A,B)\Pf(B,C)=\Pf(A,C),
\\[2pt]
\Pf(A,B)^{2}=\det(AB^{-1}),\qquad \Pf(WAW^{T},VBV^{T})=\Pf(A,B)\det(V^{-1}W);
\\[2pt]
\text{if}\,\ A^{2}=B^{2}=-1,\quad\ \text{then}\,\ \Pf(A,B)=\pm 1.
\end{gather}

\subsubsection{Dimension one: cutting obstruction and the relative Majorana
number} Now, consider a matrix that describes the ground state of Majorana
fermions on an infinite chain. From a formal point of view, it suffices to say
that $B=(B_{jk}: j,k\in\ZZ)$ is quasidiagonal, real skew-symmetric, and that
$B^{2}=-1$. Suppose that we cut the chain between sites $\eta-1$ and
$\eta$. Is it possible to find a new matrix $B'$ that contains no cross
elements (i.e., has the form $B'_{--}\oplus B'_{++}$), satisfies the condition
$(B')^{2}=-1$, and coincides with $B$ far away from the cut? The following
quantity represents an obstruction for the construction of such a matrix:
\begin{equation}
\calM_{\eta}(B)=\Pf\bigl(V(\eta)BV(\eta),\,B\bigr)=\pm 1,\qquad\quad
\text{where}\,\ V(\eta)=1-2\Pi^{[\eta,\infty)}.
\end{equation}
Recall that $\Pi^{[\eta,\infty)}$ denotes the projector onto sites $\eta$ and
higher; $\bigl(V(\eta)BV(\eta)\bigr)_{jk}=\pm B_{jk}$, where the minus sign
occurs when $j$ and $k$ lie on different sides of the cut.

We claim that the conditions on $B'$ cannot be met if $\calM_{\eta}(B)=-1$.
Indeed,
\begin{equation}
\begin{aligned}
\calM_{\gamma}(B)\,\calM_{\eta}(B)^{-1}\,
&=\,\Pf\Bigl(V(\gamma)BV(\gamma),\,V(\eta)BV(\eta)\Bigr)
\\[2pt]
&=\,\det\bigl(V(\eta)^{-1}V(\gamma)\bigr) \,=\, (-1)^{\gamma-\eta}.
\end{aligned}
\end{equation}
The same is true for $B'$, hence the \emph{relative Majorana number},
\begin{equation}
\calM(B,B')\,\bydef\, \calM_{\gamma}(B)\,\calM_{\gamma}(B')
\end{equation}
does not depend on $\gamma$. However, $\calM_{\gamma}(B')=\calM_{\gamma}(B)$
if $|\gamma-\eta|$ is large enough (since $B$ and $B'$ agree far away from the
cut), while $\calM_{\eta}(B')=1$ (because $B'=B'_{--}\oplus B'_{++}$) and
$\calM_{\eta}(B)=-1$ (by assumption). We have arrived at a contradiction.

From the physical point of view, the impossibility to construct a suitable
matrix $B'$ indicates the presence of unpaired Majorana modes on both sides of
the cut. This situation may be described by matrices $B'_{--}$ and $B'_{++}$
having one zero eigenvalue.

The cutting obstruction and the relative Majorana number can also be expressed
as Fredholm determinants. To this end, consider the quasidiagonal matrix
\begin{equation}
W=V(0)\exp(\pi X),\qquad
\text{where}\,\ X=\frac{1}{2}\Bigl(\Pi^{(x)}B+B\Pi^{(x)}\Bigr), \quad
\Pi^{(x)}\bydef\Pi^{[0,\infty)}.
\end{equation}
Clearly, $X$ is real skew-symmetric and commutes with $B$. Hence $W$ is
orthogonal, and $WBW^{T}=V(0)BV(0)$. On the other hand, $X$ is equal to
$B$ at $+\infty$ and vanishes at $-\infty$, therefore $W$ is equal to $1$ at
$\pm\infty$. One can actually show that $W-1$ has a well-defined trace. It
follows that
\begin{equation}
\calM_{0}(B)\,=\,\Pf\left(WBW^{T},B\right)\,=\, \det W.
\end{equation}
Similarly, one can define $X'$, $W'$ as functions of $B'$. Thus
\begin{equation}\label{Majnum1}
\calM(B,B')\,=\, \det\left(W^{-1}W'\right)
\,=\,\det\bigl(e^{-\pi X}e^{\pi X'}\bigr).
\end{equation}

\subsubsection{Applications to two dimensions} Eq.~(\ref{Majnum1})
may be used as a more general definition of the relative Majorana number,
which works even if $\calM_{\eta}(B)$ and $\calM_{\eta}(B')$ do not exist. For
example, let $B$ be a quasidiagonal matrix in two dimensions and
$B'=(1-2\Pi^{(y)})B(1-2\Pi^{(y)})$. In other words, $B'_{jk}=\pm B_{jk}$,
where the minus sign occurs when $j$ and $k$ lie on different sides of the
$x$-axis. If we regard the system as one-dimensional (collapsing it in the $y$
direction), the Majorana number $\calM(B,B')$ is well-defined and can be
expressed in terms of the projector $P=\frac{1}{2}(1-iB)$. The calculation
follows.

First, we obtain an expression for the operator $X$:
\[
X=i\Bigl(P\Pi^{(x)}P-(1-P)\Pi^{(x)}(1-P)\Bigr).
\]
Then we use the relation $1-2\Pi^{(y)}=\exp\bigl(\pm i\pi\Pi^{(y)}\bigr)$:
\[
\calM(B,B')\,=\,
\det\Bigl(e^{-\pi X}e^{-i\pi\Pi^{(y)}}e^{\pi X}e^{i\pi\Pi^{(y)}}\Bigr)
\,=\, \exp\Bigl(i\pi^{2}\Tr\bigl[X,\Pi^{(y)}\bigr]\Bigr).
\]
Finally, we compute the trace of the commutator:
\begin{align*}
\Tr\bigl[X,\Pi^{(y)}\bigr]
\,&=\,\Tr\Bigl(\bigl(P^{2}+(1-P)^{2}\bigr)\bigl[X,\Pi^{(y)}\bigr]\Bigr)
\,=\,\Tr\Bigl(P\bigl[X,\Pi^{(y)}\bigr]P+(1-P)\bigl[X,\Pi^{(y)}\bigr](1-P)\Bigr)
\\[2pt]
&=\,i\Tr\Bigl[P\Pi^{(x)}P,\,P\Pi^{(y)}P\Bigr]
\,-\,i\Tr\Bigl[(1-P)\Pi^{(x)}(1-P),\,(1-P)\Pi^{(y)}(1-P)\Bigr]
\\[2pt]
&=\, \frac{1}{2\pi}\bigl(\nu(P)-\nu(1-P)\bigr) \,=\, \frac{1}{\pi}\nu(P),
\end{align*}
because $\nu(1-P)=-\nu(P)$. Thus
\begin{equation}\label{MajChern}
\calM(B,B')=(-1)^{\nu(P)}.
\end{equation}

Equation~(\ref{MajChern}) can be applied to the geometry shown in
Fig.~\ref{fig_annuli}. In this case, $B$ and $B'$ correspond to two different
$\ZZ_{2}$-field configurations: the first is regarded as vortex-free and the
second has a vortex at the origin. The specific assumptions are as follows:
(i) $B$ and $B'$ are real skew-symmetric quasidiagonal matrices in two
dimensions; (ii) $B$ and $B'$ coincide, except that $B'_{jk}=-B_{jk}$ for
those links which cross the ray $r$; (iii) $B^{2}=-1$ (therefore $(B')^{2}=-1$
far away from the vortex). Note that $\calM_{\eta}(B)$ and $\calM_{\eta}(B')$
may be defined for a sufficiently large loop $\eta$ enclosing the vortex. Let
$|\eta|$ be the number of sites inside the loop. Then the \emph{absolute
Majorana numbers} $\calM(B)\bydef (-1)^{|\eta|}\calM_{\eta}(B)$ and
$\calM(B')\bydef (-1)^{|\eta|}\calM_{\eta}(B')$ do not depend on $\eta$ and
indicate the presence of unpaired Majorana modes in $B$ and $B'$,
respectively. However, an unpaired mode cannot exist in $B$ since $B^{2}=-1$;
therefore $\calM(B,B')=\calM(B')$.  We conclude that
\begin{equation}
\calM(B')=(-1)^{\nu(P)}.
\end{equation}

\section{Some remarks on the chiral central charge}
\label{sec_anomaly}

This appendix is an attempt to understand the physical and mathematical
meaning of the chiral central charge beyond the CFT framework. Recall that
chiral central charge is just the coefficient $c_{-}$ in the edge energy
current formula~(\ref{heatcurrent}). This definition will be refined in
Sec.~\ref{sec_current}. It turns out that the edge energy current is a
\emph{property of the bulk ground state}. The edge current is related to a
bulk $2$-current. Although the theory is in its embryonic stage, it looks like
a second-quantized version of the Chern number for quasidiagonal matrices
(cf.\ Sec.~\ref{sec_locChern}).

In Sec.~\ref{sec_Dehn} we discuss modular transformations of the partition
function on a space-time torus. In particular, the phase factor $e^{-2\pi
ic_{-}/24}$ appears in the description of the Dehn twist along a time circle.
However, other modular transformations are difficult to define because
space and time are not physically equivalent.

\subsection{The edge energy current and a bulk 2-current}\label{sec_current}

\subsubsection{Energy current in the Hamiltonian formalism}
\enlargethispage{\baselineskip}

Let us represent the Hamiltonian as a sum of local terms:
\begin{equation}\label{locHam}
H=\sum_{j}H_{j},
\end{equation}
where $H_{j}$ is a Hermitian operator acting only on spins in some
neighborhood of the point $j$. Note that the decomposition into local terms is
not unique. We also assume that the ground state $|\Psi\rangle$ is separated
from excited states by an energy gap; thus equal-time spin correlators decay
exponentially with distance~\cite{Hastings04}. We may slightly extend the
notion of locality so that $H_{j}$ acts on all spins but
\begin{equation}\label{weak_locality}
\bigl\|[H_{j},\sigma_{k}^{\alpha}]\bigr\| \,\le\, u\bigl(|j-k|\bigr),
\end{equation}
where $u$ is some function with fast decay at infinity (faster than any
power). Then the equal-time correlators also decay faster than any power,
provided the spectrum is gapped.

Microscopic energy current can be defined for any system with
Hamiltonian~(\ref{locHam}) at finite temperature. It is convenient to fix the
temperature at $T=1$ and vary the Hamiltonian instead. The thermal average of
an operator $X$ is defined in the standard way:
\begin{equation}
\langle X\rangle=\Tr(\rho X),\qquad
\text{where}\quad \rho=Z^{-1}e^{-H},\quad Z=\Tr e^{-H}.
\end{equation}

According to the Heisenberg equation,
\begin{equation} \label{currentop}
\frac{dH_{j}}{dt}\,=\, -i[H_{j},H]\,=\,\sum_{k}\hat{f}_{jk},\qquad
\text{where}\quad \hat{f}_{jk}=-i[H_{j},H_{k}].
\end{equation}
Thus the operator $\hat{f}_{jk}$ characterizes the energy current from $k$ to
$j$. We denote its thermal average by $f_{jk}$:
\begin{equation} \label{microencurr}
f_{jk}= f(H_{j},H_{k}),\qquad\quad \text{where}\quad
f(A,B)\bydef\bigl\langle-i[A,B]\bigr\rangle.
\end{equation}
(Note that this definition depends on the decomposition of the Hamiltonian
into local terms.)  Of course, the energy current is conserved:
\begin{equation}
(\partial f)_{k}\,\bydef\,\sum_{j}f_{jk}\, =\, 0.
\end{equation}
Indeed,\vspace{-2pt}
\[
(\partial f)_{k}\,=\, f(H,H_{k}) \,=\,
\Tr\bigl(-i\rho[H,H_{k}]\bigr)
\,=\, \Tr\bigl(-i[\rho,H]H_{k}\bigr)\,=\,0,
\]
since $\rho$ and $H$ commute.

Let us try to apply the general formula~(\ref{microencurr}) to a specific
geometry. Let $H^{(\infty)}$ be the Hamiltonian of an infinite two-dimensional
system with a ground state $|\Psi\rangle$ and gapped excitations. To simulate
an edge, we introduce another Hamiltonian,
\begin{equation} \label{edgelocHam}
\figbox{1.0}{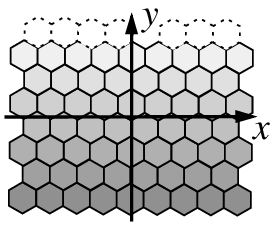}\qquad
H=\sum_{j}H_{j},\qquad
H_{j}\,= \left\{\begin{array}{c@{\quad}l}
0 & \text{if}\ y(j)\to+\infty,
\\[3pt]
\beta_{j}H_{j}^{(\infty)},\,\ \beta_{j}\to\infty & \text{if}\ y(j)\to-\infty,
\end{array}\right.
\end{equation}
where $y(j)$ is the $y$-coordinate of the point $j$. Thus the lower part of
the system is ``bulk material'' characterized by the state
$|\Psi\rangle\langle\Psi|$, whereas the upper part is ``empty space''
(actually, a set of noninteracting spins in the maximally mixed state). By
definition, the chiral central charge is $12/\pi$ times the total energy
current in the negative $x$-direction.

\subsubsection{Representing the ground state by local constraints}

At this point, we have encountered a technical difficulty: the energy current
does not necessarily vanish in the bulk. The problem can be avoided if we
describe the ground state by a set of local constraints. (A \emph{local
constraint} is a local operator that annihilates the ground state.)
Hamiltonians that are explicitly written as sums of local constraints, e.g.,
the Rokhsar-Kivelson model~\cite{RokhsarKivelson88,MoessnerSondhi00}, have
provided a lot of insight into properties of quantum many-body systems.

\begin{Proposition}\label{prop_loccons}
Any gapped local Hamiltonian $H=\sum_{j}H_{j}$ whose ground state
$|\Psi\rangle$ has zero energy can be represented as a sum of Hermitian local
operators $\widetilde{H}_{j}$ such that $\widetilde{H}_{j}|\Psi\rangle=0$.
\end{Proposition}

\begin{proof}[Proof sketch.]
Without loss of generality we may assume that
$\langle\Psi|H_{j}|\Psi\rangle=0$. (If not, change $H_{j}$ by a constant.) Let
all excited states have energy greater or equal to $\Delta$. Choose a smooth
function $\widehat{w}(\eps)$ such that
\[
\widehat{w}(-\eps)=\widehat{w}(\eps)^{*},\qquad\qquad
\widehat{w}(0)=1,\qquad\qquad
\widehat{w}(\eps)=0\,\ \text{for}\,\ |\eps|\ge\Delta.
\]
Its Fourier transform, $w(t)=\int\widehat{w}(\eps)\,e^{-\eps
t}\,\frac{d\eps}{2\pi}$ decays faster than any power of $t$ though more slowly
than $e^{-\gamma|t|}$ (because $\widehat{w}$ is smooth but not analytic). We
set
\begin{equation}
\widetilde{H}_{j}\,=\,\int_{-\infty}^{\infty}e^{iHt}H_{j}e^{-iHt}\,w(t)\,dt.
\end{equation}
It is easy to check that $\sum_{j}\widetilde{H}_{j}=H$ and
$\widetilde{H}_{j}|\Psi\rangle=0$. A locality condition of the
form~(\ref{weak_locality}) can be established using the bound of Lieb and
Robinson~\cite{LiebRobinson} on correlation propagation
(cf.~\cite{Hastings04}).
\end{proof}

For our purposes, we can actually use an arbitrary gapped Hamiltonian
$X=\sum_{j}X_{j}$ made up of local constraints for the given state
$|\Psi\rangle$. It turns out that \emph{the chiral central charge depends only
on the ground state}. Indeed, one can interpolate between different sets of
local constraints for the same state. Consider the situation where the
constrains are $X_{j}$ in close proximity of the edge, gradually changing to
$X'_{j}$ deeper in the bulk. No energy current is associated with the
transition between $X$ and $X'$.
\smallskip

As an aside, we conjecture that \emph{there is a set of local constraints
$Y_{j}$ such that the Hamiltonian $Y=\sum_{j}Y_{j}^{\dag}Y_{j}$ is
gapped}. Thus excitations can be efficiently detected locally, by coupling
each $Y_{j}$ to an external indicator:
$H_{\mathrm{detector}}=\sum_{j}\bigl(Y_{j}\otimes\bigl(|1\rangle\langle
0|\bigr)_{j}+Y_{j}^{\dag}\otimes\bigl(|0\rangle\langle 1|\bigr)_{j}\bigr)$. A
stronger version of this conjecture asserts that \emph{the original
Hamiltonian $H$ can be represented as $\sum_{j}Y_{j}^{\dag}Y_{j}$, where
$Y_{j}$ are local constraints}. This is true for the quadratic fermionic
Hamiltonian~(\ref{quadH}), which needs to be offset by its smallest
eigenvalue~(\ref{quadHenergy}) so that the ground state energy becomes
zero. It is easy to see that the matrix $|iA|$ is symmetric and that
$\sum_{j,k}\bigl(|iA|\bigr)_{jk}c_{j}c_{k}$ is equal to $\Tr|iA|$ times the
identity operator. Hence
\[
H\,=\, \frac{1}{4}\sum_{j,k}
\Bigl(iA_{jk}+\bigl(|iA|\bigr)_{jk}\Bigr)c_{j}c_{k} \,=\,\sum_{j,k} D_{jk}
c_{j}c_{k}, \qquad
\text{where}\,\ D\,=\,\frac{1}{4}\bigl(iA+|iA|\bigr)\ge\, 0
\]
(meaning that $D$ is positive semidefinite). The matrix $\sqrt{D}$ is
quasidiagonal, provided $A$ is quasidiagonal and has a spectral gap around
zero. Thus we can represent the Hamiltonian in the desired form:
\begin{equation}
H=\sum_{m}Y_{m}^{\dag}Y_{m},\qquad\quad \text{where}\quad\
Y_{m}\,=\,\sum_{k}\Bigl(\sqrt{D}\Bigr)_{mk}c_{k}.
\end{equation}

\subsubsection{Bulk $2$-current}

In the disk geometry (i.e., when interaction is strong near the origin and
vanishes at infinity), the chiral central charge is $12/\pi$ times the energy
current in the counterclockwise direction. Let us divide the disk into three
sectors, $A$, $B$, and $C$ in the counterclockwise order (see picture in
Eq.~(\ref{Chern}), or in Eq.~(\ref{sectanom}) below). In this setting, the
energy current is $\sum_{k\in B}\sum_{l\in A}f_{kl}$.

Let us try to represent the chiral central charge $c_{-}$ in a form that would
explicitly depend on the bulk Hamiltonian rather than the edge. To this end,
we construct a $2$-current $h$ such that
\begin{equation}
f=\partial h, \qquad\quad \text{where}\quad
(\partial h)_{kl}\bydef\sum_{j}h_{jkl}.
\end{equation}
(A \emph{$2$-current} is an antisymmetric function of three lattice sites that
decays fast enough as the distance between any two sites increases.) Then in
the three-sector geometry we have:
\[
\sum_{k\in B}\sum_{l\in A}f_{kl} \,=\,
\sum_{k\in B}\sum_{l\in A}\left(\sum_{j}h_{jkl}\right) \,=\,
-\sum_{j\in A}\sum_{k\in B}\sum_{l\in C} h_{jkl}.
\]
Thus,
\begin{equation} \label{sectanom}
\figbox{1.0}{sectors.eps}\qquad\qquad
c_{-}\,=\, -\frac{12}{\pi}\sum_{j\in A}\sum_{k\in B}\sum_{l\in C} h_{jkl}.
\qquad
\end{equation}
The last sum is dominated by a neighborhood of the triple contact point
between $A$, $B$, and $C$.

Unfortunately, there is no canonical expression for $h_{jkl}$. Instead, we can
define a canonical object $g$ which is a $2$-current on the lattice and a
$1$-form on the space of local Hamiltonians. It satisfies the condition
\begin{equation} \label{2-current_form}
\partial g=df.
\end{equation}
A suitable $2$-current $h$ is obtained by integrating $g$ over an arbitrary
path $H(\beta):\,\beta\in[0,\infty)$ in the space of spin Hamiltonians. Here
we assume that $H(\beta)$ is a sum of local terms $H_{j}(\beta)$, the
corresponding thermal state has short-range correlators, $H(0)=0$, and
$H(\beta)\approx\beta H^{(\infty)}$ as $\beta\to\infty$. Thus we implicitly
accept the following conjecture.

\begin{Conjecture}
The ground state of a gapped local Hamiltonian on a two-dimensional lattice
can be transformed into the maximally mixed state without a phase transition.
\end{Conjecture}
(In this formulation, $\beta$ changes from $\infty$ to $0$.) Note that usual
symmetry-related phase transitions are not a problem because $H(\beta)$ is not
required to be symmetric. Rather, we should worry about a transition from a
topological phase to a non-topological one. However, in two dimensions the
long-range topological order in the ground state is destroyed at any finite
temperature. Indeed, all topological excitations are point-like and not
confined, therefore they have a finite density at any non-zero
temperature. This argument does not apply to three-dimensional systems, where
string-like excitations are possible.

Let us give an explicit formula for $g$. For any two operators $A$ and $B$, we
define the Matsubara time-ordered average, as well as its ``truncated''
version:\footnote{This notation is reminiscent of the formula $A(\tau)=e^{\tau
H}Ae^{-\tau H}$, but the latter may be problematic to use because $e^{\tau
H}\approx\exp(\tau\beta H^{(\infty)})$ diverges as $\beta\to\infty$. Thus the
expression $\bigl\langle A(\tau)B(0)\bigr\rangle$ should be viewed as a
whole.}
\begin{align}
\bigl\langle A(\tau)&B(0)\bigr\rangle = 
Z^{-1}\Tr\bigl(e^{-(1-\tau)H}Ae^{-\tau H}B\bigr),\qquad
\text{where}\quad Z=\Tr e^{-H},\quad 0\le\tau\le 1;
\\[3pt]
\bdlangle A(\tau)&B(0)\bdrangle = 
\bigl\langle A(\tau)B(0)\bigr\rangle
-\langle A\rangle\langle B\rangle.
\end{align}
Next, we define a function of three operators:
\begin{equation}
\mu(A,B,C)\,=\, i\int_{0}^{1}
\bdlangle A(\tau)\,[B,C](0) \bdrangle\, d\tau.
\end{equation}
Finally, we assume that the Hamiltonian $H$ and the operators $A$, $B$, $C$
depend on some parameters, so that we can differentiate them. Now we can
define $g$:
\begin{equation}
g(A,B,C)=\mu(dA,B,C)+\mu(dB,C,A)+\mu(dC,A,B),\qquad
g_{jkl}=g(H_{j},H_{k},H_{l}).
\end{equation}

Let us verify Eq.~(\ref{2-current_form}). Note that
$\mu(X,Y,H)=-i\bigl\langle[X,Y]\bigr\rangle$ because $\bigl\langle X(\tau)\,
[Y,H](0)\bigr\rangle =\frac{d}{d\tau}\bigl\langle X(\tau)Y(0)\bigr\rangle$ and
$\bigl\langle[Y,H]\bigr\rangle=0$. We proceed as follows:
\begin{align*}
\partial g(B,C)\,&=\, g(H,B,C) \,=\, \mu(dH,B,C)+\mu(dB,C,H)+\mu(dC,H,B)
\\[2pt]
&=\, \Bigl(-iZ^{-1}\Tr\bigl(de^{-H}[B,C]\bigr)
-i\langle dH\rangle\langle[B,C]\rangle\Bigr)
-i\langle[dB,C]\rangle+i\langle[dC,B]\rangle
\\[2pt]
&=\,
d\Bigl(-iZ^{-1}\Tr\bigl(e^{-H}[B,C]\bigr)\Bigr)\,=\, df(B,C).
\end{align*}

\subsection{Modular transformations beyond CFT} \label{sec_Dehn}

The modular invariance~\cite{Verlinde} is usually formulated in the CFT
framework. In this section, we demonstrate some modular properties of the
partition function of edge excitations, not assuming conformal, rotational, or
any other symmetry. First, let us generalize the transformation of the vacuum
character\footnote{Some care should be taken to make a proper connection
between the physics of edge modes and the the CFT formalism. If all edge modes
have the same chirality, characters correspond to representations of some
chiral algebra (as is usual in CFT). In general, the vacuum character
$\chi_{1}$ is defined as a sum over all fields with trivial monodromy whereas
$\chi_{a}$ includes all fields that occur in the two-dimensional theory when
an anyon of type $a$ emerges on the surface (see Fig.~\ref{fig_cylinder}b on
page~\pageref{fig_cylinder}).}  under the Dehn twist:
\begin{equation}\label{Dehn_conf1}
\chi_{1}(w+1)\,=\,e^{-2\pi ic_{-}/24}\chi_{1}(w),
\end{equation}
where $w$ is the modulus of the torus and the subscript $1$ refers to the
vacuum sector.

In fact, the derivation of Eq.~(\ref{Dehn_conf1}) does not require much more
than the knowledge of the energy-momentum tensor, or just its $\tau
x$-component, which is exactly the energy current $I$. (We use imaginary time,
$\tau=it$.) In the thermodinamic limit, the coordinate
transformation\footnote{For lattice models, one should consider the
transformation $(j,\tau)\to(j,\,\tau+\xi_{j})$. It should be possible to give
an exact quantum-mechanical interpretation of such a reparametrization by
coupling the discrete derivative $\eta_{jk}=\xi_{j}-\xi_{k}$ to the energy
current operator $\hat{f}_{jk}=-i[H_{j},H_{k}]$, where $H_{j}$, $H_{k}$ are
local terms in the Hamiltonian. Such coupling is unambiguously defined only in
the first order in $\eta_{jk}$, but that does not affect the thermodynamic
limit.}  $(x,\tau)\to(x,\,\tau+\xi(x))$ induces the action change $\delta
S=iI\int \frac{d\xi}{dx}dx\,d\tau$. If the imaginary time is cyclic, then
twisting it by a full circle results in the action change $\delta
S=i\frac{\pi}{12}c_{-}$ and the multiplication of the partition function by
$e^{-S}=e^{-2\pi ic_{-}/24}$. We now describe a particular setting in which
this result is applicable.

Let us consider the partition function $Z=\Tr e^{-H/T}$ of a spin system on a
disk. As is usual, we think of the evolution over the time period $1/T$,
making the edge into a torus and the disk itself into the three-dimensional
manifold $M=D^{2}\times S^{1}$. Let us suppose that the temperature $T$ is
much smaller than the energy gap in the bulk, so that $Z$ is mostly determined
by edge excitations. In fact, the partition function depends on full detail of
the edge Hamiltonian as well as \emph{some} conditions in the bulk. For
example, we may place an anyon of type $a$ somewhere in the disk, which
slightly changes the edge excitation spectrum. In other words, we insert a
particle worldline $\ell$ into the manifold $M$, or act by the time-like
Wilson loop operator $W_{a}(\ell)$. Thus a partition function $Z_{a}$ is
defined. If all edge modes propagate with the same velocity $v$ and can be
described by a CFT, then $Z_{a}=\chi_{a}(-1/w)$, where $w=iLT/v$. In the limit
$w\to\infty$,
\begin{equation} \label{S_estimate}
Z_{a}\,=\,\chi_{a}(-1/w)\,=\,\sum_{b}\calS_{ab}\chi_{b}(w)
\,\approx\,\calS_{a1}\chi_{1}(w),
\end{equation}
because all characters $\chi_{b}(w)$ for $b\not=1$ are exponentially smaller
than the vacuum character $\chi_{1}(w)$.

The following argument does not depend on CFT. As already mentioned, twisting
the torus by $\xi$ along a time circle leads to the the action change by
$iI\xi/T$, or the multiplication of the partition function by $e^{-iI\xi/T}$.
(The sign in the exponent depends on the direction of the
twist; we choose it to be consistent with the CFT formula.)  For the full
twist, $\xi=1/T$, we get this equation:
\begin{equation} \label{Dehn0}
Z_{a}'\approx e^{-2\pi ic_{-}/24}Z_{a}.
\end{equation}
Note that the finite-size quantization of edge modes might affect the energy
current. We therefore assume that the disk circumference $L$ is much larger
than the edge correlation length $l_{\mathrm{edge}}\sim v/T$, where $v$ is the
maximum group velocity of any excitation in the system.\footnote{A rigorous
and completely general upper bound for the group velocity was obtained by Lieb
and Robinson~\cite{LiebRobinson}. It is consistent with the rough estimate
$v\lesssim rJ$, where $J$ is the inter-spin interaction strength and $r$ is
its range.} Small corrections proportional to $\exp(-L/l_{\mathrm{edge}})$ are
expected. Thus, Eq.~(\ref{Dehn0}) and its estimated precision are in perfect
agreement with the CFT formulas~(\ref{Dehn_conf1}) and~(\ref{S_estimate}).

With some more work, one can also derive an analogue of this CFT result:
\begin{equation}\label{Dehn_conf}
\chi_{b}(w+1)\,=\,e^{-2\pi ic_{-}/24}\theta_{b}\,\chi_{b}(w),
\end{equation}
where $\theta_{b}$ is the topological spin of $b$. In the non-conformal
setting, the role of the characters $\chi_{b}(w)$ is played by linear
combinations of the partition functions:
\begin{equation}\label{S_part}
\widetilde{Z}_{b}\,=\,\sum_{a}s_{b\anti{a}}Z_{a},
\end{equation}
where $s_{ab}$ are the entries of a \emph{topological S-matrix} defined in
terms of anyonic braiding, see Eq.~(\ref{sab}) on page~\pageref{sab}. (We
distinguish it form the modular matrix $\calS$, though they actually coincide
when the latter is defined.) A calculation based on the insertion of a
\emph{space-like} Wilson loop $W_{b}(\widetilde{\ell})$ shows that
\begin{equation}
\widetilde{Z}_{b}'\,\approx\, e^{-2\pi ic_{-}/24}\theta_{b}\,\widetilde{Z}_{b}.
\end{equation}
This equation is indeed similar to~(\ref{Dehn_conf}) with one important
difference: unlike $\chi_{a}(w)$ and $\chi_{a}(-1/w)$, the numbers
$\widetilde{Z}_{a}$ and $Z_{a}$ are \emph{not} obtained by evaluating the same
function. Another problem is that the Dehn twist can only be performed along a
time circle. It is therefore not clear whether the full modular invariance can
be established without conformal symmetry, or at least a $90^{\circ}$-rotation
symmetry of the space-time manifold.

\section{Algebraic theory of anyons} \label{sec_algth}

This appendix is an attempt to present an existing but difficult and somewhat
obscure theory in an accessible form, especially for the reader without
extensive field theory knowledge. (But we do assume some mathematical
background and the willingness to follow abstract arguments.) The presently
available resources may be divided into three categories: the original
field-theoretic papers where the relevant mathematical structure was
discovered~\cite{MooreSeiberg,Witten}, local field theory theory
papers~\cite{FRS89,FrohlichGabbiani90} (see Haag's book~\cite{Haag} for a
general reference on this subject), and purely mathematical
expositions~\cite{Turaev,BakalovKirillov,Kassel}. Unfortunately, all these
texts have different focus and/or are difficult to read. A nice elementary
introduction can be found in Preskill's lecture notes on quantum
computation~\cite{Preskill}, but more detail is needed for our purposes.

The theory described below is applicable to two-dimensional many-body systems
with short-range interactions and an energy gap. Using some intuition about
local excitations in such systems, we characterize the properties of anyons by
a set of axioms and derive some corollaries. This approach is somewhat similar
to what has been done rigorously in local field theory. However, we avoid many
difficulties by keeping the discussion at the physical level until all
essential properties are cast into a finite algebraic form, at which point we
enter the realm of mathematics.

More specifically, we claim that finite-energy excitations are classified by
\emph{superselection sectors}: each sector consists of states that can be
transformed one to another by local operations. (We do not care about rigorous
definition here.) It is assumed that the full classification can be
established by considering a small neighborhood of an arbitrary point (of size
compared to the correlation length). Another assumption is that local
excitations can be moved from one place to another by applying an operator
along an arbitrary path. Note that the quasiparticle transport need not be
adiabatic, nor we require translational symmetry. Topological properties, such
as braiding rules, have invariant meaning independent of possible disorder,
cf.\ Fig.~\ref{fig_exch} on page~\pageref{fig_exch}. Using some additional
arguments, we will arrive at a theory that may be called \emph{unitary braided
fusion category} (UBFC) ($=$~unitary ribbon category) in abstract
language.\footnote{Note for experts (the general reader shouldn't worry about
this): a \emph{fusion category} is a $k$-linear semisimple rigid tensor
category with finite-dimensional morphism spaces and finitely many simple
objects such that $\mathop{\mathrm{End}}(\unit)=k$, see~\cite{ENO}. We take
$k=\CC$. A unitary fusion category is automatically pivotal, spherical, and
nondegenerate. In the presence of braiding, these properties imply the
existence of a ribbon twist.}

\begin{figure}
\centerline{\begin{tabular}{c@{\qquad\qquad\qquad}c}
$\figbox{1.0}{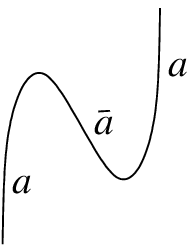}\ =\ \figbox{1.0}{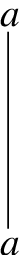}$ &
$\figbox{1.0}{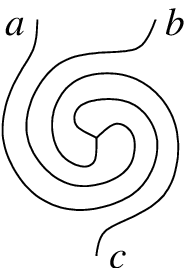}\ =\ \figbox{1.0}{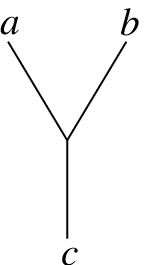}$
\\
a) & b)
\end{tabular}}
\caption{Some ``obvious'' identities.}
\label{fig_obvious}
\end{figure}

With this approach, we get a weaker set of axioms compared to an analogous
algebraic structure in conformal field theory. Therefore some ``obvious''
properties require a proof. (A similar situation occurs in local field theory,
but our exposition is different in some details and hopefully simpler.) For
nonrelativistic quantum-mechanical models, the space and time are not
equivalent even if we use the imaginary time formalism. Thus the notion of an
antiparticle being a particle propagating back in time must be applied with
caution. The identity in Fig.~\ref{fig_obvious}a illustrates a standard use of
this notion. (Shown in the figure are space-time diagrams; time goes up.) On
the left-hand side of the identity a particle $a$ annihilates with its
antiparticle $\anti{a}$ that was created as part of an $\anti{a}a$ pair. If
space and time were related by a rotational symmetry, then the zigzag on the
particle worldline could simply be removed. In our formalism, this equation is
satisfied by a suitable choice of normalization factors for the creation and
annihilation operators of particle-antiparticle pairs. However, the relation
in Fig.~\ref{fig_obvious}b is a nontrivial theorem; it follows from the
positivity of Hermitian inner product, see Section~\ref{sec_aniparticle}. (In
the categorical language, this result reads: ``Any unitary fusion category
admits a pivotal structure''.)

Having formulated fusion and braiding axioms, we will touch upon another
question: What UBFCs can be realized by a Hamiltonian with given symmetry
properties? The simplest case is where there is no built-in symmetry or
associated conservation laws. That is, we assume that the system consists of
spins (or other \emph{bosonic} degrees of freedom) and the Hamiltonian is in
\emph{generic position}. (Fermions are excluded because their number is always
conserved \latin{modulo}~$2$.)  Thus, any effective conservation law for
quasiparticles must have topological nature. In Sec.~\ref{sec_modularity} we
formulate this condition as a certain nondegeneracy property of the braiding
rules. Theories satisfying this axiom are called \emph{unitary modular
categories} (UMC).\footnote{The term ``modular'' refers to the possibility to
define a Hilbert space associated with a torus, on which the modular group
acts. More generally, one can construct a so-called \emph{modular functor} and
a \emph{topological quantum field theory}~\cite{Walker,Turaev,BakalovKirillov}
(but we do not cover these topics).}
\enlargethispage{3pt}

Unfortunately, UMC does not capture one important robust characteristic of the
physical system, namely the chiral central charge $c_{-}$. However, there is a
beautiful relation~\cite{FrohlichGabbiani90,Rehren90} that fixes\, $c_{-}\bmod
8$:
\begin{equation}\label{anomaly}
\calD^{-1}\sum_{a}d_{a}^{2}\theta_{a}\,=\, e^{2\pi ic_{-}/8}, \qquad\quad
\text{where}\quad \calD = \sqrt{\textstyle\sum_{a}d_{a}^{2}},
\end{equation}
$d_{a}$ is the \emph{quantum dimension} of the superselection sector $a$ (see
Sec.~\ref{sec_aniparticle}) and $\theta_{a}$ is the \emph{topological spin}
(see Sec.~\ref{sec_braiding}). The left-hand side of Eq.~(\ref{anomaly}) is
denoted by $\Theta$; its algebraic meaning will be explained in
Sec.~\ref{sec_modularity}. Roughly, $\Theta=S^{-1}TSTST$, where $S$ is a
so-called \emph{topological $S$-matrix} and $T$ is the diagonal matrix with
entries $\theta_{a}$. We will see that $\Theta$ is a root of unity, hence
Eq.~(\ref{anomaly}) implies that $c_{-}$ is rational.

The standard proof of Eq.~(\ref{anomaly}) is based on the assumption that the
system admits conformally invariant edge modes under suitable boundary
conditions. Then one can employ the modular invariance of the partition
function on a space-time torus, which is established in the CFT
framework~\cite{Verlinde}. It is not known whether the conformality hypothesis
is true in general. One may, however, hope to find a proof that would not
depend on conformal symmetry. In Sec.~\ref{sec_Dehn} we made partial progress
toward this goal. Specifically, we showed that the matrix $\calT=e^{-2\pi
ic_{-}/24}T$ corresponds to the Dehn twist along a time circle. While $S$ can
also be regarded as a modular transformation in a suitable mathematical
theory, it is not clear whether that theory is applicable in the same context
as the expression for $\calT$.

In the remaining part of the appendix we do not discuss the chiral central
charge. Our primary goal is to define a unitary braided fusion category (UBFC)
and its special case, unitary modular category (UMC). The definition is rather
long, so we break it into several parts. Each part contains some physical
motivation, formal axioms in terms of \emph{basic data} and \emph{equations},
as well as some corollaries. Note that our formulation of the axioms does not
involve the mathematical notion of category, so we call the corresponding
objects ``theories'' (e.g., ``fusion theory'' instead of ``fusion
category''). Only later do we introduce the powerful but heavy language of
categories and functors (also known as ``abstract nonsense''). It is surely
overkill for most of the problems considered in this paper, but the concept of
tensor functor may actually be useful in the study of phase transitions in
anyonic systems, e.g., Bose-condensation of spinless particle and weak
symmetry breaking. (The last topic is considered in
Appendix~\ref{sec_symbreak}.)

\subsection{Fusion theory} \label{sec_fusion}

In this section we consider anyons on a line (though the physical system is
two-dimensional). The motivation for the definitions has been provided in
Sec.~\ref{sec_assoc} (esp.\ in~\ref{sec_fusgen}). Each particle is
characterized by a superselection sector label (describing its ``anyonic
charge''). The position and local degrees of freedom may be ignored, but the
order of particles is important.  The main element of the the fusion theory is
the space $V_{c}^{ab}$ --- the state space of particles $a$ and $b$ restricted
to have total anyonic charge $c$. More exactly, vectors in this space
correspond to different ways of splitting $c$ into $a$ and $b$, or equivalence
classes of local operators that effect the splitting.

\subsubsection{Splitting and fusion operators}
While $V_{c}^{ab}$ may be called ``splitting space'', the corresponding fusion
space is denoted by $V^{c}_{ab}$. If $\psi\in V_{c}^{ab}$ is a splitting
operator, then $\psi^\dag\in V^{c}_{ab}$ is a fusion operator:
\begin{equation}
\text{splitting:}\quad \figbox{1.0}{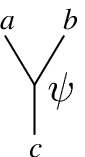}\;, \qquad\qquad
\text{fusion:}\quad \figbox{1.0}{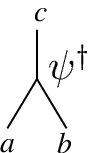}\;,
\end{equation}
where $\psi$ and $\psi^{\dag}$ label the corresponding vertices. One may also
use Dirac's notation, i.e., write $|\psi\rangle$ instead of $\psi$ and
$\langle\psi|$ instead of $\psi^\dag$. By definition, the Hermitian inner
product corresponds to the operator multiplication. Specifically, if
$\xi,\eta\in V_{c}^{ab}$, then
\begin{equation} \label{inprod}
\figbox{1.0}{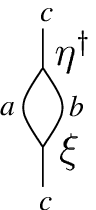}\;=\,\langle\eta|\xi\rangle\;\figbox{1.0}{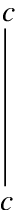}\;,
\qquad\quad \text{i.e.,}\qquad\quad
\eta^{\dag}\xi\,=\, \langle\eta|\xi\rangle\,\id_{c}.
\end{equation}

Similarly, one can define splitting and fusion of multiple particles. The most
general process is one transforming particles $b_{1},\dots,b_{m}$ into
$a_{1},\dots,a_{k}$. It can be performed in two steps: first, fusing all the
particles into one, and then splitting it as required. Therefore
\begin{equation} \label{fusplitspace}
V_{b_{1}\dots b_{m}}^{a_{1}\dots a_{k}}\,=\,
\bigoplus_{c}V_{c}^{a_{1}\dots a_{k}}\otimes V_{b_{1}\dots b_{m}}^{c}.
\end{equation}
For example, the identity operator acting on particles $a$ and  $b$ can be
decomposed as follows:
\begin{equation} \label{partidentity}
\figbox{1.0}{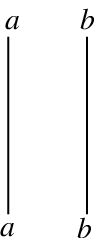}\ =\ \sum_{c} \sum_{j}\figbox{1.0}{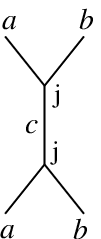}\;,
\end{equation}
where $j$ stands for both $\psi_{j}$ and $\psi_{j}^{\dag}$ --- basis vectors
in $V_{c}^{ab}$ and $V^{c}_{ab}$, respectively.

The description of splitting and fusion will be complete when we understand
relations between different spaces $V_{c}^{a_{1}\dots a_{k}}$. That is the
topic of the following subsection.

\subsubsection{Basic data and equations} The basic data include a set of
superselection sectors, fusion rules, and associativity relations. The latter
satisfy so-called pentagon and triangle equations. An additional condition
(requiring the nondegeneracy of certain operators) will be formulated in
Sec.~\ref{sec_aniparticle}.

\paragraph{Superselection sectors \textmd{(also called ``particle types'', or
``labels'')}} They form a finite set $M$.

\paragraph{Fusion rules} For any combination of $a,b,c\in M$ there is a fixed
finite-dimensional Hilbert space $V_{c}^{ab}$. The numbers $N^{c}_{ab}=\dim
V^{c}_{ab}=\dim V_{c}^{ab}$ are called \emph{fusion multiplicities}. The
choice of a special element $1\in M$ (the \emph{vacuum sector}) is also
considered part of the fusion rules.

\paragraph{Associativity relations} For each $a,b,c,d\in M$ there is a
canonical unitary isomorphism between two Hilbert spaces:
\begin{equation} \label{assoc}
\figbox{1.0}{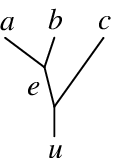} \:\xrightarrow{F_{u}^{abc}}\:
\figbox{1.0}{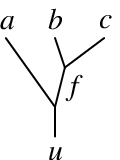} \qquad\qquad\quad
F_{u}^{abc}:\, \bigoplus_{e}V_{e}^{ab}\otimes V_{u}^{ec}
\,\to\, \bigoplus_{f}V_{u}^{af}\otimes V_{f}^{bc}.
\end{equation}
Note that both spaces in question are simply different representation of the
same physical space $V_{u}^{abc}$--- that of the particles $a$, $b$, and $c$
restricted to have total anyonic charge $u$. Therefore one may regard the
isomorphism $F_{u}^{abc}$ as equality. This view is formally justified by the
pentagon equation and MacLane's coherence theorem, which will be formulated
later. At a cruder, combinatorial level associativity may be expressed as
follows: $\sum_{e}N^{e}_{ab}N^{u}_{ec}\,=\,\sum_{f}N^{u}_{af}N^{f}_{bc}$.

From the physical perspective, splitting off (or fusing with) the vacuum
sector is trivial. In any splitting diagram all lines labeled by $1$ may be
simply erased. To guarantee the consistency of this procedure, we require that
the spaces $V_{a}^{a1}$ and $V_{a}^{1a}$ are not only one-dimensional, but
\emph{canonically isomorphic} to $\CC$. In other words, these spaces are
endowed with fixed unit vectors $|\alpha_{a}\rangle\in V_{a}^{a1}$ and
$|\beta_{a}\rangle\in V_{a}^{1a}$. The canonical isomorphisms are given by the
formulas
\begin{equation} \label{units}
\alpha_{a}:\CC\to V_{a}^{a1}\,:\quad
z\mapsto z|\alpha_{a}\rangle, \qquad\qquad
\beta_{a}:\CC\to V_{a}^{1a}\,:\quad
z\mapsto z|\beta_{a}\rangle,
\end{equation}
where $z\in\CC$. These isomorphisms must satisfy so-called triangle equations
(see below).

\begin{figure}[t]
\centerline{\begin{tabular}{@{}c@{\qquad\qquad}c@{}}
$\figbox{1.0}{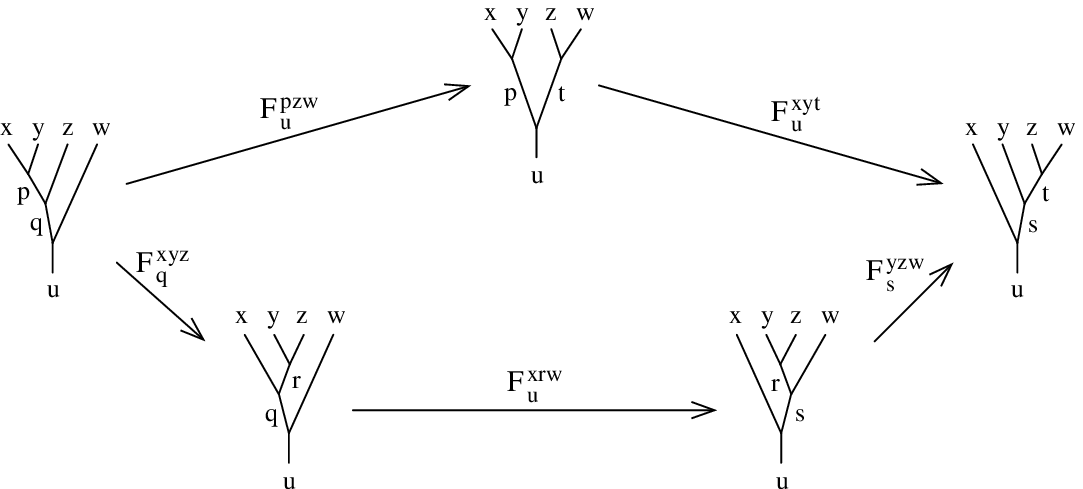}$ & $\figbox{1.0}{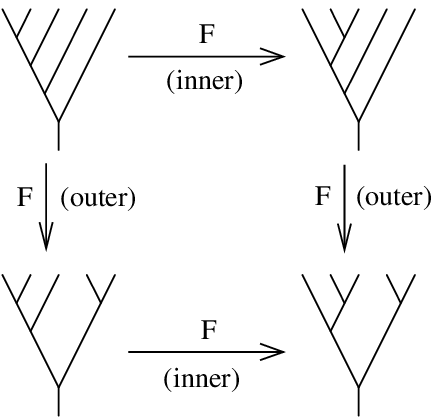}$
\\
a) & b)
\end{tabular}}
\caption{The pentagon equation (a) and a ``quadrilateral identity'' (b). (The
latter is satisfied automatically.)}
\label{fig_pentquadr}
\end{figure}

\paragraph{Pentagon equation}
The graphic representation of this axiom is shown in
Fig.~\ref{fig_pentquadr}a. Its exact meaning is this: for any $x,y,z,w,u\in M$
the following diagram commutes:
\begin{equation}
\begin{xy}
(-55,0)*+{\bigoplus_{p,q}V_{p}^{xy}\otimes V_{q}^{pz}\otimes V_{u}^{qw}}="1",
(0,14)*+{\bigoplus_{p,t}V_{p}^{xy}\otimes V_{u}^{pt}\otimes V_{t}^{zw}}="2",
(55,0)*+{\bigoplus_{s,t}V_{u}^{xs}\otimes V_{s}^{yt}\otimes V_{t}^{zw}}="3",
(-30,-24)*+{\bigoplus_{q,r}V_{q}^{xr}\otimes V_{r}^{yz}\otimes V_{u}^{qw}}="4",
(30,-24)*+{\bigoplus_{r,s} V_{u}^{xs}\otimes V_{r}^{yz}\otimes V_{s}^{rw}}="5",
\ar^-{F_{u}^{pzw}} (-4,0)+"1"; (-4,0)+"2"
\ar^-{F_{u}^{xyt}} (4,0)+"2"; (4,0)+"3"
\ar^{F_{q}^{xyz}} "1";"4"
\ar^{F_{u}^{xrw}} "4";"5"
\ar^{F_{s}^{yzw}} "5";"3"
\end{xy}
\end{equation}
Arrow labels are abbreviated: for example, the arrow on the left (labeled by
$F_{q}^{xyz}$) actually designates the map $\sum_{q}F_{q}^{xyz}\otimes
\id_{V_{u}^{qw}}$, where $\id_{V_{u}^{qw}}$ is the unit operator on
$V_{u}^{qw}$.

\begin{figure}[t]
\centerline{\begin{tabular}{@{}c@{\quad}c@{}}
$\figbox{0.8}{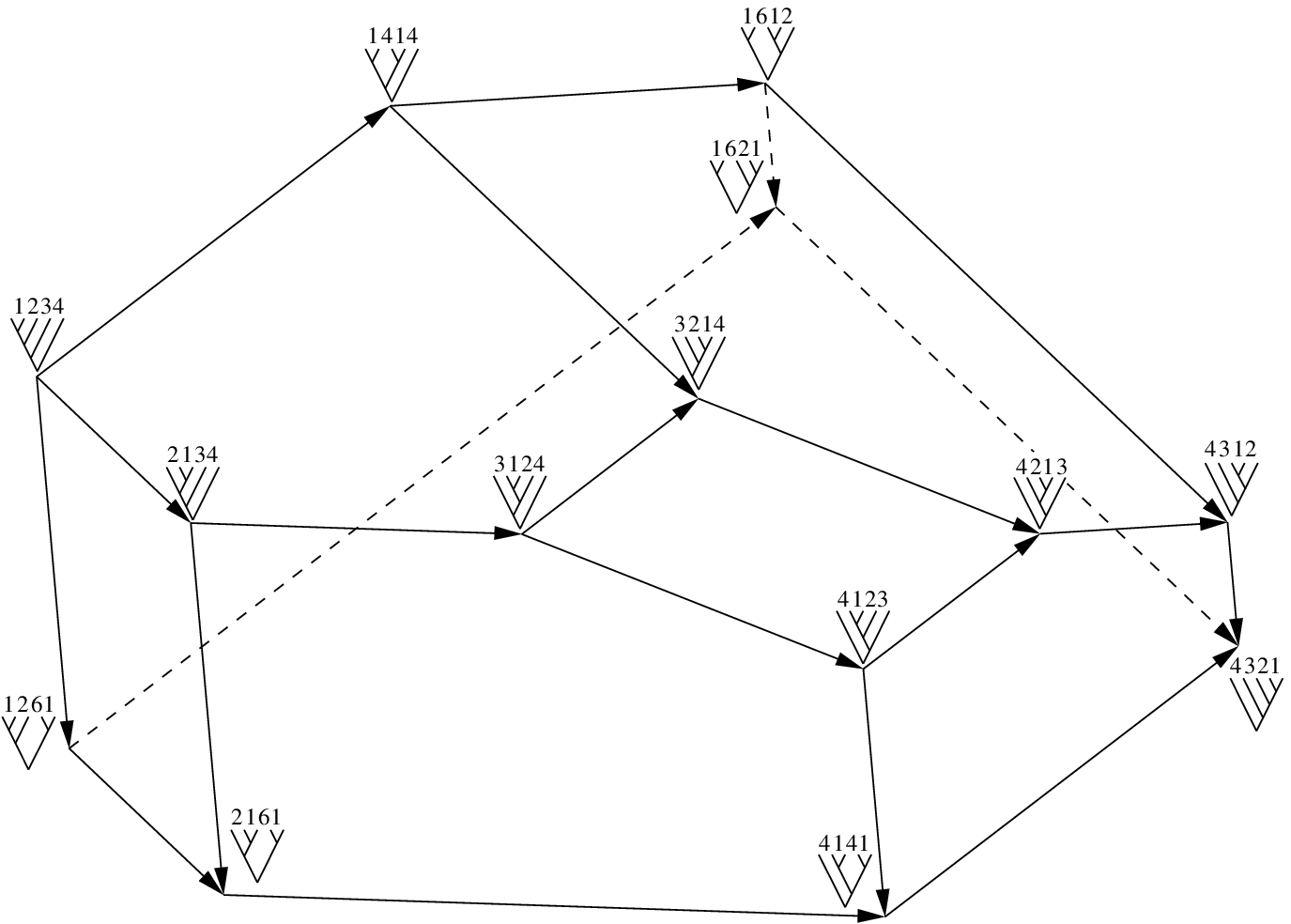}$ &
\begin{tabular}{@{}c@{}}
\hskip-2cm $\figbox{0.8}{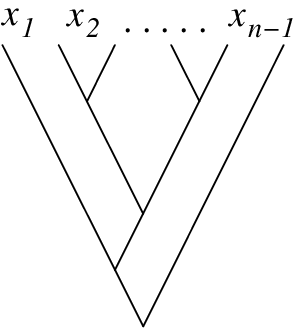}$
\parbox{4.5cm}{\noindent\hfill
$\displaystyle x_{j}=\sum_{u\in L_{j}}\sum_{v\in R_{j}}\mu_{uv}$,\hfill
\smallskip\newline where $\mu_{uv}$ are fixed positive numbers associated with
pairs of leaves (we have used $\mu_{uv}=1$).}
\vspace{25pt}\\
$\figbox{0.8}{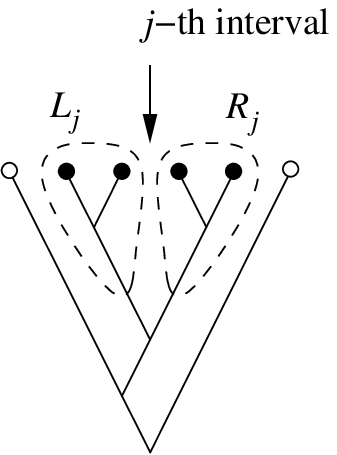}$
\end{tabular}
\\
a) & b)
\end{tabular}}
\caption{The graph of $F$-moves between binary planar trees with $n=5$ leaves
(a) and the procedure used to assign coordinates $x_{1},\dots,x_{n-1}$ to each
tree (b).}
\label{fig_stash}
\end{figure}

According to \emph{MacLane's coherence theorem}~\cite{MacLane}, any sequence
of $F$-moves between two given trees results in the same isomorphism between
the corresponding spaces: the equality of all such isomorphisms follows from
the pentagon equation. (Actually, the theorem is a bit more general and also
includes $\alpha$- and $\beta$-moves, but we ignore them for the moment.) This
result may be regarded as a combinatorial statement. Indeed, let us consider
the graph $\Gamma_{n}$ whose vertices are binary planar trees with $n$ leaves
and whose edges correspond to $F$-moves. (The graph $\Gamma_{5}$ is shown in
Fig.~\ref{fig_stash}a.) Then the theorem says that any cycle in the graph can
be filled with pentagons and quadrilaterals. The latter correspond to the
obvious fact that independent $F$-moves commute; such moves can be nested as
in Fig.~\ref{fig_pentquadr}b or occur in disjoint branches. MacLane actually
shows that any two directed paths between the same pair of vertices can be
transformed one to the other using the pentagon and quadrilateral equations;
the proof is by induction on a suitable parameter.

Let us also mention a beautiful geometric proof of the coherence theorem: the
graph $\Gamma_{n}$ together with the set of $2$-cells (i.e., the pentagons and
quadrilaterals) can be realized by the vertices, edges, and $2$-faces of some
convex polytope $K_{n}$ of dimension $n-2$. This polytope appeared in the work
of Stasheff~\cite{Stasheff} and now bears his name; it is also called
\emph{associahedron}. More exactly, this term refers to a certain
combinatorial type of a polytope while its exact shape may vary. The faces of
$K_{n}$ in all dimensions are associated with general (not necessarily binary)
planar trees. In particular, the edges correspond to trees with one degree-$3$
vertex, the quadrilaterals correspond to trees with two degree-$3$ vertices,
and the pentagons correspond to trees with one degree-$4$ vertex. Described in
Fig.~\ref{fig_stash}b is a concrete geometric realization of $K_{n}$ (cf.\
Ref.~\cite{stash_poly}). It is fairly easy to show that the convex hull of the
points $(x_{1},\dots,x_{n-1})\in\RR^{n-1}$ obtained this way is characterized
by one equation and $\frac{n(n-1)}{2}-1$ inequalities:
\begin{equation}
\sum_{j=1}^{n-1}x_{j}\,=\,\eta_{1n},\qquad
\sum_{j=k}^{l-1}x_{j}\,\ge\,\eta_{kl}\quad (1\le k<l\le n),\qquad\quad
\text{where}\quad
\eta_{kl}\bydef\sum_{k\le u<v\le l}\!\mu_{uv}.
\end{equation}

\begin{figure}[t]
\centerline{\begin{tabular}{c@{\qquad\qquad\qquad}c@{\quad\qquad}c}
$\figbox{1.0}{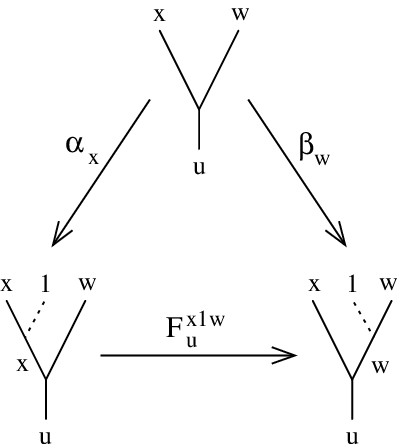}$ &
$\figbox{1.0}{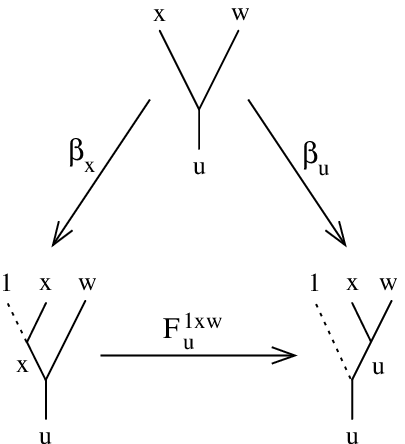}$ & $\figbox{1.0}{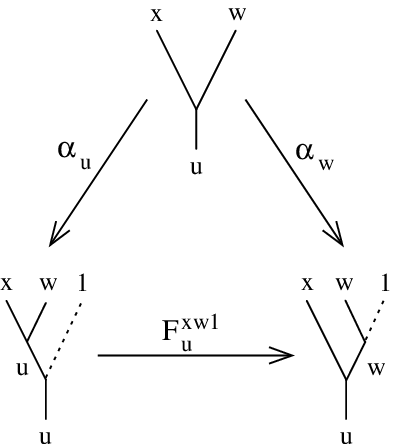}$
\\
a) & b) & c)
\end{tabular}}
\caption{The fundamental triangle equation (a) and its corollaries (b and c).}
\label{fig_triang}
\end{figure}

\paragraph{Triangle equations}
The three equations in Fig.~\ref{fig_triang} guarantee that adding or removing
trivial lines commutes with $F$-moves. Let us write the first equation in a
more conventional form:
\begin{equation}
\begin{xy}
(0,8)*+{V_{u}^{xw}}="1",
(-20,-12)*+{V_{x}^{x1}\otimes V_{u}^{xw}}="2",
(20,-12)*+{V_{u}^{xw}\otimes V_{w}^{1w}}="3",
\ar_{\alpha_{x}} "1";"2"
\ar^{\beta_{w}} "1";"3"
\ar^{F_{u}^{x1w}} "2";"3"
\end{xy}
\hskip-25pt \text{commutes,} \qquad\text{i.e.}\quad
\begin{array}{l}
F_{u}^{x1w}\bigl(|\alpha_{x}\rangle\otimes|\psi\rangle\bigr)\,=\,
|\psi\rangle\otimes|\beta_{w}\rangle
\medskip\\
\text{for any}\ |\psi\rangle\in V_{u}^{xw}.
\end{array}
\end{equation}

\begin{Lemma}\label{lem_alphabeta}\,
$\alpha_{1}=\beta_{1}$.
\end{Lemma}
\begin{proof}
Let $x=u$ and $w=1$. Then first and the third triangle equations coincide,
except that the right arrow is $\beta_{1}$ in Fig.~\ref{fig_triang}a and
$\alpha_{1}$ in Fig.~\ref{fig_triang}c. But the right arrow is the composition
of the other two, therefore it is the same in both cases.
\end{proof}

\begin{Lemma}[\normalfont cf.~\cite{Kassel}, Lemma XI.2.2]
\label{lem_triang}
The first triangle equation (together with the pentagon and quadrilaterals)
implies the second and the third.
\end{Lemma}
The statement of the lemma refers to formal tree calculus. Moves between
trees are bidirectional (therefore we may represent them by lines rather then
arrows). The quadrilaterals express the commutativity of disjoint $F$-,
$\alpha$-, and $\beta$-moves.

\begin{proof} Let us prove the equation in Fig.~\ref{fig_triang}b.

\figboxright{First, we join an additional trivial branch to the trunk of
each tree and show that the resulting equation is equivalent to the original
one. In the diagram on the right, the old and the new equation constitute the
top and the bottom of a triangular prism, respectively. The sides of the prism
are commutative quadrilateral, therefore the top triangle commutes if and only
if the bottom one does.}{1.0}{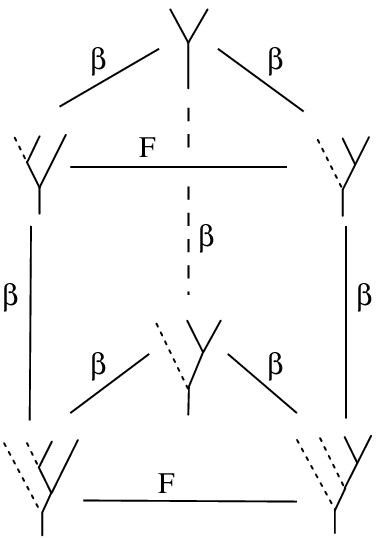}

\figboxright{We now consider the bottom of the prism, which may be identified
with triangle \mbox{4-7-5} in the new diagram. In this part of the argument it
is not important that the leftmost branch of each tree is trivial, so we
represent it by a solid line. The outline of the diagram corresponds to the
pentagon equation. Triangles \mbox{3-1-6} and \mbox{4-2-5} are instances of
the equation in Fig.~\ref{fig_triang}a, which is true by hypothesis. The
quadrilaterals \mbox{1-2-4-3} and \mbox{6-3-4-7} also commute. Thus the
required triangle \mbox{4-7-5} commutes as well.}{1.0}{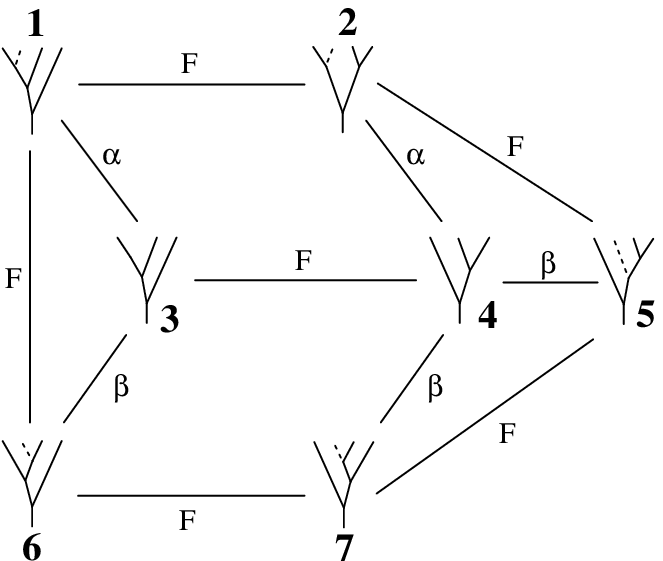}

The equation in Fig.~\ref{fig_triang}c is proved analogously.
\end{proof}

\subsubsection{Examples}

Several concrete theories have been presented in the main text, see tables on
pages~\pageref{tab_nonabelian}, \pageref{tab_0mod4}, \pageref{tab_2mod4}.  Let
us now describe some general constructions leading to infinite series of
examples.
\begin{enumerate}
\item Let the label set $M$ be a finite group and the fusion rules correspond
to the group multiplication. That is, the space $V^{x,y}_{z}$ is
one-dimensional (with a basis vector $\psi^{x,y}_{z}$) if $z=xy$, otherwise
$V^{x,y}_{z}=0$. The associativity relations are trivial. Note that if the
group $M$ is non-Abelian, this fusion theory does not admit braiding.

\item\label{ex_H3}
The associativity constraints in the above example can be deformed as follows:
\begin{equation}
F^{x,y,z}_{xyz}\,\figbox{1.0}{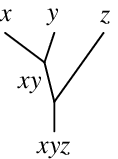}\;=\;
f(x,y,z)\,\figbox{1.0}{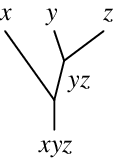}\,,
\end{equation}
where the vertices of the trees correspond to basis vectors and
$f(x,y,z)$ are some phase factors. In this case the pentagon equation
reads:
\begin{equation}\label{defgroup}
f(xy,z,w)\,f(x,y,zw)\,=\,
f(x,y,z)\,f(x,yz,w)\,f(y,z,w).
\end{equation}
This is a cocycle equation, i.e., $f$ is a $3$-cocycle on $M$ with values
in $\UU(1)$. One can show that the triangle equations do not put any
additional restriction on $f$ but rather define $\alpha$ and $\beta$:
\begin{equation}
\alpha_{x}=\zeta\,f(x,1,1)^{-1}\psi^{x,1}_{x},\qquad
\beta_{w}=\zeta\,f(1,1,w)\psi^{1,w}_{w},\qquad
\text{($\zeta$ is an arbitrary phase)}.
\end{equation}
The basis change $\psi^{x,y}_{xy}\to\,u(x,y)\psi^{x,y}_{xy}$ will result in a
new cocycle $g$ such that
\begin{equation}
f(x,y,z)^{-1}\,g(x,y,z)\,=\,
u(x,y)\,u(x,yz)^{-1}\,u(xy,z)\,u(y,z)^{-1}.
\end{equation}
The right-hand side of this equation is the coboundary of the $2$-cochain $u$,
i.e., $g$ and $f$ belong to the same cohomology class
$\widetilde{f}$. Thus, the associativity relations are classified by
$H^{3}(M,\UU(1))$.

For a concrete example, let $M=\ZZ_{2}=\{1,a\}$. It is known that
$H^{3}(\ZZ_{2},\UU(1))=\ZZ_{2}$, i.e., Eq.~(\ref{defgroup}) has one trivial
and one nontrivial solution. The nontrivial solution is given by
$f(a,a,a)=-1$, the other seven values being equal to $1$. This particular
fusion theory admits two braidings: the counterclockwise exchange of two
$a$-particles is characterized by either $+i$ or $-i$. Such particles are
called \emph{semions}. Recall that Case~2 in the table on
page~\pageref{tab_0mod4} represents two independent species of semions.

\item Let $M$ be the set of irreducible representations of a finite group $G$
and let $V^{ab}_{c}=\Hom_{\CC[G]}(c,a\otimes b)$. In other words, elements of
$V^{ab}_{c}$ are intertwiners between $c$ and $a\otimes b$, i.e., linear maps
that commute with the group action. The associativity relations are given by
$6j$-symbols. This theory admits trivial braiding.

\item Examples~3 and~1 (with $M=G$) can be combined in what is called
\emph{representation theory of Drinfeld's quantum
double}~\cite{Drinfeld,Kassel}. Models with anyons described by this theory
were proposed in~\cite{BDWP-1} (gauge-symmetric Lagrangian) and
in~\cite{Kitaev97} (lattice Hamiltonian not constrained by an external
symmetry). This construction can be deformed by an arbitrary cohomology class
$\widetilde{f}\in H^{3}(G,\UU(1))$, see Refs.~\cite{DPR91,BDWP-2}.

\item A very interesting set of fusion theories is based on a
\emph{Temperley-Lieb category}, see e.g.\ Chapter~XII in book~\cite{Turaev}.
These theories are also known as ``representations of quantum $\SU(2)$''.
\end{enumerate}

\subsubsection{Calculations with planar graphs}

The rules we have described allow us to work not only with trees but with
arbitrary oriented planar graphs. In particular, loops can be removed using
Eq.~(\ref{inprod}). If we encounter a subgraph like shown in
Eq.~(\ref{inprod}) but with different labels at the top and at the bottom
(say, $c'$ and $c''$), then its value is zero.  Let us illustrate that by a
concrete example, using the fusion theory from Table~\ref{tab_nonabelian} on
page~\pageref{tab_nonabelian} for $\FS=+1$:
\[
\figbox{1.0}{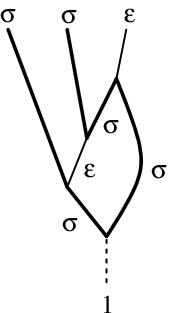}\enspace=\;
\figbox{1.0}{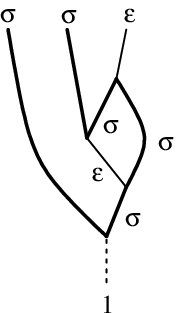}\enspace=\enspace
\frac{1}{\sqrt{2}}\underbrace{\figbox{1.0}{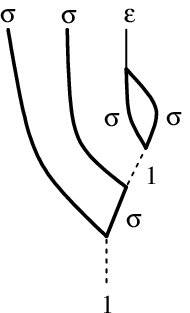}}_{0}\:
-\:\frac{1}{\sqrt{2}}\figbox{1.0}{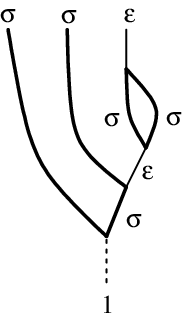}\enspace=\enspace
-\:\frac{1}{\sqrt{2}}\figbox{1.0}{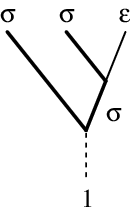}\:.
\]

In general, we consider planar graphs that satisfy the following
conditions.
\begin{enumerate}
\item \label{sourcetarget}Vertices are associated with fusion/splitting spaces
$V^{u_{1},\dots,u_{s}}_{l_{1},\dots,l_{r}}$; the indices are represented by
prongs. There may also be external labels (terminals) at the bottom and the top
of the graph. Bottom terminals and upper indices are called \emph{sources},
lower indices and top terminals are called \emph{targets}.
\item \label{regedge}Each edge connects a source to an identically labeled
target. (In pictures we use a single label for the whole edge.)
\item \label{transvedge}The edges are transversal to the horizontal direction
and oriented upwards. (We do not show the orientation in drawing because it is
obvious.)
\end{enumerate}

To each vertex we also assign an element of the corresponding space
$V^{u_{1},\dots,u_{s}}_{l_{1},\dots,l_{r}}$. Using associativity relations and
Eqs.~(\ref{inprod}), (\ref{partidentity}), we can compute the \emph{value} of
the graph, $X\in V^{a_{1},\dots,a_{k}}_{b_{1},\dots,b_{m}}$, where
$a_{1},\dots,a_{k}$ and $b_{1},\dots,b_{m}$ are the external labels. Note that
we may freely add or remove edges labeled with~$1$, thanks to the triangle
equations.

A general calculation strategy is based on the idea that $X$ can be
represented by a set of maps $X_{c}:V^{b_{1},\dots,b_{m}}_{c}\to
V^{a_{1},\dots,a_{k}}_{c}$ for each label $c$; this representation is closely
related to the decomposition of identity~(\ref{partidentity}). Thus we apply
Eq.~(\ref{partidentity}) first, and then start shrinking and removing loops as
illustrated above. For example:
\[
\figbox{1.0}{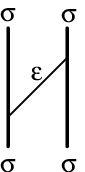}\enspace=\enspace
\figbox{1.0}{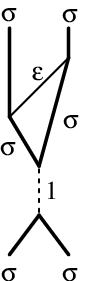}\;+\;\figbox{1.0}{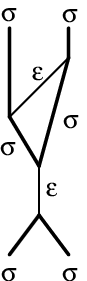}\enspace=\enspace
\figbox{1.0}{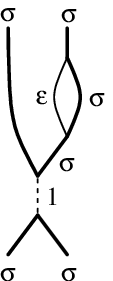}\;-\;\figbox{1.0}{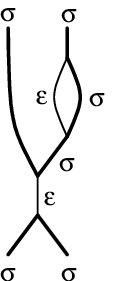}\enspace=\enspace
\figbox{1.0}{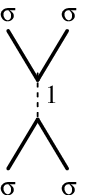}\;-\;\figbox{1.0}{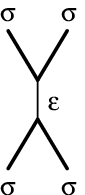}\:.
\]

We will see that for theories with particle-antiparticle duality,
condition~\ref{transvedge} can be dispensed with. The formalism will be
revised so that edges can bend, changing their direction from upward to
downward and back, and vertices can rotate by $360^{\circ}$ as in
Fig.~\ref{fig_obvious}b.

\subsection{Particle-antiparticle duality} \label{sec_aniparticle}

In this section we complement the fusion theory by the following condition.
\begin{remark}[Duality axiom.]\label{ax_duality}\itshape
For each label \,$a$\, there is some label \,$\anti{a}$\, and vectors
\,$|\xi\rangle\in V^{a\anti{a}}_{1}$,\: $|\eta\rangle\in V^{\anti{a}a}_{1}$\,
such that
\[
\bigl\langle\alpha_{a}\otimes\eta\big|F^{a\anti{a}a}_{a}\big|
\xi\otimes\beta_{a}\bigr\rangle \not=0.
\]
\end{remark}
Note that the matrix element in question corresponds to a physical process in
which the $\anti{a}$-particle from an $a\anti{a}$-pair annihilates with a
different copy of $a$:
\begin{equation}\label{cranop0}
\figbox{1.0}{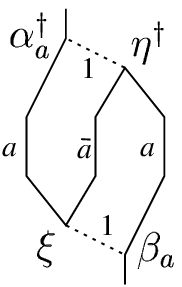}\;=\;
\bigl\langle\alpha_{a}\otimes\eta\big|F^{a\anti{a}a}_{a}\big|
\xi\otimes\beta_{a}\bigr\rangle\;
\figbox{1.0}{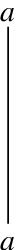}\:.
\end{equation}
(The dotted lines and the associated labels, i.e., $1$, $\alpha_{a}$,
$\beta_{a}$ may be ignored.)

Let $\langle\alpha_{a}\otimes\eta|F^{a\anti{a}a}_{a}|
\xi\otimes\beta_{a}\rangle\,=u$, assuming that $|\xi\rangle$ and
$|\eta\rangle$ are unit vectors. We will see that the spaces
$V^{a\anti{a}}_{1}$ and $V^{\anti{a}a}_{1}$ are one-dimensional, therefore
$|u|$ is uniquely defined. The number $d_{a}=|u|^{-1}$, called \emph{quantum
dimension}, plays an important role in the theory. In particular, $d_{a}=1$
for Abelian particles (i.e., such that $a\times\anti{a}=1$), otherwise
$d_{a}>1$. If $\anti{a}=a$, then we may set $|\xi\rangle=|\eta\rangle$ so that
$u$ itself has an invariant meaning. Specifically,
\begin{equation} \label{qdimFS}
\figbox{1.0}{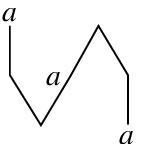}\;=\:
\frac{\FS_{a}}{d_{a}}\:\figbox{1.0}{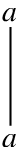}\:,
\end{equation}
where $\FS_{a}$ is a uniquely defined phase, which is actually equal to
$\pm 1$ (see below). This number is called \emph{Frobenius-Schur indicator};
in the present context it was introduced by Fredenhagen, Rehren, and
Schroer~\cite{FRS92}.

\subsubsection{Physical motivation for the duality axiom}

The existence of antiparticles follows from a locality principle mentioned at
the beginning of this appendix: a quasiparticle can be moved from one place to
another by applying an operator that acts on spins along a path connecting the
given points. (We are still considering anyons on a line, therefore the path
is unique.) The action on different spins can be performed at once or in any
particular order. For example, to move a particle from point~$3$ to point~$1$
on its left, it may be natural to move it first to some middle point $2$, and
then to the final destination. But it is also possible to act on the interval
$[1,2]$ \emph{before} $[2,3]$. The intermediate state must be a
particle-antiparticle pair.

For a slightly more rigorous argument, let $X$ be an operator that moves a
particle of type~$a$ from $3$ to $1$, acting on spins along the interval. We
can represent it as follows:
\begin{equation}
\hskip-3cm\figbox{1.0}{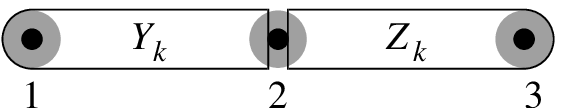}\qquad\qquad\quad
X=\sum_{k}Y_{k}Z_{k},
\end{equation}
where $Y_{k}$ and $Z_{k}$ act on disjoint sets of spins in some regions around
$[1,2]$ and $[2,3]$, respectively. Let $\Pi^{x}_{[s]}$\, ($s=1,2,3$) be the
projector onto states having particle $x$ at point $s$; such operators can be
realized locally (to act in the gray circles in the above picture) if we only
allow states that are not excited away from the given three points. Finally,
we define new versions of $X$, $Y_{k}$, and $Z_{k}$:
\begin{equation}
\begin{gathered}
X'\,=\, \Pi^{a}_{[1]}\Pi^{1}_{[2]}\Pi^{1}_{[3]}\,X\,
\Pi^{1}_{[1]}\Pi^{1}_{[2]}\Pi^{a}_{[3]}
\,=\,\sum_{k}Z'_{k}Y'_{k},\\[3pt]
Z'_{k}\,=\,\Pi^{1}_{[2]}\Pi^{1}_{[3]}\,Z_{k}\,\Pi^{a}_{[3]},\qquad\quad
Y'_{k}\,=\,\Pi^{a}_{[1]}\,Y_{k}\,\Pi^{1}_{[1]}\Pi^{1}_{[2]}.
\end{gathered}
\end{equation}
The operator $X'$ still moves $a$ from $2$ to $3$, but $Y'_{k}$ and $Z'_{k}$
overlap geometrically and therefore no longer commute. It is clear that each
operator $Y'_{k}$ creates an $a$-particle at point~$1$ and some particle at
point~$2$, whereas $Z'_{k}$ removes the second particle as well as the
original particle at point $3$. Thus each product $Z'_{k}Y'_{k}$ effects the
process shown in Eq.~(\ref{cranop0}), up to an overall factor. This factor is
nonzero for at least one value of $k$.

\subsubsection{Some properties and normalization conventions}
\label{sec_normconv}

\begin{Lemma}\label{lem_duality}
For each label $a$ the corresponding label $\anti{a}$ is unique. Moreover,
$N^{1}_{ab}=N^{1}_{ba}=\delta_{b\anti{a}}$.
\end{Lemma}

\begin{proof}
Recall that $N^{1}_{ab}\bydef\dim V_{ab}^{1}=\dim V^{ab}_{1}$. We first show
that this number is equal to one if $b=\anti{a}$ and zero otherwise. Let
$\xi$, $\eta$ be as in the duality axiom, and let $u$ be the corresponding
matrix element. For an arbitrary element $\psi\in V^{ab}_{1}$ we have:
\[
\figbox{1.0}{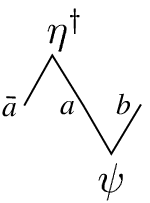}\;=\;
\figbox{1.0}{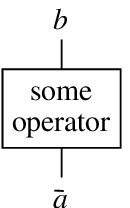}\;=\:
\left\{\begin{array}{cl}
\gamma\id_{\anti{a}}\ \text{for some}\ \gamma\in\CC & \text{if}\ b=\anti{a}
\smallskip\\
0 & \text{if}\ b\not=\anti{a}.
\end{array}\right.
\]
Therefore
\[
\psi\;\mathrel{\mathop{=}\limits_{\text{by duality axiom}}}
\enspace u^{-1}\;\figbox{1.0}{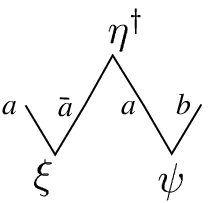}\enspace
\mathrel{\mathop{=}\limits_{\text{due to the above}}}\enspace
\left\{\begin{array}{cl}
u^{-1}\gamma\,\xi & \text{if}\ b=\anti{a}
\smallskip\\
0 & \text{otherwise}.
\end{array}\right.
\]

To prove that $N_{ba}^{1}=\delta_{b\anti{a}}$, we do a similar calculation
with $\psi'\in V^{ba}_{1}$, putting $(\psi')^{\dag}$ on the left of~$\xi$.
\end{proof}

\begin{Corollary}
$\;\anti{1}=1$\: and\/\: $\anti{\anti{a}}=a$.
\end{Corollary}

Let us now set new normalization conventions. In the definition of $d_{a}$ we
assumed that $|\xi\rangle$ and $|\eta\rangle$ are unit vectors, i.e.,
$\xi^{\dag}\xi=\eta^{\dag}\eta=\id_{1}$. However, it is more convenient to
multiply $\xi$ and $\eta$ by $\sqrt{d_{a}}$ (and some suitable phase factors)
so that the zigzag on the left-hand side of equation~(\ref{cranop0}) could be
simply removed. The so normalized operators for the creation of
particle-antiparticle pairs will be represented as smooth ``cups'' with a
triangle mark at the bottom. The adjoint operators are ``caps'' obtained by
flipping the pictures about a horizontal line:\footnote{This notation is not
standard, but it is convenient for calculations. In the notation used
in~\cite{Turaev,BakalovKirillov,Kassel}, cups and caps are not decorated but
each particle has a fictitious degree of freedom: it is considered as either
$a$ going up or $\anti{a}$ going down. This prevents noninvariant phases from
appearing in formulas.}
\begin{equation}
\figbox{1.0}{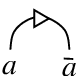}\:=\,\biggl(\figbox{1.0}{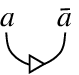}\biggr)^{\dag},
\qquad\qquad
\figbox{1.0}{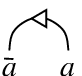}\:=\,\biggl(\figbox{1.0}{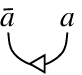}\biggr)^{\dag}
\end{equation}
In all four cases, the triangle points from $a$ to $\anti{a}$ as we follow the
line. The four operators are defined up to a single phase and satisfy the
following relations:
\begin{equation}\label{zigzag0}
\figbox{1.0}{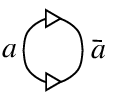}\:=\:d_{a}\:=\:\figbox{1.0}{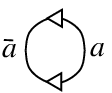}\:,
\end{equation}
\begin{equation}\label{zigzag1}
\figbox{1.0}{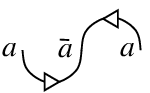}\:=\,\id_{a},\qquad
\figbox{1.0}{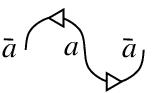}\:=\,\id_{\anti{a}},\qquad\quad
\figbox{1.0}{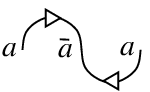}\:=\,\id_{a},\qquad
\figbox{1.0}{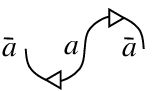}\:=\,\id_{\anti{a}}.
\end{equation}
Eq.~(\ref{zigzag0}) and the first equality in~(\ref{zigzag1}) are true by
definition. The second equality can be proved analogously to
Lemma~\ref{lem_duality}; the last two equalities are obtained by passing to
adjoint operators. Roughly, the normalization conditions may be described as
follows: \emph{oppositely oriented triangles cancel each other}.
\enlargethispage{\baselineskip}

So far the label $a$ was fixed. Repeating the same procedure for $\anti{a}$,
we get a new set of basis vectors in $V^{a\anti{a}}_{1}$, $V^{\anti{a}a}_{1}$,
$V_{a\anti{a}}^{1}$, $V_{\anti{a}a}^{1}$, which must be related to the old
ones:

\begin{equation}\label{lam0}
\begin{aligned}
\figbox{1.0}{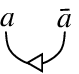}\:&=\,
\FS_{a}\,\figbox{1.0}{cupr.eps}\,,\qquad\quad
&\figbox{1.0}{cupl.eps}\:&=\,
\FS_{\anti{a}}\,\figbox{1.0}{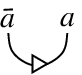}\,,
\\[3pt]
\figbox{1.0}{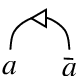}\:&=\,
\FS_{a}^{*}\,\figbox{1.0}{capr.eps}\,,\qquad\quad
&\figbox{1.0}{capl.eps}\:&=\,
\FS_{\anti{a}}^{*}\,\figbox{1.0}{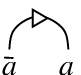}\,.
\end{aligned}
\end{equation}
Combining these relations with Eqs.~(\ref{zigzag0}), (\ref{zigzag1}), we get:
\[
|\FS_{a}|^{2}\,=\,d_{\anti{a}}/d_{a}\,=\,
|\FS_{\anti{a}}|^{-2},\qquad\qquad
\FS_{a}(\FS_{\anti{a}}^{*})^{-1}\,=\,1.
\]
Thus,
\begin{equation} \label{dlambda}
d_{\anti{a}}=d_{a},\qquad\quad  \FS_{\anti{a}}=\FS_{a}^{*},\qquad\quad
|\FS_{a}|=1.
\end{equation}
In general, the number $\FS_{a}$ depends on the arbitrary phases in the
definition of cups and caps.  However, if $a=\anti{a}$, then $\FS_{a}$
is defined uniquely and coincides with the Frobenius-Schur
indicator.\footnote{In category theory, $\FS_{a}$ is not a scalar but rather a
morphism from $a$ to $\anti{\anti{a}}$, which are regarded as different
(albeit isomorphic) objects. Thus the arbitrariness goes away, but the
Frobenius-Schur indicator has to be defined in a more complicated fashion.} In
this case, $\FS_{a}=\pm1$.

Now we are in a position to relax the requirement that edges are transversal
to the horizontal direction.
\def\ThmText{Definition-Proposition}\begin{Definition} A \emph{line} is an
alternating sequence of $2n$ cups and caps without triangle marks. It
may be open or closed. Such an object has a canonical normalization given by
$n$ triangles pointing forward and $n$ triangles pointing backward (relative
to a chosen path direction). All such decorations are equivalent.
\end{Definition}

Let us also renormalize the inner product on splitting spaces:
\begin{equation}\label{newinprod}
\dlangle\eta|\xi\drangle\,\bydef\,\sqrt{\frac{d_{c}}{d_{a}d_{b}}}\,
\langle\eta|\xi\rangle\,=\:
\frac{1}{\sqrt{d_{a}d_{b}d_{c}}}\,\figbox{1.0}{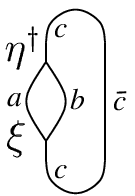}\:,\qquad\qquad
\text{where}\quad \xi,\eta\in V_{c}^{ab}.
\end{equation}
(Triangles may be added in one of the two consistent ways; one may also flip
the $c$-loop to the left.) This renormalization does not violate the unitarity
of associativity relations. The inner product on fusion spaces is defined
through the antilinear isomorphism $V^{ab}_{c}\to
V_{ab}^{c}:\,\psi\mapsto\psi^{\dag}$. When computing either type of inner
product, we stack two trees with branches touching each other and connect the
roots by a loop. Stacking the trees root to root might produce a different
result, but it turns out to be the same since we can actually rotate vertices
by $360^{\circ}$, see below.

Finally, we rewrite Eq.~(\ref{partidentity}) in a form that is consistent with
the new rules:
\begin{equation} \label{partidentity1}
\figbox{1.0}{idab.eps}\ =\ 
\sum_{c}\sum_{j}
\sqrt{\frac{d_{c}}{d_{a}d_{b}}}\,\figbox{1.0}{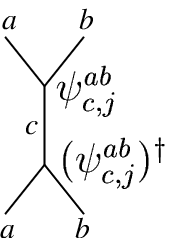}\;,\qquad\quad
\text{where}\quad\dlangle\psi^{ab}_{c,j}|\psi^{ab}_{c,k}\drangle=\delta_{jk}
\end{equation}
(the elements $\psi^{ab}_{c,j}\in V^{ab}_{c}$ form a complete basis).

\subsubsection{Raising and lowering of indices}

Let us define the following linear maps:
\begin{equation}
A^{ab}_{c}:\,V^{ab}_{c}\to V^{b}_{\anti{a}c}:\enspace
\figbox{1.0}{psiabc1.eps}\:\longmapsto\:\figbox{1.0}{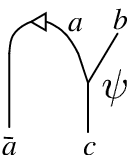}\:,
\qquad\qquad
B^{ab}_{c}:\,V^{ab}_{c}\to V^{a}_{c\anti{b}}:\enspace
\figbox{1.0}{psiabc1.eps}\:\longmapsto\,\figbox{1.0}{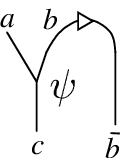}\:.
\end{equation}
They are obviously invertible, hence $N_{ab}^{c}=N_{\anti{a}c}^{b}
=N_{b\anti{c}}^{\anti{a}} =N_{\anti{b}\anti{a}}^{\anti{c}}
=N_{\anti{c}a}^{\anti{b}} =N_{c\anti{b}}^{a}$.  Unfortunately, this does not
save us from the need to to distinguish between lower and upper indices
because the simultaneous raising and lowering of indices on the two ends of a
line results in the factor $\FS_{a}$. For example:
\begin{equation}
\figbox{1.0}{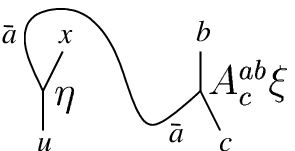}\;\:=\;\:\FS_{a}\;\figbox{1.0}{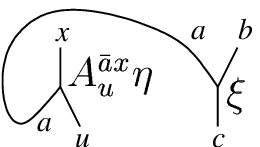}\:,
\qquad\qquad \text{where}\quad
\eta\in V^{\anti{a}x}_{u},\quad \xi\in V^{ab}_{c}.
\end{equation}
However, as the following theorem shows, we do not have to tie the types of
indices to the orientation of the corresponding vertex. This way, one obtains
an isotopy-invariant calculus for planar graphs, see Table~\ref{tab_iso}. We
will continue using the previous notation, though.

\begin{table}[t]
\[
\begin{array}{|l|c|c|}
\hline
& \text{Notation used in this paper} & \text{Isotopy-invariant calculus}
\\
\hline
\text{Vertices} &
\figbox{1.0}{psiabc1.eps}\:,\qquad
\begin{aligned}
\figbox{1.0}{cuplb.eps}\:&=\,\FS_{a}\,\figbox{1.0}{cupr.eps}
\\
\figbox{1.0}{caplb.eps}\:&=\,\FS^{*}\,\figbox{1.0}{caplb.eps}
\end{aligned} &
\figbox{1.0}{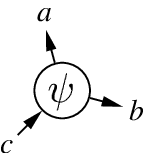}\:,\qquad
\begin{aligned}
\figbox{1.0}{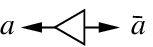}\:&=\,\FS_{a}\,\figbox{1.0}{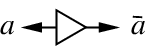}
\\[5pt]
\figbox{1.0}{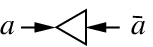}\:&=\,\FS_{\anti{a}}\,\figbox{1.0}{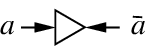}
\end{aligned}
\\
\hline
\text{Lines} &
\figbox{1.0}{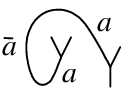},\qquad\quad \figbox{1.0}{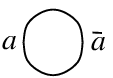} &
\figbox{1.0}{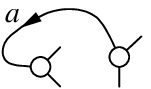}\:,\qquad\quad
\figbox{1.0}{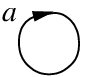}\:=\figbox{1.0}{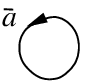}
\\
\hline
\text{Cancellation rules} &
\figbox{1.0}{zigzag1.eps}\:=\:\figbox{1.0}{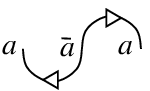}\:=\,\id_{a}
&
\figbox{1.0}{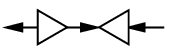}\:=\:\figbox{1.0}{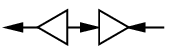}\:=\:\figbox{1.0}{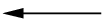}
\\
\hline
\end{array}
\]
\caption{Notation summary.}
\label{tab_iso}
\end{table}

\begin{Theorem}\label{th_raiselower}
The maps $A^{ab}_{c}$, $B^{ab}_{c}$ are unitary with respect to the inner
product $\dlangle\cdot|\cdot\drangle$, and the following diagram (in which the
arrows may be traversed in both directions) commutes:
\begin{equation}\label{raiselower}
\begin{xy}
(0,16)*+{V^{ab}_{c}}="1",
(27,9)*+{V^{a}_{c\anti{b}}}="2",
(27,-9)*+{V^{\anti{c}a}_{\anti{b}}}="3",
(0,-16)*+{V^{\anti{c}}_{\anti{b}\anti{a}}}="4",
(-27,-9)*+{V^{b\anti{c}}_{\anti{a}}}="5",
(-27,9)*+{V^{b}_{\anti{a}c}}="6",
\ar^{B^{ab}_{c}} "1";"2"
\ar_{A^{\anti{c}a}_{\anti{b}}} "3";"2"
\ar^{B^{\anti{c}a}_{\anti{b}}} "3";"4"
\ar_{A^{ab}_{c}} "1";"6"
\ar^{B^{b\anti{c}}_{\anti{a}}} "5";"6"
\ar_{A^{b\anti{c}}_{\anti{a}}} "5";"4"
\end{xy}
\end{equation}
\end{Theorem}
Note that the commutativity of this diagram is equivalent the ``pivotal
identity'' in Fig~\ref{fig_obvious}b. The proof of the theorem is preceded by
two lemmas.

\begin{Lemma}\label{lem_raiselower1}
The composition of any two adjacent arrows in Eq.~(\ref{raiselower}) is
unitary.
\end{Lemma}
\begin{proof}
Due to symmetry, it suffices to consider just one particular case, e.g., the
arrows connected at the upper left corner. Let
$X=(B^{b\anti{c}}_{\anti{a}})^{-1}A^{ab}_{c}:\,V^{ab}_{c}\to
V^{b\anti{c}}_{\anti{a}}$. For arbitrary elements
$\xi\in V^{ab}_{c}$,\, $\eta\in V^{b\anti{c}}_{\anti{a}}$ we have:
\[
X\xi\,=\:\figbox{1.0}{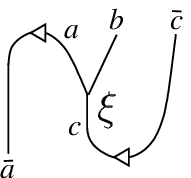},\qquad\qquad
X^{-1}\eta\,=\:\figbox{1.0}{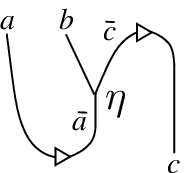}\:;
\medskip
\]
\[
\sqrt{d_{a}d_{b}d_{c}}\,\dlangle\eta|X|\xi\drangle\:=\;
\figbox{1.0}{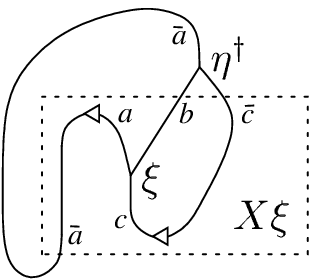}\;=\;
\figbox{1.0}{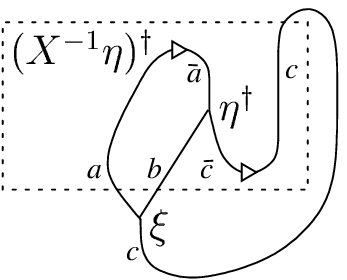}\;=\:
\sqrt{d_{a}d_{b}d_{c}}\,\dlangle X^{-1}\eta|\xi\drangle.
\]
Therefore $X^{-1}=X^{\dag}$.
\end{proof}

Now consider the two paths from top to bottom of the
hexagon~(\ref{raiselower}):
\begin{equation}
U_{l},U_{r}:\,V^{ab}_{c}\to V_{\anti{b}\anti{a}}^{\anti{c}},\qquad\qquad
U_{l}:\;\psi\,\mapsto\:\figbox{1.0}{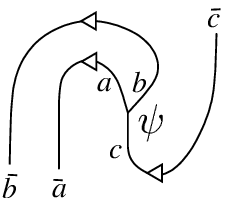}\:,\qquad\quad
U_{r}:\;\psi\,\mapsto\:\figbox{1.0}{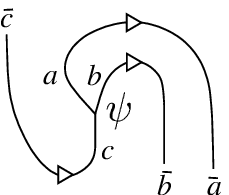}\:.
\end{equation}
From the physical point of view, these operators (which are actually equal)
correspond to the CPT symmetry.

\begin{Lemma}\label{lem_raiselower2}\; $U_{l}^{\dag}=U_{r}^{-1}$.
\end{Lemma}
\begin{proof}
Let $\xi\in V^{ab}_{c}$ and $\eta\in V_{\anti{b}\anti{a}}^{\anti{c}}$. Then
\[
\sqrt{d_{a}d_{b}d_{c}}\,\dlangle\eta|U_{l}|\xi\drangle\:=\;
\figbox{1.0}{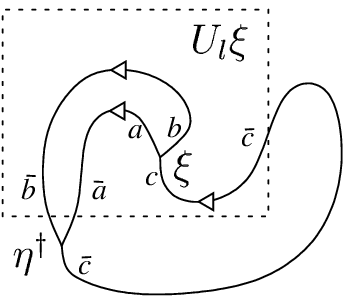}\;=\;\figbox{1.0}{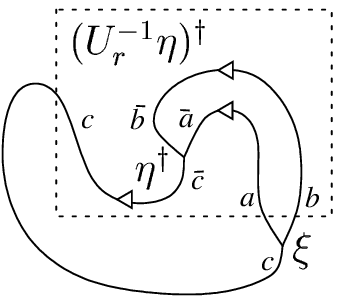}\;=\;
\sqrt{d_{a}d_{b}d_{c}}\,\dlangle U_{r}^{-1}\eta|\xi\drangle.
\]
\end{proof}

\begin{proof}[Proof of Theorem~\ref{th_raiselower}.]
Let us consider the path around the diagram~(\ref{raiselower}) in the
counterclockwise direction, $W=U^{-1}_{r}U_{l}$. This operator is a
composition of six arrows, therefore it is unitary (due to
Lemma~\ref{lem_raiselower1}). On the other hand, Lemma~\ref{lem_raiselower2}
implies that $W=U_{l}^{\dag}U_{l}$ is Hermitian and positive semidefinite. It
follows that $W$ is the identity operator, i.e., the diagram commutes.

Thus, $U_{l}=U_{r}$; let us denote this operator simply by $U$. It is unitary
by Lemma~\ref{lem_raiselower2}. Any arrow in the diagram is a composition of
$U$ (or $U^{-1}$) and some number of arrow pairs. Therefore all the arrows are
unitary.
\end{proof}

\subsubsection{Quantum dimension and fusion rules}
This is a key identity: 
\begin{equation}\label{qdimmult}
d_{a}d_{b}\,=\,\sum_{c}N^{c}_{ab}d_{c}.
\end{equation}
To prove it, we use Eq.~(\ref{partidentity1}), the pivotal property, and
Eq.~(\ref{newinprod}):
\[
d_{a}d_{b}\:=\:
\figbox{1.0}{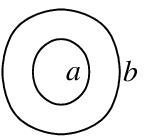}\;=\:
\sum_{c,j}\sqrt{\frac{d_{c}}{d_{a}d_{b}}}\,\figbox{1.0}{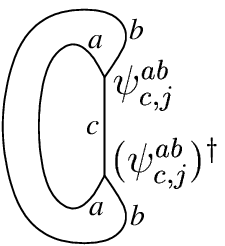}\:
\mathrel{\mathop{=}\limits_{\substack{\mathstrut
\text{rotating}\ \psi_{c,j}^{ab}\;\\
\text{counterclockwise}}}}\:
\sum_{c,j}\sqrt{\frac{d_{c}}{d_{a}d_{b}}}\,\figbox{1.0}{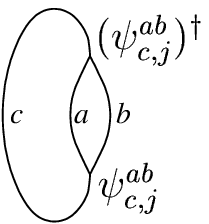}\!=\:
\sum_{c}N^{c}_{ab}d_{c}.
\]

Eq.~(\ref{qdimmult}) implies that the matrix $\widehat{N}(a)=(N^{c}_{ab}:\,
b,c\in M)$ has eigenvector $v=(d_{c}:c\in M)$, and the corresponding
eigenvalue is $d_{a}$. Note that all the entries of $v$ are positive.
According to the Perron-Frobenius theorem (about matrices with nonnegative
entries), all eigenvalues of $\widehat{N}(a)$ satisfy the inequality
\begin{equation}
|\lambda_{j}(a)|\le d_{a}.
\end{equation}
Thus, the quantum dimension $d_{a}$ is the largest eigenvalue of the matrix
$\widehat{N}(a)$.

The proof of Eq.~(\ref{qdimmult}) also motivates a definition of \emph{quantum
trace}. It is a number assigned to any element $X\in V^{a_{1}\dots
a_{n}}_{a_{1}\dots a_{n}}$. Such an element acts as an operator in the space
$V^{a_{1}\dots a_{n}}_{c}$ for each $c$; we denote this action by $X_{c}$. The
quantum trace is defined as follows:
\begin{equation}
\QTr X\:\bydef\: \sum_{c}d_{c}\Tr X_{c}\,=\:\figbox{1.0}{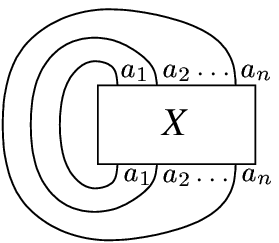}\:.
\end{equation}

\subsection{Braiding and topological spin} \label{sec_braiding}

Braiding is an additional piece of algebraic structure characterizing
anyons. It is defined by a set of elements $R_{ab}\in V^{ba}_{ab}$ that
represent transposition of two particles:
\begin{equation}
R_{ab}\,=\:\figbox{1.0}{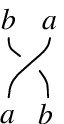},\qquad\qquad
(R_{ab})^{-1}=(R_{ab})^{\dag}\,=\:\figbox{1.0}{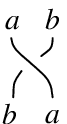}\,.
\end{equation}
They must satisfy the Yang-Baxter equation, i.e., a line can be moved over or
under a crossing between two other lines:
\begin{equation}\label{YaBa}
\figbox{1.0}{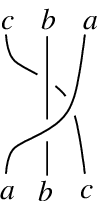}\;=\;\figbox{1.0}{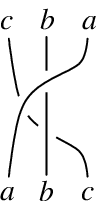}\:.
\end{equation}
More generally, a line can slide over or under an arbitrary vertex. It is
sufficient to postulate this property for three-prong vertices representing
splitting and fusion and for lines tilted left and right. Furthermore,
splitting and fusion are related by the decomposition of the
identity~(\ref{partidentity1}), whereas changing the slope of the intersecting
line is equivalent to replacing $R$ with $R^{-1}$. Thus the number of
independent conditions is reduced to two:
\begin{equation}\label{brfucons}
\figbox{1.0}{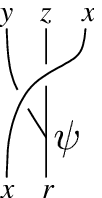}\:=\:\figbox{1.0}{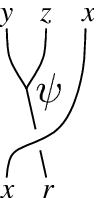}\,,\qquad\qquad
\figbox{1.0}{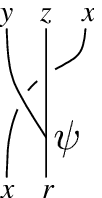}\:=\:\figbox{1.0}{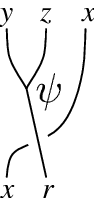}\,.
\end{equation}

\begin{figure}[t]
\centerline{\begin{tabular}{c@{\qquad\qquad}c} $\figbox{1.0}{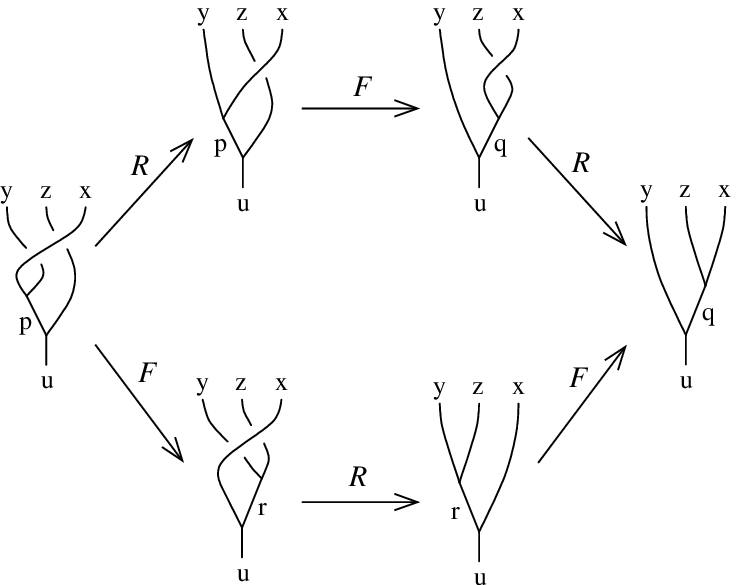}$ &
$\figbox{1.0}{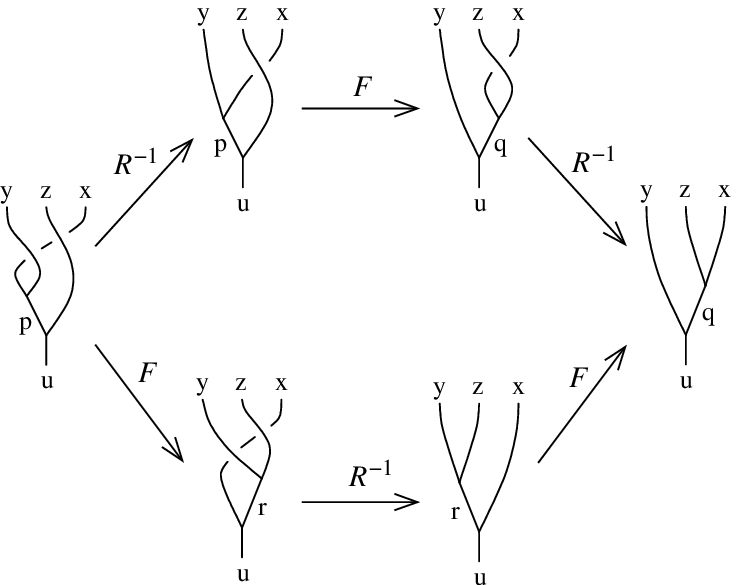}$ \\ a) & b)
\end{tabular}}
\caption{The hexagon equations.}
\label{fig_hexagon}
\end{figure}

The description of braiding in terms of basic data amounts to specifying the
action of $R_{ab}$ on splitting spaces:
\begin{equation}\label{Rabc}
\figbox{1.0}{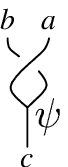}\:=\,R^{ab}_{c}\,\psi,\qquad\qquad
R^{ab}_{c}:\,V^{ab}_{c}\to V^{ba}_{c}.
\end{equation}
Note that $R^{ab}_{c}$ is a unitary map, therefore $N_{ab}^{c}=N_{ba}^{c}$.

To express Eq.~(\ref{brfucons}) in terms of $R^{ab}_{c}$, we join the two
lines at the bottom of each graph and perform equivalence
transformations. These include $F$-moves as well as \emph{$R$-moves} ---
absorbing a line crossing by a vertex. Thus we obtain the diagrams in
Fig.~\ref{fig_hexagon}; the bottommost arrow in each of them combines an
$R$-move with the first or the second equation in question. We may now forget
about the topological meaning of braiding and only keep track of the linear
maps involved:
\begin{equation}\label{hex1}
\begin{xy}
(-52,0)*+{\bigoplus_{p}V^{xy}_{p}\otimes V^{pz}_{u}}="1",
(-32,20)*+{\bigoplus_{p}V^{yx}_{p}\otimes V^{pz}_{u}}="2",
(32,20)*+{\bigoplus_{q}V^{yq}_{u}\otimes V^{xz}_{q}}="3",
(52,0)*+{\bigoplus_{q}V^{yq}_{u}\otimes V^{zx}_{q}}="4",
(-32,-20)*+{\bigoplus_{r}V^{xr}_{u}\otimes V^{yz}_{r}}="5",
(32,-20)*+{\bigoplus_{r}V^{yz}_{r}\otimes  V^{rx}_{u}}="6",
\ar^-(0.4){\bigoplus_{p}R^{xy}_{p}\otimes\id_{V^{pz}_{u}}} "1";"2"
\ar^-{F^{yxz}_{u}} "2";"3"
\ar^-(0.6){\bigoplus_{q}\id_{V^{yq}_{u}}\otimes R^{xz}_{q}} "3";"4"
\ar^-(0.6){F^{xyz}_{u}} "1";"5"
\ar^-{\bigoplus_{r}(\id_{V^{yz}_{r}}\otimes R^{xr}_{u})\,\cdot\,
\mathrm{swap}} "5";"6"
\ar^-(0.4){F^{yzx}_{u}} "6";"4"
\end{xy}
\bigskip
\end{equation}
\begin{equation}\label{hex2}
\begin{xy}
(-52,0)*+{\bigoplus_{p}V^{xy}_{p}\otimes V^{pz}_{u}}="1",
(-32,20)*+{\bigoplus_{p}V^{yx}_{p}\otimes V^{pz}_{u}}="2",
(32,20)*+{\bigoplus_{q}V^{yq}_{u}\otimes V^{xz}_{q}}="3",
(52,0)*+{\bigoplus_{q}V^{yq}_{u}\otimes V^{zx}_{q}}="4",
(-32,-20)*+{\bigoplus_{r}V^{xr}_{u}\otimes V^{yz}_{r}}="5",
(32,-20)*+{\bigoplus_{r}V^{yz}_{r}\otimes  V^{rx}_{u}}="6",
\ar^-(0.4){\bigoplus_{p}(R^{yx}_{p})^{-1}\otimes\id_{V^{pz}_{u}}} "1";"2"
\ar^-{F^{yxz}_{u}} "2";"3"
\ar^-(0.6){\bigoplus_{q}\id_{V^{yq}_{u}}\otimes (R^{zx}_{q})^{-1}} "3";"4"
\ar^-(0.6){F^{xyz}_{u}} "1";"5"
\ar^-{\bigoplus_{r}(\id_{V^{yz}_{r}}\otimes (R^{rx}_{u})^{-1})\,\cdot\,
\mathrm{swap}} "5";"6"
\ar^-(0.4){F^{yzx}_{u}} "6";"4"
\end{xy}
\end{equation}
These commutative diagrams are known as \emph{hexagon equations}. They
actually look nicer in the tensor category formalism, see Eqs.~(\ref{brcat1})
and (\ref{brcat2}).

Note that braiding with label $1$ is trivial:
\begin{equation}
\figbox{1.0}{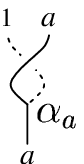}\:=\:\figbox{1.0}{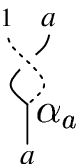}\:=\:
\figbox{1.0}{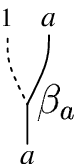}\,,\qquad\quad \text{i.e.,}\quad
R^{a1}_{a}\alpha_{a}\,=\,(R^{1a}_{a})^{-1}\alpha_{a}\,=\,\beta_{a}.
\end{equation}
(The proof is analogous to that of Lemma~\ref{lem_triang}.) Due to this
property, we need not worry about lines labeled by $1$, e.g., ones that are
implicitly attached to each cup and cap.

To each label $a$ we associate a complex number $\theta_{a}$, called
\emph{topological spin}:
\begin{equation}\label{topspin}
\theta_{a}\:\bydef\;d_{a}^{-1}\:\figbox{1.0}{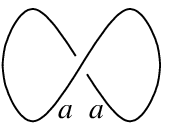}\;=\:
d_{a}^{-1}\QTr R_{aa}\:=\:d_{a}^{-1}\sum_{c}d_{c}\Tr R^{aa}_{c}.
\end{equation}
It may also be characterized by any of the following relations:
\begin{equation}
\figbox{1.0}{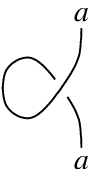}\:=\:\theta_{a}\,\figbox{1.0}{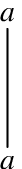}\:=\:
\figbox{1.0}{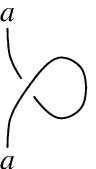}\:,\qquad\qquad
\figbox{1.0}{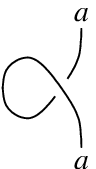}\:=\:\theta_{a}^{*}\,\figbox{1.0}{topspin0.eps}\:=\:
\figbox{1.0}{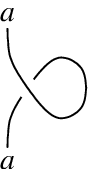}\:.
\smallskip
\end{equation}
Note the following properties of the topological spin:\smallskip
\begin{equation}
|\theta_{a}|=1,\qquad\qquad \theta_{\anti{a}}=\theta_{a}.
\end{equation}
Indeed,
\[
d_{a}\theta_{a}\theta_{a}^{*}\:=\:\figbox{1.0}{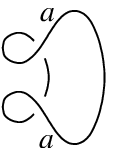}\:=\:d_{a}\:,
\qquad\qquad\quad
d_{a}\theta_{\anti{a}}\:=\:\figbox{1.0}{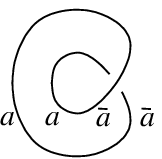}\:=\:
\figbox{1.0}{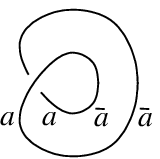}\:=\:d_{a}\theta_{a}.
\]
This is yet another expression for the topological spin:
\begin{equation}
\figbox{1.0}{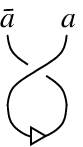}\:=\:\theta_{a}^{*}\,\figbox{1.0}{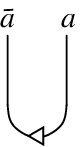}\,.
\end{equation}
(To prove it, we need to put a left-oriented cap on both sides of the equation
and rotate the resultant figure-eight on the left-hand side by $90^{\circ}$.)
In particular, if $\anti a=a$, then we have the following representation for
the invariant scalar $R^{aa}_{1}$:
\begin{equation}\label{tsFS}
R^{aa}_{1}\,=\,\theta_{a}^{*}\FS_{a}.
\end{equation}

\begin{Remark}
Eq.~(\ref{tsFS}) provides a simple physical interpretation for the
Frobenius-Schur indicator $\FS_{a}$. Suppose that the system is rotationally
invariant, so that not only the topological spin but also the usual spin
$s_{a}$ has physical meaning. Of course, $s_{a}$ may take different values
subject to the constraint $e^{2\pi is_{a}}=\theta_{a}$. Let us consider a pair
of identical particles with trivial total topological charge. What values does
the total angular momentum of this composite object take? This question may be
answered as follows. Assuming that both particles are in the same spin state
$s_{a}$ and barring additional ``isospin'' degrees of freedom, the
$180^{\circ}$ rotation is characterized by the phase factor $e^{i\pi
s_{a}}e^{i\pi s_{a}}R^{aa}_{1}=\FS_{a}$. Thus, the total angular momentum is
even if $\FS_{a}=1$ and odd if $\FS_{a}=-1$.
\end{Remark}

It is interesting that the effect of moving one particle around another is
fully characterized by the topological spin:
\begin{equation}\label{R2topspin}
R^{ba}_{c}R^{ab}_{c}\,=\,
\frac{\theta_{c}}{\theta_{a}\theta_{b}}\id_{V^{ab}_{c}}.
\end{equation}
\begin{proof}
Consider the following element of the space $V^{ab}_{ab}$:
\begin{equation}\label{brtwist}
\figbox{1.0}{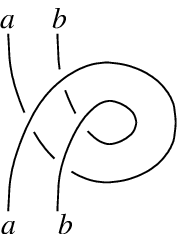}\;=\;\figbox{1.0}{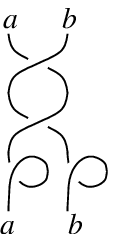}\;=\:
R_{ba}R_{ab}\theta_{a}\theta_{b}.
\end{equation}
Acting by it as an operator on the space $V^{ab}_{c}$, we obtain the required
identity. 
\end{proof}

\begin{Theorem}[Vafa~\cite{Vafa}]\label{th_Vafa}
The topological spins are roots of unity. More specifically,
$\theta_{a}^{n}=1$, where the integer $n\not=0$ depends only on the fusion
multiplicities.
\end{Theorem}
\begin{proof}
Let us use this fact from linear algebra: for any integer matrix $X$ there is
an integer matrix $Y$ (the adjugate) such that $YX=(\det X)I$. Therefore, if
$\prod_{b}\theta_{b}^{X_{ab}}=1$ for all $a$, then $\theta_{b}^{\det X}=1$ for
all $b$.

To find multiplicative relations between topological spins, we consider
determinants of $R$- and $F$-moves. Such determinants generally depend on the
choice of basis in the source and the target space. So, let us fix a basis in
each space $V^{ab}_{c}$ and replace the linear maps in the hexagon
equations~(\ref{hex1}) and~(\ref{hex2}) by their determinants. Dividing the
first equation by the second, we get:
\[
\det\left(\bigoplus_{p}(R^{xy}_{p}R^{yx}_{p})\otimes\id_{V^{pz}_{u}}\right)
\det\left(\bigoplus_{q}\id_{V^{yq}_{u}}\otimes(R^{xz}_{q}R^{zx}_{q})\right)
\:=\:
\det\left(\bigoplus_{r}\id_{V^{yz}_{r}}\otimes(R^{xr}_{u}R^{rx}_{u})\right).
\] 
Note that the determinants here are actually invariant and can be expressed in
terms of topological spins using Eq.~(\ref{R2topspin}):
\[
\prod_{p}\left(\frac{\theta_{p}}{\theta_{x}\theta_{y}}\right)
^{N^{p}_{xy}N^{u}_{pz}}
\prod_{q}\left(\frac{\theta_{q}}{\theta_{x}\theta_{z}}\right)
^{N^{u}_{yq}N^{q}_{xz}}
\:=\:
\prod_{r}\left(\frac{\theta_{u}}{\theta_{x}\theta_{r}}\right)
^{N^{r}_{yz}N^{u}_{xr}},
\medskip
\]
\begin{equation}
\prod_{p}\theta_{p}^{N^{p}_{xy}N^{\anti{p}}_{zv}+
N^{p}_{xz}N^{\anti{p}}_{yv}+N^{p}_{yz}N^{\anti{p}}_{xv}}\,=\,
\bigl(\theta_{x}\theta_{y}\theta_{z}\theta_{v}\bigr)^{N_{xyzv}},\qquad\quad
N_{xyzv}\bydef\,\sum_{q}N^{q}_{xy}N^{\anti{q}}_{zv},
\end{equation}
where we have substituted $v$ for $\anti{u}$.

For our purposes, it is sufficient to consider the case where $x=y=a$ and
$z=v=\anti{a}$. Thus we obtain a system of equations described by a square
matrix:
\begin{equation}
\prod_{b}\theta_{b}^{X_{ab}}\,=\,1,\qquad\quad \text{where}\quad
X_{ab}\,=\,4N_{aa\anti{a}\anti{a}}\,\delta_{ab}
-2N^{b}_{a\anti{a}}N^{\anti{b}}_{a\anti{a}}
-N^{b}_{aa}N^{\anti{b}}_{\anti{a}\anti{a}}.
\end{equation}
It remains to show that $X$ is nondegenerate. For this purpose, we may divide
the $a$-th row of $X$ by $N_{aa\anti{a}\anti{a}}$ so that the matrix becomes
$4I-Y$; the eigenvalues of $Y$ are bounded by a suitable multiplicative matrix
norm:
\[
Y_{ab}\,=\, (N_{aa\anti{a}\anti{a}})^{-1}
\bigl(2N^{b}_{a\anti{a}}N^{\anti{b}}_{a\anti{a}}
+N^{b}_{aa}N^{\anti{b}}_{\anti{a}\anti{a}}\bigr), \qquad\quad
\bigl|\mathrm{eigenvalue}_{j}(Y)\bigr|\le\max_{a}\sum_{b}|Y_{ab}|=3.
\]
Thus the matrix $4I-Y$ cannot have a zero eigenvalue.
\end{proof}

\subsection{Verlinde algebra and topological $S$-matrix}

In this section we investigate some properties of fusion, based on the axioms
stated above. First, let as consider a fusion theory with
particle-antiparticle duality, but without braiding. The \emph{Verlinde
algebra} is an associative $*$-algebra spanned by elements
$\mathbf{e}_{a}:a\in M$ which satisfy the following relations:
\begin{equation}
\mathbf{e}_{a}\mathbf{e}_{b}\,=\,\sum_{c}N^{c}_{ab}\mathbf{e}_{c},
\qquad\qquad
\mathbf{e}_{a}^{\dag}=\mathbf{e}_{\anti{a}}.
\end{equation}
The basis element $\mathbf{e}_{a}\in\Ver$ may be represented by the matrix
$\widehat{N}(a)=(N^{c}_{ab}:\, b,c\in M)$; this representation is
faithful. Since $\widehat{N}(\anti{a})=\widehat{N}(a)^{\dag}$, the algebra
operation $\dag$ corresponds to taking the adjoint matrix. Therefore $\Ver$ is
actually a finite-dimensional $C^{*}$-algebra. Equation~(\ref{qdimmult}) says
that the linear map $\Ver\to\CC:\:\mathbf{e}_{a}\mapsto d_{a}$ is a
homomorphism.

If braiding is defined, then $N_{ab}^{c}=N_{ba}^{c}$, therefore the Verlinde
algebra is commutative. A finite-dimensional commutative $C^{*}$-algebra
splits into several copies of $\CC$. Therefore we have $|M|$ distinct
homomorphisms from $\Ver$ to $\CC$, which we refer to as \emph{fusion
characters}:
\begin{equation}\label{fuchar}
\lambda_{j}:\Ver\to\CC,\qquad\qquad
\lambda_{j}(a)\lambda_{j}(b)\,=\,\sum_{c}N^{c}_{ab}\lambda_{j}(c).
\end{equation}
The vector $v_{j}=(\lambda_{j}(c):c\in M)$ is a common eigenvector of the
matrices $\widehat{N}(a)$, the eigenvalues being equal to
$\lambda_{j}(a)$. Note that the matrices $\widehat{N}(a)$ are normal,
therefore eigenvectors corresponding to distinct eigenvalues must be
orthogonal. Thus the following orthogonality condition holds:
\begin{equation}
\sum_{a}\lambda_{j}(a)\lambda_{k}(a)^{*}\,=0\qquad\quad
\text{if $j\not=k$}.
\end{equation}

We will now construct a map from the set of labels to the set of fusion
characters. Specifically, we are to show that some of the characters are given
(up to a constant factor) by the columns of the \emph{topological $S$-matrix}
$S=(s_{ab}:a,b\in M)$, where
\begin{equation}\label{sab}
s_{ab}\,\bydef\:\frac{1}{\calD}\,\figbox{1.0}{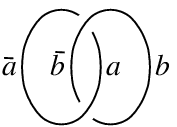}\:=\:
\frac{\QTr(R_{\anti{b}a}R_{a\anti{b}})}{\calD}\:=\:
\frac{1}{\calD}\sum_{c}d_{c}
\Tr(R^{\anti{b}a}_{c}R^{a\anti{b}}_{c})\:=\:
\frac{1}{\calD}\sum_{c}N^{c}_{a\anti{b}}
\frac{\theta_{c}}{\theta_{a}\theta_{b}}d_{c}.
\end{equation}
The normalization factor $1/\calD$ with $\calD=\sqrt{\sum_{a}d_{a}^{2}}$ is
chosen so that $S$ is unitary under certain conditions.

Let us study properties of the numbers $s_{ab}$. First, we observe these
symmetries:
\begin{equation}
s_{ab}=s_{ba}=s_{\anti{a}\anti{b}}=s_{\anti{b}\anti{a}}=
s_{\anti{a}b}^{*}=s_{b\anti{a}}^{*}=s_{a\anti{b}}^{*}=s_{\anti{b}a}^{*}.
\end{equation}
Indeed, up to normalization we have:
\[
s_{ab}\sim\figbox{1.0}{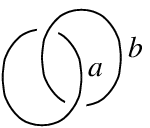}=\figbox{1.0}{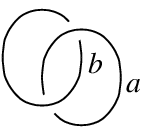}
\sim s_{ba},
\qquad
s_{ab}\sim\,\figbox{1.0}{sab.eps}\,=\figbox{1.0}{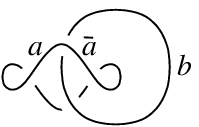}\,=\,
\theta_{a}^{\phantom{*}}\theta_{a}^{*}\,\figbox{1.0}{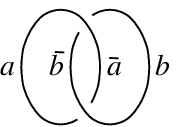}\,
\sim s_{\anti{a}b}^{*}.
\]
Factors of the form $s_{ax}/s_{1x}$ often arise via this equation:
\begin{equation}
\figbox{1.0}{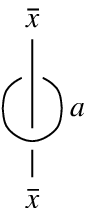}\;=\:\frac{s_{ax}}{s_{1x}}\,\figbox{1.0}{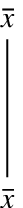}\,,
\qquad\qquad\quad
s_{1x}\,=\,\frac{d_{x}}{\calD}.
\end{equation}
For example, let us repeat the calculation done in proof of
Eq.~(\ref{qdimmult}), but now with an additional line passing through the
loops:
\[
\figbox{1.0}{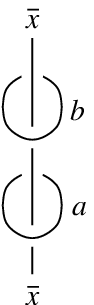}\;=\;\figbox{1.0}{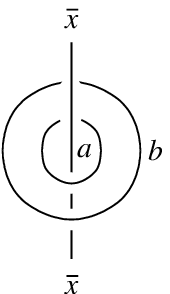}\;=\:
\sum_{c,j}\sqrt{\frac{d_{c}}{d_{a}d_{b}}}\,\figbox{1.0}{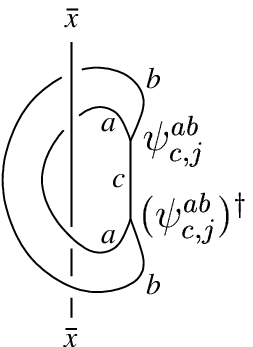}\;=\:
\sum_{c}N^{c}_{ab}\,\figbox{1.0}{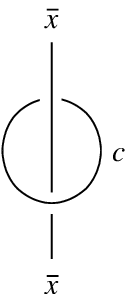}\:.
\]
The first and the last expression are equal to
$\displaystyle\frac{s_{ax}s_{bx}}{s_{1x}^{2}}\id_{\anti{x}}$\, and\,
$\displaystyle\sum_{c}N^{c}_{ab}\frac{s_{cx}}{s_{1x}}\id_{\anti{x}}$,\,
respectively. Thus,
\begin{equation}\label{Verlinde0}
\frac{s_{ax}s_{bx}}{s_{1x}}\,=\,\sum_{c}N^{c}_{ab}s_{cx}.
\end{equation}
Comparing this equation with Eq.~(\ref{fuchar}), we conclude that
\begin{equation}
\frac{s_{ax}}{s_{1x}}\,=\,\lambda_{j}(a)\qquad
\text{for some $j=j(x)$.}
\end{equation}
In other words, for each $x$ the map $a\mapsto s_{ax}/s_{1x}$ is a fusion
character.

In the next section we will see that if braiding is sufficiently nontrivial,
then the $S$-matrix is unitary. In this case, the map from labels to fusion
characters is one-to-one, and Eq.~(\ref{Verlinde0}) may be cast into a form
known as Verlinde formula:\footnote{Eqs.~(\ref{Verlinde0}), (\ref{Verlinde1})
were first obtained by Pasquier for statistical models and conformal theories
associated with Dynkin diagrams~\cite{Pasquier}. A general derivation in the
CFT framework is due to Verlinde~\cite{Verlinde}.}
\begin{equation}\label{Verlinde1}
N^{c}_{ab}\,=\,\sum_{x}\frac{s_{ax}s_{bx}s_{\anti{c}x}}{s_{1x}}.
\end{equation}

\subsection{Braiding nondegeneracy~$=$~modularity} \label{sec_modularity}

Let us recall the definition of a superselection sector: it is a class of
states that can be transformed one to another by local operators. From the
physical perspective, the operators we use must respect any unbroken symmetry
present in the Hamiltonian, e.g., the $\UU(1)$ symmetry associated with the
conservation of electric charge. Another example is the number of fermions
\latin{modulo}~$2$, if that number is not conserved as an integer. (Here we
speak about actual fermions forming the system rather than effective Majorana
modes obtained by a nonlocal transformation.)  However, in the model studied
in this paper the superselection sectors are stable with respect to \emph{all}
local operators. More generally, we may consider an arbitrary system built of
spins or other \emph{bosonic} degrees of freedom, and ask for properties that
are stable to a generic perturbation. In this case, the superselection sectors
have purely topological nature: some signature of a nontrivial excitation
$x\not=1$ will be preserved even if we cut out a piece of material containing
the quasiparticle. It is quite reasonable to assume that the presence of such
an excitation can actually be detected by an Aharonov-Bohm measurement, i.e.,
by moving a test particle $a$ around $x$. Thus we arrive at the condition that
the braiding is nondegenerate, which must be true for any anyonic system not
complicated by external symmetries.

\begin{Definition}\label{def_modularity} Braiding is said to be
\emph{nondegenerate} if for each label $x\not=1$ there is some label $a$ such
that the operator $R_{ax}R_{xa}$ is not identity.
\end{Definition}

\begin{Theorem}\label{th_Sunitary}
In a theory with nondegenerate braiding, the following operator $S_{z}$ acting
in the space $\calL_{z}=\bigoplus_{b}V^{bz}_{b}$ is unitary:
\begin{equation}
S_{z}\figbox{1.0}{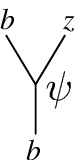}\:\bydef\;
\frac{1}{\calD}\sum_{a}d_{a}\,\figbox{1.0}{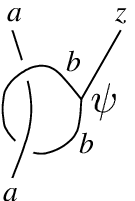}\:,\qquad\qquad
\text{where}\quad \psi\in V^{bz}_{b}.
\end{equation}
\end{Theorem}
Note that the standard $S$-matrix corresponds to $z=1$. The operator
$S_{z}^{\dag}$ differs from $S_{z}$ by the orientation of crossings. (This is
easy to show by considering a matrix element of $S_{z}$.) The converse of
Theorem~\ref{th_Sunitary} is also true and can be proved easily: if the
standard $S$-matrix is unitary, then the braiding is nondegenerate.

\begin{Lemma}\label{lem_trivbraid} The operator
$R_{ax}R_{xa}$ is trivial for all labels $a$ if and only if $x$ is mapped to
the trivial fusion character $j(x)=j(1)$, i.e., if $s_{ax}/s_{1x}=d_{a}$ for
all $a$.
\end{Lemma}
\begin{proof}
The triviality of $R_{ax}R_{xa}$ means that for all $c$,\,
$R^{ax}_{c}R^{xa}_{c}=\id^{xa}_{c}$, i.e.,
$\frac{\theta_{c}}{\theta_{x}\theta_{a}}=1$ whenever $N^{c}_{xa}\not=0$. We
calculate $s_{x\anti{a}}=s_{ax}^{*}$ using Eq.~(\ref{sab}):
\[
\calD\,s_{x\anti{a}}\,=\,\sum_{c}N^{c}_{xa}
\frac{\theta_{c}}{\theta_{x}\theta_{a}}d_{c}\,=\,
\sum_{c}N^{c}_{xa}d_{c}\,=\,d_{x}d_{a}.
\]
Thus, $s_{ax}/s_{1x}=d_{a}$. Conversely, if the operator $R_{ax}R_{xa}$ is
nontrivial, then
\[
\calD\,\Re s_{x\anti{a}}\,=\,\sum_{c}N^{c}_{xa}
\Re\left(\frac{\theta_{c}}{\theta_{x}\theta_{a}}\right)d_{c}\,<\,
\sum_{c}N^{c}_{xa}d_{c}\,=\,d_{x}d_{a},
\]
therefore $s_{ax}/s_{1x}\not=d_{a}$.
\end{proof}

The proof of Theorem~\ref{th_Sunitary} is based on the following equation,
which is useful on its own right:
\begin{equation} \label{Sproj}
\frac{1}{\calD^{2}}\sum_{a}d_{a}\,\figbox{1.0}{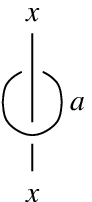}\;=\:
\delta_{j(x),j(1)}\:\figbox{1.0}{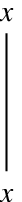}\,.
\end{equation}
Indeed, the condition $j(x)=j(1)$ means that the fusion character
$\lambda_{j(x)}:a\mapsto s_{ax}/s_{1x}$ is trivial, i.e.,
$s_{ax}/s_{1x}=d_{a}$. On the other hand, if $\lambda_{j(x)}$ is nontrivial,
then it is orthogonal to the trivial character, i.e.,
$\sum_{a}d_{a}s_{ax}^{*}=0$. The above equation combines both cases.

\begin{proof}[Proof of Theorem~\ref{th_Sunitary}.] 
By Lemma~\ref{lem_trivbraid}, the nondegeneracy condition implies that
$j(x)=j(1)$ if and only if $x=1$. Thus, the linear combination of loops in
Eq.~(\ref{Sproj}) is a projector onto label $1$. We apply it as follows:
\begin{equation}\label{proj2lines}
\frac{1}{\calD^{2}}\sum_{a}d_{a}\,\figbox{1.0}{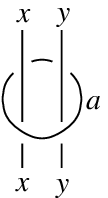}\;=\:
\frac{1}{d_{x}}\,\delta_{\anti{x}y}\:\figbox{1.0}{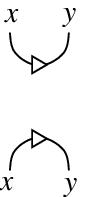}\;;
\medskip
\end{equation}
\[
S_{z}^{\dag}S_{z}\psi\:=\:
\frac{1}{\calD^{2}}\sum_{x,a}d_{x}d_{a}\,\figbox{1.0}{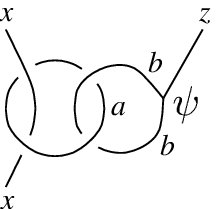}\:=\:
\sum_{x}d_{x}\frac{\delta_{xb}}{d_{x}}\,\figbox{1.0}{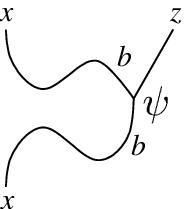}\:=\:
\psi.
\]
\end{proof}

Theories with a unitary $S$-matrix are often called \emph{modular} due to
their relation to the group of diffeomorphisms of the two-dimensional torus
considered up to topological equivalence (namely, isotopy). Let us discuss
this subject at a speculative level. The first thing to note is that a system
of spins or electrons can in principle be put on a torus whether or not the
braiding is nondegenerate. However, as indicated earlier, degeneracy generally
occurs due to some external symmetry. If it is a gauge symmetry, then putting
the system on the torus involves some choice. For example, in the case of
$\UU(1)$ symmetry, one may run an arbitrary magnetic flux through each basic
circle of the torus. Those fluxes can be detected by Aharonov-Bohm
measurements, but they cannot be changed by intrinsic operations, which
include splitting, fusion, and moving particles around the
torus. Mathematically, such operations form a so-called \emph{skein algebra},
which may be defined in terms of graphs on the torus up to equivalence
transformations. (Recall that a planar graph without external lines can be
transformed to a number, therefore the skein algebra of the plane is simply
$\CC$.) The skein algebra is a finite-dimensional $C^{*}$-algebra; in general
it is a direct sum of several blocks, each block being isomorphic to the
algebra of operators on some finite-dimensional Hilbert space. Physically,
different blocks correspond to different values of external parameters such as
the magnetic fluxes. Modular theories are special in that there is only one
block. In other words, the torus is characterized by a single
finite-dimensional space.

The Hilbert space of the torus is actually $\calM=\CC^{M}$. A basis in this
space may be associated with effective anyonic charge that is detected by an
Aharonov-Bohm measurement along some circle. Different circles correspond to
different bases. The $S$-matrix effects a transition between two bases. Other
important matrices are $C=(\delta_{\anti{a}b}: a,b\in M)$ (which corresponds
to a $180^{\circ}$-degree rotation) and $T=(\theta_{a}\delta_{ab}:a,b\in M)$
(which corresponds to a Dehn twist). We will justify this description by
showing that $S$, $C$, and $T$ obey the modular relations up to a phase factor
$\Theta$:
\begin{equation}\label{modrel}
(ST)^{3}=\Theta C,\qquad S^{2}=C,\qquad C^{2}=I, 
\end{equation}
where
\begin{equation}\label{Theta}
\Theta=\calD^{-1}\sum_{a}d_{a}^{2}\theta_{a}.
\end{equation}
As a corollary, note that $\Theta$ is a root of unity. Indeed, $S^{4}=I$,
therefore $\Theta^{4|M|}=(\det T)^{12}$ is a product of topological spins,
which are roots of unity by Theorem~\ref{th_Vafa}.

Let us actually study relations between more general operators acting in the
space $\calL_{z}=\bigoplus_{b}V^{bz}_{b}$ (which corresponds to a punctured
torus, with $z$ being the anyonic charge of the puncture). The operator
$S_{z}$ has been already defined, whereas
\begin{equation}
C_{z}\figbox{1.0}{psibzb.eps}\:\bydef\:
\theta_{b}^{*}\figbox{1.0}{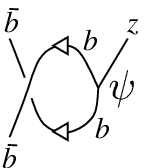}\:,\qquad\qquad
T_{z}\figbox{1.0}{psibzb.eps}\:\bydef\:\theta_{b}\figbox{1.0}{psibzb.eps}\,.
\end{equation}

\begin{Theorem}
The operators $S_{z}$, $C_{z}$, and $T_{z}$ satisfy the following modular
relations:
\begin{equation}
S_{z}^{\dag}T_{z}S_{z}\,=\,\Theta\,T_{z}^{\dag}S_{z}^{\dag}T_{z}^{\dag},
\qquad\quad
S_{z}=S_{z}^{\dag}C_{z},\qquad\quad
C_{z}^{2}=\theta_{z}^{*}.
\end{equation}
\end{Theorem}
\begin{proof}
The calculation of $S_{z}^{\dag}T_{z}S_{z}$ parallels the proof of
Theorem~\ref{th_Sunitary}, but the nondegeneracy condition is not necessary.
First, let us apply the $S$-matrix to the row
vector with entries $d_{a}\theta_{a}$:
\[
\sum_{a}d_{a}\theta_{a}s_{a\anti{x}}\,=\,
\frac{1}{\calD}\sum_{a,c}d_{a}N_{ax}^{c}
\frac{\theta_{c}}{\theta_{x}}d_{c}\,=\,
\frac{1}{\calD}\sum_{c}d_{x}d_{c}\frac{\theta_{c}}{\theta_{x}}d_{c}\,=\,
\Theta\, d_{x}\theta_{x}^{*}.
\]
The result may be written as follows:
\begin{equation}
\frac{1}{\calD}\sum_{a}d_{a}\theta_{a}\,\figbox{1.0}{sax.eps}\;=\:
\Theta\,\figbox{1.0}{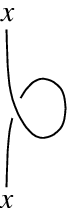}\:.
\end{equation}
Then we replace the single line by two lines and simplify the result (cf.\
Eq.~(\ref{brtwist})):
\begin{equation}
\frac{1}{\calD}\sum_{a}d_{a}\theta_{a}\,\figbox{1.0}{saxy.eps}\;=\:
\Theta\,\figbox{1.0}{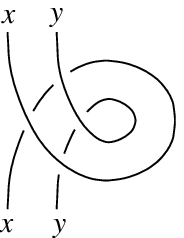}\;=\:
\Theta\,\theta_{x}^{*}\,\figbox{1.0}{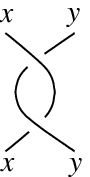}\,\theta_{y}^{*}.
\end{equation}
Attaching a graph representing $\psi\in V^{\anti{y}z}_{\anti{y}}$ to the first
and the last expression, we get $S_{z}^{\dag}T_{z}S_{z}\psi$ and
$\Theta\,T_{z}^{\dag}S_{z}^{\dag}T_{z}^{\dag}\psi$, respectively.

Now we show that $S_{z}=S_{z}^{\dag}C_{z}$. If $\psi\in V^{bz}_{b}$, then
\[
S_{z}\psi\:=\:
\frac{1}{\calD}\sum_{a}d_{a}\,\figbox{1.0}{sabz.eps}\:=\:
\frac{1}{\calD}\sum_{a}d_{a}\,\figbox{1.0}{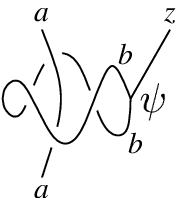}\:=\:
\frac{1}{\calD}\sum_{a}d_{a}\theta_{b}^{*}\,\figbox{1.0}{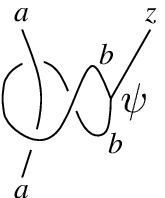}\:=\:
S_{z}^{\dag}C_{z}\psi.
\]
The formula $C_{z}^{2}\psi=\theta_{z}^{*}\psi$ follows from~(\ref{brtwist}).
\end{proof}

\subsection{Gauge freedom and Ocneanu rigidity} \label{sec_Ocneanu}

In the preceding sections, we defined a theory of anyons as a solution to a
certain system of algebraic equations for $F^{abc}_{u}$, $\alpha_{a}$,
$\beta_{a}$, and $R^{ab}_{c}$. Specifically, these are the pentagon equation,
the triangle and hexagon equations as well as unitarity conditions. (The
unitarity conditions are algebraic in the real and imaginary part of the
corresponding matrices.) The set of solutions is some real algebraic
variety. This description, however, does not take into account a gauge degree
of freedom. Indeed, two solutions are physically equivalent if they are
related to each other by a simultaneous basis change in the spaces
$V^{ab}_{c}$. In the examples studied in this paper, quotienting over such
transformations makes the set of solutions discrete. In other words, the
algebraic structure is \emph{rigid}, with the physical consequence that the
properties of anyons are stable to small perturbations of the Hamiltonian. Is
it true in general?  The affirmative answer was obtained by Ocneanu, but he
did not publish his proof. The only written proof I know of is due to Etingof,
Nikshych, and Ostrik~\cite{ENO}, but it is more general and hence complex; in
particular, it does not depend on the unitarity or the pivotal property.
Meanwhile, the proof of Ocneanu rigidity for unitary theories is not very
difficult and may be interesting to a mathematical physicist.

To prove that a solution to an equation system is rigid, one needs to study
infinitesimal deformations. Let us call a deformation of $F$, $\alpha$,
$\beta$, $R$ \emph{permissible} if it satisfies the equations in the first
order of Taylor expansion. The deformation is called \emph{trivial} if it can
be obtained by a change of basis, also in the first order. (By restricting our
attention to the first order, we potentially make the space of permissible
deformations larger and the space of trivial deformation smaller.) Our goal is
to calculate the quotient, permissible \latin{modulo} trivial, and show that
it vanishes. The analysis consists of several steps.

First, we only consider $F$ and the pentagon equation. The resulting
deformation problem resembles the definition of $H^{3}(G,\RR)$ for a group
$G$. In fact, it is exactly that in example~\ref{ex_H3} on
page~\pageref{ex_H3}. Associated with an arbitrary fusion theory is a cochain
complex defined by Crane and Yetter~\cite{CraneYetter} and independently by
Davydov~\cite{Davydov}. It is a sequence of real spaces and linear maps
\begin{equation}
C^{1}\xrightarrow{\delta^{1}}
C^{2}\xrightarrow{\delta^{2}}C^{3}\xrightarrow{\delta^{3}}\cdots
\end{equation}
whose third cohomology space $H^{3}(C)\bydef\Ker\delta^{3}/\Im\delta^{2}$
classifies the nontrivial deformations of the theory. (The first and the
second cohomology spaces have natural interpretation too.) We will use the
term ``tangent cohomology''\footnote{Etingof, Nikshych, and Ostrik call it
``Yetter cohomology'' in recognition of Yetter's further work in this area.}
proposed by Davydov.

It is well known that any finite group has trivial real cohomology in all
dimensions $n\ge 1$. Likewise, the tangent cohomology vanishes for an
arbitrary unitary fusion theory (actually, any fusion theory). The proof of
this statement is based on the same idea as the corresponding proof for
groups. In the latter case, one uses averaging over the group, which is
well-defined because the group is finite. For a unitary fusion theory, one
needs to take the quantum trace (actually, a partial quantum trace) and
average it over the label set with weight $d_{a}/\calD^{2}$.

The next step is to include $\alpha$, $\beta$ and to consider the triangle
equations. This is just a technical detail that involves trivial changes to
the deformation theory~\cite{Yetter01}. (The argument is quite general and
applicable even if the label set is infinite, in which case the cohomology may
not vanish.) The rigidity of braiding follows from the vanishing of $H^{2}(C)$
in conjunction with Vafa's theorem (see Theorem~\ref{th_Vafa}).

\subsubsection{Gauge freedom in the description of anyons}

An isomorphism between two fusion theories, $\calA$ and $\calA'$ is given by a
bijection between the label sets (we simply assume that they are equal) and a
collection of unitary maps\footnote{The direction of these maps is chosen to
be consistent with the definition of a tensor functor (see
Definition~\ref{def_tensfun}).}  $\Gamma^{ab}_{c}:(V')^{ab}_{c}\to
V^{ab}_{c}$; these data are enough to relate one system of associativity
constraints to the other:
\begin{equation} \label{1-iso}
F_{u}^{abc},(F')^{abc}_{u}:\, 
\figbox{1.0}{abcu1.eps} \,\longrightarrow\,
\figbox{1.0}{abcu2.eps}\,,\qquad
F^{abc}_{u} \Bigl(\sum_{e}\Gamma_{e}^{ab}\otimes\Gamma_{u}^{ec}\Bigr)\,=\,
\Bigl(\sum_{f}\Gamma_{u}^{af}\otimes\Gamma_{f}^{bc}\Bigr) (F')^{abc}_{u}.
\end{equation}
For example, let $\calA'$ be obtained from $\calA$ by changing left and right,
i.e., $(V')^{ab}_{c}=V^{ba}$ and $(F')^{abc}_{u}=(F^{cba}_{u})^{-1}$. If
theory $\calA$ has braiding $R$, then $\Gamma^{ab}_{c}=(R^{ab}_{c})^{-1}$ is
an isomorphism (for a quick demonstration, apply the Yang-Baxter
braid~(\ref{YaBa}) to both trees above). However, we will be mostly interested
in the case where $(V')^{ab}_{c}=V^{ab}_{c}$ so that $\Gamma^{ab}_{c}$ may be
called ``basis change''; examples can be found in Sections~\ref{sec_assoc}
and~\ref{sec_16}.

By analogy with Eq.~(\ref{1-iso}), one can write similar equations for
$\alpha_{a}$ vs.\ $\alpha_{a}'$ and $\beta_{a}$ vs.\ $\beta_{a}'$. However,
the physical meaning of $\alpha$ and $\beta$ is rather elusive, so it is not
clear whether we obtain the most general form of equivalence this way. A
reasonable notion of isomorphism between two theories, namely \emph{tensor
functor} arises naturally in the categorical formalism (see
Sec.~\ref{sec_tenscatdef}). To state it in concrete terms, we just need to
supplement the basis change with an overall phase factor $\gamma$ so that
\begin{equation} \label{1-iso1}
\alpha_{a}\,=\,\Gamma^{a1}_{a}\gamma\,\alpha'_{a},\qquad\qquad
\beta_{a}\,=\,\Gamma^{1a}_{a}\gamma\,\beta'_{a}.
\end{equation}

Finally, the braiding rules are changed as follows:
\begin{equation}
R^{ab}_{c}\Gamma^{ab}_{c}\,=\,\Gamma^{ba}_{c}(R')^{ab}_{c}.
\end{equation}

Thus we have defined the family of theories $\calA'$ that are isomorphic to a
given theory $\calA$. However, their parametrization by $\Gamma^{ab}_{c}$ and
$\gamma$ is redundant. In particular, $F$, $\alpha$, $\beta$, $R$ do not
change if we replace $\Gamma^{ab}_{c}$ and $\gamma$ with $\Phi^{ab}_{c}$ and
$\ph$ such that
\begin{equation} \label{2-iso}
\Phi^{ab}_{c}=\,\frac{h_{c}}{h_{a}h_{b}}\Gamma^{ab}_{c},\qquad\qquad
\ph=h_{1}\gamma,
\end{equation}
where $h_{x}$ are arbitrary phase factors. In the categorical language, the
tensor functors defined by $(\Gamma,\gamma)$ and by
$(\Phi,\ph)$ are isomorphic.

Thus we have entities of dimension $0$, $1$, and $2$: theories of anyons,
$1$-isomorphisms (i.e., isomorphisms between theories), and $2$-isomorphisms
(i.e., isomorphisms between $1$-isomorphisms). Let us define three sets
indexed by a complementary dimension:
\begin{enumerate}
\item $2$-automorphisms $h$ of a given $1$-isomorphism $(\Gamma,\gamma)$;
\item Equivalence classes of $1$-automorphisms $(\Gamma,\gamma)$ of a given
theory $\calA$ up to $2$-isomorphisms;
\item Equivalence classes of theories with given fusion multiplicities up to
$1$-isomorphisms.
\end{enumerate}
The first set is an Abelian group, the second is a group (see
Appendix~\ref{sec_symbreak}), the third does not have any special structure.

\subsubsection{Infinitesimal deformations and the vanishing of tangent
cohomology} Let us define infinitesimal analogues of sets 1, 2, 3; we will
eventually prove that they are trivial. The constructions involved are typical
to cohomology theory, following a pattern that may already be recognized by
studying set~1. We are eventually interested in set~3, which classifies
solutions of the pentagon equation up to a basis change. In this subsection we
ignore braiding and all attributes of the vacuum sector (i.e., $\gamma$,
$\alpha$, $\beta$, and the triangle equations); these things will be
considered later.

For a fixed fusion theory with associativity constraints $F^{abc}_{u}$, let
\begin{equation}
h_{a}\approx\,1-iX_{a},\qquad\quad
\Gamma^{ab}_{c}\,\approx\,\id_{V^{ab}_{c}}-iY^{ab}_{c},\qquad\quad
(F')^{abc}_{u}\,\approx\,F^{abc}_{u}\bigl(\id_{V^{abc}_{u}}-iZ^{abc}_{u}\bigr),
\end{equation}
where $X_{a}$ is an infinitely small real number and $Y^{ab}_{c}$,
$Z^{abc}_{u}$ are infinitely small Hermitian operators acting in $V^{ab}_{c}$
and $V^{abc}_{u}$, respectively.\footnote{To be rigorous, $Z^{abc}_{u}$ acts
in the space $\bigoplus_{e}V_{e}^{ab}\otimes V_{u}^{ec}$, the representation
of $V^{abc}_{u}$ corresponding to the $((ab)c)$ label grouping. But that is
not so important as we always identify different groupings ($=$~trees) using
$F$.} Now, let us substitute the expressions for $h$ and $\Gamma$ into
Eq.~(\ref{2-iso}), assuming that $\Phi^{ab}_{c}=\id_{V^{ab}_{c}}$. We get:
\begin{equation}
Y^{ab}_{c}\,=\,\bigl(\delta^{1}X\bigr)^{ab}_{c}\,\bydef\,
(X_{b}-X_{c}+X_{a})\id_{V^{ab}_{c}}.
\end{equation}
An infinitesimal analogue of set~1 is the space of deformations $X=(X_{a}:a\in
M)$ such that $\delta^{1}X=0$. One can deal with Eq.~(\ref{1-iso}) and the
pentagon equation in an analogous way.

To present the results in a more convenient form, let us get rid of the lower
index $c$. (In a categorical formulation of the theory, it does not appear at
all.) Note that the set of operators $Y^{ab}_{c}$ for all $c$ represents the
action of a single element $Y_{ab}\in V^{ab}_{ab}$, which may be constructed
as follows:
\begin{equation}
\figbox{1.0}{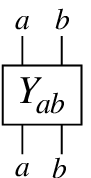}\:=\:
\sum_{c,j}\sqrt{\frac{d_{c}}{d_{a}d_{b}}}\,\figbox{1.0}{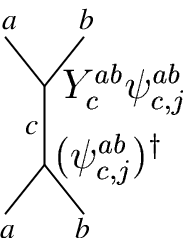}\:,
\qquad\quad
\text{where}\quad\dlangle\psi^{ab}_{c,j}|\psi^{ab}_{c,k}\drangle=\delta_{jk}
\end{equation}
(cf.\ Eq.~\ref{partidentity1}). For $\bigl(\delta^{1}X\bigr)^{ab}_{c}$
this procedure yields:
\begin{equation}
\bigl(\delta^{1}X\bigr)_{ab}\:=\;
\figbox{1.0}{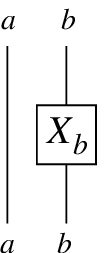}\:-\:
\sum_{c,j}\sqrt{\frac{d_{c}}{d_{a}d_{b}}}\,\figbox{1.0}{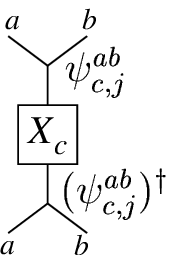}\,+\:
\figbox{1.0}{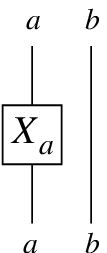}\:.
\end{equation}
More generally, we can use the following definition.

\begin{Definition}
Let $C^{n}$\, be the set of self-adjoint elements in
$\bigoplus_{a_{1},\dots,a_{n}}V^{a_{1}\dots a_{n}}_{a_{1}\dots a_{n}}$. The
\emph{tangent complex} of the fusion theory is the sequence of $\RR$-linear
maps
\begin{equation}
C^{0}\xrightarrow{\delta^{0}}C^{1}\xrightarrow{\delta^{1}}
C^{2}\xrightarrow{\delta^{2}}C^{3}\xrightarrow{\delta^{3}}\cdots,
\qquad\qquad
\delta^{n}\,=\,\sum_{k=0}^{n+1}(-1)^{k}f^{n}_{k},
\end{equation}
where the maps $f^{n}_{k}:\,C^{n}\to C^{n+1}$ are defined as follows:
\begin{equation}
\begin{gathered}
\bigl(f^{n}_{0}X\bigr)_{a_{1}\dots a_{n+1}}\,=\:
\figbox{1.0}{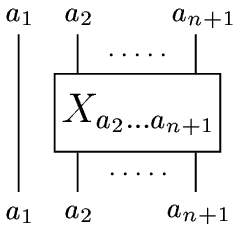}\,,\qquad\qquad\qquad
\bigl(f^{n}_{n+1}X\bigr)_{a_{1}\dots a_{n+1}}\,=\:
\figbox{1.0}{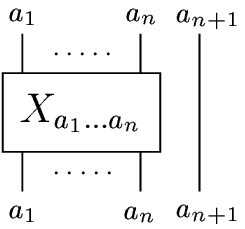}\,,
\\
\bigl(f^{n}_{k}X\bigr)_{a_{1}\dots a_{n+1}}\,=\:
\sum_{c,j}\sqrt{\frac{d_{c}}{d_{a_{k}}d_{a_{k+1}}}}\,
\figbox{1.0}{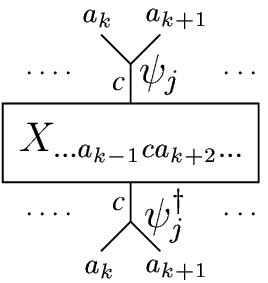}\:\qquad\quad
\text{for}\enspace k=1,\dots,n.
\end{gathered}
\end{equation}
(For $n=0$ we use $V^{\varnothing}_{\varnothing}\bydef V^{1}_{1}=\CC$, hence
$C^{0}=\RR$. Correspondingly,
$f^{0}_{0},f^{0}_{1}:\,1\mapsto\sum_{a}\id_{V^{a}_{a}}$, therefore
$\delta^{0}=f^{0}_{0}-f^{0}_{1}=0$.)
\end{Definition}

Note that $C$ is indeed a cochain complex, i.e.,
$\delta^{n+1}\delta^{n}=0$, which follows from this easily verifiable
identity:
\begin{equation}
f^{n+1}_{k}f^{n}_{m}=f^{n+1}_{m+1}f^{n}_{k}\qquad\quad
\text{for}\enspace 0\le k\le m\le n+1.
\end{equation}
(It is part of structure that makes $C$ into a \emph{cosimplicial
space}.) 

The reader may check that Eq.~(\ref{1-iso}) becomes
$Z=\delta^{2}Y$. Therefore the infinitesimal analogue of set~2 is given by
solutions to the equation $\delta^{2}Y=0$ \latin{modulo} elements of the
form $Y=\delta^{1}X$. Similarly, the pentagon equation may be written as
$\delta^{3}Z=0$, and the solutions should be considered \latin{modulo}
$\delta^{2}Y$. Thus, low-dimensional cohomology of the tangent complex has
the following meaning:
\begin{itemize}
\item $H^{0}(C)=\RR$;
\item $H^{1}(C)$ classifies infinitesimal $2$-automorphisms of the identity
$1$-automorphism;
\item $H^{2}(C)$ classifies infinitesimal $1$-automorphisms (i.e., basis
changes that leave the associativity constraints invariant) up to
$2$-isomorphisms;
\item $H^{3}(C)$ classifies infinitesimal deformations of the fusion theory
up to $1$-isomorphisms (i.e., arbitrary basis changes).
\end{itemize}

\begin{Theorem}\label{th_tancoh}
$\,H^{n}(C)=0\,$ for all $n>0$.
\end{Theorem}
\begin{proof} We will use a standard method of proving vanishing cohomology
results, namely \emph{contracting homotopy}. Let
\begin{equation}
\chi^{n}:C^{n}\to C^{n-1},\qquad\qquad
\bigl(\chi^{n}X\bigr)_{a_{1}\dots a_{n-1}}\:=\:
\frac{1}{\calD^{2}}\sum_{c}d_{c}\,\figbox{1.0}{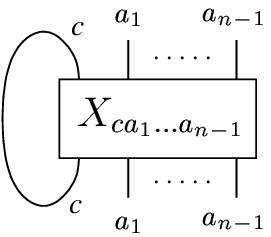}\:.
\end{equation}
We will show that $\delta\chi+\chi\delta=1$ or, more exactly,
\begin{equation} \label{contraction}
\chi^{n+1}\delta^{n}+\delta^{n-1}\chi^{n}\,=\,\id_{C^{n}}
\qquad\quad \text{for}\enspace n>0.
\end{equation}
If this is true, then any $X\in\Ker\delta^{n}$ can be represented as
$\delta^{n-1}\chi^{n}X\in\Im\delta^{n-1}$, hence $H^{n}(C)=0$.

Equation~(\ref{contraction}) is an immediate corollary of this identity:
\begin{equation}
\chi^{n+1}f^{n}_{k}\,=\,
\left\{\begin{array}{l@{\quad}l}
\id_{C^{n}} & \text{if}\ k=0,\\[3pt]
f^{n-1}_{k-1}\chi^{n} & \text{if}\ k,n>0.
\end{array}\right.
\end{equation}
Let us rewrite it using graphic notation:
\[
\begin{array}{l@{\qquad\quad}l}
k=0: & \displaystyle
\frac{1}{\calD^{2}}\sum_{c}d_{c}\,\figbox{1.0}{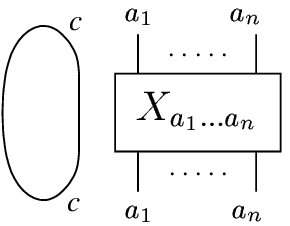}\;=\:
X_{a_{1}\dots a_{n}}\,,
\\
k=1: & \displaystyle
\frac{1}{\calD^{2}}\sum_{a,c,j}d_{a}
\sqrt{\frac{d_{c}}{d_{a}d_{a_{1}}}}\,\figbox{1.0}{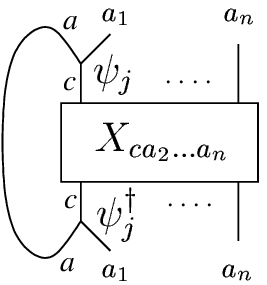}\;=\:
\frac{1}{\calD^{2}}\sum_{c}d_{c}\,\figbox{1.0}{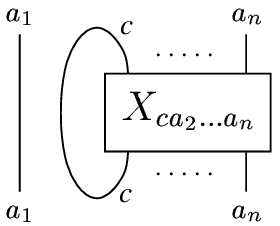}
\end{array}
\]
(the case $k>1$ is trivial). The equation for $k=0$ is also obvious, whereas
the one for $k=1$ follows from this identity:
\[
\sum_{j}\,\figbox{1.0}{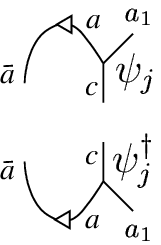}\;=\:\sum_{j}\,\figbox{1.0}{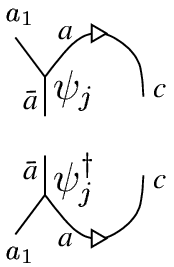},
\]
where we have used Lemma~\ref{lem_raiselower1}. (Note that we did not use the
pivotal property or the positivity of the inner product; this sheds some light
on why the result holds in a more general setting~\cite{ENO}.)
\end{proof}

\subsubsection{Technicalities related to the unit and braiding}

Let us now take into account additional structure that was neglected in the
above analysis. We will get more variables and more equations, but the old
equations will not change. The new degrees of freedom are characterized by
infinitely small real numbers $g$, $A_{a}$, $B_{a}$ and Hermitian operators
$W^{ab}_{c}$ which are defined as follows:
\begin{equation}
\gamma\approx\,1-ig,\qquad\quad
\alpha_{a}'\approx\,(1-iA_{a})\alpha_{a},\qquad
\beta_{a}'\approx\,(1-iB_{a})\beta_{a},
\smallskip
\end{equation}
\begin{equation}
(R')^{ab}_{c}\approx\,R^{ab}_{c}(\id_{V^{ab}_{c}}-iW^{ab}_{c}).
\end{equation}
(Recall that $\gamma$ is part of the definition of a $1$-isomorphism.) It has
been previously shown that any infinitesimal deformation of the associativity
constraints $F$ can be compensated by a suitable basis change
$\Gamma$. Although $\alpha$, $\beta$, and $R$ may still remain deformed, the
problem is reduced to the case where $F$ is fixed. Furthermore, any
infinitesimal basis change not affecting $F$ has the form~(\ref{2-iso}), hence
$\alpha$, $\beta$, or $R$ are not altered either.  The only parameter to tune
is $\gamma\approx 1-ig$.

Let us first show that any permissible deformation of $\alpha$ and $\beta$ is
trivial, where ``permissible'' means satisfying the triangle equations and
``trivial'' means satisfying Eq.~(\ref{1-iso1}). Taken to an infinitesimal
form, the first triangle equation (see Fig.~\ref{fig_triang}a) and
Eq.~(\ref{1-iso1}) read:
\begin{equation}
A_{x}=B_{w}\quad\text{for all $x,w$};\qquad\qquad
g+A_{a}=g+B_{a}=0\quad\text{for all $a$}.
\end{equation}
Clearly, the first condition implies the second if we put $g=-A_{1}$.

Thus we may assume that the whole fusion theory, i.e., $F$, $\alpha$, and
$\beta$ are fixed. It remains to show that the braiding deformation $W$
vanishes, provided it satisfies an infinitesimal version of the hexagon
equations. Instead of using the hexagon equations directly, we will rely on
the fact that braiding defines an isomorphism between the given fusion theory
$\calA$ and the theory $\calA'$ in which left and right are changed. Thus
$(R^{ab}_{c})^{-1}(R')^{ab}_{c}$ is an automorphism of $\calA$. It follows
that $\delta^{2}W=0$, hence
\[
W^{ab}_{c}\,=\,\bigl(\delta^{1}X\bigr)^{ab}_{c}\,=\,
(X_{b}-X_{c}+X_{a})\id_{V^{ab}_{c}}.
\]
for some $X\in C^{1}$. It is important that the right-hand side is a scalar
times the identity, which enables us to calculate the deformation of
$R_{c}^{ba}R_{c}^{ab}=\frac{\theta_{c}}{\theta_{a}\theta_{b}}$ easily:
\[
\frac{\theta_{c}'}{\theta_{a}'\theta_{b}'}\approx\,
\frac{\theta_{c}}{\theta_{a}\theta_{b}}\bigl(1-2i(X_{b}-X_{c}+X_{a})\bigr).
\]
But the topological spin is rigid due to Vafa's theorem, hence
$X_{b}-X_{c}+X_{a}=0$.

\subsection{Categorical formalism (aside)} \label{sec_tenscat}

Categories and functors are the language used by mathematicians to describe
fusion, braiding, and related concepts. I originally tried to write an
exposition of the theory of anyons using this formalism, but found it too
awkward. I still think that functors are necessary for the understanding of
phase transitions and other advanced properties of anyonic systems, but most
things can be explained in more elementary terms. This section is a remainder
of the abandoned plan. Please be warned that these notes are very incomplete,
e.g., there is no discussion of duality and related concepts: rigid, pivotal,
and spherical categories (not to mention that we focus on semisimple
categories --- this restriction is natural for the intended applications).

Let us outline the main elements of the theory. An abstraction called
\emph{tensor category} generalizes the notion of fusion theory. While anyonic
fusion has a compact description in terms of basic data, a category is a huge
collection of ``objects'' related by ``morphisms''.  However, these relations
form a regular structure that does not leave much freedom of choice. A rather
trivial example is the category $\Hilb$, whose objects are all possible
finite-dimensional Hilbert spaces and the morphisms are all linear maps. A
more interesting category $\Rep(G)$ is defined as follows: the objects are
finite-dimensional unitary representations of a compact group $G$ and the
morphisms are intertwiners. We may also think of a fusion theory as a
category: the objects are finite sequences of particle labels, and the
morphisms between $(b_{1},\dots,b_{m})$ and $(a_{1},\dots,a_{k})$ are
fusion/splitting operators, i.e., elements of the space $V_{b_{1}\dots
b_{m}}^{a_{1}\dots a_{k}}$. Another way to turn a fusion theory into a
category will be described later.

By definition, a tensor category is equipped with an operation $\tens$ that is
analogous to fusion. The role of the vacuum sector is played by a special
object $\unit$. In the categories $\Hilb$ and $\Rep(G)$, the operation $\tens$
is the usual tensor product and $\unit$ is the set of complex numbers $\CC$
regarded as a one-dimensional space or the trivial representation of $G$.

Yet another example: the matrix category $\Matr_{m,m}$. In this
category,\footnote{The more scientific name for this creature is
$\CC^{m}$-$\mathrm{mod}$-$\CC^{m}$, the category of finite-dimensional
bimodules over the algebra $\CC^{m}$ (the product of $m$ copies of
$\CC$).}  objects are ${m\times m}$ matrices whose entries are
finite-dimensional complex linear spaces, morphisms between matrices are
entrywise linear maps, and
\begin{equation} \label{matcatmult}
(A\tens B)_{jl}\,=\,\bigoplus_{k}A_{jk}\otimes B_{kl},\qquad\qquad
\unit_{jk}\,=
\left\{\begin{array}{ll}
\CC & \text{if}\ j=k,\smallskip\\
0 & \text{if}\ j\not=k.
\end{array}\right.
\end{equation}
(If the matrix entries are finite-dimensional Hilbert spaces, we use the
notation $\Matr_{m,m}^{\dag}$ because in this case each morphism has a
Hermitian adjoint.) Note that $(A\tens B)\tens C=A\tens(B\tens C)$ and
$A\tens\unit=A=\unit\tens A$; tensor categories with this property are called
\emph{strict}. In general, the equalities are replaced by isomorphisms $F$,
$\alpha$, and $\beta$ satisfying the pentagon and triangle equations.

A \emph{tensor functor} is a map from one tensor category to another, a
classic example being the embedding $\Rep(G)\to\Hilb$. Tensor functors have
some physical applications, such as transformations of particles by global
symmetries (see Appendix~\ref{sec_symbreak}). They are also related to the
gauge freedom in the description of anyonic fusion, cf.\ Eqs.~(\ref{1-iso}),
(\ref{1-iso1}), and~(\ref{2-iso}). However, our main goal is to understand the
status of the isomorphisms $F$, $\alpha$, $\beta$. To this end, we will find
an embedding of an arbitrary tensor category into a strict one by a tensor
functor that preserves distinction between morphisms (such functors are called
\emph{faithful}). In particular, any unitary fusion theory with label set $M$
embeds into the matrix category $\Matr^{\dag}_{M,M}$. The embedding theorem
implies
\begin{remark}[MacLane's coherence theorem:]
\emph{All morphisms composed of $F$, $\alpha$, $\beta$ and having the same
source and target are equal.}
\end{remark}
Indeed, in the strict category the source and the target are the same object,
and any composition of $F$, $\alpha$, and $\beta$ is the identity morphism
(i.e., equality). Thus all such compositions in the original category are
mapped to the same morphism. Since the functor preserves distinction between
morphisms, the original morphisms are also equal.

\subsubsection{The basics of category theory}
A \emph{category} is a collection of objects and morphisms. The set of
morphisms from $A$ to $B$ in a category $\calC$ is denoted by $\calC(A,B)$.

\emph{Morphisms} are anything that can be composed: if $f:A\to B$ and $g:B\to
C$, then $gf:A\to C$. It is only required that $(hg)f=h(gf)$ and that every
object $A$ has an \emph{identity morphism} $\id_{A}$ such that
$f\id_{A}=\id_{B}f=f$ for any $f:A\to B$. (Example: objects are vertices of a
given graph, morphisms are paths of arbitrary length, identity morphisms are
paths of length $0$.)

A morphism $f:A\to B$ is called an \emph{isomorphism} if there is a morphism
$f^{-1}:B\to A$ such that $f^{-1}f=\id_{A}$ and $ff^{-1}=\id_{B}$. Notation
for isomorphic objects: $A\cong B$.

For applications in quantum mechanics, the basic example is the category
$\Hilb$ of finite-dimensional Hilbert spaces. Among its objects are the set of
complex numbers $\CC$ and the one-dimensional space corresponding to the first
level of a harmonic oscillator. These two spaces are isomorphic but
different. (The isomorphism maps the complex number $1$ to a normalized
wavefunction $\psi$, but this map is not canonical because there is no reason
to prefer $\psi$ over $-\psi$.)  So, it does not generally make sense to ask
whether two given objects are \emph{equal} (unless they are the same by
definition). Morphisms between two given objects may be compared for equality
though. For two objects, $X$ and $Y$, a reasonable question is whether $X\cong
Y$. Of course, two spaces are isomorphic if and only if they have the same
dimension. But we also want to keep track of isomorphisms, for they may not
commute: a sequence of isomorphisms may result in a nontrivial automorphism
$u:X\to X$,\, $u\not=\id_X$. (For Hilbert spaces, it is natural to consider
unitary isomorphisms.)

The next example has motivation in superselection theory. Let us consider an
infinite spin system which is almost in the ground state, with excitations
being allowed only in some finite region. Quantum states of such a system are
classified by superselection sectors. For each sector $a$ the state belongs to
some finite-dimensional Hilbert space $X_{a}$, which depends on specific
constraints on the excitations. We are going to consider different sets of
constraints (called ``objects'') and transformations from one object to
another (called ``morphisms'').

\begin{Definition}
Let $M$ be some set. The category $\Vect^{\dag}_{M}$ is defined as follows.
\begin{itemize}
\item An \emph{object} $X\in\Vect^{\dag}_{M}$ is a collection of
finite-dimensional Hilbert spaces $(X_{a}:a\in M)$, of which only finitely
many are nonzero.
\item A \emph{morphism} $f:X\to Y$ is a collection of linear maps
$f_{a}:X_{a}\to Y_{a}$ (for each $a\in M$). The \emph{identity morphism}
$\id_{X}:X\to X$ consists of the unit operators acting in $X_{a}$.
\item Additional structure: The set $\Vect^{\dag}_{M}(X,Y)$ of morphisms from
$X$ to $Y$ is a complex linear space. For any $f:X\to Y$ we define the
\emph{adjoint morphism} $f^{\dag}:Y\to X$ such that
$(f^{\dag})_{a}=(f_{a})^{\dag}$.
\end{itemize}
A \emph{unitary isomorphism} is an isomorphism $f$ such that
$f^{-1}=f^{\dag}$.
\end{Definition}
An simplified version of this construction is the category $\Vect_{M}$: we use
complex linear spaces instead of Hilbert spaces and do not consider adjoint
morphisms.\smallskip

For each $a\in M$ we define the object $\refobj{a}$ such that
$\refobj{a}_{a}=\CC$ and $\refobj{a}_{b}=0$ for $b\not=a$. Objects isomorphic
to $\refobj{a}$ are called \emph{simple}. For any object $Y$,
\begin{equation}\label{refobj}
Y_{a}\,=\,\Vect^{\dag}_{M}\bigl(\refobj{a},Y\bigr).
\end{equation}
From the physical perspective, simple objects correspond to quantum states
(e.g., a particle pinned to a point), whereas $\refobj{a}$ is a reference
state in the given superselection sector. For a more formal example, consider
the category $\Rep(G)$ whose objects are finite-dimensional unitary
representations of a compact group $G$ and whose morphisms are
intertwiners. Let $M$ be the set of irreps considered up to isomorphism, and
let $\refobj{a}$ be a particular irrep in the given isomorphism class. Then we
may identify the category $\calA=\Rep(G)$ with $\Vect^{\dag}_{M}$ this way:
$Y_{a}\bydef\calA(\refobj{a},Y\bigr)$.

\begin{Remark}
Our definition of $\Vect_{M}$ and $\Vect^{\dag}_{M}$ resembles the
construction of a vector space using coordinates. An invariant
characterization is also possible, though somewhat complicated. When the label
set $M$ is unspecified, $\Vect_{M}$ is called a \emph{semisimple $\CC$-linear
category}, and $\Vect^{\dag}_{M}$ is called a \emph{unitary category}.
\end{Remark}

The concepts of \emph{functor} and \emph{natural morphism} are central in
category theory. They are extremely general and therefore hard to grasp. We
will try to illustrate them by simple examples, which are still rather
abstract. More meaningful (though less direct) examples can be found in the
next subsection.

\begin{Definition}
A \emph{functor} $\calF:\calA\to\calB$ maps each object $A$ of the category
$\calA$ to an object $\calF(A)$ of the category $\calB$ and each morphism
$f:A\to A'$ to a morphism $\calF(f):\calF(A)\to\calF(A')$ such that
\[
\calF(f_{2}f_{1})\,=\,\calF(f_{2})\,\calF(f_{1}),\qquad\qquad
\calF(\id_{A})=\id_{\calF(A)}.
\]
A \emph{morphism} (or \emph{natural transformation}) between two functors
$\calF,\calG:\calA\to\calB$ is a collection $h$ of morphisms
$h_{A}:\calF(A)\to\calG(A)$ such that for any $f:A\to A'$ this diagram
commutes
\begin{equation}\label{natmor}
\vcenter{\xymatrix@+5pt{
{\calF(A)}  \ar[d]_{h_{A}} \ar[r]^{\calF(f)} &
{\calF(A')} \ar[d]^{h_{A'}} \\
{\calG(A)}  \ar[r]^{\calG(f)} &
{\calG(A')}
}}
\end{equation}
\end{Definition}

We have already mentioned one example of a functor, namely the embedding
$\calF:\Rep(G)\to\Hilb$. Acting on morphisms, it maps the set of intertwiners
to the set of all linear maps between two representation spaces. A natural
morphism $h:\calF\to\calF$ may be constructed as follows: $h_{A}$ is the
action of some fixed group element on the representation space of $A$. Indeed,
for this particular choice of the functors $\calF=\calG$ the commutative
diagram~(\ref{natmor}) simply says that $\calF(f)$ is an intertwiner.

\begin{Example}
Let $H$ be a finite-dimensional Hilbert space. We define a unitary functor
$\calH:\Hilb\to\Hilb$ by tensoring with $H$ on the left (notation:
$\calH=\lefttens{H}$):
\[
\calH(A)\bydef H\otimes A,\qquad\qquad \calH(f)\bydef \id_{H}\otimes f.
\]
A linear map $u:H\to G$ between two spaces defines a natural morphism
$U=\lefttens{u}$ between the corresponding functors:
\[
U_{A}\bydef u\otimes \id_{A}.
\]
\end{Example}

The functor $\calH=\lefttens{H}$ has some special properties, namely the map
$f\mapsto\calH(f)$ is linear and $\calH(f^{\dag})=(\calH(f))^{\dag}$. Such
functors are called \emph{unitary}.

\begin{Proposition} \label{prop_Hilbfunc}
\begin{enumerate}
\item
Any unitary functor $\calF:\Hilb\to\Hilb$ is naturally isomorphic to the
left-tensoring functor $\lefttens{H}$ for some space $H$, namely
$H=\calF(\CC)$. This isomorphism is unitary.
\item
Any natural morphism $U:\,\lefttens{G}\to\lefttens{H}$ has the form
$\lefttens{u}$, where $u=U_{\CC}$.
\end{enumerate}
\end{Proposition}
\begin{proof}
The idea is very simple: any space is a direct sum of one-dimensional spaces,
any one-dimensional space is isomorphic to $\CC$, therefore $\calF$ and $U$
are completely characterized by their action on $\CC$. Let us go through the
detail to see how the formalism works.

1.\, Let $A$ be an arbitrary finite-dimensional Hilbert space. Elements of $A$
may be associated with morphisms $\CC\to A$. An orthonormal basis corresponds
to a set of morphisms $\psi_{j}:\CC\to A$ such that
$\psi_{k}^{\dag}\psi_{j}=\delta_{kj}\id_{\CC}$ and
$\sum_{j}\psi_{j}\psi_{j}^{\dag}=\id_{A}$. Since $\calF$ is a unitary functor,
\[
\calF(\psi_{k})^{\dag}\,\calF(\psi_{j})\,=\,\delta_{kj}\id_{\calF(\CC)},
\qquad\quad
\sum_{j}\calF(\psi_{j})\,\calF(\psi_{j})^{\dag}\,=\,\id_{\calF(A)}.
\]
The same is true for the functor $\calH=\lefttens{\calF(\CC)}$. We define a
morphism $h_{A}:\calF(A)\to\calH(A)$ as follows:
\[
h_{A}\,=\,\sum_{j}\calH(\psi_{j})\,h_{\CC}\,\calF(\psi_{j})^{\dag},
\]
where $h_{\CC}:\calF(\CC)\to\calH(\CC)=\calF(\CC)\otimes\CC$ is an equality
(indeed, $X=X\otimes\CC$ for any space $X$). It is obvious that
$h_{A}^{\dag}h_{A}=\id_{\calF(A)}$ and
$h_{A}h_{A}^{\dag}=\id_{\calH(A)}$, hence $h_{A}$ is a unitary
isomorphism. To show that $h$ is natural, consider another object $A'$ with an
orthonormal basis $\{\psi'_{j}\}$ and a linear map
$f=\sum_{j,k}c_{jk}\psi'_{j}\psi_{k}^{\dag}$, where $c_{jk}\in\CC$. It is easy
to check that
\[
h_{A'}\,\calF(f)\,=\,
\sum_{j,k} c_{jk}\,\calH(\psi'_{j})\,h_{\CC}\,\calF(\psi_{k})^{\dag}
\,=\,\calH(f)\,h_{A}.
\]

2.\, Let $\calG=\lefttens{G}$ and $\calH=\lefttens{H}$. For an arbitrary
object $A$ with an orthonormal basis $\{\psi_{j}\}$ we have
$U_{A}\,\calG(\psi_{j})=\calH(\psi_{j})\,U_{\CC}$ (due to the naturality of
$U$). Hence
\[
U_{A}\,=\, U_{A}\id_{\calG(A)} \,=\,
U_{A} \left(\sum_{j}\calG(\psi_{j})\calG(\psi_{j}^{\dag})\right)
\,=\, \sum_{j}\calH(\psi_{j})\,U_{\CC}\,\calG(\psi_{j}^{\dag})
\,=\, U_{\CC}\otimes\id_{A}.
\]
\end{proof}

Adding another level of abstraction, we can reformulate
Proposition~\ref{prop_Hilbfunc} as follows: \emph{The category
$\Func^{\dag}(\Hilb,\Hilb)$ of unitary functors from $\Hilb$ to $\Hilb$ is
isomorphic to the category $\Hilb$}. (One may replace $\Hilb$ by the category
of finite-dimensional complex linear spaces, omitting the unitarity
condition.) By analogy, we obtain the following result.

\begin{Proposition} \label{prop_matrixcat}
The category $\Func^{\dag}(\Vect^{\dag}_{M},\Vect^{\dag}_{N})$ is isomorphic
to the category $\Matr^{\dag}_{N,M}$ whose objects are $N\times M$ matrices of
finite-dimensional Hilbert spaces with a finite number of nonzero entries in
each column and whose morphisms are entrywise linear maps between such
matrices. (If $M$ is finite, then
$\Matr^{\dag}_{N,M}\cong\Vect^{\dag}_{N\times M}$.)
\end{Proposition}

\subsubsection{Fusion in categorical terms} Fusion theory may be formulated as
additional structure on the category $\calC=\Vect^{\dag}_{M}$. Let us assume
that the the label set $M$ contains a special element $1$ and that
$V^{ab}_{c}$, $F^{abc}_{u}$, $\alpha_{a}$, $\beta_{a}$ are defined and satisfy
the pentagon and triangle equations as well as unitarity conditions. If in
addition $M$ is finite and the duality axiom on page~\pageref{ax_duality}
holds, then the resulting construction is called \emph{unitary fusion
category}. (For a general definition of fusion category see Ref.~\cite{ENO}.)

The use of categorical formalism has mathematical as well as physical
motivation. Let $A$ and $B$ be spatially confined excitations such that $A$ is
located on the left of $B$. Each excitation may be described by an object in
the category $\calC$. If we consider both excitations together, we will obtain
a new object, $A\tens B$, which may be called ``physical tensor product''. If
$A$ and $B$ do not have local degrees of freedom, i.e., $A\cong\refobj{a}$,\,
$B\cong\refobj{b}$, then $(A\tens B)_{c}\cong V_{c}^{ab}$. A nice property of
the physical tensor product is that $(A\tens B)\tens C=A\tens(B\tens C)$
(provided $A$, $B$, and $C$ are arranged on the line in that particular
order).  Mathematically, the operation $\tens$ is only associative up to a
canonical isomorphism, but we will find some abstract representation of
objects in which the isomorphism becomes an equality.

The tensor product of two objects, $A,B\in\calC$ is defined by the equation
\begin{equation}\label{tensprod}
\left(A\tens B\right)_{c}\,=\, \bigoplus_{a,b}
V_{c}^{ab}\otimes A_{a}\otimes B_{b}.
\end{equation}
This operation is neither commutative nor \latin{a priory} associative.
Different ways to multiply several objects may be described by parenthesis
structures, or trees. For example,
\[
\figbox{1.0}{abcu1.eps}\;:  \qquad\qquad\quad
\Bigl(\bigl(\refobj{a}\tens\refobj{b}\bigr)\tens\refobj{c}\Bigr)_{u}
\ =\ \bigoplus_{e}V_{e}^{ab}\otimes V_{u}^{ec}.
\]

The operation $\tens$ is a \emph{functor}, meaning that for any morphisms
$f:A\to A'$ and $g:B\to B'$ there is a morphism $f\tens g:A\tens B\to A'\tens
B'$ such that
\begin{equation}\label{functorprop}
(f_{2}f_{1})\tens (g_{2}g_{1})
=(f_{2}\tens g_{2})(f_{1}\tens g_{1}), \qquad\qquad
\id_{A}\tens\id_{B}=\id_{A\tens B}.
\end{equation}
(The tensor product of morphisms is defined in the obvious way: $(f\tens
g)_{c}=\sum_{a,b}\id_{V_{c}^{ab}}\otimes f_{a}\otimes g_{b}$.)

We may define a unitary isomorphism between $(A\tens B)\tens C$ and
$A\tens(B\tens C)$:
\begin{equation}\label{Ffusion}
F_{A,B,C}: (A\tens B)\tens C\to A\tens(B\tens C),\qquad
\bigl(F_{A,B,C}\bigr)_{u}=\,
\sum_{a,b,c}F_{u}^{abc}\otimes
\id_{A_{a}}\otimes \id_{B_{b}}\otimes \id_{C_{c}},
\end{equation}
where $F_{u}^{abc}$ is the associativity map~(\ref{assoc}). Similarly,
\begin{equation}
\alpha_{A}: A\to A\tens\refobj{1},\qquad\qquad\quad
\beta_{A}: A\to \refobj{1}\tens A
\end{equation}
(the definition is obvious).  It is important that the isomorphisms
$F_{A,B,C}$, $\alpha_{A}$, $\beta_{A}$ are \emph{natural}, i.e., for any
morphisms $f:A\to A'$,\, $g:B\to B'$, and $h:C\to C'$ we have
\begin{equation}\label{naturality}
\begin{array}{c}
F_{A',B',C'}\,\bigl((f\tens g)\tens h\bigr)\,=\,
\bigl(f\tens(g\tens h)\bigr)\,F_{A,B,C},
\medskip\\
\alpha_{A'}f\,=\,(f\tens\id_{\refobj{1}})\,\alpha_{A}, \qquad
\beta_{A'}f\,=\,(\id_{\refobj{1}}\tens f)\,\beta_{A}.
\end{array}
\end{equation}

\begin{Proposition} \label{prop_quadr}
The functor property~(\ref{functorprop}) and the naturality~(\ref{naturality})
imply all quadrilateral identities of the tree calculus.
\end{Proposition}
Indeed, there are two kinds of such identities. In the example shown in
Fig.~\ref{fig_pentquadr}b, one of the commuting moves occurs inside a subtree
descending from the other. The inner move plays the role of $f$ in the
naturality condition. If both moves occur in disjoint subtrees, one should use
the functor property.

\subsubsection{Main definitions} \label{sec_tenscatdef}

Among things we are going to define, \emph{tensor category} is just a general
formulation of properties of the operation $\tens$. \emph{Tensor functor} is a
new concept though. For an elementary but important example, let the label
types and fusion spaces be fixed but the associativity constraints vary. Given
two sets of associativity constraints, we may ask whether they are equivalent
up to a basis change, cf.\ Eqs.~(\ref{1-iso}) and~(\ref{1-iso1}). This type of
equivalence is a special case of tensor functor. Note that two basis
transformations may differ by trivial factors, cf.\ Eq.~(\ref{2-iso}); the
corresponding tensor functors are said to be \emph{naturally isomorphic}.

\begin{Definition}
A \emph{tensor category} $\hat{\calC}$ is a category $\calC$ endowed with a
functor $\tens:\calC\times\calC\to\calC$, a special object $\unit\in\calC$,
and natural isomorphisms
\begin{equation} \label{tenscat}
F_{A,B,C}: (A\tens B)\tens C\to A\tens(B\tens C),\qquad\quad
\alpha_{A}: A\to A\tens\unit,\qquad
\beta_{A}: A\to \unit\tens A
\end{equation}
such that the following diagrams (variants of the pentagon equation in
Fig.~\ref{fig_pentquadr}a and the triangle equation in Fig.~\ref{fig_triang}a)
commute:
\begin{equation} \label{tenscat_unit}
\begin{xy}
/u 10pt/ \xymatrix@C=20mm@R=7mm{
{((X\tens Y)\tens Z)\tens W}
\ar[r]^{F_{X\tens Y,Z,W}}
\ar[dd]^{F_{X,Y,Z}\tens\id_{W}} &
{(X\tens Y)\tens (Z\tens W)}
\ar[rd]^{F_{X,Y,Z\tens W}}
\\
& & X\tens(Y\tens(Z\tens W))
\\
{(X\tens(Y\tens Z))\tens W}
\ar[r]^{F_{X,Y\tens Z,W}} &
{X\tens((Y\tens Z)\tens W)}
\ar[ru]^*+<2pt>{\scriptstyle\id_{X}\tens F_{Y,Z,W}}
}
\end{xy}
\medskip
\end{equation}
\begin{equation}
\vcenter{\xymatrix@C=14mm@R=11mm{
{} & {}\save ;[l] **{}?(.5)
*+{X\tens W}
\ar"2,1"_*+<2pt>{\scriptstyle\alpha_{X}\tens\id_{W}}
\ar"2,2"^*+<2pt>{\scriptstyle\id_{X}\tens\beta_{W}}
\restore
\\
{(X\tens\unit)\tens W}  \ar[r]^{F_{X,\unit,W}} &
{X\tens(\unit\tens W)}
}}
\end{equation}
The tensor category $\hat{\calC}$ is called \emph{strict} if $F_{A,B,C}$,
$\alpha_{A}$, $\beta_{A}$ are equalities. It is called \emph{semisimple} if
the base category $\calC$ is $\Vect_{M}$. It is called \emph{unitary} if
$\calC=\Vect^{\dag}_{M}$, the functor $\tens$ is unitary, and the isomorphisms
$F$, $\alpha$, $\beta$ are unitary.
\end{Definition}

Semisimple tensor categories admit a compact description in terms of ``basic
data'', which is just slightly more general than that of a fusion
theory. Indeed, by analogy with Propositions~\ref{prop_Hilbfunc}
and~\ref{prop_matrixcat} one can show that any functor of two arguments,
$\tens:\Vect_{M}\times\Vect_{M}\to\Vect_{M}$ is given by a set of
finite-dimensional spaces $V_{c}^{ab}$. However, the unit object is not
necessarily simple, an example being the category $\Matr_{M,M}$ or
$\Matr_{M,M}^{\dag}$ with the multiplication rule and the
unit~(\ref{matcatmult}). (Note that these tensor categories are strict.) In
physical terms, the nonsimplicity of the unit means that the vacuum is not
unique; thus simple objects in $\Matr_{M,M}^\dag$ are boundaries between
vacua.

\begin{Definition}\label{def_tensfun}
Let $\hat{\calC}=(\calC,\tens,\unit,F,\alpha,\beta)$ and
$\hat{\calC}'=(\calC',\tens',\unit',F',\alpha',\beta')$ be tensor
categories. A \emph{tensor functor} $\hat{\calG}:\hat{\calC}\to\hat{\calC}'$
is a functor $\calG$ between the corresponding base categories, plus two
natural isomorphisms,\footnote{The direction of arrows is a convention, which
is slightly inconvenient for our purposes.}
\begin{equation}
\Gamma_{A,B}:\, \calG(A)\tens'\calG(B)\to\calG(A\tens B),
\qquad\qquad
\gamma:\, \unit'\to\calG(\unit),
\end{equation}
such that the the following diagrams commute:
\begin{equation}\label{tensfunc}
\vcenter{\xymatrix@C=7pt@R+5pt{
{\calG\bigl((X\tens Y)\tens Z\bigr)} 
\ar[d]^{\calG(F_{X,Y,Z})}
& \leftAR{}{\Gamma_{X\tens Y,Z}} &
{\calG(X\tens Y)\tens'\calG(Z)} 
& \leftAR{}{\Gamma_{X,Y}\tens'\id_{\calG(Z)}} &
{\bigl(\calG(X)\tens'\calG(Y)\bigr)\tens'\calG(Z)}
\ar[d]^{F'_{\calG(X),\calG(Y),\calG(Z)}}
\\
{\calG\bigl(X\tens (Y\tens Z)\bigr)}
& \leftAR{}{\Gamma_{X,Y\tens Z}} &
{\calG(X)\tens'\calG(Y\tens Z)}
& \leftAR{}{\id_{\calG(X)}\tens'\Gamma_{Y,Z}} &
{\calG(X)\tens'\bigl(\calG(Y)\tens'\calG(Z)\bigr)}
}}
\medskip
\end{equation}
\begin{equation}\label{tensfunc_unit}
\vcenter{\xymatrix@C=6pt@R+2pt{
{\calG(X)}
\ar[d]_{\calG(\alpha_{X})}
& \rightAR{}{\alpha'_{\calG(X)}} &
{\calG(X)\tens'\unit'}
\ar[d]^{\id_{\calG(X)}\tens'\gamma}
\\
{\calG(X\tens\unit)}
& \leftAR{}{\Gamma_{X,\unit}} &
{\calG(X)\tens'\calG(\unit)}
}}
\qquad\qquad\quad
\vcenter{\xymatrix@C=6pt@R+2pt{
{\calG(X)}
\ar[d]_{\calG(\beta_{X})}
& \rightAR{}{\beta'_{\calG(X)}} &
{\unit'\tens'\calG(X)}
\ar[d]^{\gamma\tens'\id_{\calG(X)}}
\\
{\calG(\unit\tens X)}
& \leftAR{}{\Gamma_{\unit,X}} &
{\calG(\unit)\tens'\calG(X)}
}}
\end{equation}
The tensor functor $\hat{\calG}=(\calG,\Gamma,\gamma)$ is called
\emph{unitary} if all three of its components are unitary.
\end{Definition}

This definition seems complicated, but basically it says the following: we may
identify $\calG(A\tens B)$ with $\calG(A)\tens'\calG(B)$ and $\calG(\unit)$
with $\unit'$ (by means of $\Gamma$ and $\gamma$) so that
\[
\calG(F_{X,Y,Z})=F'_{\calG(X),\calG(Y),\calG(Z)},\qquad
\calG(\alpha_{X})=\alpha'_{\calG(X)},\qquad
\calG(\beta_{X})=\beta'_{\calG(X)}.
\]

\begin{Remark}\label{laxfunc}
A \emph{lax tensor functor} is defined likewise, but $\Gamma_{A,B}$ and
$\gamma$ are not required to be isomorphisms.  In physics, this construction
describes a situation in which theory $\hat{\calC}$ is obtained from
$\hat{\calC}'$ as a result of Bose-condensation, cf. note~\ref{Boselaxfunc} on
page~\pageref{Boselaxfunc}.
\end{Remark}

\begin{Definition}
A \emph{morphism} between two tensor functors,
$\hat{\calG}=(\calG,\Gamma,\gamma)$ and $\hat{\calH}=(\calH,\Phi,\ph)$ is a
natural morphism $h:\calG\to\calH$ satisfying these additional commutation
relations:
\begin{equation}\label{tensmor}
\begin{array}{@{}c@{\qquad\qquad\qquad}c@{}}
\vbox{\xymatrix@C=6pt@R+5pt{
{\calG(X\tens Y)}  \ar[d]_{h_{X\tens Y}}
& \leftAR{}{\Gamma_{X,Y}} &
{\calG(X)\tens'\calG(Y)} \ar[d]^{h_{X}\tens'h_{Y}} \\
{\calH(X\tens Y)}
& \leftAR{}{\Phi_{X,Y}} &
{\calH(X)\tens'\calH(Y)}
}\hbox{}}
&
\vbox{\xymatrix@R+5pt{
{\calG(\unit)}
\save ; [d] **{}?(0.5) +(16,0) *+{\unit'}
\ar_{\gamma} "1,1" \ar^{\ph} "2,1"
\restore
\ar[d]_{h_{\unit}}\\
{\calH(\unit)}
}\hbox{}}
\end{array}
\end{equation}
The morphism $h$ is called a (unitary) \emph{isomorphism} if all the maps
$h_{X}$ are (unitary) isomorphisms.
\end{Definition}

Let us see how these definitions work for fusion theories. Let $\calC$ and
$\calC'$ have label sets $M$ and $M'$, respectively. We already know that an
arbitrary functor $\calG:\calC\to\calC'$ is described by some matrix
${(G^{a}_{a'}:\, a\in M,\,a'\in M')}$ of finite-dimensional Hilbert spaces.
Then the supplementary components of the tensor functor, $\Gamma$ and $\gamma$
are characterized by some unitary linear maps
\begin{equation}
\Gamma^{ab}_{c'}:\: 
\bigoplus_{a',b'\in M'} (V')^{a'b'}_{c'}\otimes G^{a}_{a'}\otimes G^{b}_{b'}
\,\to\, \bigoplus_{c\in M} V^{ab}_{c}\otimes G^{c}_{c'},
\qquad\qquad \gamma:\CC\to G^{1}_{1'}.
\end{equation}
Note that since the original $\gamma$ (the map form $\unit'$ to
$\calG(\unit)$) is an isomorphism, $G^{1}_{a'}=0$ if $a'\not=1'$.
Eqs.~(\ref{tensfunc}) and~(\ref{tensfunc_unit}) impose certain algebraic
relations on $\Gamma^{ab}_{c'}$ and $\gamma$. A unitary isomorphism between
two tensor functors is described by a set of unitary maps
$h^{a}_{a'}:G^{a}_{a'}\to H^{a}_{a'}$ satisfying the equations that follow
from Eq.~(\ref{tensmor}).

Returning to the example of basis transformations, let $\calC=\calC'$,\,
$\tens=\tens'$, but possibly $F\not=F'$,\, $\alpha'\not=\alpha$,\,
$\beta'\not=\beta$. Thus we fix the label types and fusion spaces while
allowing different associativity relations. Let us further assume that $\calG$
is the identity functor, i.e., $G^{a}_{a}=\CC$ and $G^{a}_{a'}=0$ if
$a'\not=a$. In this case, $\Gamma^{ab}_{c}$ is just a unitary operator acting
in the space $V^{ab}_{c}=(V')^{ab}_{c}$, and $\gamma$ is a phase factor.  Then
Eqs.~(\ref{tensfunc}) and~(\ref{tensfunc_unit}) become~(\ref{1-iso})
and~(\ref{1-iso1}), respectively. In this setting, an isomorphism between two
tensor functors is described by unital complex numbers $h_{a}$ satisfying
Eq.~(\ref{2-iso}).

\subsubsection{The embedding theorem}

Now, we will see the ``abstract nonsense'' in action. While we manipulate with
definitions, some combinatorial magic happens behind the scenes. Basically, we
turn pentagons into rectangles; this process may be viewed as a proof that the
$2$-skeleton of the Stasheff polytope is simply connected.

\begin{Theorem}
For any tensor category $\hat{\calC}=(\calC,\tens,\unit,F,\alpha,\beta)$ there
exists a strict tensor category $\hat{\calC'}=(\calC',\tens',\unit')$ and a
faithful tensor functor $\hat{\calG}=(G,\Gamma,\gamma)$ from $\hat{\calC}$ to
$\hat{\calC'}$ (where ``faithful'' means that the images of distinct morphisms
are also distinct).
\end{Theorem}

\begin{proof}[Proof {\normalfont (borrowed from Ref.~\cite{Rosenberg})}.]
Let $\calC'$ be the category of functors from $\calC$ to itself, $\tens'$ the
operation of composing two functors (usually denoted by $\circ$), and $\unit'$
the identity functor. We assume that functors act on the left, i.e., $(X\circ
Y)(W)\bydef X(Y(W))$. For any object $X$ we set $\calG(X)=\ltens{X}$, i.e.,
$\calG(X)(W)\bydef X\tens W$. The isomorphism $\Gamma_{X,Y}$ between two
functors may also be defined in terms of its action on a test object $W$:
\[
\Gamma_{X,Y}:\,\ltens{X}\circ\ltens{Y}\to\ltens{(X\tens Y)},\qquad\quad
\Gamma_{X,Y}(W)\,\bydef\,F_{X,Y,W}^{-1}:\,
X\tens (Y\tens W)\to (X\tens Y)\tens W.
\]
Finally, we define
\[
\gamma:\id_{\calC}\to\ltens{\unit},\qquad\quad
\gamma(W)\,\bydef\,\beta_{W}:\,W\to\unit\tens W.
\]

Now we simply rewrite Eqs.~(\ref{tensfunc}) and~(\ref{tensfunc_unit}),
applying them to $W$:
\[
\vcenter{\xymatrix@C=5pt@R+3pt{
{((X\tens Y)\tens Z\bigr)\tens W} 
\ar[d]^{F_{X,Y,Z}\tens\id_{W}}
& \rightAR{}{F_{X\tens Y,Z,W}} &
{(X\tens Y)\tens(Z\tens W)} 
& \rightAR{}{F_{X,Y,Z\tens W}} &
{X\tens(Y\tens(Z\tens W))}
\ar@{=}[d]
\\
{(X\tens (Y\tens Z))\tens W}
& \rightAR{}{F_{X,Y\tens Z,W}} &
{X\tens((Y\tens Z)\tens W)}
& \rightAR{}{\id_{X}\tens F_{Y,Z,W}} &
{X\tens(Y\tens(Z\tens W))}
}}
\medskip
\]
\[
\vcenter{\xymatrix@C=5pt{
{X\tens W}
\ar[d]_{\alpha_{X}\tens\id_{W}}
& \rightAR{@{=}}{} &
{X\tens W}
\ar[d]^{\id_{X}\tens\beta_{W}}
\\
{(X\tens\unit)\tens W}
& \rightAR{}{F_{X,\unit,W}} &
{X\tens(\unit\tens W)}
}}
\qquad\qquad\quad
\vcenter{\xymatrix@C=6pt@R+2pt{
{X\tens W}
\ar[d]_{\beta_{X}\tens\id_{W}}
& \rightAR{@{=}}{} &
{X\tens W}
\ar[d]^{\beta_{X\tens W}}
\\
{(\unit\tens X)\tens W}
& \rightAR{}{F_{\unit,X,W}} &
{\unit\tens(X\tens W)}
}}
\]
These are the conditions we need to check. But the first and the second of
them are identical to Eqs.~(\ref{tenscat}) and~(\ref{tenscat_unit}),
respectively. The third condition is a different type of triangle equation,
which is shown in Fig.~\ref{fig_triang}b.  It follows from the standard one by
Lemma~\ref{lem_triang}.

To see that the functor $\calG$ is faithful, we set $W=\unit$ and observe that
a morphism $h:X\to Y$ is mapped to
$\calG(h)(\unit)=h\tens\id_{\unit}:\,X\tens\unit\to Y\tens\unit$. Thus,
\[
\alpha_{Y}^{-1}\,\bigl(\calG(h)(\unit)\bigr)\,\alpha_{X}\,=\,h
\]
(due to the naturality of $\alpha$). It follows that distinct morphisms remain
distinct.
\end{proof}

Note that if $\calC=\Vect^{\dag}_{M}$, then $\calC'=\Matr^{\dag}_{M,M}$. The functor $\calG$
takes a simple object $a$ to the matrix whose $[b,c]$ entry is $V^{ab}_{c}$.

\subsubsection{Braiding}

Let us be brief and just give some definitions.

\begin{Definition} A \emph{braiding} in a tensor category
$\hat{\calC}=(\calC,\tens,\unit,F,\alpha,\beta)$ is a collection of
isomorphisms
\begin{equation}
R_{A,B}:\,A\tens B\to B\tens A
\end{equation}
which are natural with respect to $A$ and $B$, i.e., for any morphisms $f:A\to
A'$,\: $g:B\to B'$ this square commutes:
\begin{equation}
\vcenter{\xymatrix@+5pt{
{A\tens B}  \ar[d]_{R_{A,B}} \ar[r]^{f\tens g} &
{A'\tens B'} \ar[d]^{R_{A',B'}} \\
{B\tens A}  \ar[r]^{g\tens f} &
{B'\tens A'}
}}
\end{equation}
It is also required that the following diagrams are commutative:
\begin{equation}\label{brcat1}
\begin{xy}
(-40,0)*+{(X\tens Y)\tens Z}="1",
(-20,17)*+{(Y\tens X)\tens Z}="2",
(20,17)*+{Y\tens(X\tens Z)}="3",
(40,0)*+{Y\tens(Z\tens X)}="4",
(-20,-17)*+{X\tens(Y\tens Z)}="5",
(20,-17)*+{(Y\tens Z)\tens X}="6",
\ar^-(0.4){R_{X,Y}\tens\id_{Z}} "1";"2"
\ar^-{F_{Y,X,Z}} "2";"3"
\ar^-(0.6){\id_{Y}\tens R_{X,Z}} "3";"4"
\ar^-(0.6){F_{X,Y,Z}} "1";"5"
\ar^-{R_{X,Y\tens Z}} "5";"6"
\ar^-(0.4){F_{Y,Z,X}} "6";"4"
\end{xy}
\bigskip
\end{equation}
\begin{equation}\label{brcat2}
\begin{xy}
(-40,0)*+{(X\tens Y)\tens Z}="1",
(-20,17)*+{(Y\tens X)\tens Z}="2",
(20,17)*+{Y\tens(X\tens Z)}="3",
(40,0)*+{Y\tens(Z\tens X)}="4",
(-20,-17)*+{X\tens(Y\tens Z)}="5",
(20,-17)*+{(Y\tens Z)\tens X}="6",
\ar^-(0.4){R_{Y,X}^{-1}\tens\id_{Z}} "1";"2"
\ar^-{F_{Y,X,Z}} "2";"3"
\ar^-(0.6){\id_{Y}\tens R_{Z,X}^{-1}} "3";"4"
\ar^-(0.6){F_{X,Y,Z}} "1";"5"
\ar^-{R_{Y\tens Z,X}^{-1}} "5";"6"
\ar^-(0.4){F_{Y,Z,X}} "6";"4"
\end{xy}
\bigskip
\end{equation}
\end{Definition}

\begin{Definition}
A \emph{braided tensor functor} is a tensor functor that commutes with
braiding. More exactly, let $\hat{\calC}=(\calC,\tens,\unit,F,\alpha,\beta)$
and $\hat{\calC}'=(\calC',\tens',\unit',F',\alpha',\beta')$ be tensor
categories furnished with braidings $R$ and $R'$, respectively. A tensor
functor $\hat{\calG}=(\calG,\Gamma,\gamma)$ from $\hat{\calC}$ to
$\hat{\calC}'$ is called braided if the following diagram commutes:
\begin{equation}
\vcenter{\xymatrix@C=6pt@R+5pt{
{\calG(A\tens B)}  \ar[d]_{\calG(R_{A,B})}
& \leftAR{}{\Gamma_{A,B}} &
{\calG(A)\tens'\calG(B)} \ar[d]^{R'_{\calG(A),\calG(B)}} \\
{\calG(B\tens A)}
& \leftAR{}{\Gamma_{B,A}} &
{\calG(B)\tens'\calG(A)}
}}
\end{equation}
\end{Definition}

\begin{Example}
Let $\hat{\calC}$ be an arbitrary braided tensor category and let
$\hat{\calC}'$ be defined as follows:
\[
\calC'=\calC,\quad A\tens'B=B\tens A,\quad \unit'=\unit,\quad
F'_{X,Y,Z}=F_{Z,Y,X}^{-1},\quad \alpha'=\beta,\quad \beta'=\alpha,\quad
R'_{A,B}=R_{B,A}.
\]
Then the identity functor $\calG=\id_{\calC}$ supplemented with
$\Gamma_{A,B}=R_{A,B}^{-1}$ and $\gamma=\id_{\unit}$ is a braided tensor
functor from $\hat{\calC}$ to $\hat{\calC}'$. (We omit the proof.)
\end{Example}

\section{Weak symmetry breaking}
\label{sec_symbreak}

A complete, yet to be discovered description of topological quantum order must
include fusion and braiding rules (characterized by a unitary braided fusion
category (UBFC), see Appendix~\ref{sec_algth}) as well as symmetries of the
underlying microscopic Hamiltonian. The no-symmetry case was considered in
Sec.~\ref{sec_modularity}. On the other hand, in our study of Abelian anyons
in the honeycomb lattice model (see Sec.~\ref{sec_Abelian}) we observed a
rather strange property: the translation by a lattice vector interchanges two
superselection sectors. This phenomenon may be called a ``weak breaking'' of
the translational symmetry.

Let us now consider an arbitrary \emph{two-dimensional} many-body system with
a symmetry group $G$. For simplicity, we assume that the system consists of
spins (or other bosonic degrees of freedom) rather than fermions. There are
several ways in which the ground state can break the symmetry that is present
in the Hamiltonian. First, the symmetry may be spontaneously broken in the
usual sense, i.e., there may be a local order parameter. If such an order
parameter does not exist, we may look for finer signs of symmetry breaking, in
particular, for nontrivial action of $G$ on superselection sectors. Such an
action is described by a homomorphism $\omega_{1}:G\to\Gamma_{1}$, where
$\Gamma_{1}$ is some finite group defined below.

But even if $G$ does not permute the superselection sectors, there is still
possibility for subtle symmetry-breaking properties, some of which were
studied by Wen~\cite{Wen-1,Wen-2} under the rubric of ``projective symmetry
groups'' (PSG). While Wen's approach is rather indirect, one may instead
investigate the action of symmetry transformations on superselection
sectors. For a simple example, let us consider the Hamiltonian~(\ref{torcode})
with $J_{m}<0$ (note the analogy with Wen's model~\cite{Wen-2}). The ground
state of this Hamiltonian may be obtained from the $J_{m}>0$ state by putting
an $m$-particle on every plaquette (so that any missing $m$-particle would now
be regarded as an excitation). Therefore an $e$-particle picks up the phase
factor $-1$ when it winds around a plaquette. It follows that the translations
by two basis lattice vectors commute up to a minus sign when we consider their
action on the nontrivial superselection sector $e$.

Mathematically, the up-to-sign commutation means that the translational group
$G=\ZZ\times\ZZ$ is replaced by a central extension $G_{e}\to G$ with kernel
$\ZZ_{2}$. If general, central extensions are classified by the cohomology
group $H^{2}(G,\UU(1))$. For each superselection sector $a$ we get an
extension $G_{a}$. The whole collection of extensions is characterized by a
cohomology class $\omega_{2}\in H^{2}(G,\Gamma_{2})$, where $\Gamma_{2}$ is
some finite Abelian group (see below).

However, this is not the end of the story. Even without anyonic excitations
(or if anyons exist but do not complicate the group action in any way), the
system may have nontrivial properties such as the integer quantum Hall
effect. The latter is described by a Chern-Simons term in the effective action
for the electromagnetic field. Dijkgraaf and Witten~\cite{DijkgraafWitten}
showed that a general topological action for a gauge field in $2+1$ dimensions
is characterized by an element of the cohomology group $H^{4}(BG,\ZZ)$, where
$BG$ is the classifying space of $G$. (The group $G$ is assumed to be
compact.) An analogue of the Chern-Simons action exists even for finite
groups. In this case, $BG=K(G,1)$ (where $K(G,n)$ is the Eilenberg-MacLane
space) and $H^{4}(BG,\ZZ)\cong H^{3}(G,\UU(1))$ (where the first $H$ refers to
the topological cohomology and the second to the group cohomology).

\begin{table}[t]
\centerline{\begin{tabular}{|c|p{6cm}|p{6cm}|}
\hline
\parbox{3.5cm}{\strut Effective dimension in which the symmetry is
broken\strut}
& \multicolumn{1}{c|}{Mathematical description}
& \multicolumn{1}{c|}{Examples} \\
\hline\hline
$0$ & A local order parameter taking values in the coset space $G/H$,
where $H$ is a subgroup of $G$ & Ferromagnet; Neel phase \\
\hline
$1$ & Nontrivial action of the symmetry group on superselection sectors, which
is characterized by a homomorphism $\omega_{1}:G\to\Gamma_{1}$
& Abelian phases $A_{x}$, $A_{y}$, $A_{z}$ in the honeycomb lattice model \\
\hline
$2$ & The action of $G$ on  each super\-selection sector $a$ is described by a
central extension $G_{a}$; the whole set of extensions is characterized by
an element $\omega_{2}\in H^{2}(BG,\Gamma_{2})$
& Hamiltonian~(\ref{torcode}) with $J_{m}<0$;\newline a number of models
in~\cite{Wen-1,Wen-2} \\
\hline
$3$ & & \\
\hline
$4$ & Effective topological action for a $G$-gauge field; such actions are
classified by $H^{4}(BG,\ZZ)$
& Integer quantum Hall effect \\
\hline
\end{tabular}}
\caption{Four levels of symmetry breaking in a two-dimensional quantum system.}
\label{tab_symbreak}
\end{table}

The four-level classification of symmetry-breaking phenomena is summarized in
Table~\ref{tab_symbreak}. It has the flavor of topological obstruction theory
(see e.g.~\cite{Hatcher}). One may hypothesize that there is a somehow
relevant topological space $Y$ such that $\pi_{1}(Y)=\Gamma_{1}$,\,
$\pi_{2}(Y)=\Gamma_{2}$,\, $\pi_{3}(Y)=0$,\, $\pi_{4}(Y)=\ZZ$, and all higher
homotopy groups vanish. We now argue that the analogy with obstruction theory
has precise mathematical meaning in dimensions~$1$ and~$2$. 

Suppose that the symmetry remains unbroken in dimension $0$, but can be
removed (if so is desired) by introducing a small perturbation to the
Hamiltonian. That is obviously possible if the system consists of spins, while
the $\ZZ_{2}$ symmetry associated with fermions can never be removed.  The
properties that survive the perturbation are described by a unitary modular
category (UMC) $\calA$ and a chiral central charge $c_{-}$, but we ignore the
latter.

The action of the symmetry group $G$ on superselection sectors can be
described algebraically as well as topologically. To begin with algebra,
consider the $2$-groupoid $\Gamma=\mathop{\mathrm{Aut}}(\calA)$. It has a
single object, the morphisms are invertible unitary braided tensor functors
$\calA\to\calA$, and the $2$-morphisms are natural isomorphisms between such
functors.\footnote{As explained to me by Ezra Getzler, higher groupoids may be
nicely defined as Kan complexes~\cite{May}; this approach also allows one to
endow the morphism sets with topology. In the definition of $\Gamma$ no
topology is assumed. It would be interesting to see what changes (if anything)
when a natural topology is included.}  If the functors are considered up to
isomorphisms, we obtain the group $\Gamma_{1}$ that was mentioned above. The
Abelian group $\Gamma_{2}$ consists of all automorphisms of the identity
functor. Note that $\Gamma_{1}$ may act on $\Gamma_{2}$ in a nontrivial
way. The $2$-groupoid $\Gamma$ is characterized by this action and a
cohomology class $h\in H^{3}(\Gamma_{1},\Gamma_{2})$.

On the topological side, one can consider the classifying space $X=B\Gamma$,
which is a topological space with a basepoint. I claim that $X$ is connected,
${\pi_{1}(X)=\Gamma_{1}}$,\, ${\pi_{2}(X)=\Gamma_{2}}$, and all higher
homotopy groups vanish. Furthermore, the action of $G$ on superselection
sectors is described by a continuous map $f:BG\to X$ defined up to a homotopy
(some explanation is given below). One may ask if $f$ is homotopic to a
constant map. The first obstruction to such a homotopy is given by a homotopy
class of maps $\omega_{1}:BG\to K(\Gamma_{1},1)$; such classes are in
one-to-one correspondence with group homomorphisms $G\to\Gamma_{1}$. If this
obstruction vanishes, one can define $\omega_{2}\in
H^{2}(BG,\Gamma_{2})=H^{2}(G,\Gamma_{2})$. If $\omega_{2}=0$, then $f$ is
homotopic to identity, or equivalently, the action of $G$ on $\calA$ is
trivial.

Note that $X$ can be represented as a fibration with base $K(\Gamma_{1},1)$
and fiber $K(\Gamma_{2},2)$. Its structure is characterized by the element
$h\in H^{3}(\Gamma_{1},\Gamma_{2})$. If $h\not=0$, then the fibration does not
have a cross section. In this case, the homomorphism
$\omega_{1}:G\to\Gamma_{1}$ must satisfy a certain constraint, namely the
inverse image of $h$ by $\omega_{1}$ must vanish. Thus, some seemingly
possible types of symmetry breaking in dimension $1$ may be forbidden due to
an obstruction in dimension $2$.

In conclusion, I conjecture a ``physical'' interpretation of the space $X$ and
the map $f$. Specifically, $X$ parametrizes quantum states in the universality
class described by the unitary modular category $\calA$. Each point of $X$ is
associated with a set of states that differ from each other locally; thus all
finite-energy excitations of a given gapped Hamiltonian are represented by the
same point. The original unperturbed system corresponds to some $x_{0}\in
X$. To define the map $BG\to X$, let us assume that $G$ is a compact Lie group
which acts on each spin independently according to a faithful representation
$G\to\UU(m)$. The Hilbert space of the spin, $\CC^{m}$ may be embedded in a
large space $\CC^{M}$. Each embedding $u$ has an associated state $g(u)\in
X$. The space $E$ of embeddings is contractible in the limit $M\to\infty$,
therefore its quotient with respect to the natural $G$-action may be regarded
as a model of $BG$. Furthermore, since $x_{0}$ is symmetric, the map $g:E\to
X$ factors through $BG$. Thus the map $f:BG\to X$ is defined.

\begin{Remark}
It is not clear how to define the ``extended classifying space'' $Y$ that has
nontrivial homotopy in dimension $4$. One may conjecture that $Y=X\times
K(\ZZ,4)$, but this definition seems contrived. It may well be the case that
the very idea to classify the Chern-Simons action as symmetry breaking in
dimension $4$ is wrong. Indeed, the Chern-Simons action is not fully described
by homotopy theory because the fundamental homology class of the source
(physical) manifold is involved.
\end{Remark}

\section*{Acknowledgments}
\addcontentsline{toc}{section}{Acknowledgments}

During the years this work was in progress I received genuine interest and
words of encouragement from John Preskill, Michael Freedman, Mikhail
Feigelman, Grigory Volovik, and many other people. I thank Andreas Ludwig for
having convinced me in the importance of the chiral central charge in the
study of topological order. A conversation with Dmitri Ivanov was essential
for understanding the difference between non-Abelian anyons and Majorana
vortices in two-dimensional $p$-wave superconductors. I also thank Matthew
Fisher, Nicholas Read, and Xiao-Gang Wen for helpful discussions. I am
especially grateful to Jean Bellissard who read a preliminary version of the
manuscript and made some valuable comments; in particular, he directed me to
Ref.~\cite{Lieb94}. This work was supported in part by the National Science
Foundation under Grant No.~EIA-0086038 and by the Army Research Office under
grant No.~\hbox{W911NF-04-1-0236}.

\end{document}